\documentclass[12pt]{book} 

\usepackage{amsfonts,graphicx, color}
\input{tcilatex}

\begin{document}

\setcounter{page}{0}

\thispagestyle{empty}

\indent \ \

\centerline{\LARGE \textsl{
Bootstrap Methods in 1+1-Dimensional}}

\vspace{0.15cm}

\centerline{\LARGE \textsl{Quantum Field Theories:}}

\vspace{0.15cm}

\centerline{\LARGE \textsl{the Homogeneous Sine-Gordon Models}}

\vspace{0.5cm}

{\centerline{\Large Olalla A. Castro Alvaredo{\large
\footnote{olalla@physik.fu-berlin.de}$^{,}$
\footnote{Current adress: Institut f\"ur Theoretische Physik,
 Freie Universt\"at Berlin, Arnimallee 14, D-14195 Berlin, Germany.}}}}

\vspace{0.5cm}

\centerline{\large \em Departamento de F\'{\i}sica de Part\'{\i}culas, 
Facultade de F\'{\i}sica,}

\vspace{0.1cm}

\centerline{\large \em Universidade de Santiago de Compostela,}

\vspace{0.1cm}

\centerline{\large \em E-15706 Santiago de Compostela, Spain}

\vspace{0.4cm}

{\centerline{\Large Abstract}}

\vspace{0.4cm}

The bootstrap program for 1+1-dimensional integrable Quantum Field Theories (QFT's)
is developed to a large extent for the Homogeneous sine-Gordon (HSG) models.  
This program can be divided into various steps, which
include  the computation of the exact S-matrix, Form Factors of
local operators and correlation functions, as well as the identification of the operator
content of the QFT and the development of various consistency checks. 
Taking as an input the fairly recent S-matrix proposal for the HSG-models, we confirm its
consistency by carrying out both a Thermodynamic Bethe Ansatz (TBA) and a Form Factor analysis, 
which allow for extracting the main characteristics of the underlying Conformal Field Theory (CFT) 
associated to these theories. In contrast to many other 1+1-dimensional integrable models
studied in the literature, the HSG-models possess two remarkable features, namely the
parity breaking, both at the level of the Lagrangian and S-matrix, as well as the presence
of unstable particles in the spectrum. These features have specific consequences in our
analysis which are given a physical interpretation.
By exploiting the Form Factor approach, we develop further the QFT advocated
to the HSG-models by evaluating  correlation functions of various local
operators of the model. We compute the renormalisation group (RG) 
flow of Zamolodchikov's c-function and  $\Delta$-functions which carry information about
the RG-flow of the operator content of the underlying CFT and provide a means for the
identification of the operator content of the QFT. For these functions, as well
as for the scaling function computed in the TBA-context, an `staircase' pattern 
which can be physically interpreted, by associating the different
plateaux to energy scales for the onset of  stable and/or unstable
particles of the model, is found. For the $SU(3)_2$-HSG model we
show how the form factors of different local operators are interrelated by means of the momentum space
cluster property. We find closed formulae for all $n$-particle form
factors of a large class of operators of the $SU(N)_2$-HSG models. These
formulae are  expressed in terms of universal building blocks which 
allow both a determinant and an integral representation. 

\newpage

\setcounter{page}{0}

\thispagestyle{empty}
\indent \ \

\vspace{3cm}

\centerline{SUPERVISOR}

\vspace{1cm}

\noindent Prof. Dr. J.L.~MIRAMONTES ANTAS,\\
 Departamento de F\'{\i}sica de Part\'{\i}culas, 
Universidade de Santiago de Compostela, Spain.
\vspace{2cm}

\centerline{THESIS COMMISSION}

\vspace{2cm}

\noindent Dr. P.E.~DOREY,\\
 Department of Mathematical Sciences, University of Durhan, England.\\

\vspace{0.1cm}

\noindent Dr. A.~FRING,\\
 Institut f\"ur Theoretische Physik, Freie Universit\"at Berlin, Germany.\\

\vspace{0.1cm}

\noindent Prof. Dr. G. MUSSARDO,\\
International School for Advanced Studies, Trieste, Italy and\\
Istituto Nazionale di Fisica Nucleare, Sezione di Trieste, Trieste, Italy.\\

\vspace{0.1cm}

\noindent Prof. Dr. J.~S\'ANCHEZ GUILL\'EN,\\
Departamento de F\'{\i}sica de Part\'{\i}culas, 
Universidade de Santiago de Compostela, Spain.\\

\vspace{0.1cm}

\noindent Prof. Dr. G.~SIERRA RODERO,\\
 Instituto de Matem\'aticas y F\'{\i}sica Fundamental, C.S.I.C., Madrid, Spain.

\newpage

\setcounter{page}{0}

\thispagestyle{empty}
\indent \ \

The original work presented in this PhD thesis, collects the results
which can be found in the following publication list:

\vspace{0.5cm}

\begin{enumerate}

\item Decoupling the
$SU(N)_2$-homogeneous sine-Gordon model.\\
O.A.~Castro-Alvaredo and  A.~Fring,\\
Accepted for publication in Phys. Rev. D.\\
hep-th/0010262. Chapter 5.\\

\item Renormalization group
flow with unstable particles.\\
O.A. Castro-Alvaredo and A.~Fring,\\
\emph{Phys. Rev.} \textbf{D63} (2001) 21701.\\
hep-th/0008208. Chapter 4.\\

\item
Identifying the
Operator Content, the Homogeneous Sine-Gordon models.\\
O.A.~Castro-Alvaredo and A.~Fring,\\ 
\emph{Nucl. Phys.} \textbf{B604} (2001) 367.\\
hep-th/0008044. Chapter 4.\\

\item
Form factors of the homogeneous sine-Gordon models.\\
 O.A.~Castro-Alvaredo, A.~Fring and  C.~Korff,\\ 
\emph{Phys. Lett.} \textbf{B484} (2000) 167.\\
hep-th/0004089. Chapter 4.\\

\item Massive symmetric space sine-Gordon soliton
theories and perturbed conformal field theory.\\
O.A.~Castro-Alvaredo  and
J.L.~Miramontes, \\
{\em Nucl. Phys.} {\bf B581} (2000) 643.\\
 hep-th/0002219.
Chapter 2\footnote{Since this
thesis is mainly concerned with the study of the  Homogeneous
sine-Gordon models, whereas this paper is devoted to the study
of the so-called symmetric space sine-Gordon models, 
we have decided to recall in this manuscript 
only a small part of the results found in this paper.}.\\

\item 
Thermodynamic Bethe ansatz of the homogeneous sine-Gordon models.\\
O.A.~Castro-Alvaredo, A.~Fring, C.~Korff and
J.L.~Miramontes,\\
{\em Nucl. Phys.} {\bf B575} (2000) 535.\\
hep-th/9912196.
Chapter 3.\\

\end{enumerate}

\newpage

\vspace{0.4cm}

\def\lab#1{\label{eq:#1}}

\tableofcontents
\listoftables
\listoffigures  

\chapter{Introduction}
\indent \ \
The study of 1+1-dimensional massive quantum field theories (QFT) has turned
out to be a very fruitful research field for almost three decades.
Various distinguished properties arising in the 1+1-dimensional context are
responsible of this success and it is our intention to begin this thesis
by providing a very brief glimpse of them through  points {\bf I-IV}:

\vspace{0.25cm}

\noindent {\bf I.}\,{\bf Conformal invariance} becomes in the 1+1-dimensional
context an extremely powerful symmetry \cite{schroer, BPZ, CFT, cardy}.
In any dimension, conformal transformations are the subset of coordinate
transformations which leave the metric invariant up to a local scale factor, namely  

$$
g_{\mu \nu} \rightarrow e^{\Lambda({x})} g_{\mu\nu}.
$$

However, it turns out that only in 1+1-dimensions the conformal symmetry 
algebra becomes infinite dimensional, which means that
there will be infinitely many conserved quantities associated
to any 1+1-dimensional conformal field theory (CFT) . Consequently,
conformal invariance becomes an extremely constraining requirement in
the 1+1-dimensional context and many problems, which for general QFT's
can only be handled with great difficulties find an exact solution
within the context of 1+1-dimensional CFT's.

\vspace{0.25cm}

\noindent {\bf II.} \,
Although it is more restrictive for 1+3-dimensional massive QFT's, 
the so-called {\bf Coleman-Mandula theorem} \cite{colman}
is one of the key properties one has to appeal to in order to unravel the origin 
of the distinguished features of 1+1-dimensional massive QFT's. 
The mentioned theorem was formulated and proven 
in 1967 by S. Coleman and J. Mandula \cite{colman}. These authors  
determined (under certain assumptions) the maximum S-matrix symmetry group
associated to a  1+3-dimensional massive  QFT. 
They found that for any local, relativistic, massive 1+3-dimensional QFT 
a symmetry group, $G$, containing the Poincar\'e group, $\mathcal{P}$,  and
an arbitrary internal symmetry group, $\mathcal{I}$, should necessarily be a 
direct product of the form
$$
G= \mathcal{I} \otimes \mathcal{P},  
$$
\noindent 
Amongst all the hypothesis  involved in the derivation of  Coleman-Mandula
theorem, it is worth emphasising  that  the symmetry group
$G$ is assumed to be a Lie group whose generators obey a 
Lie algebra based on commutators.  Alternatively, the  theorem can be
formulated by stating that, under the mentioned assumptions \cite{colman},
space-time and internal symmetries can not be combined in any but 
a trivial way for 1+3-dimensional massive QFT's.

\vspace{0.25cm}

\noindent {\bf III.}\, 
As mentioned above, the Coleman-Mandula theorem explicitly refers to the 1+3-dimensional
domain. Nonetheless it has also
a crucial counterpart in the 1+1-dimensional context.
The result we are referring to can be formulated in very 
close spirit to the Coleman-Mandula theorem: 
the combination of space-time and internal symmetries
in a non-trivial way is possible 
for 1+1-dimensional massive QFT's. Equivalently, in
1+1-dimensions and assuming a purely massive particle spectrum, 
 the existence of conserved quantities different from the energy,
the space momentum, and the charges associated to internal symmetries
does not force the S-matrix to be trivial. However,
the S-matrix is constrained in a very severe way whenever
the mentioned additional symmetries are present in the theory. In that case, 
the corresponding QFT is said to be {\bf integrable} and the
precise constraints we refer to are the following: 

\vspace{0.25cm}
{\bf Absence of particle production} in any scattering process.

\vspace{0.25cm}

`Strict' {\bf momentum conservation} i.e., equality of the sets of 
 momenta of incoming and outgoing particles.

\vspace{0.25cm}

{\bf Factorisability} of all $n$-particle scattering amplitudes into $2$-particle ones.

\vspace{0.25cm}

\noindent  Denoting by $S_{a_1 \cdots a_n}^{b_1 \cdots b_k} (p_{a_1},\cdots, p_{a_n}; p_{b_1}, \cdots, p_{b_k})$
 the scattering amplitude of a process with $n$ incoming particles of quantum numbers $a_1, \cdots, a_n$ and 
momenta $p_{a_1}, \cdots, p_{a_n}$ and $k$ outgoing particles of quantum numbers
$b_1, \cdots, b_k$ and momenta $p_{b_1}, \cdots, p_{b_k}$, the previous constraints can be expressed as 

$$
S_{a_1 \cdots a_n}^{b_1 \cdots b_k} (p_{a_1},\cdots, p_{a_n}; p_{b_1}, \cdots, p_{b_k})
 \propto 
 \delta_{n,k} \prod_{i=1}^{n} \delta(p_{a_i}-p_{b_i}) \equiv 
S_{a_1 \cdots a_n}^{b_1 \cdots b_n} (p_{1},\cdots, p_{n}),
$$
$$
S_{a_1 \cdots a_n}^{b_1 \cdots b_n} (p_{1},\cdots, p_{n})=
\prod_{ i<j, l<k} S_{a_i, a_j}^{b_k, b_l} (p_i, p_j).
$$

Such severe constraints to the form of the scattering
amplitudes for integrable models were originally observed in the context of
the study of various concrete 1+1-dimensional QFT's \cite{smatrix}. In particular, the 
investigation
of these models lead the different authors to
conjecture the factorisability property mentioned above. Remarkably, a very
analogous property had been already encountered much earlier in the
non-relativistic framework (see e.g. \cite{facto}). The properties itemized above
were thereafter reviewed in more generality by R. Shankar and E. Witten
 \cite{SW}
and by D. Iagolnitzer in \cite{sm2}. The arguments presented
 in those articles relied
on the assumption of the presence of infinitely
 many conserved quantities in the QFT,
in order to derive the outlined S-matrix constraints. 
Later, S. Parke refined the arguments in \cite{SW, sm2} by showing
 in \cite{parke} that actually the presence of just
two non-trivial conserved quantities in the theory leads to the same
conclusions drawn in \cite{SW, sm2}. The main arguments presented in
\cite{SW, sm2, parke} will be reviewed in chapter 2 of this thesis.

\noindent In the previous equations, 
 the particles $a_i, b_i$, for all values of $i=1, \cdots, n$ are
particles belonging to the same mass multiplet. The conservation of the set of momenta of incoming
and outgoing particles still allows for a possible exchange of quantum
numbers between particles in the $in$- and $out$-states. This is possible
whenever the particle spectrum is degenerate, namely there is more
than one particle in each particle multiplet. In that case the S-matrix 
is said to be non-diagonal. However, for
many interesting theories (in particular, the ones
studied in this thesis)  the mentioned degeneration does not
occur and  the two-particle scattering 
amplitudes can be written as
$$
S_{a_i, a_j}^{b_k, b_l} (p_i, p_j)= \delta_{i}^{k} \delta_{j}^{l} S_{i, j}(p_i, p_j),
$$
\noindent which means that the corresponding  S-matrix is {\bf diagonal} and
usually simplifies its explicit construction.

The constraints arising from integrability are also in the origin of the so-called 
Yang-Baxter \cite{yb} and bootstrap equations \cite{Boot} which together with
the physical requirements of  unitarity,  crossing symmetry, 
Hermitian analyticity, and  Lorentz invariance of the scattering amplitudes 
\cite{Boot,KTTW,ELOP, Wein, Zalgebra, sinegor}
allow in many cases for the exact calculation of the corresponding S-matrix. 
The main steps involved in  such  a construction  as well as
the nature of the properties summarised above will be analysed  in detail
 in the next chapter. 

\vspace{0.25cm}

\noindent {\bf IV.}\, A fundamental link between the properties stated in {\bf I} and
{\bf III} was established by A.B. Zamolodchikov in \cite{Pertcft}. This result has been
extensively exploited in the study of 1+1-dimensional integrable QFT's over the
last decade and can be summarised as follows: a 1+1-dimensional QFT may be viewed as
a  perturbation of a CFT by means of a particular operator of the CFT itself.
Equivalently, one could write formally the action functional associated to a 1+1-dimensional
QFT as
$$
S=S_{\text{CFT}} + \lambda \int d^2 x \, \Phi (x,t),
$$
\noindent for $S_{\text{CFT}}$ to be the action of the unperturbed CFT, 
$\lambda$ a coupling constant and $\Phi (x,t)$ a local field which in 
the ultraviolet limit corresponds to a local field of the CFT. In this fashion the unperturbed or
{\bf underlying CFT} is recovered in the ultraviolet limit of the
 {\bf perturbed CFT}. Equivalently,
the perturbation of the CFT breaks the original conformal invariance
 by taking the CFT away 
from its associated renormalisation group critical fixed point.
Nonetheless, being the original conformal invariance extremely powerful it is to
be expected that it has some `remaining' counterpart in the massive QFT. 
Indeed, it was proven in \cite{Pertcft} that
for suitable choices of the perturbing field (we will see what this means in the next chapter),
 we will end up with a massive QFT which is not conformally 
invariant but still possesses an infinite number of conserved
quantities and is therefore integrable in the sense of {\bf III}. In order to prove
the integrability of the model at hand, those conserved
quantities can be explicitly constructed by doing perturbation theory around the original CFT or
their existence may be proven by appealing to the so-called  {\bf counting-argument} 
presented in \cite{Pertcft}.

\vspace{0.25cm}
As a summary of the previous points {\bf I-IV}, let us consider a 1+1-dimensional
CFT for which, if we are fortunate, a lot of information will be available. We may perturb it
by means of a certain local field of the CFT itself and construct this way a 
1+1-dimensional massive QFT.
Thereafter, by exploiting the methods mentioned in {\bf IV}, we might
 be able to establish whether or not
the perturbed CFT is integrable. In case the answer is positive the properties stated in 
{\bf III} are automatically fulfilled namely, the latter QFT will be described by a factorisable
S-matrix and there will be no particle production in any scattering process.
 Furthermore, the scattering
amplitudes are forced to satisfy other requirements already mentioned in {\bf III}, which
 may allow for the exact construction of the S-matrix associated to the massive QFT
 by carrying out the so-called {\bf bootstrap program} \cite{Boot}.
It is important for later purposes to mention that, the outlined
procedure often involves certain assumptions and ambiguities which are ultimately justified by the
self-consistency of the results obtained. An example of the former is the extrapolation
of semi-classical results to the QFT and, concerning the latter, any S-matrix proposal will be
always determined up to certain factors, the so-called {\bf CDD-factors} \cite{CDD}.
 These factors are
functions which do not add any physical information
 to the S-matrix proposal and satisfy trivially
all the requirements summarised in {\bf III}.

\vspace{0.25cm}

In the light of the previous observations, once a certain S-matrix has
been constructed by means of the bootstrap program and assumed to describe the scattering
theory associated to a certain 1+1-dimensional integrable QFT, 
it is highly desirable to develop tools which allow for consistency checks of this
 S-matrix proposal,
that is, approaches which permit a definite one-to-one identification
 between the S-matrix constructed and
the particular QFT under consideration. If the massive QFT has been
constructed along the lines summarised in the preceding paragraph, its ultraviolet limit
should lead to the original unperturbed CFT, whose main characteristics are 

\vspace{0.25cm}
\begin{center}
{\bf Virasoro central charge}
 
\vspace{0.25cm}

{\bf Conformal dimension of the  perturbing field}

\vspace{0.25cm}

{\bf Local  operator content}

\vspace{0.25cm}
\end{center}
\noindent Therefore, having a certain perturbed CFT at hand, for which an S-matrix proposal has
been constructed, we want to develop methods which taking this S-matrix proposal as an input
allow for checking if it really corresponds to the specific massive QFT under study. Moreover, since
the massive QFT has been constructed as perturbed CFT, such consistency checks may exploit the knowledge
of the characteristics of the underlying CFT  itemized above. 
These sorts of tools or approaches are commonly referred to as {\bf Bootstrap methods} \cite{Boot}. 
Amongst them, the {\bf thermodynamic Bethe ansatz} (TBA) originally proposed by C.N. Yang
and C.P. Yang in \cite{Yang} and formulated in the present form by A.B. Zamolodchikov \cite{TBAZam1}, 
and the {\bf form factor approach}, pioneered in the late seventies by the Berlin group of the 
Freie Universit\"at \cite{Kar}, 
constitute  prominent examples. The purpose of the work presented in this
thesis will be the application of these methods to the study of a concrete family of 1+1-dimensional
massive integrable QFT's together with the further investigation of  the mentioned approaches 
themselves.

\vspace{0.25cm}
We do not want to describe in detail now the formulation of these two approaches, 
which will be done in subsequent chapters. Nonetheless, it is worth noticing here that
both the TBA-  and form factor approach take as an input the knowledge
of the exact S-matrix associated to a certain 1+1-dimensional QFT and allow in principle
for computing both the Virasoro central charge of the underlying CFT and the 
conformal dimension of the perturbing field. Moreover, in the 
TBA-context, the {\bf finite size scaling function} \cite{scaling} can be computed (usually
numerically). The latter function can be understood for unitary CFT's as a sort of `off-critical' 
Virasoro central charge which measures the amount of effective light degrees of
 freedom present in the theory at each energy scale. Such  a function has a counterpart which
is expressible in terms of {\bf correlation functions} involving different components 
of the energy momentum tensor and is known as {\bf Zamolodchikov's $c$-function} \cite{ZamC}.
It carries the same physical information
as the finite size scaling function and both functions turn out to be qualitatively very similar,
despite the fact,  that their precise relationship is still an outstanding problem. The computation
of Zamolochikov's $c$-function  will be possible within the form factor framework, 
since the knowledge of
the form factors associated to a certain local operator allows for the computation of its
 two-point correlation function.

\vspace{0.25cm}
Moreover, it must be emphasised that the form factor approach goes, at least at present, 
beyond the previous applications and, in contrast to the TBA-analysis, allows also for
the further development of the QFT advocated to a certain model. In particular, as noticed
above, the knowledge of the form factors associated to any local operator of the QFT allows for
the computation of correlation functions involving such operator. 
The latter use of form factors can be exploited, for instance, in
re-constructing at least a large part of the local operator content of the underlying CFT
(apart from the perturbing field)
by assuming a one-to-one correspondence between the local operator content of the 
unperturbed  and perturbed CFT. Such correspondence can be established 
by evaluating the ultraviolet conformal dimensions of local operators of the massive QFT, that is,
the conformal dimensions of those primary fields of the underlying CFT 
which are identified as their counterpart in the UV-limit. In order to carry out
this identification we can consider the UV-limit of the two-point functions of
local operators of the QFT, and extract thereafter the associated conformal dimension. 
Alternatively, in the form factor framework,
{\bf $\Delta$-sum rules}, like the one proposed in \cite{DSC} which requires the 
knowledge of the two-point function of the local operator at hand and the trace of 
the energy momentum tensor, can be numerically
evaluated for the ultraviolet conformal dimensions of certain local fields of the QFT. 

The identification of the conformal dimensions of certain local operators of the underlying
CFT, different from the perturbing field, can also be carried out in the TBA-context. This
requires the re-formulation of this approach in order to extract the energies of
excited states of the QFT, instead of the ground state energy available
in the standard TBA-framework. These energies can be  related thereafter to
the conformal dimensions of certain operators of the underlying CFT. Work in
this direction was first carried out in \cite{marfen}, where models whose ground state becomes
degenerated for large volume were studied. Later in \cite{zamdota},  the energies of
excited states have been found to be obtainable via the  analytical continuation to the
complex plane of the parameters entering the standard TBA-equations. Unlike as in 
the form factor context described above, the latter
{\bf ``excited TBA''} analysis still does not provide a direct 
mechanism which allows for matching
the operator contents of the perturbed and unperturbed CFT. Instead,
this analysis allows, in principle, for reconstructing the Hilbert 
space of the theory at different energy scales (system sizes), providing
therefore a map between energy eigenstates in the UV- and IR-limits.
In the form factor context, the $\Delta$-sum rule proposed in \cite{DSC} can  be modified
by introducing a dependence on the RG-parameter, as shown in \cite{CF2} in
such a way that we can now reconstruct the operator content of the theory at different
energy scales, as we show in particular for the HSG-models in this thesis. 
In that context, we will compute quantities which we could name as 
``off-critical'' conformal dimensions
$\Delta(r)$,  whose variation in terms of  the RG-parameter from the UV-
to the IR-regime, reproduces the renormalisation group flow of the operator content of
the theory in terms of the RG-energy scale.

\vspace{0.25cm}

Having now introduced the main ideas entering the study we will present in this thesis, what
is left is the description of the concrete theories for which the TBA- and form factor analysis
will be carried out. These theories are particular examples of the large family
of {\bf Toda field theories} \cite{classtoda} which arise as field generalisations of one
of the simplest classically integrable models: the {\bf Toda lattice} \cite{lattice, lattice2}.
 The Toda lattice is a discrete mechanical system consisting of a set of $n$ particles located along
a line and characterised by non-linear interactions. 
In fact, as shown in the picture at the end of this chapter, it is usual to
distinguish between the Toda lattice and Toda molecule, to indicate that in the $n$-particle
system mentioned above the two particles on the extremes are coupled to a solid wall (lattice) 
or that this wall is `taken to infinity' (molecule)  and, in that sense, 
there are no boundaries\footnote{Notice that the reference
to boundaries we make here must be understood as we just explained. 
Therefore there is no relation to  boundary integrable QFT's.}.
A precursor of these models was studied 
 by E. Fermi, J. Pasta and S. Ulam in \cite{FPU}
 in the course of a numerical simulation to study heat conductivity in
solids and their field
generalisations gave rise to the large family of theories whose main features shall
be itemize below.
The mentioned generalisation is carried out by adding to 
the original discrete coordinates $q_i(t)$, 
$i=1, \cdots, n$, a space dependence of the form $q_i (x,t)$. Thereafter, the latter functions
can be naturally arranged into a field, $h(x,t)$  which takes values in a certain Lie algebra, $g$. 
If the field takes values in the Cartan subalgebra of $g$,
the corresponding theories are known as {\bf abelian Toda field theories}.
These theories  have been extensively studied in the literature and can be further classified 
into the three following groups: 

\vspace{0.25cm}
\noindent
If the Lie algebra $g$  is finite, the corresponding abelian Toda field theories are known as
{\bf conformal Toda field theories} \cite{classtoda}. As their name indicates, they are conformally invariant.
 The simplest example of this type of theories is  Liouville's theory.

\vspace{0.25cm}

\noindent If the Lie algebra is an affine Kac-Moody algebra (affine extension of
 $g$ without central extension), the resulting
theories are known as {\bf affine Toda field theories} (ATFT). These are massive QFT's
which may be understood, following point {\bf IV}, as perturbations of the class
of conformally invariant models introduced in the previous paragraph. 
Amongst these theories, we encounter two very different groups depending whether 
the associated coupling constant, $\beta$ is real or purely imaginary.
 In the former case, 
the corresponding models do not possess solitonic solutions at classical level and their
description, although well established in the infrared limit, 
is problematic in the 
ultraviolet regime. 
These theories constitute
examples of 1+1-dimensional QFT's which have been most extensively studied in 
the literature, first on a case-by-case basis, both for simply laced Lie algebras 
\cite{Roland, eeight, ATFTS, dis, mussrev}
 and for the non simply-laced case \cite{ATFTNS}. Eventually,  
universal representations for the S-matrices valid for all simply laced ATFT's 
with real coupling constant in form of hyperbolic functions \cite{uni} 
and integral representations valid for all ATFT's \cite{Oota,FKS2} based on 
purely algebraic quantities, were formulated thereafter (see also \cite{ravanini}).
However, a rigorous proof
of the equivalence between the S-matrix representation in terms of hyperbolic
functions and the integral representation for all ATFT's 
was first carried out  in \cite{FKS2}. 
The simplest example of ATFT's is the sinh-Gordon model \cite{SSG, SSG2}.
Concerning ATFT's associated to purely imaginary coupling constant, one of its 
 most relevant features is the existence of solitonic solutions. The infrared description
of such theories involves in general non-unitary S-matrices but their ultraviolet limit
is better established than for the ATFT's with real coupling constant mentioned before.
The simplest and best known example of this class of models is the sine-Gordon model
 ($SU(2)$-ATFT) even though scattering matrices corresponding to other choices of the 
Lie algebra have been also constructed in \cite{imATFT}.

\vspace{0.25cm}
\noindent  Finally, the last example of abelian Toda field theories are the so-called
{\bf conformal affine Toda field theories} \cite{babelon},
which may be obtained from the affine Toda field theories by introducing auxiliary
fields $\eta, \mu$ in order to restore conformal invariance. They can also be defined as those 
conformal theories whose spontaneous symmetry breaking leads
 to the affine Toda field theories defined above. 

\vspace{0.25cm}
\noindent On the other hand, when the field $h(x,t)$ takes values in a  non-abelian Lie algebra $g$, 
a new rich family of models emerges. These models are known as
{\bf non-abelian Toda field theories} (NAAT) \cite{nt} and a particular subset of them
will be the final object of our study. The NAAT-theories were originally constructed as
non-abelian  generalisations of the abelian Toda field theories described above. 
Such a generalisation can be carried out in different ways. The original construction
due to A.N.~Leznov and M.V.~ Saveliev (see first reference in \cite{nt}) associates
different types of equations of motion to each possible embbeding of $sl(2, \Bbb{C})$
into the non-abelian  Lie algebra $g$. However, a systematic and more general description
of the NAAT-theories was carried out in the two last references in \cite{nt} and in 
\cite{HSG}. 
The mentioned construction 
associates different types of equations of motion to each possible  gradation
of an affine Lie algebra $\hat{g}$ associated to a finite semisimple Lie algebra $g$. 
This gradation is induced by a finite order automorphism of $g$, usually denoted by $\sigma$ \cite{HSG}. 
However, it is worth mentioning that the explicit use of affine Lie algebras
is not needed but when one attempts to explicitly construct higher spin conserved
quantities. 

Although all NAAT-theories
are classically integrable, it was shown in \cite{HSG} that at quantum level only two
particular subsets of NAAT-theories can be associated with unitary and massive QFT's. These
two families of theories were named in \cite{HSG} as {\bf homogeneous sine-Gordon models} (HSG)
and {\bf symmetric space sine-Gordon models} (SSSG). The latter two subsets of NAAT-theories 
are of special interest since, remarkably, they simultaneously possess classical solitonic solutions,
their ultraviolet description is well established and they are expected to admit also an infrared 
description in terms of unitary S-matrices, which have been explicitly constructed in \cite{HSGS} for
the HSG-models corresponding to simply-laced Lie algebras.

The HSG-models have been
studied to quite a large extent by members of the Theory group of the University of Santiago
de Compostela (Spain) along the last  years \cite{HSG, HSG2, HSGsol, HSGS, gallasphd, pousaphd}.
This study gave rise ultimately to an S-matrix proposal for all the HSG-models associated to
simply-laced Lie algebras \cite{HSGS} which constitutes the starting point of the whole analysis
presented in this thesis. 
The development of consistency checks for this S-matrix proposal has been one
of the main aims of the work presented here.

The action associated both to the HSG- and SSSG-models may be written in the general form

$$
S=S_{WZNW} + \frac{m^2}{\pi \beta^2} \int d^2 x
\, \langle \Lambda_{+}, h^{\dagger}\Lambda_{-}h \rangle,
$$
\noindent where the first term is the action of a Wess-Zumino-Novikov-Witten (WZNW) \cite{wzw, wzw2}
coset model, associated to a coset of the general form $G_0/U(1)^{p}$. $G_0$ is the Lie group associated
to a compact finite and semisimple Lie algebra $g_0 \subset g$ 
and $p$ is  an integer whose value depends on
the particular type of theories under study. The WZNW-models \cite{wzw, wzw2} are conformally invariant
theories so that the HSG- and SSSG-models may be understood as perturbed CFT's in the sense described in
{\bf IV}. The perturbing field is identified in that case to be
$\Phi=\langle \Lambda_{+}, h^{\dagger}\Lambda_{-}h \rangle$
where $\Lambda_{\pm}$ are two semisimple elements of $g$ and $\langle \,,\,\rangle$ denotes
a bilinear form in $g$.

\vspace{0.25cm}

For the HSG-models $g_0=g$ and the integer $p$  equals the rank of the Lie algebra
$g$, which we will denote by $\ell$. Therefore, they are perturbations of WZNW-coset models associated
to cosets of the form $G_k/U(1)^\ell$ which are also known in the literature as
 {\bf $G_k$-parafermion theories}.
Here we introduced an integer $k$ that is called the `level' \cite{wzw, wzw2} and in terms of which
the coupling constant $\beta^2$  gets quantized for the quantum theory to be well-defined.
The main characteristics of these CFT's have been studied in \cite{Gep, GepQ, DHS} and, in particular,
their local operator content is well classified, a fact which we might exploit in the context of our form
factor analysis. Finally, the elements $\Lambda_{\pm}$ characterising the perturbing field take
values in the Cartan subalgebra of $g$. 
The simplest example amongst the HSG-models is associated to $g=su(2)$ and can be
identified with the so-called complex sine-Gordon model \cite{CSGold, Park}.
This theory corresponds to the perturbation of the usual
$\mathbb{Z}_k$-parafermions \cite{paras} by the first thermal operator~\cite{CSGBAK}, whose exact
factorisable scattering matrix is the minimal one associated to
$A_{k-1}$~\cite{paras, Roland}.

As mentioned above, recently an S-matrix proposal for all HSG-models related 
to simply-laced Lie algebras
has been provided in \cite{HSGS}. These S-matrices
include some novel features with respect to many
other integrable QFT's which is worth emphasising already at this point:
they break parity invariance, they have resonance poles which may interpreted as the trace of the
presence of unstable particles in the model and  they posses
at the same time a well-defined
Lagrangian description. Although in \cite{brazhnikov}
the SSSG-models associated to the symmetric space $SU(3)/SO(3)$ were first
investigated and found to be quantum 
integrable and to possess the above mentioned properties, an
S-matrix proposal for these concrete theories is absent for the time being. 
Therefore, the S-matrices constructed
in \cite{HSGS} for the HSG-models still provide the first examples
of scattering amplitudes which having a non-trivial rapidity dependence,
incorporate consistently  parity breaking together with
the presence of unstable particles in their spectrum.

\vspace{0.25cm}

Concerning the SSSG-models, some of their characteristics have been studied in \cite{cm,tesinha}
despite the fact that little information is known in comparison to the above mentioned HSG-models. 
In particular,
the development of an S-matrix proposal is an open problem
 for the time being for all SSSG-theories. However, part
of their classical soliton spectrum was constructed in \cite{cm} and also the quantum integrability
of a subset of them proven. We will devote part of the next chapter to a more detailed description
of the results obtained in \cite{cm} for these models. 

The SSSG-theories are in one-to-one correspondence with the compact
symmetric spaces $G/G_0$ \cite{helgas} which also means that the Lie algebra $g$ 
associated to $G$ admits a decomposition
of the form $g=g_0 \oplus g_1$. Here 
$g_0$ is a subalgebra of $g$ whereas $g_1$ is a certain 
subspace of $g$  in which
the elements $\Lambda_{\pm}$ take values. These theories are perturbations of WZNW-coset models
associated to cosets of the form $G_0 / U(1)^p$, where the value of $p$ is now not fixed but can take
a certain range of values which is determined by the properties of the elements $\Lambda_{\pm}$.
In particular there are two specially interesting situations:

\vspace{0.25cm}
\noindent $p=0$ : in that case the corresponding SSSG-model is just an integrable perturbation of the
WZNW-theory associated to the Lie group $G_0$. These models have been called {\bf split models}
in \cite{cm, tesinha}.

\vspace{0.25cm}

\noindent $p=\ell_0$: in that case the corresponding model is a perturbation of a WZNW-coset theory
associated to a coset of the form $G_0/U(1)^{\ell_0}$, for $\ell_0$ to be the rank of $G_0$. 
Therefore we have new integrable perturbations of $G_0$-parafermion theories, different from the
ones provided by the HSG-models.

\vspace{0.25cm}
The simplest examples of SSSG-models are the sine-Gordon model, which corresponds to
the symmetric space $G/G_0=SU(2)/SO(2)$ and again the complex sine-Gordon theory which
is now related to the symmetric space $Sp(2)/U(2)$. Also the SSSG-theories related to
the symmetric space $SU(3)/SO(3)$ have been  studied by V.A. Brazhnikov in \cite{brazhnikov}. 
In particular, for $p=1$, the perturbed CFT associated to the latter coset corresponds again
to the perturbation of the usual $\mathbb{Z}_k$-parafermion theories \cite{paras}, 
in this case by the second thermal operator.

\vspace{0.25cm}
Having now presented the key ideas entering the work
we will carry out in this thesis, as well as the main defining
characteristics of the homogeneous sine-Gordon models \cite{HSG, HSG2},
we will now summarise the content of each
of the subsequent chapters:

\vspace{0.25cm}
{\it In chapter \ref{ntft} we will provide a fairly detailed revision of the
most relevant properties of 1+1-dimensional QFT's}. In particular, since
these properties are closely linked to the powerful nature of conformal invariance
in the 1+1-dimensional context, we will start the chapter with a review of some of
the most important properties of 1+1-dimensional CFT's. Thereafter, we will enter
the analysis of the properties of massive QFT's constructed as perturbed CFT's along
the lines of \cite{Pertcft}. We will also summarise the arguments presented in
\cite{SW, parke}, concerning the distinguished features of the scattering amplitudes
of 1+1-dimensional integrable QFT's mentioned in {\bf III}. We will analyse
the constraints any scattering amplitude is subject to in virtue of Lorentz invariance,
unitarity, Hermitian analyticity and crossing symmetry 
and describe the link between the pole 
structure of the S-matrix and the stable and unstable particle content of the QFT 
\cite{Boot,KTTW,ELOP, Wein, Zalgebra, sinegor}.
After this general part we will enter the analysis of the main properties of the
non-abelian affine Toda field theories (NAAT) \cite{nt} paying special attention to
the description of the  classical and quantum aspects of the subset of these
theories known under the name of homogeneous sine-Gordon models (HSG) \cite{HSG, HSG2}.
In particular, we will present in detail the semi-classical 
construction of their stable particle spectrum carried out in \cite{HSGsol}, 
as well as the S-matrix proposal of  \cite{HSGS}. These data will constitute
the starting point of the analysis carried out in the next chapters.
We also present in this chapter a revision of some of the most relevant
aspects of the symmetric space sine-Gordon models (SSSG) investigated in {\bf \cite{tesinha, cm}}. Finally, we provide the reader with a brief overview of the characteristics
of the $g|\tilde{g}$-theories proposed in \cite{FK}, whose S-matrices have been constructed as
generalisations of the HSG-model and minimal ATFT \cite{Roland} S-matrices and,
 therefore, contain the
latter as particular examples.

\vspace{0.25cm}
{\it In chapter \ref{tba} we will introduce the fundamental ideas entering the thermodynamic
Bethe ansatz approach (TBA)  \cite{Yang, TBAZam1} and carry out a TBA-analysis for the
HSG-models \cite{HSG, HSG2, HSGsol, HSGS}}. Our TBA-analysis will permit the identification of the 
Virasoro central charge of the underlying CFT for all the HSG-models. Also,
the conformal dimension of the perturbing field will be identified by  conjecturing its relation
to the periodicities of the so-called $Y$-systems \cite{TBAZamun}.
The finite size scaling function and $L$-functions entering the TBA-equations 
will be numerically computed for some particular examples of the $SU(3)_k$-HSG models,
corresponding to $k=2,3$ and 4 and different values of the resonance parameter
characterising the mass of the unstable particles present in the model. The original 
results presented in this chapter can be found in {\bf \cite{CFKM}} (see also \cite{KK, MM, FF}).

\vspace{0.25cm}
{\it In chapter \ref{ffs} we will present the fundamental properties and applications of form factors
to the study of 1+1-dimensional QFT's and carry out a form factor analysis for the $SU(3)_2$-HSG model}.
We will introduce the consistency equations derived originally 
in \cite{Kar}, whose solution leads to the exact computation of the form factors associated
to a certain local operator of the QFT, that is, matrix elements of the mentioned operator between the
vacuum state and an $n$-particle $in$-state. Thereafter, we will discuss in total generality the
applications of these form factors to the computation of different interesting  quantities:
the Virasoro central charge of the underlying CFT, Zamolodchikov's $c$-function \cite{ZamC},
the conformal dimensions of various local operators of the underlying CFT, the renormalisation
group flow of the operator content of the underlying CFT, etc...After this general introduction
we will carry out a detailed form factor analysis for the $SU(3)_2$-HSG model. We shall construct
all $n$-particle form factors associated to a large class of local operators of the model in terms of
general building blocks which admit both a determinant and an integral
 representation. We will identify
the ultraviolet conformal dimensions of these operators by exploiting the knowledge of
the operator content of the underlying CFT. We will also compute the Virasoro central charge of the
unperturbed CFT and identify the conformal dimension of the perturbing field. We will numerically
determine Zamolodchikov's $c$-function \cite{ZamC} and generalise the $\Delta$-sum rule proposed in
\cite{DSC} to the `off-critical' situation. We shall also analyse in detail the so-called
momentum space cluster property of form factors for the model at hand. The original results presented
in this chapter collect the work published in {\bf \cite{CFK, CF1, CF2}} (see also \cite{FF}).

\vspace{0.25cm}
{\it In chapter \ref{ffs2} we will generalise the study of the previous chapter to all
$SU(N)_2$-HSG models}. We shall also construct all $n$-particle form factors associated to
a large class of operators of the model finding again the same sort of building blocks
encountered in the $SU(3)_2$-case. We will compute the Virasoro central charge, Zamolodchikov's
$c$-function and the conformal dimension of the perturbing field for several concrete values of $N$.
We will also study the renormalisation group flow of the operator content of the underlying CFT and
define what we have called $\beta$-like functions in order to have a clear-cut identification of
the different fixed points both the $c$- and $\Delta$-function surpass in their flow from the
ultraviolet to the infrared regime. The original work presented in this chapter may be found in
{\bf \cite{CF3}}.

\vspace{0.25cm}
{\it In chapter \ref{conclu} we will summarise
the main conclusions of the work carried
out in this thesis and state some open problems
which are left for future investigations.}

\vspace{0.25cm}
{\it In appendix A we collect some useful properties
of elementary symmetric polynomials.}

\vspace{0.25cm}
{\it In appendix B we present the explicit expressions
of the form factors associated to a large
class of operators of the $SU(3)_2$-HSG model
up to the 8-particle form factor.}

\begin{figure}
\begin{center}
\includegraphics[width=14.65cm,height=20.3cm]{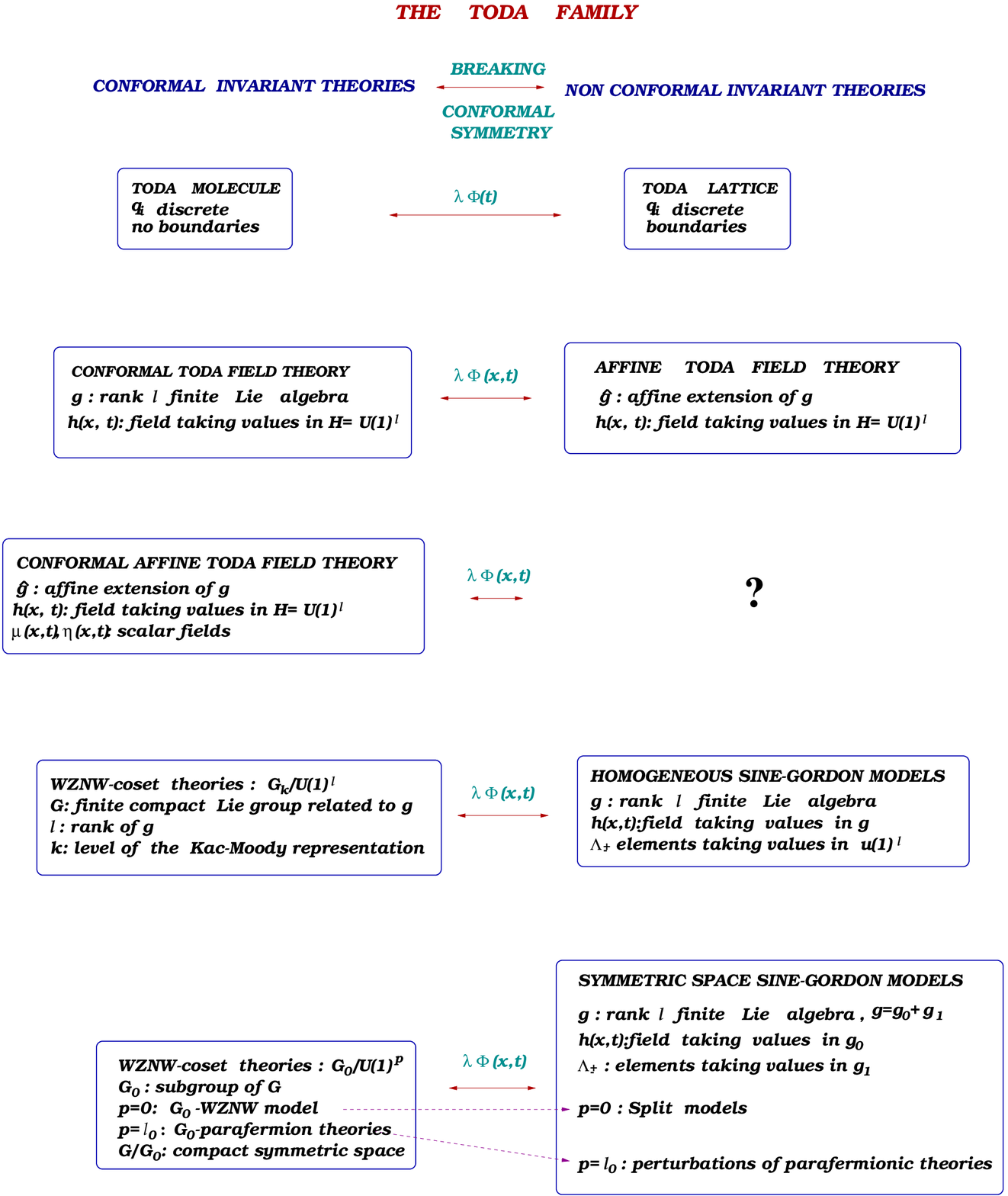}
\label{Toda}
\end{center}
\end{figure}

\chapter{1+1-dimensional integrable massive quantum field theories}
\label{ntft}
\indent \ \
In the previous introduction we provided a first glimpse of 
the distinguished properties of 1+1-dimensional integrable quantum field theories (QFT's). 
Our aim
was to furnish an introductory justification for 
the enormous interest that these sort of models
have achieved over the last 30 years. In particular, we introduced  the Toda field theories 
as a especially prominent
class of  QFT's included  in the previous category. We also reported very briefly the main
properties of  1+1-dimensional integrable QFT's and emphasised their consequences,
in particular, in what concerns the construction of exact S-matrices. 

\vspace{0.25cm}

The aim of the present chapter will be first, a more detailed analysis and derivation
of the properties of 1+1-dimensional integrable QFT's: their construction,
the special characteristics of their S-matrices,
and the general procedure  which may 
allow  for the exact construction of these S-matrices
on the basis of a series of physical requirements together with the
concrete constraints due to integrability.
Second, we want to use the previous general results and
techniques for the study of a concrete family of 1+1-dimensional
integrable QFT's, a subset of 
the non-abelian affine Toda (NAAT) field theories \cite{nt, HSG}.
In particular, we will pay special attention to a subclass of
the latter theories, the homogeneous sine-Gordon (HSG) models \cite{HSG2}, 
whose study, has been carried out to a large extent by the Theory group
of the University of Santiago de Compostela (Spain), over the last years
\cite{HSG, HSG2, HSGsol, HSGS, gallasphd, pousaphd}. The development
of non-perturbative tests of the S-matrix proposal provided in \cite{HSGS}
for those  theories is one of the main objectives of the work we
will present in this thesis. However, it must be emphasised that 
our results constitute also a valuable
contribution to the understanding of several aspects 
related to the thermodynamic Bethe ansatz and form factor approach themselves.

\vspace{0.25cm}

In addition, we will also provide in this chapter original results \cite{cm}
concerning the quantum properties of a second family of theories, the symmetric space
sine-Gordon (SSSG) models \cite{HSG, tesinha}, 
which are also particular examples of massive NAAT-theories and
whose study,  for the time being, has not been carried out to such an extent as for the
HSG-theories. Even the construction of  exact  S-matrices related to the
SSSG-models is still a completely open problem. However, since the main objectives
of this thesis are the ones stressed in the previous paragraph, we will not present here 
in detail the results found in \cite{cm}. 

\vspace{0.4cm}
More concretely, the structure of the present chapter will be the following:

\vspace{0.25cm}

In section \ref{cftqft} we shall present a brief overview of the main
properties of conformally invariant theories \cite{CFT, cardy},
 necessary for the 
understanding of the more relevant characteristics of perturbed CFT's. Recall
that it was pointed out by A.B. Zamolodchikov in \cite{Pertcft} that a 1+1-dimensional
QFT may be viewed as a perturbation of a CFT which takes the latter away from its
associated renormalisation group fixed point. The results provided in \cite{Pertcft}
will be reviewed
 in subsection  \ref{PCFT}. Thereafter, in section \ref{exactS},
we will analyse the specific properties of 1+1-dimensional integrable QFT's, paying special
 attention to their consequences in what concerns the exact computation of  S-matrices.
After the introduction of the so-called Zamolodchikov's algebra \cite{Zalgebra} in
subsection \ref{Zamalgebra} as a means for representing the asymptotic states in a 1+1-dimensional
QFT, we will present the definition and properties of the so-called 
higher spin 
conserved charges. We shall explain how their existence 
in 1+1-dimensional QFT's 
leads to the conclusion that the corresponding
S-matrix factorises into products of two particle S-matrices and
 that there is no particle
production in any scattering process \cite{SW, sm2, parke}. 
In section \ref{analit} we shall
summarise the specific properties of
 two-particle scattering amplitudes in 1+1-dimensional
QFT's \cite{Boot, KTTW, ELOP, Wein, Zalgebra, sinegor}. 
These properties are Lorentz invariance, 
Hermitian analyticity, unitarity and crossing symmetry together with two sets
of highly non-trivial equations known as Yang-Baxter \cite{yb} and bootstrap
equations \cite{Boot, KTTW}. The first set of properties have their origin 
in physically motivated  requirements 
whereas the latter two equations exploit the specific consequences of quantum
 integrability or the existence of higher spin conserved charges in the QFT.
 All these constraints allow
in many cases for the exact computation of  the S-matrix associated to a 1+1-dimensional
QFT by means of the bootstrap program, originally proposed  in \cite{Boot}. We will also
pay attention in section \ref{analit} to the pole structure of  the two-particle scattering amplitudes,
and report its intimate connection with the  stable and unstable particle spectrum of the model at hand.
Once the general framework and techniques have been reported, we will turn in section
\ref{ntft1} to the description of the specific theories we will focus our interest on: 
the non-abelian affine Toda (NAAT) field theories \cite{nt}. After a very brief review on classical
integrability, we will go through the quantum properties of all NAAT-theories, studied
in \cite{HSG, HSG2, HSGsol, Park, HMQH, gallasphd, pousaphd, HSGS}, 
exploiting the general results of sections \ref{cftqft}, \ref{exactS} and \ref{analit}. 
In particular, we will describe in detail the characteristics of
the two families of unitary and massive NAAT-theories found in \cite{HSG}: the symmetric
space (SSSG) and homogeneous sine-Gordon (HSG) theories. Since the latter  models
have been studied to a larger extent than the former and most of the original
results presented in this thesis are related  to the HSG-models, we will pay more attention to the
description of the properties of these theories, and ultimately report the S-matrix
proposal  for the HSG-models related to simply-laced Lie algebras derived in \cite{HSGS}.
We will also dedicate subsection \ref{SSSG} to a brief description of the properties 
of the SSSG-models, 
report the most prominent results obtained in \cite{cm} and provide arguments in order to motivate
the interest of their further investigation.
Finally, we will recall some of the features of a new type of S-matrices proposed in \cite{FK} which
contain the HSG-models \cite{HSG2} and minimal ATFT \cite{Roland, eeight}
as particular distinguished examples and whose underlying CFT was studied in \cite{DHS}.
These are the $g|\tilde{g}$-theories proposed by A. Fring and C. Korff in \cite{FK} for
$g$, $\tilde{g}$ to be simply laced Lie algebras and recently
generalised by C. Korff in \cite{KK2} for the case when $\tilde{g}$ is non-simply laced.

\vspace{0.4cm}
\section{From conformal field theory to massive quantum field theory}
\label{cftqft}
\indent \ \
As outlined above,  since in 1989 A.B. Zamolodchikov \cite{Pertcft}
pointed out that a 1+1-dimensional integrable QFT
may be formally viewed as a  perturbation of a CFT by means
of a certain  relevant field of the CFT itself, the latter approach has 
been successfully exploited by many authors  in the course 
of the construction, classification and characterisation of
 1+1-dimensional QFT's.
In the spirit of  \cite{Pertcft}, 
the original ultraviolet CFT plays the role of 
starting point in the construction of a 1+1-dimensional
 massive  integrable QFT. 
The key idea is that a CFT can always be thought of 
as a renormalisation group (RG)  critical fixed point  
(see e.g. \cite{cardy}).  
Therefore,  its perturbation by means of any relevant
operator  amounts to ``moving''  the CFT away from its 
associated RG-fixed point 
and consequently, to  breaking the initial conformal invariance. For arbitrary choices
of the perturbing field  one would expect to end up with a 1+1-dimensional QFT
which, in general,  may be neither conformal invariant nor integrable in the sense of possessing
an infinite number of integrals of motion. However,  the combination of 
a suitable choice of the perturbation together with the fact that conformal symmetry is
an extremely high symmetry which provides every  CFT with an infinite number 
of local conserved quantities, allows for  the construction of 1+1-dimensional
QFT's which, although breaking conformal invariance still have associated
an infinite number of conserved quantities. These conserved quantities 
arise as particular  combinations of those of the original unperturbed CFT
and may even be explicitly constructed along the lines of \cite{Pertcft}.
This construction has been carried out for instance
for the mentioned HSG- and some SSSG-models in \cite{HSG2} and 
\cite{cm} respectively, aiming to prove their integrability.
However, it is worth mentioning that the integrability of the perturbed CFT, 
is guaranteed by the existence of such quantities. This means that their
explicit construction is not necessary in order to prove the integrability
of the model  whenever their existence can be established  by other means. 
In this direction, it is possible to resort to the so-called 
``counting-argument'' developed also by A.B. Zamolodchikov in  \cite{Pertcft},
which will be described in detail later. The mentioned argument provides
a sufficient condition for the existence of conserved quantities in a
perturbed CFT and can be worked out provided the characters of the irreducible
representations of the Virasoro algebra (see subsection \ref{cftbo}) 
associated to the unperturbed CFT are
known. As we said this criterium is sufficient to prove the existence 
of conserved charges  in the massive QFT. To our knowledge,
the mentioned  characters are not known up to now for the generality of
the underlying CFT's related to the HSG- and SSSG-models, which is the
reason why the explicit construction of some conserved quantities 
has been necessary for establishing their integrability. 

The above qualitative arguments justify the important role
CFT plays in the construction of 1+1-dimensional integrable QFT's,
of which the non-abelian affine Toda field theories \cite{nt} at hand
are particular examples.  We will devote the next subsection to the
introduction of some basic notions on 
conformal field theory necessary for the subsequent understanding of
the most relevant features of perturbed conformal field theory. Some of these
properties will also  be recovered both in the thermodynamic Bethe
ansatz and form factor context since the application of any of these approaches
to the study of 1+1-dimensional integrable models allows ultimately for
the identification of the main data  characterising the CFT 
which describes the ultraviolet behaviour of the QFT at hand.

\subsection{Conformal field theory: a brief overview}
\label{cftbo}
 \indent \ \
In the light of the previous paragraph and to achieve self-consistency 
we will  now introduce some basic notions on 1+1-dimensional
conformal field theory. A more exhaustive discussion and derivation  
of these properties may be found
for instance in \cite{CFT, cardy}.

\subsubsection{The classical conformal group}
\indent \ \
The conformal group in arbitrary dimension $d$ is the subset of coordinate transformations which
leave the metric $g_{\mu \nu}$ invariant up to a scale transformation, namely
\begin{equation}
g_{\mu\nu} \rightarrow e^{\Lambda(x)} g_{\mu \nu}.
\end{equation}
It can be proven  by considering initially infinitesimal transformations of the
type $x_{\mu} \rightarrow x_{\mu}+ \epsilon_{\mu}$, 
that conformal symmetry requires 
\begin{equation}
\partial_{\mu} \epsilon_{\nu}+\partial_{\nu} \epsilon_{\mu} = \frac{2}{d}
\partial_{\rho} \epsilon^{\rho} \, \eta_{\mu \nu}
\end{equation}
\noindent provided we consider a flat metric $g_{\mu \nu}=\eta_{\mu\nu}$.
In this thesis we will be interested in the 1+1-dimensional case.
Hence, the latter equation gives 
\begin{equation}
\partial_{1}\epsilon_{1}=\partial_{2}\epsilon_{2}, \,\,\,\,\,\,\text{and}\,\,\,\,\,\,
\partial_{1}\epsilon_{2}= -\partial_{2}\epsilon_{1},
\end{equation}
\noindent which, according to the Cauchy-Riemann theorem, suggests the definition
of two new functions  $\epsilon(z)$ and $\bar{\epsilon}(\bar{z})$ depending upon the complex
coordinates $z, \bar{z}=x^0 \pm i x^1$ as follows,
\begin{equation}
 \epsilon(z)= \epsilon_1 + i   \epsilon_2,\,\,\,\,\,\,\text{and}\,\,\,\,\,\, 
\bar{\epsilon}(\bar{z})= \epsilon_1 - i   \epsilon_2.
\end{equation}
\noindent This result amounts to the conclusion that 1+1-dimensional conformal
transformations are just analytic coordinate transformations in the complex plane
of the form,
\begin{equation}
 z \rightarrow  f(z) ,\,\,\,\,\,\, \,\,\,\,\,\,\text{and}\,\,\,\,\,\, \,\,\,\,\,\,
\bar{z} \rightarrow \bar{f}(\bar{z}).
\label{cg2}
\end{equation}
\noindent The algebra which generates the sort of transformations (\ref{cg2})
is infinite dimensional and the corresponding infinitesimal generators are
found to be 
\begin{equation}
l_n=-z^{n+1} \partial ,\,\,\,\,\,\, \,\,\,\,\,\,\text{and}\,\,\,\,\,\, \,\,\,\,\,\,
\bar{l}_n=-\bar{z}^{n+1} \bar{\partial},
\label{cftgen}
\end{equation}
\noindent with $\partial:=\partial/{\partial z}$ and  $\bar{\partial}:=\partial/{\partial \bar{z}}$
and $n \in \mathbb{Z}$. These generators satisfy the {\bf Witt algebra},
\begin{equation}
\left[ l_n, l_m \right]=(m-n) l_{n+m}, \,\,\,\,\,\,
\left[ \bar{l}_n, \bar{l}_m \right]=(m-n) \bar{l}_{n+m},
\label{calgebra}
\end{equation}
\noindent with $\left[ l_n, \bar{l}_m \right]=0$ for any value of $n, m$.
 Therefore, the conformal algebra is the direct product
of two isomorphic subalgebras generated by the  $l$'s and  the $\bar{l}$'s. At the quantum level,
 these commutation relations acquire an additional constant contribution on the r.h.s., 
 giving rise to a so-called Virasoro algebra.  

Having  (\ref{cftgen}) at hand one easily observes that in the limits $z \rightarrow 0, \infty$ only the 
infinitesimal generators $l_0, l_{\pm 1}$ are globally well defined and similarly for their anti-holomorphic
counterpart. Moreover, Eq. (\ref{calgebra}) shows that their commutation algebra closes and is isomorphic
to $sl(2, \mathbb{C})/\mathbb{Z}_2$. It might also be easily derived that $l_{-1}, \bar{l}_{-1}$  are the generators of translations ($z \rightarrow z+a$) whereas $l_0 +\bar{l}_0$ and $i(l_0-\bar{l}_0)$ generate dilatations 
($z \rightarrow \lambda z$)  and rotations ($z \rightarrow e^{i \theta}z$)  respectively.

\subsubsection{Conformal symmetry at quantum level}
\indent \ \
Over the last 30 years, conformal field theory has  became one of the most
active and fruitful research fields in the context of mathematical physics. 
The explanation
of this success relies on the fact that conformal invariance turns 
out to be an extremely 
powerful symmetry in 1+1-dimensions since, only in that case
 it is associated to an
infinite number of independent generators. Consequently, many problems 
which can only be handled with great
difficulties for general QFT's find in this context an exact solution.  
In the framework of physical systems characterised by local interactions,
conformal invariance can be understood as an immediate consequence of
 scale invariance. This observation was originally
made by A.M. Polyakov \cite{poly}. Thereafter there have been 
various works elaborating on
these ideas, e.g. \cite{schroer}. However, the key work which
 really initiates the modern study of 
conformal invariance in 1+1-dimensions dates back to 1984 
and is due to A.A. Belavin, A.M. 
Polyakov and A.B. Zamolodchikov \cite{BPZ}. In \cite{BPZ} 
the authors showed how to construct
completely solvable CFT's, the minimal models, which thereafter 
have been extensively studied in the literature \cite{FQS}.
 In particular they were able to
formulate differential equations (Ward identities) satisfied by correlation functions.

In the light of the previous paragraph, we want to devote 
 this subsection to a review of some
of the most important features of  1+1-dimensional CFT's.  
We will not give here all the details of the 
quantization procedure which matches the results of the 
preceding subsection with
the ones we want to present now. To keep it brief we start, 
at classical level, with a formulation
of the theory by means of the coordinates $\sigma^0$, $\sigma^1$ and
 introduce, as usual  in Euclidean space, the complex coordinates
 $\omega, \bar{\omega}=\sigma^0 \pm i \sigma^1$. 
The first step in the quantization
procedure is the compactification of  the space dimension:
 $\sigma^1\equiv \sigma^1+ 2\pi$. 
Therefore we end
up with a theory formulated in an infinitely long  cylinder whose
 circumference is identified as the compactified space dimension. 
 Thereafter, the introduction of  the conformal map $z = e^{\omega}$,  
which  allows for  defining   the QFT  in the $z$-plane, transforms
the problem in what is usually referred to as {\bf radial quantization}.

The conserved charges associated to the QFT in 
the $z$-plane are generated by the {\bf energy
momentum tensor}  $T_{\mu\nu}$ which is always symmetric and
in conformally invariant theories, also traceless ($T^{\mu}_{\mu}:= \Theta=0$). 
It is usually more convenient to express the components of the
energy momentum tensor in terms of the $z, \bar{z}=x^0\pm i x^1$ coordinates.
 We obtain
\begin{eqnarray}
T_{zz}&=&\frac{1}{4}(T_{00}-2 iT_{10}-T_{11}),\nonumber\\
T_{\bar{z}\bar{z}}&=&\frac{1}{4}(T_{00}+2 iT_{10}-T_{11}),\nonumber\\
T_{z\bar{z}}&=&T_{\bar{z}z}=\frac{1}{4}(T_{00}+T_{11})=\frac{\Theta}{4}.
\end{eqnarray}
\noindent The conservation of the energy momentum tensor 
amounts to the imposition of the following constraints,
\begin{equation}
\bar{\partial} T_{zz}=\partial T_{\bar{z}\bar{z}}=0,
\label{tt}
\end{equation}
\noindent which justify the definitions $T(z):=T_{zz}$ and $\bar{T}(\bar{z}):= T_{z\bar{z}}$. Consequently, local conformal
 transformations in the complex $z$-plane are generated by the
holomorphic and anti-holomorphic components 
of the energy momentum tensor defined before. In fact, Eq. (\ref{tt}) suggests  the introduction of an infinite set of generators $L_n$, $\bar{L}_n$
which arise as the `coefficients' of the Laurent 
expansion of the holomorphic and anti-holomorphic
components of the energy momentum tensor,
\begin{equation}
T(z)=\sum_{n\in \mathbb{Z}} z^{-n-2} L_n \iff
 L_n=\oint_{z} \frac{d \omega}{2\pi i} (\omega-z)^{n+1} T(\omega),
\label{expan}
\end{equation}
\noindent  and act on the space of local fields of the CFT.
A similar mode expansion  can be performed for the anti-holomorphic component $\bar{T}(\bar{z})$ in
terms of  modes $\bar{L}_n$. In order to compute now 
the algebra of commutators satisfied by these modes
it is required the evaluation of commutators of contour integrals of the type $\left[\oint dz, \oint d\omega\right]$ together with the computation 
of {\bf operator product expansions} (OPE) of the 
holomorphic and anti-holomorphic components of the energy momentum tensor.
These OPE's 
characterise the leading order behaviour in the 
limit $z \rightarrow \omega$ and they can be
easily computed once the QFT has been formulated in the plane by means of 
the radial quantization  procedure summarised  before. 
In 1+1-dimensions and in the Euclidean regime
we can exploit our  knowledge about contour integrals 
and complex analysis, in particular when evaluating 
short distance expansions and we refer the reader to
 \cite{CFT} for a more complete
description of these applications. For the energy 
momentum tensor we have the following relevant  OPE
\begin{equation}
T(z) T(\omega)=\frac{c/2}{(z-\omega)^4}+\frac{2 T(\omega )}{(z-\omega)^2}+\frac{\partial T(\omega)}{(z-\omega)}.
\label{ttope}
\end{equation}
\noindent The constant $c$ arising in the $\mathcal{O}((z-\omega)^{-4})$-term is the so-called 
{\bf central charge} of the CFT and depends on 
the particular theory considered being one of  its  most characteristic data. 
The latter OPE has a completely analogous counterpart when
 considering the anti-holomorphic component of the energy
 momentum tensor and allows for the computation of the algebra satisfied by the
 generators $L_n$ above introduced. The mentioned algebra has the form
\begin{equation}
\left[L_n, L_m\right]\,=\,(n-m)L_{n+m}\,+\,\frac{c}{12}\,n(n^2 -1)\delta_{n+m,0},
\label{VV}
\end{equation}
\noindent and is known as {\bf Virasoro algebra} although the central extension 
 was originally found by J. Weis (see note added in proof of \cite{FV}).
Consequently, the central charge $c$ is usually referred
to also as {\bf Virasoro central charge}. Notice that the algebra (\ref{VV}) is a sort of  `extension' 
of the classical algebra (\ref{calgebra}) which is still recovered for the 
generators $L_n$, with  $n= 0, \pm 1$. Constant terms of the type ${c}\,n(n^2 -1)\delta_{n+m,0}/12$
which arise at quantum level and have the effect of providing additional constant 
contributions to the classical commutation relations are generically called {\bf central extensions}.

In summary, if at classical level one has an algebra  of symmetry transformations
of the type (\ref{calgebra}), at quantum level the commutation relations are expected to acquire
quantum corrections (typically  $\mathcal{O}(\hbar^2)$) which should give rise
to a new symmetry algebra still compatible with the Jacobi identities.
This is easily achieved whenever the mentioned quantum corrections are proportional to an element
of the symmetry algebra whose commutator with all the remaining generators vanishes.
In that case the proportionality coefficient is  referred to as a central extension, 
as explained for instance in  \cite{olive}.

\subsubsection{Fields and correlation functions in CFT}
\indent \ \
Let us now consider a conformal mapping of the form $z \rightarrow f(z)$ and 
$\bar{z} \rightarrow \bar{f}(\bar{z})$ and a local field of the CFT, say $\phi(z, \bar{z})$,  
which under this map transforms as
\begin{equation} 
\phi (z, \bar{z}) \rightarrow ( \partial f )^{\Delta}
 (\bar{\partial} \bar{f} )^{\bar{\Delta}} \phi (f(z), \bar{f}(\bar{z})).
\label{primaryf}
\end{equation}
\noindent This sort of transformation is
very similar to the transformation law of a tensor of the form 
$\phi_{z\cdots z \bar{z} \cdots \bar{z}} (z, \bar{z})$ 
with $\Delta$ lower $z$-indices and
$\bar{\Delta}$ lower $\bar{z}$-indices. 
Such transformation property defines what is
known as a {\bf primary field} of the CFT of {\bf conformal dimensions}
 $(\Delta, \bar{\Delta})$. However, 
there will be many fields in a CFT which do not have this sort of transformation property,
for instance, the energy momentum tensor introduced above. They are referred to as {\bf secondary 
or descendant fields}. 
There is an especially relevant subclass of secondary fields which are known as
{\bf quasi-primary fields}. Quasi-primary fields satisfy  (\ref{primaryf}) but only when
the mapping $z, \bar{z} \rightarrow f(z), \bar{f}(\bar{z})$ is generated by the globally
defined Virasoro generators $L_0, L_{\pm 1}$ namely, they are primary fields under global
conformal transformations. The holomorphic and anti-holomorphic components of the 
energy momentum tensor are particular examples of quasi-primary fields of conformal
dimensions (2, 0) and (0, 2) respectively. It is also clear from the previous 
definitions that a primary field is automatically quasi-primary. 

By using the transformation property (\ref{primaryf}) and  exploiting  the fact that the theory
is conformally invariant, it is possible to establish very restrictive constrains for  the
general form of any correlation function involving quasi-primary fields. In particular, the
two-point function of a quasi-primary field $\phi (z, \bar{z})$ must necessarily have
the form
\begin{equation}
\langle \phi(z, \bar{z}) \phi (\omega, \bar{\omega}) \rangle =\frac{\mathcal{C}}
{(z-\omega)^{2 \Delta}(\bar{z}-\bar{\omega})^{2 \bar{\Delta}}},
\label{uvbeh}
\end{equation}
\noindent for $\mathcal{C}$ being a constant. In particular it is common to define
$s:=\Delta-\bar{\Delta}$ and $d:=\Delta+\bar{\Delta}$ as the {\bf spin} and 
{\bf scale dimension} of the field under consideration. Thus, for spinless fields ($s=0$) the
two-point function reduces to a simpler form
\begin{equation}
\langle \phi(z, \bar{z}) \phi (\omega, \bar{\omega}) \rangle =\frac{\mathcal{C}}
{|z-\omega|^{4 \Delta}},
\label{uvbeh2}
\end{equation}
\noindent which we will exploit in the form factor context (see chapters \ref{ffs} and \ref{ffs2})
in order to extract  the conformal dimensions of  part of the quasi-primary fields of the unperturbed CFT. 
This can be done once the assumption that there is a one-to-one correspondence between
the field content of the perturbed and unperturbed CFT is made (we will provide
arguments which support this belief  in the next sections). Consequently,
we will be able to extract the conformal dimensions of  primary fields of  
the underlying or unperturbed CFT by studying the ultraviolet behaviour 
of the two-point functions of their corresponding counterparts in the perturbed CFT,
which are in principle available within the form factor approach.
The mentioned ultraviolet behaviour is therefore expected to be of 
the form (\ref{uvbeh2}).

Similarly, conformal invariance together with the transformation law (\ref{primaryf}) also restrict
severely the possible form of the 3- and 4-point functions. However, since we will not require
their behaviour in what  follows we will not report them in here.

Another aspect which is relevant concerning the properties of primary fields is the form
of their OPE's with the holomorphic and anti-holomorphic components of the 
energy momentum tensor. They turn out to be
\begin{eqnarray}
T(z) \phi (\omega, \bar{\omega})&=&\frac{\Delta \, \phi(\omega, \bar{\omega})}{(z-\omega)^2}+
\frac{\partial \phi (\omega, \bar{\omega})}{z-\omega},\nonumber\\
\bar{T}(\bar{z}) \phi (\omega, \bar{\omega})&=&\frac{\bar{\Delta} \, \phi(\omega, \bar{\omega})}
{(\bar{z}-\bar{\omega})^2}+\frac{\bar{\partial} \phi (\omega, \bar{\omega})}
{\bar{z}-\bar{\omega}},
\label{topes}
\end{eqnarray}
\noindent meaning that once the previous OPE's are known, the conformal dimensions of
a primary field can be identified by looking at the proportionality constant characterising 
the $\mathcal{O}((z-\omega)^{-2})$ and $\mathcal{O}((\bar{z}-\bar{\omega})^{-2})$ terms.
It must be emphasised once more that the OPE's (\ref{topes})  characterise only primary fields,
therefore they do not have the same form for quasi-primary fields like, for instance, the energy
momentum tensor itself. In fact, this is clear from the OPE (\ref{ttope}) which shows in
that case that the leading order behaviour when $z \rightarrow \omega$ is governed by 
the term containing  the Virasoro central charge, term which does not have a counterpart
for primary fields.  However, if  we ignore
the $\mathcal{O}((z-\omega)^{-4})$ contribution to  (\ref{ttope})  the remaining terms
are entirely analogous to the ones encountered in (\ref{topes}) and confirm the previous assertion
that the conformal dimensions of the holomorphic and anti-holomorphic components of the
energy momentum tensor are indeed $(2,0)$ and $(0,2)$ respectively.

Combining now Eqs. (\ref{expan}) and (\ref{topes}) for a purely holomorphic
primary field $\phi(z)$ one can easily derive
\begin{equation}
\left[L_n, \phi(z) \right]= \Delta (n+1) z^n \phi(z) + z^{n+1}\partial \phi,
\label{fff}
\end{equation}
\noindent which means that $\left[L_n, \phi(0) \right]=0$ for $z=0$ and $n>0$ and
$\left[L_0, \phi(0) \right]=\Delta \phi (0)$. The latter property is of great relevance in
what concerns the definition of the asymptotic states in conformally invariant QFT's,
which will be constructed by means of the successive 
action of a primary field on the vacuum state $|0\rangle$.

\vspace{0.25cm}

 Let  $|0\rangle$ denote the vacuum state of the theory.  If we require the regularity of 
$T(z) |0\rangle$ at $z=0$ it follows from the expansion (\ref{expan}) that
\begin{equation}
 L_n |0\rangle =0, \,\,\,\,\,\,\,\,\,\text{for} \,\,\,\,\,\,\,\,\, n\geq -1.
\label{nn1}
\end{equation}
\noindent On the other hand, we may use the convention
$L^{\dagger}_{n}=L_{-n}$  and similarly
for the anti-holomorphic generators,
 which is consistent with the hermitian character of the 
holomorphic and anti-holomorphic components of the energy momentum tensor,
 $T(z)$, $\bar{T}(\bar{z})$. 
Therefore, (\ref{nn1}) is equivalent to
\begin{equation}
 \langle 0| L_n =0, \,\,\,\,\,\,\,\,\,\text{for} \,\,\,\,\,\,\,\,\, n\leq 1.
\label{nn2}
\end{equation}
\noindent Notice that again the subalgebra $\{L_0, L_{\pm 1}\}$ appears to play
a distinguished role since
 only these generators are common to the sets (\ref{nn1}), (\ref{nn2})
namely, they simultaneously annihilate the $|0\rangle$ and $\langle 0|$ states.

\subsubsection{Integrals of motion  in CFT}
\label{motion}
\indent \ \
Once the vacuum state has been defined satisfying the conditions (\ref{nn1}), (\ref{nn2}) one is
in the position to construct  highest weight states, namely eigenstates of  the Virasoro generators
$L_n, \bar{L}_n$ generating  a highest weight representation of the Virasoro
algebra (\ref{VV}). The construction of this sort of representations starts with a single 
primary field $\phi(z)$ of conformal dimension $(\Delta, 0)$, {\em i.e.} let us consider the generic
asymptotic state
\begin{equation}
|\Delta\rangle = \phi (0) |0 \rangle,
\end{equation}
\noindent created by the holomorphic field $\phi(z)$. Following (\ref{fff}) and the subsequent discussion 
we derive 
\begin{equation}
L_0 |\Delta \rangle= \Delta | \Delta \rangle, \,\,\,\,\,\,\,\,L_n |\Delta \rangle =0, \,\,\,\,\,\,n>0.
\end{equation}
\noindent Any state satisfying the latter conditions is referred to as a {\bf highest weight state}. 

The combination of the constraints (\ref{nn1}) and (\ref{nn2}) 
with the Virasoro algebra (\ref{VV}) and the previous definition 
of a highest weight state
gives, for $n>0$,  the interesting relationship

\begin{equation}
{|| L_{-n} | \Delta \rangle || }^2=
\langle \Delta| \left[L_{n}, L_{-n}\right]|\Delta \rangle =
(2n \Delta + cn (n^2-1)/12) ||\, |\Delta \rangle \, ||^2,
\label{unitary}
\end{equation}
\noindent from which we infer that, if we assume the norm
of the states $|\Delta\rangle$ in the Hilbert space to
be positive, taking $n=1$ and using the fact that $L_{-1} | 0 \rangle =0$,
we obtain the condition $\Delta \geq 0$ whereas for $n$ to
be very large, we get the constraint $c>0$. 
Therefore,  for {\bf unitary CFT's} the conformal dimensions of 
fields and the Virasoro central charge must be non-negative.

All the states constructed by the successive  application of Virasoro generators $L_{-n}$ with
$n>0$ to a highest weight state are referred to as {\bf descendant states} and have the generic form,
\begin{equation}
L_{-n_1}L_{-n_2}\cdots L_{-n_p}|\Delta \rangle,\,\,\,\,\,\,\,n_i >0, \,\,\,\,i=1, \cdots, p.
\label{set}
\end{equation}
\noindent  They are also eigenstates of $L_0$ with weight or eigenvalue 
$n=\Delta + \sum_{i=1}^{p}n_i$.
Therefore, starting with a highest weight state $|\Delta \rangle$, it is possible to construct
a `tower' of states (\ref{set}) which  is usually referred to as a {\bf Verma module}.
However, it is not guaranteed that this collection of states are all independent from each
other and frequently, depending on the concrete values of $\Delta$ and the central charge $c$,
one can find vanishing combinations of states of the same weight. These combinations
are known as {\bf null states} and an irreducible representation of the Virasoro algebra built up from
the initial state $|\Delta \rangle $ is ultimately constructed by removing all the null states of
the Verma module.

Another important concept to be introduced in order to identify the integrals of
motion characterising a CFT are the so-called {\bf descendant or secondary fields}, already
mentioned at the beginning of the previous subsection. As we have seen
the highest weight representations of the Virasoro algebra are obtained starting with
a primary field. The remaining fields of the representation can be obtained from this
initial one by the commutation of the Virasoro generators $L_{-n}$ with the initial
primary field. In other words the descendant states (\ref{set}) can be seen also as 
\begin{equation}
 L_{-n_1}L_{-n_2}\cdots L_{-n_p}|\Delta \rangle=
 L_{-n_1}L_{-n_2}\cdots L_{-n_p} \phi (0)|0 \rangle\equiv \psi(0) |0\rangle,
\end{equation}
\noindent where $\psi(0)= L_{-n_1}L_{-n_2}\cdots L_{-n_p} \phi (0)$ would be
a descendant field of conformal dimensions $(\Delta+\sum_{i=1}^{p}n_i , 0)$.

A relevant example of a descendant field is the energy momentum tensor. By using 
(\ref{expan}) and denoting by $\mathcal{I}$ the identity operator we find
\begin{equation}
(L_{-n} \mathcal{I})(z)=\frac{\partial^{n-2}T(z)}{(n-2)!},
\label{construction}
\end{equation}
\noindent which means that for $n=2$ we obtain the energy momentum tensor
$T(z)=(L_{-2}\mathcal{I})(z)$. All the descendant fields of the identity are
composite fields build up from the holomorphic component of the energy momentum
tensor and its derivatives. They span an infinite dimensional space which we shall
denote by $\mathcal{D}$ and which admits a decomposition
 \begin{equation}
\mathcal{D}=\bigoplus_ {s=-\infty}^{\infty} \mathcal{D}_s,
\label{decomp}
\end{equation}
\noindent in term of subspaces $\mathcal{D}_s$ spanned by holomorphic
fields of spin $s$, namely conformal dimensions $(s, 0)$. Equivalently
\begin{equation}
L_0\mathcal{D}_s\,=\,s\mathcal{D}_s, \,\,\,\,\,  \,\,\,\,\, \bar{L}_0 \mathcal{D}_s\,=\,0.
\label{auto}
\end{equation}
\noindent It is clear from (\ref{tt}) and (\ref{construction}) that the fields in $\mathcal{D}$
are analytic or chiral, namely
\begin{equation}
\bar{\partial}{\mathcal{D}}=0,
\end{equation}
\noindent similarly to the field $T(z)$. 

Notice that the fields constructed in (\ref{construction}) are not all  linearly independent. 
All the fields arising for $n>2$ are in fact  total derivatives and it is convenient for our analysis
to eliminate them from the space $\mathcal{D}$. In other words we define the new space
\begin{equation}
\hat\mathcal{D}= \mathcal{D}/{L_{-1}\mathcal{D}},
\end{equation}
\noindent where we take out the total derivatives $ L_{-1}\mathcal{D }$. The  subspace of
$\mathcal{D}$ denoted by $ \hat\mathcal{D}$ can be also decomposed  similarly to (\ref{decomp})
in terms of subspaces $\hat\mathcal{D}_s$  which also satisfy the relations (\ref{auto}). 

Let us now denote by  $\mathcal{T}_s$ any field belonging to the subspace
 $\hat\mathcal{D}_s$\footnote{In order to simplify the notation, 
 we will label each field with only one index denoting the spin. However, it must be kept in mind that the dimension of the subspace $\hat\mathcal{D}_s$ may be higher than one.}.
As usual, these fields admit a mode expansion of the form
\begin{equation}
\mathcal{T}_s=\sum_{n\in \mathbb{Z}} z^{-n-s} \mathcal{L}_{s,n}
 \iff \mathcal{L}_{s,n}=\oint_{z}\frac{d\omega}{2 \pi i} (\omega-z)^{n+s-1}\mathcal{T}_s,
\label{lll}
\end{equation}
\noindent in terms of the modes $\mathcal{L}_{s,n}$. Let now $\xi(z, \bar{z})$  be a local
field of the CFT. The  operators
\begin{equation}
(\mathcal{L} _{s,n}\xi)(z,\bar{z})\,=\,{\oint_z}{\frac{d\omega}{2\pi i} {(\omega-z)}^{n+s-1} 
\mathcal{T}_s (w) \xi (z,\bar{z})},
\label{ope1}
\end{equation}
\noindent with $n=\pm1, \pm2, ...$ are an infinite  set of linearly independent integrals of motion associated to any CFT.
The key result in order to construct integrable perturbed CFT's is that
for suitable choices of the perturbation certain combinations of the fields (\ref{ope1}) may
 remain conserved, even after the CFT has been perturbed. Therefore, we have now all the ingredients
required for the study of perturbed CFT. Before we enter this study, we will now report
 some basic notions concerning the formulation of a CFT on the cylinder. These
results will be used later within the context of the thermodynamic Bethe ansatz analysis.

\subsubsection{Conformal field theory on the cylinder}
\indent \ \

We already pointed out before in this section 
that the components $T(z), \bar{T}(\bar{z})$ of the energy momentum tensor do not transform tensorially under conformal transformations. In other words,  the energy momentum is not a primary but
a quasi-primary field of the CFT. In fact, under a conformal 
transformation $ z, \bar{z} \rightarrow \omega, \bar{\omega}$, the 
holomorphic component  of the energy momentum satisfies,
\begin{eqnarray}
T(z) \rightarrow T(\omega)&=& \Bigg( \frac{\partial{z}}{\partial{\omega}}\Bigg) ^2 T(z) 
+ \frac{c}{12} S(z, \omega), \,\,\,\, \,\,\,\, \textnormal{with}\,\,\,\, \,\,\,\,\nonumber \\
 \Bigg(\frac{\partial z}{\partial \omega} \Bigg) ^2 S(z, \omega)&=&
\frac{\partial z}{\partial w}\frac{\partial^3 z}{\partial \omega^3}-\frac{3}{2} 
\Bigg(\frac{\partial^2 z}{\partial \omega^2} \Bigg) ^2,
\label{tbart}
\end{eqnarray}
\noindent instead of (\ref{primaryf}),
and analogously for the anti-holomorphic component, $\bar{T}(z)$. The function 
$S(z, \omega)$ is usually named as the {\bf  Schwartzian derivative},
and the quantity $c$ is the  Virasoro central charge of the conformal field theory. 

\vspace{0.25cm}
Let us consider now a CFT defined on an infinitely long cylinder, with periodic boundary conditions
defined in terms of the coordinates  $-\infty < \sigma^0 < \infty$ and  $0 \leq \sigma^1 \leq R$ which
are  related to the $z$-plane by means of the conformal mapping 
\begin{equation}
z = e^{{2 \pi \omega}/{R}}=e^{{2 \pi (\sigma^0 + i \sigma^1)}/R},
\label{this}
\end{equation}
\noindent then we may perform the transformation (\ref{tbart}) and obtain the expression
of the holomorphic component of the energy  momentum tensor,
\begin{equation}
T(w)_{cylinder}= \Bigg( \frac{2 \pi}{R}\Bigg) ^2 \Big( z^2 T(z)_{plane}-\frac{c}{24}\Big),
\label{wz}
\end{equation}
\noindent  and analogously for the anti-holomorphic part $\bar{T}( \bar{z})$.
By substituting the mode expansion (\ref{expan}) and its anti-holomorphic counterpart
in (\ref{wz}) we obtain the following expression
\begin{equation}
T(w)_{cylinder}=  \Bigg( \frac{2 \pi}{R}\Bigg) ^2 
\Big( \sum_{n \in \mathbb{Z}} z^{-n}L_n- \frac{c}{24}\Big)=
\Bigg( \frac{2 \pi}{R}\Bigg) ^2 \sum_{n \in \mathbb{Z}}
 \Big( z^{-n}L_n- \frac{c}{24}\delta_{n,0}\Big).  
\end{equation}
\noindent Therefore, the generator  $(L_0)_{cylinder}$ on the cylinder 
is given in terms of  the $L_0$ generator in the plane as
\begin{equation}
 (L_0)_{cylinder}=  \Bigg( \frac{2 \pi}{R}\Bigg) ^2 \Big( L_0- \frac{c}{24}\Big), 
\label{lolo}
\end{equation}
\noindent and the same for its anti-holomorphic counterpart. Now, the last step towards a derivation of
the hamiltonian of the CFT in the new geometry,  is to notice that the combination 
$L_0+\bar{L}_0$ generates dilatations in the plane, namely transformations of the type
$z \rightarrow \lambda z$. These transformations  are mapped  via (\ref{this}) 
into  time  translations in the cylinder, namely 
\begin{equation}
z \rightarrow \lambda z \,\,\,\leftrightarrow \,\,\, \omega \rightarrow \omega + \frac{R}{2 \pi} \ln \lambda.
\end{equation}
\noindent Therefore, the combination $(L_0)_{cylinder}+ (\bar{L}_0)_{cylinder}$ can be identified as  the generator of  time  translations in the cylinder which in other words means that, apart from a constant factor, it gives the hamiltonian of the system in the new cylindrical geometry. Accordingly we can finally write, 
\begin{equation}
H_{cylinder}= \frac{2 \pi}{R} \Big( L_0 +\bar{L}_0 -\frac{c}{12}\Big)
\label{hcyl}
\end{equation}
\noindent where the latter expression is obtained after integration of the energy density over
the space dimension, which cancels out  one of the factors $\frac{2 \pi}{R}$ present in
(\ref{lolo}).

\subsection{Perturbed conformal field theory: conserved densities}
\label{PCFT}
\indent \ \
As mentioned at the beginning of this section, 1+1-dimensional  QFT's can
be understood as particular perturbations of 1+1-dimensional CFT's \cite{Pertcft}. Therefore
the action describing a 1+1-dimensional QFT can be written as
\begin{equation}
S=S_{CFT}+ \lambda \int d^2 x \Phi (x^0,x^1),
\label{pss}
\end{equation}
\noindent where $S_{CFT}$ is the action of the original unperturbed CFT, 
$\lambda$ is a coupling constant and $\Phi(x^0,x^1)$ is the perturbing field,
a primary field of the original CFT which is taken to have conformal dimensions 
$(\Delta, \bar{\Delta})$.  Here, we simplify (\ref{pss}) by considering a single 
perturbing field although in the
most general case one could have a sum of terms involving different perturbations
and coupling constants. However, the models we will treat in this thesis  are described
by actions of the type (\ref{pss}) and consequently, it will be sufficient for our purposes
to consider this simplified case.
 
\vspace{0.25cm}

We will assume that the conformal dimensions of the perturbation are positive,
which always holds for unitary CFT's like the ones we will study later. Furthermore,
we will consider that both ``right'' and ``left'' dimensions are equal, namely the
field $\Phi (x^0,x^1)$ is spinless and has scale dimension $d=2\Delta$. Moreover,
the perturbation must be relevant, meaning that $\Delta <1$. In fact, we will see
below that the study of  perturbed CFT's gets considerably simplified if $\Delta$ is
taken to be smaller than $1/2$, which is  the condition of  super-renormalisability at first order. 
Dimensionality arguments indicate that the coupling constant must have dimensions
$(1-\Delta, 1-\Delta)$ in order to guarantee the dimensionless character of the action, $S$.

\vspace{0.25cm}

Aiming towards the construction of conserved densities or integrals of motion
associated to the perturbed CFT  we  start by making the 
fundamental assumption that the 
local field content of  the original CFT is enough to describe also the perturbed
CFT provided the latter is super-renormalisable. In \cite{Pertcft} a qualitative 
argument which supports this assumption was provided. In general, the local
fields of the original CFT have to be renormalised when the CFT is perturbed but,  if
the resulting perturbed CFT is super-renormalisable, this ensures that each
field acquires under renormalisation a finite set of additional terms which involve
local fields of  lower conformal dimensions. Therefore one ends up with  the same
local field content  of the original theory. 

Following  the previous argument, we assume that, 
whenever we consider a  field 
$\mathcal{T}_s \in \hat\mathcal{D}_s$, which before the CFT has been perturbed 
satisfies  $\bar{\partial} \mathcal{T}_s=0$, in the perturbed CFT
 we will have 
\begin{equation}
\bar{\partial}\mathcal{T}_s= \lambda \mathcal{R}_{s-1}^{(1)} + 
\lambda^2 \mathcal{R}_{s-1}^{(2)}+\cdots+\lambda^n \mathcal{R}_{s-1}^{(n)}+\cdots,
\label{deforme}
\end{equation}
\noindent where $\mathcal{R}_{s-1}^{(n)}$ is a local field of the
 original CFT which has
conformal dimensions $(s-n(1-\Delta), 1-n(1-\Delta))$ and therefore spin $s-1$.
Taking into account that we are considering unitary CFT's, all
the fields on the r.h.s. of (\ref{deforme}) must have non-negative
 conformal dimensions
\begin{equation}
  s-n(1-\Delta) \geq 0, \,\,\,\,\,\,\,\,\,\,\,\,\,1-n(1-\Delta) \geq 0,
\end{equation}
\noindent which means that we can always find a value of $n$ high enough
such that the right conformal dimension of the field $\mathcal{R}_{s-1}^{(n)}$
becomes negative or equivalently,  the term $\lambda^n \mathcal{R}_{s-1}^{(n)}$ is vanishing.
This $n$ is the smallest integer satisfying
\begin{equation}
n  > \frac{1}{1-\Delta}.
\label{dime}
\end{equation}
\noindent Therefore, 
we conclude that the amount of terms on the r.h.s. of (\ref{deforme})
is always finite for unitary CFT's.  In the simplest case 
only the $\mathcal{O}(\lambda)$-term will arise in (\ref{deforme}), which
reduces to
\begin{equation} 
\bar{\partial}\mathcal{T}_s= \lambda \mathcal{R}_{s-1}.
\label{situation}
\end{equation}
\noindent In what follows we will focus our discussion on this
particular situation. Although (\ref{situation}) seems to be a very
special and restrictive case,  it can be directly deduced from (\ref{dime})
that we will encounter that situation whenever the perturbing field is
chosen in such a way that  
\begin{equation}
0 <\Delta < \frac{1}{2}.
\label{super}
\end{equation}
 \noindent In this case we say that
the perturbed CFT is {\bf super-renormalisable at first order}.
It was proven in  \cite{HSG2, cm} that 
the constraint  (\ref{super}) holds in particular for the HSG- and 
SSSG-models for values of the level of the Kac-Moody
representation, $k$, higher than a certain minimum value
which depends on the particular model under consideration.

In summary, the conservation laws of the unperturbed CFT
are mapped  into the new equations (\ref{situation}) whenever
the theory is perturbed by means of a primary, relevant and spinless
fields of the original CFT having scale dimension smaller than 1.

\vspace{0.25cm}

Clearly, the next step would be the explicit identification of the field
$\mathcal{R}_{s-1}$ for which we need to perform conformal perturbation
theory (CPT) around the unperturbed CFT. Any correlation function
involving the field $\mathcal{T}_s$ will have the form
\begin{equation}
\langle \mathcal{T}_s \cdots \rangle = \langle \mathcal{T}_s \cdots \rangle_{CFT}+
\lambda \int {d \omega} \int {d \bar{\omega}}
\langle \Phi (\omega, \bar{\omega}) \mathcal{T}_s (z) \cdots \rangle_{CFT},
\end{equation}
\noindent for $\langle \cdots \rangle_{CFT}$ to be the correlation functions
computed in the original CFT and  $\Phi (\omega, \bar{\omega})$ the perturbing
field. In particular we can use the OPE 
\begin{equation}
\mathcal{T}_s (z) \Phi (w,\bar{w})=\sum_{n \in \mathbb{Z}} {(z-\omega)}^{n-s}
 (\mathcal{L} _{s,-n} \Phi)(w,\bar{w}),
\label{ope2}
\end{equation}
\noindent  which is easily derived from  (\ref{construction}) 
and   (\ref{ope1}).
The combination of the OPE (\ref{ope2}) with the formal identity\footnote{This formula only needs
to be proven  for $m=0$, since the rest of the cases can be obtained from that one via successive
derivation with respect to $z$ or $\omega$. Hence, let us consider the case $m=0$ which gives

$$\bar{\partial} y^{-1}=2 \pi i \delta(y) \delta(\bar{y})$$

\noindent  for $y=z-\omega$ and 
$\bar{y}=\bar{z}-\bar{\omega}$. Instead of trying to prove the latter relation
it is easier to demonstrate the relation obtained when  we multiply first the mentioned
equation by an arbitrary holomorphic function, say $A(z)$, and carry out thereafter
the integration in  the complex variables $z$, $\bar{z}$. The integrals are considered
in an arbitrary region of the complex plane, say $D \subset \Bbb{C}$,
which contains the point $\omega$.
Proceeding in that way, we find for the r.h.s. containing the $\delta$-functions

$$
A(\omega)= \int_{D} {dz} d\bar{z} \, \delta(z-\omega)\,\delta(\bar{z}-\bar{\omega}) A(z),
$$
\noindent and for the l.h.s. we have
$$
A(\omega)=\frac{1}{2\pi i} \, \int {dz} \int d\bar{z} \,A(z) \bar{\partial}\, 
\Big( \frac{1}{\omega-z}\Big)=\frac{1}{2\pi i} \,   \oint_{\Gamma_{\omega}} d{z}
\frac{A(z)}{\omega-z}, 
$$
\noindent where in the last equation we have used Green's and Cauchy's theorems and, in
the final contour integral, $\Gamma_{\omega}= \partial D$ denotes the boundary
of the region $D$.}

\begin{equation}
\bar{\partial}{(w-z)}^{-m-1}=-\frac{2\pi i}{m!} \partial^m \delta^{(2)} (z-w),
\label{formula}
\end{equation}
\noindent gives the following crucial relation
\begin{equation}
\mathcal{R}_{s-1}(z, \bar{z})= 
\oint_{z}\frac{d\omega}{2 \pi i}\Phi (\omega, \bar{z}) \mathcal{T}_s (z).
\label{cruc}
\end{equation}
\noindent Therefore, we have identified the field arising on the r.h.s. of  (\ref{situation}) in
terms of the perturbing field and the conserved quantities of the unperturbed CFT.
It is now clear that a chiral field $\mathcal{T}_s$, that is, a field conserved in the original CFT, 
will remain conserved only if (\ref{cruc}) is a total derivative, that is
\begin{equation}
\mathcal{R}_{s-1}= \partial \Theta_{s-2},
\end{equation}
\noindent for $\Theta_{s-2}$ to be a local field of spin $s-2$ of the initial CFT. Thus,
 the conservation laws in the perturbed CFT acquire the form,
\begin{equation}
\bar{\partial} \mathcal{T}_s=\partial \Theta_{s-2}.
\label{conslaws}
\end{equation}

Therefore, in order to prove the quantum integrability of a 1+1-dimensional 
QFT constructed as shown in (\ref{pss}), one starts with one of the chiral fields 
of the original CFT of a certain spin $s$ and computes the OPE
occurring in (\ref{cruc}) in the usual fashion. In case we are fortunate, 
 the evaluation of   the residue of this OPE  as indicated on the r.h.s. of (\ref{cruc}) 
may turn out to give a total derivative. Provided this is the case, we can conclude
that the quantity $\mathcal{T}_s$ is also one of the integrals of motion
of the perturbed CFT.  However, this procedure does not seem to be very 
effective if we do not have any guess for the values of the spin $s$ at which
we expect to find conserved quantities. In this direction,
there exists a {\bf ``counting-argument''} due also to A.B. Zamolodchikov  
\cite{Pertcft} which is derived  as follows:
We know from subsection \ref{motion} that   $\mathcal{T}_s \in \hat\mathcal{D}_s$
 whereas the field $\mathcal{R}_{s-1} \in \hat\mathcal{D}_{s-1}$.
We can now re-interpret Eq. (\ref{situation}) as a map ${{f}}_{s}$ of
the form
\begin{equation}
{f}_{s}: \hat{\mathcal{D}}_s \,\, \rightarrow \,\, 
\hat{\mathcal{D}}_{s-1},
\end{equation}
\noindent  As usual, we can associate to this map a kernel, 
$\textnormal{Ker}{f}_{s}$,
which will contain the fields in $\hat{\mathcal{D}}_s$ satisfying
$\bar\partial {\mathcal{T}_s}=0$ modulo total derivatives, that is, 
those fields which remain conserved in the perturbed CFT.
Consequently, the existence of spin $s$ conserved
quantities in the perturbed CFT will be guaranteed provided 
\begin{equation}
\dim \textnormal{Ker}f_{s}  \neq 0,
\label{sufne}
\end{equation}
\noindent which means Eq. (\ref{sufne}) is a necessary and sufficient
condition for the existence of spin $s$ conserved quantities 
in the massive QFT.

Associated to the map  ${f}_{s}$ we will also find an image, 
$\textnormal{Im} {f}_{s}$, defined as the subspace of $\hat\mathcal{D}_{s-1}$ 
containing those fields $\mathcal{R}_{s-1}$
which arise on the r.h.s. of Eq. (\ref{situation}).
Obviously,

\begin{equation}
\dim \hat{\mathcal{D}}_s= \dim \textnormal{Ker}{{f}}_{s}+
\dim \textnormal{Im}{{f}}_{s}.
\end{equation}
\noindent
Since the image contains always fields in the subspace $\hat\mathcal{D}_{s-1}$, namely
$\dim \textnormal{Im}{{f}}_{s} \leq \dim{\hat\mathcal{D}_{s-1}}$,
whenever the condition
\begin{equation}
\dim {\hat{\mathcal{D}}_s} > \dim{\hat\mathcal{D}_{s-1}},
\label{count} 
\end{equation}
\noindent is satisfied, we can surely claim it does exist some spin $s$ 
conserved charge of the underlying CFT which remains conserved in the
 perturbed CFT, since the constraint (\ref{count}) ensures that
 (\ref{sufne}) is fulfilled. However, the opposite statement is not true in general, since
(\ref{sufne}) could hold even if (\ref{count}) does not.
Therefore, the counting-argument \cite{Pertcft} provides a sufficient condition which 
allows for proving the quantum integrability of a 
perturbed CFT by making only use of the knowledge of the dimensionalities of the subspaces 
$\hat{\mathcal{D}}_s$ and $\hat\mathcal{D}_{s-1}$ and without the need of explicitly computing
the corresponding conserved charges. Such dimensionalities are available once the
characters of the irreducible representations of the Virasoro algebra associated
to the unperturbed CFT are known \cite{Pertcft}. As we mentioned before, the counting-argument 
has not been used for the HSG- and SSSG-models. To our knowledge, 
the outlined characters have not been computed for the underlying CFT's related
to these models. For that reason, the integrability of the HSG- and part of the SSSG-models
was established in \cite{HSG2, cm} via the explicit construction of certain higher rank conserved
charges.

\vspace{0.25cm}

Still, there is the question of how many of these
quantities need to be identified  in order to conclude the quantum integrability
of the perturbed CFT. Although the quantum integrability 
of a 1+1-dimensional massive
QFT possessing an infinite number of quantum conserved charges 
was established in the light of the results found in \cite{smatrix}
whitin the study of concrete models
and argued in more generality in 
\cite{SW, sm2}, the answer to the question posed at 
the beginning of this paragraph was given by
S. Parke \cite{parke} who demonstrated that really, only the existence of
two of these quantities different from the energy momentum tensor and
having different spin from each other needs to be proven  in order to conclude the quantum
integrability of the theory. We report the main arguments leading
to this important conclusion as well as the key consequences integrability has
concerning the exact computation of  S-matrices in the next section.

\section{Exact S-matrices: Factorisability and absence of particle production}
\label{exactS}
\indent \ \
In 1967 S. Coleman and J. Mandula \cite{colman} demonstrated that, under certain assumptions,
 the existence of any conserved charge which transforms under Lorentz transformations
like a tensor of spin higher than one in a QFT formulated in more
than one space dimension is sufficient to conclude that its  S-matrix is trivial, 
namely particles do not interact amongst each other. This observation is usually
referred to as {\bf Coleman-Mandula theorem}. Amongst other assumptions which are explained
in detail in the original paper, it is especially important to
mention that the S-matrix symmetry group is assumed to be a Lie group 
whose generators satisfy an algebra based on commutators. 
This assumption turns out to be crucial in the derivation of the theorem.
Very different results are obtained when the presence of anticommutators
in the S-matrix symmetry algebra is allowed \cite{susy}, that is,  
the latter algebra is a supersymmetry algebra. 

This pioneering result immediately suggests that if the existence of higher spin conserved
quantities turns out to be such a constraining condition in more than one space dimension,
forcing the whole S-matrix to be $S=\pm 1$,  in 1+1-dimensions
 it should at least restrict severely the 
form and properties of  the mentioned  S-matrix. Indeed, in the light of 
 the result of  S. Coleman and J. Mandula \cite{colman}, later
 investigations have shown that this conjecture is justified, which 
 is on the basis of the enormous success the study of
1+1-dimensional integrable QFT's has achieved over the last 30 years. 

The pioneer works which pointed out the drastic consequences the existence
of higher spin conserved quantities has in 1+1-dimensional QFT where
concerned with the investigation of the scattering matrices of
concrete QFT's \cite{smatrix}. The severe constraints to the form
of the S-matrices observed in \cite{smatrix} were later reviewed  
by R. Shankar and E. Witten \cite{SW} and by D. Iagolnitzer 
\cite{sm2} in 1977.  By exploiting model-independent  
arguments which we will report 
later in more detail, these authors shown that
the existence of an infinite number of higher
 spin conserved quantities associated
to a 1+1-dimensional QFT  has two immediate consequences:

\vspace{0.3cm}
{\bf i)} There is no particle production in any scattering process, namely the number
of particles in the $in$- and $out$-states is the same. Moreover, the set of momenta
associated to incoming and outgoing particles coincide.

\vspace{0.3cm}

{\bf ii)} The S-matrix associated to any scattering process always factorises into
a product of two-particle scattering matrices.

\vspace{0.3cm}

As anticipated before, properties {\bf i), ii)} show that 
quantum integrability in 1+1-dimensions turns out to
be a very powerful property  in what concerns the form of the S-matrix and
the construction of exact S-matrices appears to be a much simpler task in that context.
 In particular,  we infer from {\bf ii)} that the scattering matrix
of a 1+1-dimensional integrable QFT is completely determined
once all the two-particle scattering amplitudes are known. This
fact, together with {\bf i)} allowed for  the exact computation
of the exact S-matrices associated to many 1+1-dimensional QFT's by carrying out the so-called
{\bf bootstrap program} originally proposed in \cite{Boot}. 
Amongst the models for which exact S-matrices have been computed, we find
 the sine-Gordon and  non-linear $\sigma$ model  \cite{sinegor, Zalgebra}, 
the supersymmetric non-linear $\sigma$ model \cite{SW}, the
sinh-Gordon model \cite{SSG2, Boot}, the affine Toda field theories 
\cite{Roland,eeight,ATFTS, dis, mussrev, uni, ATFTNS, Oota, FKS2} etc...
Fairly recently also the S-matrices
of the HSG-models related to simply laced Lie algebras have been determined 
\cite{HSGS} via the bootstrap program and the extrapolation of semi-classical  
results \cite{HSGsol}. Exact S-matrices have also been constructed in \cite{FK} for
the  $g|\tilde{g}$-theories, $g$ and $\tilde{g}$ being simply-laced
Lie algebras, as a generalisation of those related to the HSG-models and
minimal ATFT's. This construction has been extended thereafter in \cite{KK2} to the case when
$\tilde{g}$ is a non-simply laced Lie algebra.

However, the results presented in  \cite{SW} require the existence
of infinitely many higher spin conserved quantities in the theory in order
 to prove
properties {\bf i)} and {\bf ii)}. On the other hand, one is in general 
able to construct explicitly or prove by other means the existence of only a certain finite 
number of conserved quantities which makes the result of  S. Parke \cite{parke}
very useful in this context.
In 1980 he established \cite{parke} that the existence of at least
two higher spin conserved quantities of different spin in a 1+1-dimensional
QFT is sufficient to conclude the quantum integrability of the latter theory.
Concretely, the proof developed  in \cite{parke} 
takes as its starting point the assumption of the existence of only
two  higher spin conserved charges, which differentiates his arguments from the ones
provided in \cite{SW} in what concerns the proof of the S-matrix factorisability. 

Let us now review in more detail the arguments contained in \cite{SW, parke}
(see e.g. \cite{mussrev, dorey} for a more recent review) and introduce the definition
and basic properties of the {\bf  higher spin conserved charges}
in 1+1-dimensional QFT's. For this purpose it is interesting 
first to review the definition of the physical states in a 1+1-dimensional QFT by
recalling what is known as {\bf Zamolodchikov's algebra} \cite{Zalgebra}.

\subsection{Zamolodchikov's algebra}
\label{Zamalgebra}
\indent \ \
As mentioned above, the existence
of an infinite number of higher spin conserved quantities associated to any
1+1-dimensional integrable QFT makes it
exhibit the two fundamental properties of 
factorisability and absence of particle production \cite{SW,parke}.
As a consequence, the task of computing the
corresponding exact S-matrices becomes much simpler. 

The construction of the two-particle S-matrices requires, as
a fundamental assumption, the existence of a set of vertex operators of
creation and annihilation type, which we will denote by
$V_{A}(\theta _{A})$ representing a particle whose quantum numbers
are labeled by the index ${A}$ and which has rapidity  $\theta _{A}$.

It is common in this context to characterise the particle states by  the rapidity variable 
$\theta_A$, which is defined by the relations
\begin{equation}
p^0_A=M_A \cosh \theta_A, \,\,\,\, \,\,\,\, \,\,\,\,
p^1_A=M_A \sinh \theta_A,
\label{rapidity}
\end{equation}
\noindent for $M_A$ to be the mass of the particle and ${p}_A^{\mu}=(p^0_A, p^1_A)$ its momentum.

Let  $p_A^{\mu}=(M_A,0)$ be the components of the momentum of the particle in the rest frame,
that is, $\theta_A=0$, and study now the transformation of the momentum components under
a  Lorentz boost characterised by a velocity $v$. If we denote by  ${p}_A^{\mu\,\,\prime}$
the momentum after the Lorentz transformation, its components will be given by
\begin{equation}
{p}_A^{\mu\,\,\prime}=\frac{M_A}{\sqrt{1-v^2}}(1, v).
\end{equation}
\noindent Recalling the relations (\ref{rapidity}), the value of the rapidity in the
new reference frame is easily found to be
\begin{equation}
\theta_A^{\prime}= \ln \Lambda, \,\,\,\,\,\,\textnormal{with}\,\,\,\,\,\,\Lambda=\sqrt{\frac{1+v}{1-v}}.
\label{rap}
\end{equation}
\noindent Therefore, the change experienced by the rapidity variable has been simply a constant
shift. Such property can be easily extended to the case when we take as our starting point
a frame different from the rest frame. Therefore, 
the rapidity difference between two particles $A, B$, usually
denoted as $\theta_{AB}:=\theta_A -\theta_B$, is a Lorentz invariant. This property is on the basis
of the common use of the rapidity variable in the study of 1+1-dimensional QFT's. 
Since the scattering amplitudes must be Lorentz invariant the two-particle S-matrices can only depend
upon the rapidity difference between the interacting  particles. In particular, as we reported at the
beginning of this section, for a 1+1-dimensional integrable QFT there is no particle production
in any scattering process. Moreover, it can be proven  that the rapidities of the incoming and 
outgoing
particles have to be the same, so that for a general  $2$ particle $ \rightarrow$ 2 particle 
 scattering process of the type 
$A + B \rightarrow C+D $ we  can write the two-particle scattering amplitude as
\begin{eqnarray}
&&S_{A_1 A_2}^{B_1 B_2}(\theta_{A_1}=\theta_{B_1}, \theta_{A_2}=\theta_{B_2}):=
\nonumber \\
&&\,\,\,\,\,\,\,\,\,\,\,\,
\,\,\,\,\,\,\,\,\,\,\,\,\,\,\,\,\,\,\,\,\,\,=_{\,\,out}\langle V_{B_1}(\theta_1) V_{B_2}(\theta_2)| 
 V_{A_1}(\theta_1)V_{A_2}(\theta_2)\rangle_{in}= 
S_{A_1 A_2}^{B_1 B_2}(\theta_{12}),
\label{Lorentz}
\end{eqnarray}
\noindent where the characterisation of the $in$- and $out$-states will be presented below.  
The vertex operators $V_{A_i}(\theta_i)$ provide a generalisation of bosonic
or fermionic algebras and allow for the definition of a space of
physical states. 
They are assumed to 
obey the following highly non-trivial algebra, involving the
S-matrix which was originally employed as an auxiliary algebra in
the construction of the S-matrix \cite{Zalgebra}
\begin{eqnarray}
V_{A_i}(\theta _{i})V_{A_{j}}(\theta _{j})&=&\sum_{B_i, B_j} 
S_{A_i A_j}^{B_i B_j}(\theta _{ij})V_{B_j}(\theta _{j})V_{B_i}(\theta _{i}),
\label{Zalg1}\\
V_{A_i}^{\dagger}(\theta _{i})V_{A_j}^{\dagger}(\theta _{j})&=&\sum_{B_i, B_j}
S_{A_i A_j}^{B_i B_j}(\theta _{ij})V_{B_j}^{\dagger}(\theta _{j})
V_{B_i}^\dagger(\theta _{i}),\label{Zalg2}\\
V_{A_i}(\theta _{i})V_{A_j}^{\dagger}(\theta _{j})&=&\sum_{B_i, B_j}
S_{A_i A_j}^{B_i B_j}(-\theta _{ij})V_{B_j}^\dagger(\theta _{j})
V_{B_i}(\theta _{i})+ 2\pi \delta_{A_i A_j}\delta(\theta _{ij}),
\label{Zalg3}
\end{eqnarray}

\noindent which is known in the literature as Zamolodchikov's algebra 
\cite{Zalgebra} and also named sometimes as
{\bf Faddeev-Zamolodchikov algebra}, since the last
term in (\ref{Zalg3}) was suggested by L.D. Faddeev in \cite{cFad}. 
A space-time interpretation for this algebra was
recently proposed by B. Schroer \cite{schroer2}.

Therefore, each commutation of these operators is interpreted as a scattering process. 
Being the S-matrix of the theory involved in (\ref{Zalg1})-(\ref{Zalg3}),
the explicit form of the vertex operators associated to a 
1+1-dimensional integrable QFT depends  very much on the
particular theory under consideration and, in fact, an
explicit realisation for such operators has not been found for many theories.
As mentioned above, once the vertex operators $V_{A_i}(\theta _{i})$
have been  introduced, they can be used  in order to define a space of physical
states. For this purpose one starts by defining the vacuum
state as the one annihilated by any vertex operator, namely 

\begin{equation}
V_{A_i}\arrowvert 0 \rangle=0=\langle 0 
\arrowvert V_{A_i}^\dagger, 
\end{equation}
\noindent thus the Hilbert space will be generated by the repeated action
of creation operators on this vacuum state

\begin{equation}
\arrowvert V_{A_1}(\theta_1) V_{A_2}(\theta_2)
 \cdots V_{A_n}(\theta_n)\rangle= 
V_{A_1}^\dagger (\theta_1)V_{A_2}^\dagger (\theta_2)
\cdots V_{A_n}^{\dagger}(\theta_n)
\arrowvert 0 \rangle. 
\end{equation}

\begin{figure}[!h]
 \begin{center}
  \leavevmode
    \includegraphics[scale=0.7]{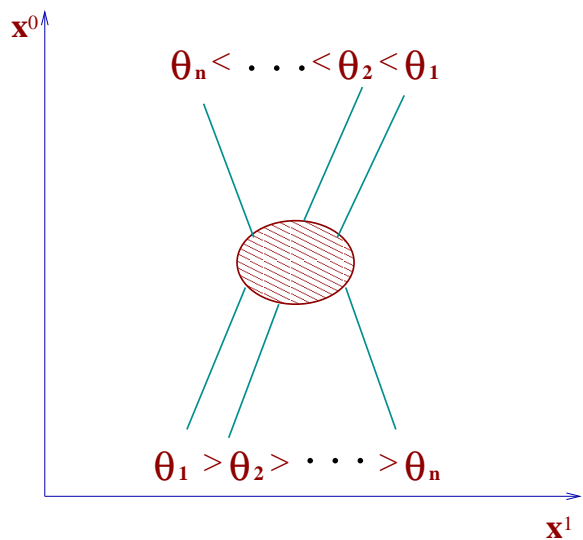}   
    \caption{$\protect{n}$-particle scattering process.}
 \label{scattering}
 \end{center}
\end{figure}

\noindent Since they obey
(\ref{Zalg1})-(\ref{Zalg3}), these states are not all of then independent and 
one has to introduce a certain prescription in order to select out a basis 
of independent or physical states. The mentioned prescription consists of
characterising the $in$- and $out$-states as follows:

An {\em{in}}-state is characterised by the fact that 
there are no further interactions when we consider the limit $t\rightarrow -\infty$.
Consequently, the particle possessing the  highest rapidity (the ``fastest''  one) 
must be on the left (see Fig. \ref{scattering}), the particle possessing the lowest
 rapidity (the ``slowest'' one)  must be on the right and all the rest should be ordered 
in between, namely for an $n$-particle {\em{in}}-state,

\begin{equation}
{\arrowvert V_{A_1}(\theta_1) V_{A_2}(\theta_2) \cdots V_{A_n}(\theta_n)\rangle}_{in}, 
\,\,\,\,\, \textnormal{with} \,\,\,\,\,  \theta_1 > \theta_2 > \cdots > \theta_n.
\end{equation}

\noindent Likewise, an $n$-particle {\em{out}}-state 
contains particles which do
not interact when $t\rightarrow \infty$. 
Then, its  natural definition is 

\begin{equation}
{\arrowvert V_{A_1}(\theta_1)V_{A_2}(\theta_2)
\cdots V_{A_n}(\theta_n)\rangle}_{out}, \,\,\,\,\, \textnormal{with}
\,\,\,\,\,  \theta_1 < \theta_2 < \cdots < \theta_n.
\end{equation}

The latter definitions of the $in$- and $out$-states allow for dropping out the subindices 
$in$ or $out$ in what follows, since the ordering of the rapidities permits a clear-cut
distinction between the set of incoming and outgoing particles without the 
need of more information. 

\subsection{Higher spin conserved charges}
\label{hrcc}
\indent \ \
Within  the study of any QFT 
we will encounter in general two types of conserved charges. First
of all, the familiar momentum and charges associated to internal symmetries of the model, which
under the Lorentz group transform as vectors and scalars, respectively. Second, we can encounter
also {\bf higher spin conserved charges} namely, objects whose Lorentz transformations read
\begin{equation}
Q_{\ell}\rightarrow \Lambda^{\ell} Q_{\ell},
\label{qq}
\end{equation}
\noindent for $\Lambda$ a Lorentz boost. In particular, for $\ell=\pm 1$ we
obtain the transformation laws of the light-cone components $p^{\pm}=p^0 \pm p^1$ of
the momentum ${p}^{\mu}=(p^0, p^1)$.

In what follows we
shall make the assumption that the charges $Q_{\ell}$ are local, meaning that they
can be expressed as integrals of local current densities
\begin{equation}
Q_{\ell} \sim \int\limits_{-\infty}^{\infty} d^2 x  \mathcal{T}_{\ell+1},
\label{nlocal}
\end{equation}
\noindent where $\mathcal{T}_{\ell+1}$ may be one of the integrals of motion
defined in the previous section, which remains conserved in the perturbed CFT. 
However, one should keep in mind that also non-local
conserved quantities can be present in the QFT at hand.
Examples of such a situation have been studied in \cite{nlocal}. 

Provided (\ref{nlocal}) holds, the quantities (\ref{qq}) satisfy 
the following algebra of commutators
\begin{equation}
\left[Q_{\ell}, Q_n \right]=0,
\label{mm}
 \end{equation}
\noindent for all $n, \ell$. Since the light-cone  components of the 
momenta and the masses of the particles are respectively conserved charges of
spins 1 and 0, Eq. (\ref{mm}) indicates that the eigenstates of higher spin
conserved quantities are linear combinations of eigenstates of the mass  namely, states
representing particles in the same mass multiplet. Consequently, each multiplet
will contain a set of one-particle states which simultaneously diagonalise the
momentum and charges $Q_{\ell}$, $\ell=\pm 1, \pm 2, ..$. These states will be
characterised as mentioned in the previous subsection by the vertex
operators,  $|V_{A}(\theta_{A})\rangle$ for $A$ to be the quantum numbers
of the particle under consideration.

\vspace{0.3cm}
The eigenvalues of the charges $Q_{\ell}$ are determined by Lorentz invariance to be
\begin{equation}
Q_{\ell} |V_A (\theta_A) \rangle= \xi_A^{\ell} (M_A e^{\theta_A})^{\ell} |V_A (\theta_A) \rangle, 
\label{oneparticle}
\end{equation}
\noindent where $\xi_A^{\ell}$ are non-vanishing Lorentz scalars.
The locality assumption (\ref{nlocal}) ensures that the generalisation of Eq. (\ref{oneparticle}) to
multi-particle states $|\theta_1 \theta_2 \cdots \theta_n \rangle$
is easily obtained as the sum of  the actions over each individual particle state
$|\theta_A \rangle$, $A=1, \cdots, n$,
\begin{equation}
Q_{\ell} |V_{A_1} (\theta_1)V_{A_2} (\theta_2) \cdots V_{A_n} (\theta_n) \rangle = 
\left[\sum_{i=1}^{n} \xi_i^{\ell} (M_i e^{\theta_i})^{\ell} \right]
 |V_{A_1} (\theta_1)V_{A_2} (\theta_2) \cdots V_{A_n} (\theta_n) \rangle 
\label{multipart}
\end{equation}
\noindent 

Let us now consider a scattering process with $k$ particles in the $in$-state 
and $n$ particles in the $out$-state. The associated scattering amplitude will be
\begin{equation} 
S_{A_1 A_2 \cdots A_k}^{B_1 B_2 \cdots B_n}:=
\langle V_{B_1}(\theta_1) V_{B_2}(\theta_2) \cdots V_{B_n}(\theta_n)|
 V_{A_1}(\theta_1)  V_{A_2}(\theta_2) \cdots V_{A_k}(\theta_k) \rangle
\label{smatrix}
\end{equation}
\noindent where the S-matrix establishes a correspondence between
 the basis of $in$- and $out$-states.
By looking at this completely general scattering process we can easily prove
property {\bf ii)} in the introduction, {\em i.e.} the absence of particle production or
the fact that necessarily $n=k$ in (\ref{smatrix}). If $Q_{\ell}$ is a  higher spin
conserved quantity satisfying (\ref{multipart}) it must remain conserved in the
scattering process (\ref{smatrix}), namely
\begin{equation}
\sum_{i=1}^k \xi_{A_i}^{\ell} (M_{A_i} e^{\theta_{A_i}})^{\ell} =
\sum_{i=1}^n \xi_{B_i}^{\ell} (M_{B_i} e^{\theta_{B_i}})^{\ell}.
\end{equation}
\noindent If we further assume that the number of conserved charges $Q_{\ell}$ is infinite,
the latter equation is really a system of infinitely many equations for different values of $\ell$ which,
for generic values of the rapidities of the particles, admits only the trivial solution $n=k$ and
\begin{equation}
\theta_{A_i}=\theta_{B_i}, \,\,\,\,\,\,\,\,\,\, \,\,\,\,\,\,\,\,\,\,
\xi_{A_i}^{\ell} (M_{A_i})^{\ell}=\xi_{B_i}^{\ell} (M_{B_i})^{\ell},
\label{parke1}
\end{equation}
\noindent for $i=1, \cdots, n$. Therefore, there is no particle production and the set of momenta
of the particles in the $in$- and $out$-states must be the same. The only freedom allowed by
the second set of equations in  (\ref{parke1})  is the possible exchange of quantum numbers between particles in the $in$- and $out$-state, in case there are more than one particle in each particle multiplet namely, the spectrum is degenerate.

The argument reported above can be found in \cite{SW, dorey} and, as we have specified, 
requires the assumption of the existence of infinitely many conserved charges $Q_{\ell}$.
This assumption allows also for proving the S-matrix factorisability, as we might see in
the next subsection. However, it is interesting to report now
 the arguments exhibited by
S. Parke in \cite{parke}, who established quantum integrability as a 
consequence of the existence
of only two of the mentioned conserved charges. These arguments can also 
be found in the review article \cite{dorey}. 
In that case, the proofs of {\bf i),ii)} are
not so straightforward but due to the relevance of this result it is 
interesting to summarise here
the key steps of Parke's argument.

\subsection{Factorisability and absence of particle production}
\label{parke}
\indent \ \
The starting point of the  argument in \cite{parke} and also of the proof 
of the S-matrix factorisability 
presented in \cite{SW} is the assumption that asymptotic one-particle states,  $|\theta_A\rangle$ 
can be represented  by means of  localised wave packets $\Psi_A (x^0,x^1)$.
Although the wave function formalism is not valid in the context
of relativistic QFT, because of particle production and annihilation,
 we can make use of it when considering asymptotic multi-particle
states ($in$- or $out$-states) namely, states describing a set of 
particles in the limits  $t \rightarrow \pm \infty$,  $t$  being  the time, in which assuming a
purely massive particle spectrum and  short
range interactions, particles become free. In that situation we can associate to each particle
in the multi-particle $in$- ($|\theta_A \rangle$) or $out$-state ($\langle \theta_A |$), 
 a  localised wave packet
\begin{eqnarray}
\Psi_A (x^0, x^1)&=&\mathcal{N}\,\int{dp^1\,{e^{f(p^1)}}}\nonumber \\
f(p^1)&=&-a{(p^1-p^1_A)}^2 \,+\,i{(p^1 (x^1-x^1_A)-p^0_A  (x^0-x^0_A))},
\label{wavef}
\end{eqnarray}
\noindent where $p^1_A$ is the mean spatial momentum of the particle $A$ and 
$p^0_A$ its energy.  $\mathcal{N}$ is a normalisation constant and $(x^0_A, x^1_A)$ are the
coordinates of the centre of the wave packet, namely the approximated time-space
position of the particle $A$. Finally, the parameter $a$ is a constant expressing
the spreading on the velocity of the wave packet.
Therefore, any multi-particle state will be
 represented by a set of wave packets
of this type whose overlapping  regions are identified to be the regions where particles
interact. The key property we want to use in the course of our argumentation
is the fact that the transformation of the wave function  (\ref{wavef}) generated by
an operator $e^{i \alpha Q_{\ell}}$, $\alpha$ being a free parameter,
amounts to shifting  the coordinates of the centre of the wave packet as follows
\begin{eqnarray}
&&x^0_A \rightarrow x^0_A + \alpha \ell \xi_A^{\ell} (M_A e^{\theta_A})^{\ell-1}, \label{time0}\\
&&x^1_A \rightarrow x^1_A +\alpha \ell \xi_A^{\ell} (M_A e^{\theta_A})^{\ell-1}\label{space0}.
\end{eqnarray}
\noindent A relevant characteristic is that
the mentioned shift is not constant in general but, on the contrary,  depends upon 
the rapidity of the particle under consideration. This is a crucial fact which 
means that when performing the same type of transformation for a multi-particle state,
the outcome will be another  multi-particle state where the mean positions of the particles differ from
the original ones  by a factor which  is different for particles with different masses and
rapidities. This observation turns out to be fundamental  in order to prove 
the factorisability of the S-matrix
into two-particle scattering matrices in massive 1+1-dimensional QFT's.

\subsubsection{Absence of particle production}
\indent \ \
The absence of particle production is a property which, as we have seen,  is easily proven  once
the assumption of the existence of infinitely many higher spin conserved quantities is made. However,
the proof becomes more involved if only two of these quantities are assumed to exist. We will not
present here the details but only summarise the main ideas involved. First of all, the scattering amplitude
must be invariant under transformations generated by the conserved charges $Q_\ell$ which
means 
\begin{equation}
S_{A_1 A_2 \cdots A_k}^{B_1 B_2 \cdots B_n}= e^{-i{\alpha}{Q_\ell}} 
S_{A_1 A_2 \cdots A_k}^{B_1 B_2 \cdots B_n}e^{i{\alpha}{Q_\ell}}.
\label{inva}
\end{equation}
In particular, Eq. (\ref{inva}) should also hold if we substitute $Q_{\ell}$ by a linear combination
of two higher spin conserved quantities of spin $m$ and $-s$
\begin{equation}
Q_{\varphi}=\frac{\cos \varphi}{m} Q_{m} - \frac{\sin \varphi}{s} Q_{-s},
\label{phichar}
\end{equation}
\noindent where $\varphi$ is a free parameter taking values in the interval $[0, 2 \pi)$.
In \cite{parke} it was proven  that any transformation generated by a conserved charge
of type (\ref{phichar}) acting on the scattering amplitude (\ref{inva}) for $k=2$ 
will lead to a violation of the {\bf macrocausality principle} 
whenever the set of rapidities of the particles in the $in$- and $out$-states are different.
The macrocausality principle states that the interaction time associated to  the two
incoming  particles, say $t_{12}$, has to be always smaller of equal than the time at which
any of the outgoing particles is produced. The key observation presented in 
\cite{parke} is that this principle could be violated if there was particle production in the scattering
process and if the set of rapidities associated to incoming and outgoing particles are different. Consequently,  the conclusion that $n$ should be always $2$ for the case at hand is immediate. 
Since in the next subsection we will see that any scattering amplitude can be ultimately
expressed as a product of two-particle amplitudes, the proof of the absence of particle production can
be easily generalised to any scattering process even when the number of incoming particles
is bigger than two. In short, we can write
\begin{equation}
S_{A_1 A_2 \cdots A_k }^{B_1 B_2 \cdots B_n}\sim
\delta_{k n}\prod_{i=1}^{n}\delta (\theta_{A_i B_i}).
\end{equation}

\subsubsection{Factorisability of the scattering amplitudes}
\begin{figure}[!h]
 \begin{center}
  \leavevmode
    \includegraphics[scale=0.7]{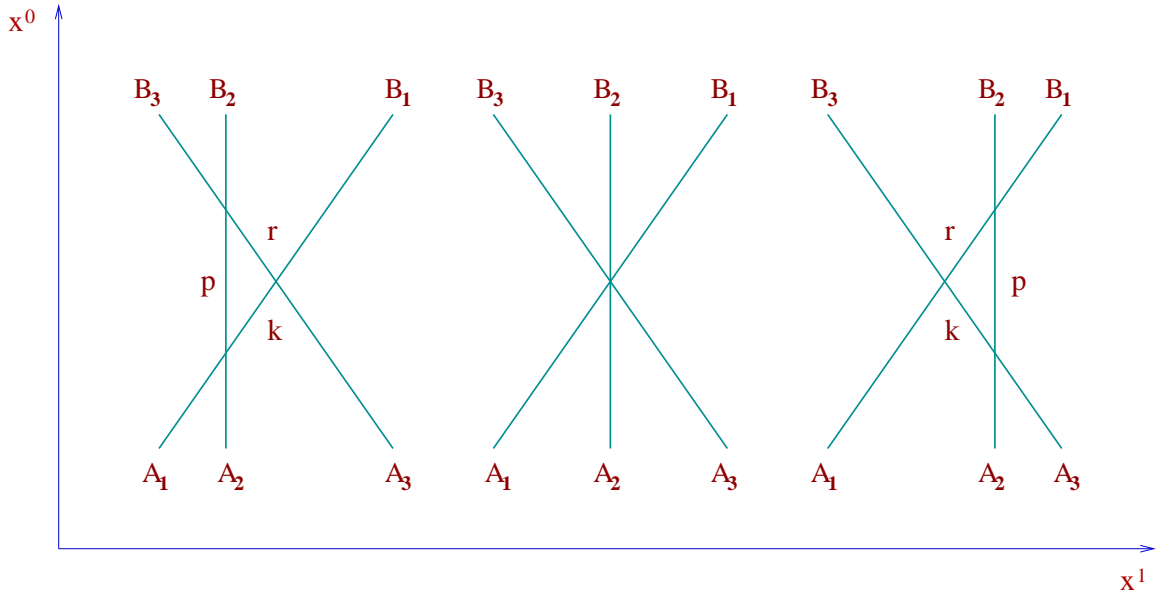}   
    \caption{Graphical representation of the  Yang-Baxter equation}
 \label{yangbaxter}
 \end{center}
\end{figure}

\indent \ \
The factorisability of any scattering amplitude in a 1+1-dimensional integrable
QFT is easily proven provided a wave function description of the type (\ref{wavef})
is used. As we have seen, the action of a conserved charge $Q_{\ell}$ on a multi-particle
$in$- or $out$-state amounts to shifting the wave packet centres by a factor which
is rapidity-dependent. The consequences of this property leading to the conclusion
of the S-matrix factorisability are typically analysed by
considering the three possible $3$ \,\,particle $\rightarrow$ $3$\,\, particle
scattering processes depicted in Fig. \ref{yangbaxter}.

In a generic QFT there will not be in principle any reason to think the scattering amplitudes
describing these three processes should have something to do with each other. However,
if the theory possesses at least  two higher spin conserved quantities, like the ones entering
Eq. (\ref{phichar}) and is 1+1-dimensional, the mentioned amplitudes are automatically forced to be
identical. The reason is that any of the three diagrams in Fig. \ref{yangbaxter} can be transformed into
each other by means of the `translation' of one of the particles.  Such a `translation'
can be generated by means of a conserved charge (\ref{phichar}) and, according to (\ref{inva}),
must leave invariant the corresponding scattering amplitude. Consequently, the processes
in Fig. \ref{yangbaxter} correspond to  the same scattering amplitude, which leads to the following
set of equations
\begin{equation}
S_{A_1 A_2}^{\,k\,\, p}(\theta_{12}) S_{k A_3}^{B_1 r}(\theta_{13}) S_{\,p \,\,r}^{B_2 B_3}(\theta_{23})= 
S_{A_1 k}^{r B_3}(\theta_{13}) S_{A_2 A_3}^{\,p \,\,k}(\theta_{23}) S_{\,r \,\,p}^{B_1 B_2}(\theta_{12}).
\label{yb}
\end{equation}

It is also evident that the same argument holds for any $n \rightarrow n$ scattering process whose
scattering matrix will then factorise into a product of $n(n-1)/2$ two-particle scattering amplitudes.
Eq. (\ref{yb}) is known in the literature as {\bf Yang-Baxter equation} \cite{yb}. 
In particular, when the number of particles in each particle multiplet is just one, {\em i.e.} the particle
spectrum is non-degenerate,  not only the rapidities but also the quantum numbers of the
particles must be the same in the $in$- and $out$-states. In that case the S-matrix
is diagonal, meaning that  it has the general form
\begin{equation}
S_{A_1 A_2}^{B_1 B_2}(\theta_{12})=
\delta_{A_1}^{B_1} \delta_{A_2}^{B_2} S_{A_1 A_2} (\theta_{12}),
\label{diag}
\end{equation}
\noindent  and consequently, the Yang-Baxter equation (\ref{yb}) becomes trivial. 
For this reason the construction of diagonal
S-matrices is in general much simpler than in the non-diagonal case. 
In this thesis we will be concerned with the diagonal case, since the
HSG S-matrices constructed in \cite{HSGS} are diagonal.

\vspace{0.3cm}

To close this section we would like to qualitatively show how the Coleman-Mandula theorem \cite{colman}
turns out to be a very natural property in the light of the arguments of this section and, in particular,
as a consequence of the properties of  higher spin conserved quantities 
exploited here. Recall that the Coleman-Mandula theorem \cite{colman} states that, under
certain assumptions, the S-matrix
of any $1+d$-dimensional QFT with $d>1$ is trivial if the theory possesses any higher
spin conserved quantity. The reason becomes clear  if we consider any of the scattering processes
in Fig. \ref{yangbaxter} but  now in more than one space dimension. The existence of a single conserved
charge $Q_{\ell}$ would allow us to `translate'  particles away from the plane where the
interaction is taking place and to separate them from each other as much as we like.
Therefore, the particles would not interact anymore and the scattering amplitudes should
necessarily be trivial.

\section{Analytical properties of  two-particle scattering amplitudes}
\label{analit}
\indent \  \
It is clear from the preceding section that the determination of the exact S-matrix
associated to a 1+1-dimensional massive integrable model is equivalent to the exact
computation of all  two-particle scattering amplitudes, corresponding
to the different $2 \rightarrow 2$ scattering processes occurring in the theory. 
Therefore, before we enter the specific description of the
non-abelian affine Toda field theories \cite{nt}, 
it is interesting to provide a general description 
of the main properties of two-particle scattering amplitudes,
paying special attention to non-parity invariant theories, since the S-matrices of
the HSG-models which will enter our analysis  in subsequent chapters 
break parity invariance. A more detailed derivation of these properties may be found
in \cite{Boot,KTTW, ELOP, Wein, Zalgebra, sinegor}.

In the context of 1+1-dimensional integrable QFT's  the calculation
of exact two-particle S-matrices is possible by  solving a set
of constraining equations.
These equations arise as a consequence of very general physical principles, in particular
as we have stressed before, quantum integrability itself imposes severe restrictions on
the two-particle scattering amplitudes which, in the non-diagonal case, 
must satisfy the highly non-trivial Yang-Baxter equation \cite{yb}
reported in (\ref{yb}). Apart from the latter equation, the two-particle scattering 
amplitudes are constrained by other requirements which we itemise below

\vspace{0.25 cm}

\begin{center}

{\bf Lorentz invariance.}

\vspace{0.25 cm}

 {\bf Analyticity.}

\vspace{0.25 cm}

 {\bf Unitarity.}

\vspace{0.25 cm}

{\bf Crossing symmetry.}

\end{center}

\vspace{0.25 cm}

\noindent Moreover, in case bound states are present in the theory, the two-particle
S-matrices have to obey  the so-called {\bf bootstrap equations} 
\cite{Boot,KTTW} which we will report later.
By solving the set of equations emerging from the previous constraints
it is possible to determine the exact S-matrix of the theory up to certain multiplicative
factors which, without adding any physical information, trivially satisfy 
all the constraining equations.  These ambiguity 
 factors are usually referred to as {\bf CDD-factors} and their
existence was pointed out in \cite{CDD} by L. Castillejo, R.H. Dalitz and F.J. Dyson. 
They are usually fixed by appealing 
to consistency requirements, although these sort of arguments are in general not rigorous enough
in order to make sure a certain S-matrix proposal is certainly correct, meaning it is really 
describing the scattering theory of a very specific model which is known through a Lagrangian
formulation or by some other physical quantities. Recall that in  the
UV-limit  one should recover the underlying CFT serving
as starting point for  our construction. For this reason it
is desirable to develop tools which allow for consistency checks 
of the outlined proposal. The thermodynamic Bethe ansatz \cite{Yang, TBAZam1}
and form factor approach \cite{Kar, Smir} are precisely  examples of this sort of tools which 
we might exploit in order to check the consistency
of the S-matrix proposal \cite{HSGS} for the HSG-models. 

\vspace{0.25cm}

Let us now commence with a more detailed discussion 
on the physical requirements summarised above. 
We already pointed out before that Lorentz invariance of the two-particle scattering amplitudes ensures
that the S-matrix  dependence  upon the momenta of the particles has to enter via Lorentz invariant 
quantities. In particular, it is well known in 1+3-dimensions that the quantities we refer
to are the so-called {\bf Mandelstam variables} $s$, $t$ and $u$, defined as
\begin{equation}
s=({p}_A+{p}_B)^2, \,\,\,\,\,t=({p}_A-{p}_C)^2
\,\,\,\,\,u=({p}_A-{p}_D)^2,
\end{equation}
\noindent for a scattering process of the type $A+B \rightarrow C+D$.
 However, in 1+1-dimensions
only one of  these three variables is really  independent, so that  the
two-particle amplitudes shall depend only on one of  Mandelstam's variables, 
say $s$.

\begin{figure}[!h]
 \begin{center}
  \leavevmode
    \includegraphics[scale=0.7]{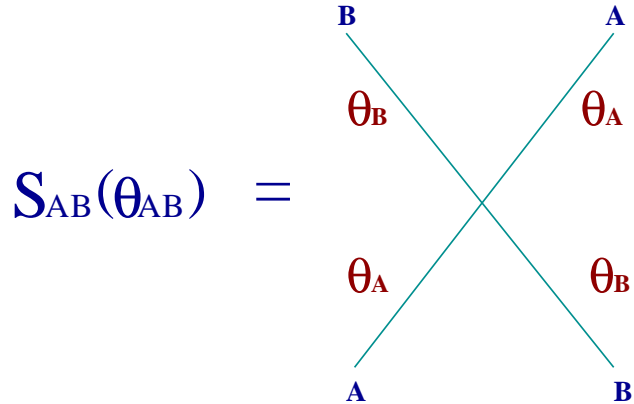}   
    \caption{Two-particle scattering amplitude.}
 \label{22}
 \end{center}
\end{figure}

The sort of theories we will be interested in are characterised by a 
non-degenerate particle spectrum which ensures that the S-matrix will be diagonal,
meaning that all two-particle scattering amplitudes have the form (\ref{diag}). Therefore, the
Yang-Baxter equation (\ref{yb}) is trivially satisfied,
 in other words, it does not introduce any new constraint
for the two-particle scattering amplitudes. Consequently, the incoming and
outgoing particles involved in any two-particle scattering process have the same momenta (rapidities), 
say $p_A, p_B$ ($\theta_A, \theta_B$), in virtue of integrability, and the same 
quantum numbers, say $A, B$, due to the non-degeneracy of the
spectrum (see Fig. \ref{22}).
We will then denote such an amplitude by $S_{AB}(\theta_{AB})$.
If the masses of the two incoming (outgoing) particles
are $M_A, M_B$, the Mandelstan variable $s$ reads
\begin{equation}
s=({p}_A + {p}_B)^2= M_A^2 + M_B^2 + 2 M_A M_B \cosh \theta,
\label{Mss}
\end{equation}
\noindent where, as usual $\theta=\theta_{AB}:=\theta_A-\theta_B$. The dependence of the $s$-variable
on the rapidity difference  is consequent with the discussion leading to  Eq. (\ref{Lorentz}). 

We can now interpret the scattering amplitude $S_{AB}(\theta)=S_{AB}(s)$ as a function
of the  variables $s$ or $\theta$,
 which can take values in all the complex plane, namely we analytically
continue the real variables $s$, $\theta$ to the complex plane. 
Once this is done, the unitarity and crossing symmetry
of the scattering amplitudes, which we will talk about a bit later, imply that the amplitude 
$S_{AB}(s)$ must have square root branch points starting at the values $s=(M_A\pm M_B)^2$
which correspond to $\theta=0, i\pi$ in the $\theta$-plane. 
These branch points give rise to branch cuts along the real axis
$s \leq (M_A-M_B)^2$ and  $s \geq (M_A+M_B)^2$ so that the two-particle amplitude,
as a function of $s$, is not a meromorphic function. However, as a function of 
the $\theta$-variable, $S_{AB}(\theta)$ is a meromorphic function since,
using the fact that $s(\theta)=s(-\theta)$, the mentioned branch cuts 
do not occur in the $\theta$-plane. In order to guarantee  that the
analytical continuation of the scattering amplitudes to the complex plane (both in
$s$ and $\theta$) gives rise to uniquely defined functions, several {\bf Riemann sheets}
have to be considered. In particular, in the diagonal case, the number of Riemann sheets in the
$\theta$-plane is just two. The physical values of the rapidity $\theta$ are constrained
to the strip,
\begin{equation}
\text{Im}\, \theta \in ( 0, \pi ).
\label{sheet}
\end{equation}
\noindent   
The region of physical values of  $\theta$ given in (\ref{sheet})
is known under the name of {\bf physical sheet} and corresponds to the first
of the two Riemann sheets arising  when diagonal scattering amplitudes 
are expressed in terms of  the rapidity variable. Obviously, 
the relationship (\ref{Mss}) between
the $s$ and $\theta$ variables implies also that, 
associated to the region (\ref{sheet}), there
is a corresponding physical sheet for the $s$-variable. 

As we shall see later in more detail,
scattering amplitudes corresponding to the creation of stable bound states will
be characterised by purely imaginary poles in the $\theta$-variable located at the physical
strip, whereas the presence of unstable particles in the spectrum will be related
to the existence of complex poles in the scattering amplitudes located in the
second $\theta$ Riemann sheet, beyond the physical sheet (\ref{sheet}).

\subsubsection{Analyticity}
\indent \ \
Being the region of physical values of  $s$ defined by  (\ref{sheet}), the  physical
values of the two-particle scattering amplitudes correspond to
\begin{equation}
S_{AB}^{\text{phys}}(s):=\lim_{\epsilon \rightarrow 0^+} S_{AB}(s+ i\epsilon),
\label{fey}
\end{equation}
\noindent for $s$ to be real, in the spirit of  Feynman's $i\epsilon$ prescription in perturbation theory. {\bf Hermitian analyticity} \cite{HERMAN, Mir, TW}
then postulates that the physical scattering amplitude $S_{AB}(s)$  and
its complex conjugated $[S_{BA}(s)]^*$ are boundary values in
opposite sides of the $s$-plane branch cut of the same analytic function, namely
\begin{equation}
[S_{BA}^{\text{phys}}(s)]^*:=\lim_{\epsilon \rightarrow 0^+} S_{AB}(s-i\epsilon).
\label{fey2}
\end{equation}
\noindent Taking furthermore into account that
\begin{equation}
\lim_{\epsilon \rightarrow 0^+} S_{AB}(s \pm i\epsilon)=S_{AB}(\pm \theta), \,\,\,\,\,\,\,\theta>0, 
\end{equation}
\noindent the  Eqs. (\ref{fey}) and (\ref{fey2}) translate into the condition
\begin{equation}
S_{AB}(\theta)=[S_{BA}(-\theta^*)]^*
\label{HS}
\end{equation}
\noindent after analytic continuation to the complex plane and using the fact that if 
 $S_{AB}(s)$ is an analytic function of $s$ the same holds for  $[S_{AB}(s^*)]^*$. 
A consequence of (\ref{HS}) is that the amplitudes $S_{AB}(\theta)$ will not be
real analytic functions unless there are additional symmetries in the theory like 
parity invariance, 
\begin{equation}
 S_{AB}(\theta)=S_{BA}(\theta).
\end{equation}
\noindent In that case the combination of (\ref{HS}) with the latter equation gives
\begin{equation}
S_{AB}(\theta)=[S_{AB}(-\theta^*)]^*,
\label{RA}
\end{equation}
\noindent which is the usual condition of {\bf real analyticity } for  two-particle scattering
amplitudes. Therefore, the two-particle S-matrices are real analytic functions
 in 1+1-dimensional QFT's only in the parity invariant case (for S to be diagonal). 
We will see later that 
some of the HSG  S-matrices \cite{HSGS} provide particular examples 
of  amplitudes which break parity invariance and therefore satisfy (\ref{HS}) instead of (\ref{RA}). 

\subsubsection{Unitarity}
\indent \ \
The S-matrix also  has to be unitary, meaning that $S S^\dagger=1$ for physical values
of the variables $s$, $\theta$. The unitarity of the S-matrix expresses the fact that
the total probability of producing an arbitrary  $out$-state from any initial $in$-state
must be one. Physical unitarity together with the Hermitian analyticity condition (\ref{HS})
lead to 
\begin{equation}
S_{AB}(\theta) S_{BA}(-\theta)=1,
\label{unit}
\end{equation}
\noindent which by analytic continuation to the complex plane can be assumed to
hold for any complex value of $\theta$.

\subsubsection{Crossing symmetry}
\indent \ \
The two-particle scattering amplitudes have to satisfy also the constraints derived from
{\bf crossing symmetry} which means they must remain invariant under the replacement  of an incoming
particle by an outgoing particle of opposite momentum. This leads to the constraint
\begin{equation}
S_{AB}(i\pi-\theta)=S_{B \bar{A}}(\theta),
\label{crossing}
\end{equation}
\noindent where $\bar{A}$ denotes the antiparticle of the particle $A$. Notice
that changing  the  momentum of the particle $A$  to $-p_A$ amounts to
changing the Mandelstam  $s$-variable to the $t$-variable,
which explains the branch cut at $(M_A-M_B)^2$. Eq. (\ref{crossing})
expresses the fact that scattering processes described in the $s$- and $t$-channels
are not independent from each other. 

\subsubsection{Bootstrap equations}
\indent \ \
All the constraints summarised above do not require any information about the particle
content of the theory, they are completely general for any 1+1-dimensional integrable
QFT and in fact, they are sufficient to establish already the general form of the two-particle 
scattering amplitudes, which in \cite{Mitra} were shown to be products
of general building blocks depending upon hyperbolic functions of the form
\begin{equation}
f_{x}(\theta):=\frac{\sinh \frac{1}{2}\Big(\theta + i\pi x\Big)}
{\sinh \frac{1}{2}\Big(\theta - i\pi x\Big)},
\label{blocks}
\end{equation}
\noindent whenever the scattering matrix is assumed to be diagonal.
The preceding building blocks (\ref{blocks}) depend on certain variables
$x$ which might  encode the information about the particle spectrum of
the theory. At this point, specific information about the QFT considered
enters the S-matrix construction. It is clear that the block (\ref{blocks}) 
has a simple pole corresponding to the value $\theta=i \pi x$. If the 
pole lies on the physical sheet $0 < \text{Im}\, (\theta) < \pi$ which
means $ 0 < x < 1$ it is assumed to be the trace of the formation of a {\bf stable bound state}.

\vspace{0.25cm}

Let us now consider again the scattering amplitude $S_{AB}(\theta)$ which,
in the light of the previous paragraph, will be a certain product 
of building blocks of the type (\ref{blocks}). Suppose that, for the particular
QFT at hand,  the mentioned scattering amplitude possesses a simple pole
$\theta=i \,u_{AB}^C$ characterising  the formation
of a stable bound state, say $C$, of mass $M_C$ in a scattering process 
of the type $A + B \rightarrow C$. By stable bound state we mean that particle $C$ is
 one of the asymptotic one-particle states present in the theory. 
Therefore, the mass of the bound state $C$ is given by,
\begin{equation}
M_C^2=M_A^2+M_B^2+ 2 M_A M_B \cos u_{AB}^C,
\label{cc}
\end{equation}
\noindent which is easily derived  from  (\ref{Mss}) by considering the formation
of  particle $C$ in the centre-of-mass collision of particles $A, B$. 
Notice that Eq. (\ref{cc})  establishes a  relationship between the masses of the stable particles present
 in the model. The values $u_{AB}^C$ are commonly referred to as {\bf fusing angles}.

\begin{figure}[!h]
 \begin{center}
  \leavevmode
    \includegraphics[scale=0.7]{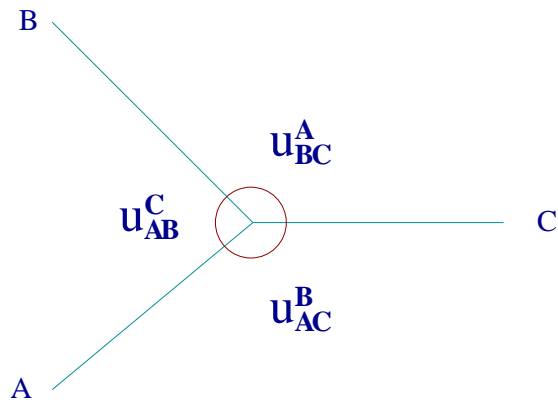}   
    \caption{Fusing angles associated to the scattering processes $\protect{A +B \rightarrow C}$, 
     $\protect{A +C \rightarrow B}$ and $\protect{B +C \rightarrow A}$.}
 \label{uabc}
 \end{center}
\end{figure}

In fact, appealing to crossing symmetry, any of the three particles
$A$, $B$ or $C$ may be seen as a bound state of the other two corresponding
to fusing angles $u_{AB}^C$, $u_{BC}^A$ and $u_{AC}^B$ respectively. 
Consequently, we could write two more equations completely analogue to (\ref{cc})
by permuting the indices $A, B, C$. The combination of these equations leads to the
constraint,
\begin{equation}
u_{AB}^C+u_{BC}^A+u_{AC}^B=2 \pi,
\end{equation}
\noindent which admits the graphic representation shown in Fig. \ref{uabc}.

\begin{figure}[!h]
 \begin{center}
  \leavevmode
    \includegraphics[scale=0.6]{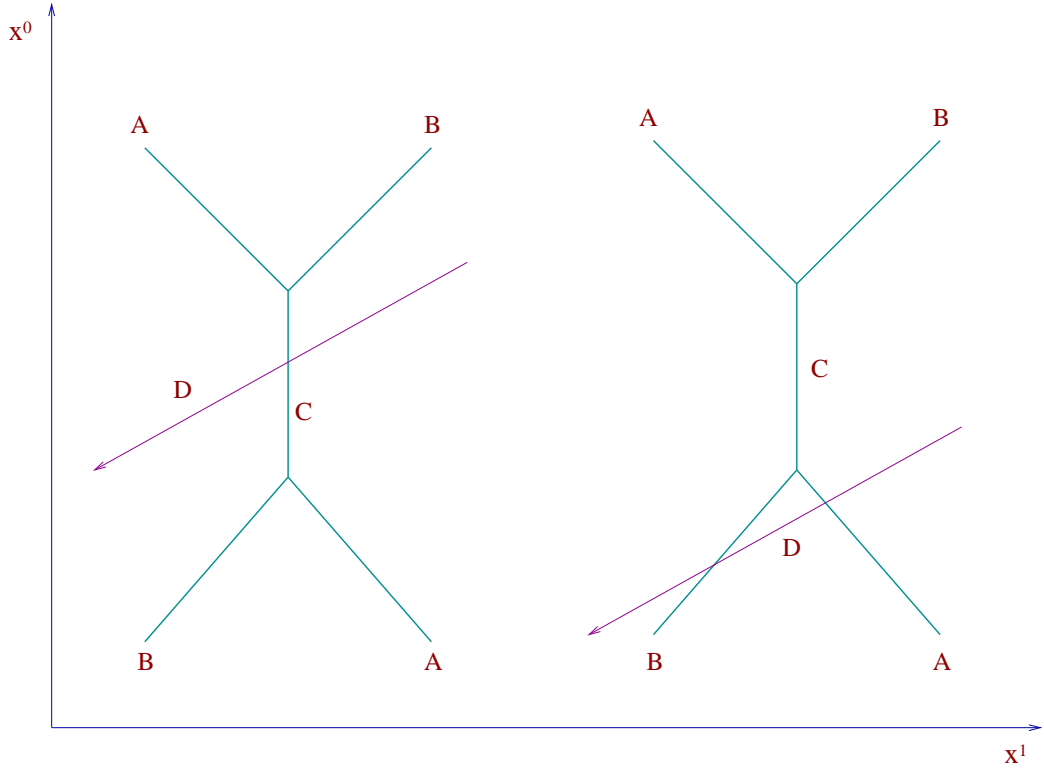}   
    \caption{Graphical representation of bootstrap equations}
 \label{boots}
 \end{center}
\end{figure}

The consideration of stable bound states as part of the asymptotic particle spectrum of the
theory together with the integrability of the model  leads to the so-called  {\bf bootstrap equations}
 \cite{Boot, KTTW},
which establish the equality of the scattering amplitudes related to the two scattering processes
depicted in Fig. {\ref{boots}}. Notice that again the representation of  one-particle asymptotic
states by means of localised wave packets together with Eqs. (\ref{time0}) and (\ref{space0})
 are on the basis of the outlined relationship.
Mathematically, this equivalence leads to the mentioned 
bootstrap equations which have the general form,
\begin{equation}
S_{AD}(\theta + i\bar{u}_{A \bar{C}}^B) S_{BD} (\theta - i\bar{u}_{B\bar{C}}^{A})=S_{CD}(\theta),
\label{bootseqs}
\end{equation}
\noindent where $\bar{u}_{AB}^C=\pi-u_{AB}^C$. 

\subsubsection{Unstable particles}
\label{unstableparticles}
\indent \ \
The discussion reported above refers exclusively to the presence of stable bound states
related to purely imaginary simple poles in the rapidity variable located  in the  physical
sheet, $0<\text{Im}\, (\theta) < \pi$. However, we might encounter a situation in which also unstable
bound states can be present in the theory. In particular, the HSG-models on which
we will mainly focus our attention later, provide examples of theories possessing unstable
particles in their spectrum, that is, particles possessing a  finite lifetime.
 Accordingly, some explanation is necessary concerning the interpretation
of unstable bound states within the scattering theory context, in other words, we wish to identify
 the trace left by the formation of an unstable particle in a two-particle scattering amplitude. 

Therefore, let us consider again a scattering process $A+B \rightarrow \tilde{C}$, where the
particle $\tilde{C}$ is now unstable i.e., it is not encountered as part of the one-particle
asymptotic spectrum which only contains particles whose lifetime is infinite. 
Similarly to the description of stable bound states, the unstable particle 
$\tilde{C}$ is expected to be produced whenever the particles
$A$, $B$ scatter at a certain centre-of-mass energy $\sqrt s$ close enough to the mass of the unstable
particle. In that case the scattering amplitude $S_{AB}(s)$ is expected to have a resonance 
 pole located at a certain complex value of the Mandelstam variable $s$ given by
\begin{equation}
s_{R}:=M_{R}^2=\left[M_{\tilde{C}}-
\frac{i\Gamma_{\tilde{C}}}{2}\right]^2,
\end{equation}
\noindent which shows that the description of unstable particles is usually carried out by 
complexifying  the physical mass of a stable particle.  In the $s$-plane, the mentioned
complexification amounts to
adding to the physical mass a complex contribution given by the decay width $\Gamma_{\tilde{C}}>0$,
whose inverse is identified as the lifetime of the unstable particle. Therefore, the form of the
two-particle S-matrix near to the resonance pole $s_{R}$ is given by the well known Breit-Wigner
resonance formula \cite{BW, ELOP, Wein}
\begin{equation}
S_{AB}(s) \approx 1-\frac{2 iM_{\tilde{c}}\Gamma_{\tilde{C}}}{s-s_R}.
\end{equation}
\noindent As mentioned in \cite{ELOP}, whenever $M_{\tilde{C}}\gg \Gamma_{\tilde{C}}$ the following approximation 
\begin{equation}
M_{R}^2 \approx  M_{\tilde{C}}^2 +iM_{\tilde{C}}\Gamma_{\tilde{C}},
\end{equation}
\noindent is justified, which allows for a clear-cut interpretation of $M_{\tilde{C}}$ as the physical
mass of the unstable particle.
The pole $s_{R}$ in the Mandelstam variable has a counterpart in the rapidity 
description which we will denote by 
\begin{equation}
\theta_{R}=\sigma_{AB}^{\tilde{C}}-
i\bar{\sigma}_{AB}^{\tilde{C}},
\label{unstableres}
\end{equation}
\noindent that is, in the $\theta$-plane, the presence of unstable particles in
the models reflects in the existence of complex poles in the second Riemann sheet or non physical sheet.
Therefore,  $\sigma_{AB}^{\tilde{C}}, \bar{\sigma}_{AB}^{\tilde{C}}>0$.
 In comparison to the description of 
 stable particles, the fusing angles $u_{AB}^C$ describing
 stable bound states, get formally complexified as follows 
\begin{equation}
u_{AB}^C \rightarrow  -\bar{\sigma}_{AB}^{\tilde{C}}- i\sigma_{AB}^{\tilde{C}},
\label{fusingu}
\end{equation}
\noindent from where it is easily inferred that, unstable particles are associated to poles
arising in the  non physical sheet, namely $\textnormal{Im}( \theta_R)$ is now negative. 
In addition, whenever the so-called {\bf resonance parameter}
$\sigma_{AB}^{\tilde{C}}$ is vanishing,  the unstable particle $\tilde{C}$ becomes a `virtual state',
meaning that $\theta_{R}$ becomes purely imaginary as for the case of stable bound states, but
still does not lie inside the physical strip. In that situation,
unstable particles become virtual states characterised by poles on the imaginary axis beyond
the physical sheet.

The particularisation  of Eq. (\ref{cc}) to the case at hand amounts to substituting $M_C$ by
$M_{R}$ and $u_{AB}^C$ as shown in (\ref{fusingu}). The identification of the real and imaginary
part of this equation leads to the following relations

\begin{eqnarray}
M_{\tilde{C}}^{2}-\frac{(\Gamma _{\tilde{C}})^{2}}{
4} &=&(M_{A})^{2}+(M_{B})^{2}+2M_{A}M_{B}\cosh \sigma_{AB}^{\tilde{C}}
 \cos \bar{\sigma}_{AB}^{\tilde{C}},  \label{BW11} \\
M_{\tilde{C}}\Gamma _{\tilde{C}}
&=&2M_{A} M_{B} \sinh \sigma_{AB}^{\tilde{C}} \sin \bar{\sigma}_{AB}^{\tilde{C}}.   \label{BW22}
\end{eqnarray}
\noindent It becomes again clear from (\ref{BW11}) and (\ref{BW22}) that,
 whenever the resonance parameter $\sigma_{AB}^{\tilde{C}}$ 
is vanishing the unstable particle becomes a virtual state.
The decay width $\Gamma_{\tilde{C}}$ vanishes in virtue of (\ref{BW22})
as would correspond to an stable particle but, as mentioned above, the 
corresponding pole is located in the non physical sheet.

\section{Non-abelian affine Toda field theories}
\label{ntft1}
\indent \ \
In the preceding sections we have summarised the most relevant features of
 1+1-dimensional integrable QFT's. We started our study by reviewing their 
construction as perturbations of conformally 
invariant QFT's and reported their most celebrated properties, which we have shown to be 
intimately linked to the powerful implications conformal symmetry has in 1+1-dimensions. 
As we have seen, the trace of these implications  is  still crucial once the CFT is driven
away from its associated RG-fixed point.  The most substantial consequence 
of the distinguished properties of 1+1-dimensional
integrable massive
 QFT's reported above is that, in many cases,  they allow for the exact computation of the
S-matrix of the theory under consideration. It is identified as the solution to a certain set of
physically-motivated consistency equations expressing the requirements of unitarity, crossing symmetry,
Hermitian analyticity and Lorentz invariance of the scattering 
amplitudes, together with the mentioned
Yang-Baxter \cite{yb} and bootstrap equations \cite{Boot, KTTW}.

\vspace{0.25cm}

Having now settled the general framework and techniques, it is interesting to exploit these 
techniques in order to  study a concrete family  of 1+1-dimensional integrable theories,  
the so-called {\bf non-abelian affine Toda (NAAT) field theories} \cite{nt},
 whose equations of motion
were originally formulated by A.N. Leznov and M.V. Saveliev  in 1983. 
The NAAT-theories are particular
examples of {\bf Toda field theories} which, as we have seen in the introduction, constitute 
field generalisations of the well known {\bf Toda lattice} \cite{FPU, lattice, lattice2}.

Clasically, the NAAT-equations~\cite{nt} are integrable
(multi-component) generalisations of the sine-Gordon equation for a bosonic
field  which takes values in a non-abelian Lie group, in contrast with the usual 
Toda field theories where the field takes values in the (abelian) Cartan
subgroup of a Lie group \cite{classtoda}. The mentioned equations
can be obtained as the equations of motion associated to an action functional
which is the sum of the  WZNW-action \cite{wzw, wzw2, olive}  associated
to a complex non-abelian Lie group $\bar{G}_0$ and a bosonic field $h(x^0,x^1)$, 
and a potential $V(h)$, depending upon this field,
\begin{equation}
S[h]\> =\> {1\over\beta^2}\> \Bigl\{ S_{WZNW}[h]\> -\> \int d^2x\>
V(h)\Bigr\}\>. 
\label{ActGen}
\end{equation}
\noindent Here $\beta$ is a coupling constant which does not play any role in the classical theory but 
gets  quantized in the quantum theory in terms of  an integer, $k$, usually referred to as  ``level'' 
\cite{wzw, olive, CFT} (see also Eq. (\ref{kacm}).

The latter structure already gives a first glimpse
concerning  the quantum formulation of these theories, which may be viewed, at least
in some cases, as perturbations of a gauged WZNW-coset action \cite{Gep, GepQ, DHS, paras}
by means of a primary field of the latter CFT. Consequently, all the properties and techniques
described before in this chapter will inspire the quantum study of some NAAT-theories. 

\vspace{0.25cm}

The most general construction of the NAAT-theories was carried out in \cite{HSG} and
takes as its starting point a semisimple, complex and finite Lie algebra denoted by $\bar{g}$,
and a finite order automorphism $\sigma$ of the latter,  which induces the following decomposition, 
\begin{eqnarray}
\bar{g}&=&\bigoplus_{\j\in \mathbb{Z}} \bar{g}_{\bar{\j}}, \,\,\,\,\,\,\,\,\left[g_{\bar{\j}},
 g_{\bar{k}}\right] \subset \bar{g}_{\overline{\j+k}}, \label{auto1}\\
\sigma(x)&=&e^{2 \pi i j/N}x, \,\,\,\,\,\,\,\,\text{for}\,\,\,\,\,\,\, x \in \bar{g}_{\bar{\j}}.
\label{auto2}
\end{eqnarray}
\noindent It is clear from (\ref{auto2}) that $\sigma$ is an order-$N$ automorphism i.e., 
$\sigma^N=1$. A decomposition of the type  (\ref{auto1})  is referred to as a 
{\bf ${\mathbb{Z}}/{N\mathbb{Z}}$-gradation}
of the Lie algebra $\bar{g}$. The subindices $\bar{\j}$ must be understood as $\bar{\j}=\j\,\, 
\text{modulo}\,\, N$. 
Furthermore, the invariant subspace under the action of the automorphism, $\bar{g}_{0}$, 
is  in virtue of (\ref{auto1}), a complex subalgebra of $\bar{g}$,
 whose associated Lie group is denoted by $\bar{G}_0$. It is this Lie group 
where the field of every NAAT-theory related to a particular pair $(\bar{g}, \sigma)$ 
takes values i.e., $h(x^0,x^1) \in \bar{G}_0$. Notice that until here we have always referred
to finite Lie algebras, although the explicit construction of classically conserved charges
requires the use of the affine or extended  Lie algebra, $\hat{g}$.  

\vspace{0.3cm}
Concerning the construction of quantum NAAT-theories, it was found 
in~\cite{HSG} (see also~\cite{HMQH}) that, although there is a large class
of interesting QFT's related at classical level to NAAT-models, 
only a certain subset of NAAT-equations lead at quantum level to  QFT's 
possessing the following remarkable properties:

\vspace{0.25cm}

{\bf i)} {\bf Unitarity}, which in that context means we are interested in theories whose
action functional takes real values. In other words, we wish to study the subset of 
theories which are expected to lead at quantum level to a
Hilbert space containing only  particle states of  non-negative norm.
In \cite{HSG} it was shown that the reality condition is satisfied for the choice of 
action functionals whose  kinetic term is positive-definite, whereas the 
potential  is real and bounded from below.

\vspace{0.25cm}

{\bf ii)} Existence of  a {\bf mass-gap} namely, we are interested in theories possessing a purely 
 massive spectrum. Notice that our discussion of the preceding sections always referred to massive
particles and accordingly, it is this sort of models which are expected to admit an S-matrix description
of the type described in subsection \ref{exactS}.

\vspace{0.25cm}

In \cite{HSG} the preceding requirements where shown to be only fulfilled by two families of
NAAT-theories which were named as {\bf Homogeneous sine-Gordon} (HSG) and {\bf Symmetric Space 
sine-Gordon} (SSSG) theories and  are in one-to-one correspondence with
 the different compact Lie groups and
compact symmetric spaces, respectively. Actually, the HSG- and SSSG-theories
of~\cite{HSG} are particular examples of the deformed coset models constructed
by Q-H. Park in~\cite{Park} and the Symmetric Space sine-Gordon models 
constructed by I. Bakas, Q-H. Park and H-J. Shin in~\cite{sssg3}, respectively, where the specific form
of the potential makes them exhibit a mass gap namely, a purely massive particle spectrum.

\subsubsection{Some remarks on classical integrability}
\indent \ \
The results presented in this thesis  are mainly concerned with  the quantum characteristics
of the previously mentioned HSG- and SSSG-theories. 
For this reason, we want to report now in this subsection only 
a very  briefly overview on the meaning and main implications of classical integrability, and 
its  specific consequences in the context of NAAT-models. For a more detailed discussion
concerning general aspects of classical integrability we refer the reader for instance 
to \cite{das, raja}. The specific
study of the classical integrability of the HSG- and SSSG-models was carried out in \cite{HSG2, cm}
respectively, where the classically conserved charges were explicitly constructed.

The {\bf classical integrability} of all NAAT-theories can be inferred from their associated
equations of motion, which admit a zero-curvature expression of the form

\begin{eqnarray}
&{\partial}_{-}(h^{-1}{\partial}_{+} h)=-m^2 \left[ \Lambda_{+}, h^{-1} \Lambda_{-} h \right]&
 \nonumber \\
&\Updownarrow& \nonumber \\
&\left[{\partial}_{+}\,+ h^{-1}{\partial}_{+} h + 
i\,m \Lambda_{+}, {\partial}_{-}\,+i\,m\, h^{-1} \Lambda_{-} h \right]=0&,
\label{zerocur}
\end{eqnarray}
\noindent where we used the light-cone coordinates $x^{\pm}=x^0 \pm x^1$. The objects $\Lambda_{\pm}$
 are semisimple elements taking values in the subspaces 
$\bar{g}_{\bar{k}}$ and $\bar{g}_{\overline{N-k}}$ defined in (\ref{auto1}) and (\ref{auto2}) 
respectively.
These elements enter into the expression of the potential $V(h)$ arising in (\ref{ActGen}) which
has the general form
\begin{equation}
V(h)=-\frac{m^2}{\pi} \langle \Lambda_{+}, h^{\dagger} \Lambda_{-} h \rangle,
\label{potential}
\end{equation}
\noindent  where $m$ is a coupling constant that has mass dimension and the brackets 
$\langle \, ,\,\rangle$ denote an invariant and non-degenerate bilinear form in $\bar{G}$ 
\cite{bau, cornwell}.
 
\vspace{0.25cm}
The classical concept of integrability
is encoded in  {\bf Liouville's theorem} which defines a classically integrable model
as a Hamiltonian system possessing at the same time a  $2n$-dimensional phase space and $n$
independent conserved charges $Q_i$, $i=1, \cdots, n$ in involution, namely
\begin{equation}
\frac{\partial Q_i}{\partial x^0}=0, \,\,\,\,\,\,\,\,\,\,\,\,\,\,\,\{Q_i, Q_j\}=0, 
\,\,\,\,\,\,\,\,\,\, i, j=1, \cdots, n,
\end{equation}
\noindent $\{\,,\,\}$ denoting  Poisson brackets.

As mentioned above, the zero-curvature expression (\ref{zerocur}) is intimately related
to the classical integrability of NAAT-theories. This is based on the celebrated {\bf Lax-pair
formalism} commonly used in the formulation of classically integrable models. A Lax-pair
$(L, B)$ is a pair of operators depending on the dynamical variables of the system at hand.
Although the classical equations of motion of integrable systems are often non-linear, 
they can be formulated in terms of Lax-pairs in the form 
\begin{equation}
\frac{\partial{L}}{\partial x^0}=\left[ L, B \right].
\end{equation}
\noindent Once a suitable Lax-pair $(L, B)$ has been constructed, 
the existence of infinitely many classically conserved charges follows immediately (see e.g. \cite{das})
and their form is easily found to be $Q_n = \text{Tr} (L^n)$. 
There are various ways to identify the most suitable
Lax-pair associated to a certain model. In particular, 
the generalised Drinfel'd-Sokolov construction of~\cite{drinfel} 
was used in \cite{HSG2, cm} to obtain the conserved densities related to (\ref{zerocur})
for the HSG- and part of the SSSG-models.  It was also observed in \cite{HSG2, cm} that 
the classically conserved charges of a certain spin are recovered from the corresponding  quantum higher 
rank conserved quantities in the limit $\hbar \rightarrow 0$. This result  provides a consistency check 
for  the quantum conserved charges constructed along the lines reported in subsection \ref{PCFT}.

\vspace{0.25cm}

It is worth emphasising that one 
 of  the most important characteristics of classically integrable theories is the fact that 
their equations of motion very often posses {\bf solitonic and multi-solitonic solutions} \cite{raja}. 
These sort of configurations turn out to be of special interest, since the classical 
interaction between solitons has, for classically integrable systems,
 very similar properties to the interaction of particles at quantum level, described 
 by means of an S-matrix. The reason is that a {\bf soliton} is defined as a non-singular solution 
to the classical non-linear equations of motion whose energy
density is localised and remains undistorted under the time evolution. Consequently, soliton
solutions have finite energy \cite{raja}. These properties suggest a relationship between
 soliton solutions and extended particles described by localised wave packets of the form (\ref{wavef}).
Indeed, very often at least part of the quantum spectrum of the model can be constructed by means of
the {\bf semi-classical quantization} of the classical solitonic solutions and there is a link
between the semi-classical limit of the S-matrix and the solitonic solutions 
of the classical scattering \cite{semi, raja}.

\vspace{0.25cm}

In particular, all NAAT-equations \cite{nt} admit solitonic solutions, 
which is in contrast with the usual affine Toda field theories \cite{Roland, eeight, ATFTS, dis, mussrev},
where the condition of having soliton
solutions (imaginary coupling constant) leads to an ill defined
action~\cite{illToda}. In particular in \cite{HSGsol} the semi-classical spectrum 
of the HSG-models was constructed. Thereafter, the S-matrix proposal pointed out in 
\cite{HSGS} made use of this result by assuming  the exact spectrum of stable particles
 at quantum level  to coincide with part of the solitonic spectrum determined 
semi-classically. We will see in later chapters that 
all our consistency checks of the S-matrix proposal for the HSG-models have
confirmed the legitimacy of the mentioned assumption. 

\vspace{0.25cm}

Let us now turn to the description of some other
general properties of NAAT-field theories and ultimately focus our attention in the homogeneous
and symmetric space sine-Gordon models \cite{HSG}.

\subsection{Quantum aspects of non-abelian affine Toda theories}
\label{hsgsssg}
\indent \ \
Classically  all NAAT-theories are integrable and make perfect sense for each possible
choice of the objects $\{ \bar{g}, \sigma, \Lambda_{\pm}\}$ characterising their action
functional (\ref{ActGen}). 
However, it was shown in \cite{HSG} that not all classical  NAAT-theories would be
expected to lead 
at quantum level to well defined QFT's in the sense described in the 
previous section (see {\bf i), ii)}). It was found also in  \cite{HSG} that every consistent
quantum NAAT-field theory will be described by an action of the form
\begin{equation}
S[h]\> =\> {1\over\beta^2}\> \Bigl\{ S_{WZNW}[h]\> +\frac{m^2}{\pi}\> \int d^2x\>
\langle \Lambda_{+}, h^{\dagger} \Lambda_{-} h \rangle
\Bigr\}\>,
\label{Actq}
\end{equation}
\noindent where now the matrix element $\langle \Lambda_{+}, h^{\dagger} \Lambda_{-} h \rangle$ 
entering the potential (\ref{potential}) is identified with a matrix element of a certain spinless
 primary WZNW-field  i.e., at quantum level the potential $V(h)$ turns out to play the role of perturbing field 
in the sense of (\ref{pss}).

Therefore, the NAAT-theories  will provide a Lagrangian formulation for
some already known integrable perturbations of CFT's and, furthermore, they will
also lead us to discovering new ones.

The unitarity and the  presence of
 a mass-gap in the theory are conditions which turn out to restrict severely
the possible choices of the objects $\{ \bar{g}, \sigma, \Lambda_{\pm}\}$ at quantum level. In particular

\vspace{0.25cm}

{\bf i)} The {\bf unitarity} of the QFT requires an action
functional whose kinetic term is positive-definite.
This can only be achieved  if the bilinear form $\langle \, , \, \rangle$ has
a definite sign. This requires that the Lie group $G_0$ is chosen to be    
a compact Lie group. In that case 
the bilinear form $\langle \, , \, \rangle$ is the 
compact real form  \cite{bau, cornwell}.
This fact additionally implies,  for $G_0$ to be non-abelian, that the field is unitary 
\begin{equation}
h ^\dagger=h^{-1}.
\end{equation}
\noindent The latter  property is only consistent with the NAAT-equations (\ref{zerocur}) if 
\begin{equation}
\Lambda_{\pm}^{\dagger}=\Lambda_{\pm} \Rightarrow  V(h) \in \mathbb{R}.
\end{equation}
\noindent The previous equation, together with the conditions $\Lambda_{+}\in g_{k}$ and
$\Lambda_{-} \in g_{N-k}$ reported in the previous section, restricts severely the possible inequivalent
choices of the automorphism $\sigma$, which can only have order $N=1$ or 2. Then, the decomposition
(\ref{auto1}) gives
\begin{eqnarray}
\sigma&=&1 \Rightarrow g=g_0,  \,\,\,\,\,\,\,\,\,\, \,\,\,\,\,\,\,\,\,\,
 \,\,\,\,\,\,\,\,\,\,\,\,\,\,\,\,\,\,\,\, \,\,\,\,\,\,\,\,\,\, \,\,\,\,\,\,\,
\left[g_0, g_0 \right] \subset g_0 \label{n1}\\
\sigma^2&=&1 \Rightarrow g=g_0 \oplus g_1, \,\,\,\,\,\,\,\,\,\,\left[g_0, g_1 \right] \subset g_1, \,\,\,
\left[g_1, g_1 \right] \subset g_0, \,\,\, \left[g_0, g_0 \right] \subset g_0. \label{n2}
\end{eqnarray}
\noindent Accordingly,
\begin{eqnarray}
&&\Lambda_{\pm}\in g_0  \,\,\,\,\,\,\,\,\,\, \,\,\,\,\,\,\,\,
\,\,\,\,\,\,\,\,\,\,\,\,\, \,\,\,\,\,\,\,\,\,\, \,\,\,\,\,\,\,\,\,\,
\text{for $\sigma=1$},\\
&&\Lambda_{\pm}\in g_1  \,\,\,\,\,\,\,\,\,\,\, \,\,\,\,\,
\,\,\,\,\,\,\,\,\,\,\,\,\,\,\, \,\,\,\,\,\,\,\,\,\, \,\,\,\,\,\,\,\,\,\,
{\text{for $\sigma^2=1$}}.
\end{eqnarray}
In this fashion we can restrict ourselves to the two families of NAAT-theories associated
to the identity automorphism $(\sigma=1, N=1)$ or to involutions $(\sigma^2=1, N=2)$.
The first family of theories are the already mentioned  Homogeneous sine-Gordon models,
 whereas the NAAT- theories associated with involutions were named as  
{\bf Symmetric space sine-Gordon models} \cite{HSG}, for the reason they are related to symmetric spaces in
the way we will see later. However, at this stage of the construction there is still an additional requirement to be fulfilled.

\vspace{0.25cm}

{\bf ii)} We wish to investigate theories with a {\bf purely massive particle spectrum}. 
This leads to additional constraints which select out the precise CFT whose perturbations
 we want to investigate. The concrete details of the derivation of these constraints can
 be found in \cite{HSG}. To put it briefly, the potential (\ref{potential}) 
can be shown to be globally invariant under certain transformations of the field 
\begin{equation}
h(x^0, x^1)\rightarrow \alpha_{+}h(x^0, x^1) \alpha_{-},
\label{transf}
\end{equation} 
\noindent for $\alpha_{\pm}\in g_{\pm}$ and $g_{\pm}$ being subalgebras of $g$.
The breaking of such a  symmetry amounts to the possibility 
of the existence of massless particles in the spectrum. In order to avoid that
situation i.e., the degeneracy of the vacuum configuration, it is necessary to identify the mentioned symmetry with a gauge symmetry of the model, so that the different vacuum configurations
are identified under gauge transformations. Therefore, one must substitute the classical action
(\ref{ActGen}) by a {\bf gauged action} for which is needed to modify the original WZNW-action
associated to the Lie group $G_0$ by
introducing certain gauge fields and select thereafter a particular gauge-fixing prescription. The outcome of the whole procedure is that
the term  $S_{WZNW}$ entering the action of  the massive
 HSG- and SSSG-models (\ref{Actq}) is a gauged 
WZNW-coset action \cite{Gep, GepQ, DHS, paras}
associated to cosets of the general form $G_0/U(1)^p$ where $g_{\pm}=u(1)^p$ is the gauge symmetry algebra where the gauge fields take values, and
\begin{eqnarray}
\,\,\,\,\,\,\,\,p=\ell, \,\,\,\,\,\,\,\,\,\,\,\,\,\,\,\,\,\,\,\,\,\,\,\,\Lambda_{\pm} \in U(1)^\ell \,\,\,\,\,\,\,\,\,&&\text{for the HSG-models},\label{hsg22}\\
0 \leq \ell -\ell_{G/G_0}  \leq p \leq {\rm min\/}[\ell_0, \ell- \nu],
 \,\,\,\,\,\,\,\,&&\text{for the SSSG-models}
\label{sssg22}
\end{eqnarray}
Here $\ell, \ell_0$ are the ranks of the compact Lie algebras $g$ and $g_0$
respectively, $\ell_{G/G_0}$ is the rank of the compact symmetric space
$G/G_0$, defined as  the dimension of the maximal abelian subspaces 
contained in $g_1$ \cite{helgas}, and $\nu=2$ or~1 depending on 
whether $\Lambda_+$ and $\Lambda_-$ are linearly independent or
not, respectively. In particular, the lower bound is reached when
$\Lambda_+, \Lambda_- \in g_1$ are regular, meaning that, the subset of elements of
$g$ which simultaneously commute with $\Lambda_{\pm}$
constitute already a maximal abelian subspace of $g$. 
It is important to mention also here that, once the gauge has been fixed, the 
action (\ref{Actq}) still possesses a remaining $U(1)^p$ global symmetry.
This symmetry is responsible for the fact that the soliton solutions we
will describe in the next subsection carry conserved  Noether charges.

\vspace{0.25cm}
It is worth emphasising that the particular form of the abelian gauge transformation
(\ref{transf}) has a crucial consequence which effects further results concerning
the properties of the HSG S-matrices. The HSG- and SSSG-models are not parity invariant in
general, unless the elements $\Lambda_{\pm}$ are not linearly independent from each other, namely
$\Lambda_{+}=\eta  \Lambda_{-}$. This is in the origin of the {\bf parity breaking} of some of
the scattering amplitudes we will encounter in the study of the HSG-theories.

\vspace{0.25cm}

{\bf iii)} Moreover, if the quantum theory is to be well
defined, the coupling constant has to be quantized: $\beta^2= 1/k$, for some
positive integer $k$ (see~\cite{wzw, olive, cm} for a more precise form of this
quantization rule). Such a quantization does not occur in the sine-Gordon
theory or the usual affine Toda theories because the field takes values in an
abelian group in those cases. An important consequence of this is that, in the
quantum theory, the $\beta^2$ will not be a continuous coupling constant.
However, the quantum theory will have other continuous coupling constants that
appear in the potential and, in particular, determine the mass spectrum. 

 \vspace{0.25cm}

The constraints stated in 
{\bf i),ii), iii)} result in a rich variety of HSG- and  SSSG-models. In summary:

\vspace{0.25cm}

{\bf I.} The HSG-theories are perturbations of WZNW-coset theories associated to
cosets of the type $G_k/U(1)^{\ell}$, for $\ell$ to be the rank of the compact 
 Lie algebra $g$  and $k$ an integer which is identified with the level of the Kac-Moody
representation of the Virasoro algebra \cite{wzw} generating the conformal symmetry
of the WZNW-models. These sort of CFT's are also named
as {\bf $G$-parafermion theories} and their properties have been studied in \cite{Gep, GepQ, DHS}.

\vspace{0.25cm}

{\bf II.} The SSSG-theories include a larger variety of models, since the value of $p$ given
by (\ref{sssg22}) is not fixed like for the HSG-models. They are associated to perturbations
of WZNW-coset models related to cosets of the form $G_0/U(1)^p$. 
There are several interesting subclasses of SSSG-models 
 that include, for $p =\ell_0$, new massive perturbations of the theory of {\bf $G_0$-parafermions}
different from those provided by the Homogeneous  sine-Gordon
theories~\cite{HSG2}. Notice that this case happens only if the symmetric
space satisfies $\ell_0\leq \ell- \nu$. Another
particularly interesting class of models occurs when $\ell=\ell_{G/{G_0}}$ and $p=0$. 
In this case, the SSSG-theory is just a massive
perturbation of the WZNW-model corresponding to $G_0$. We have named
this particular subclass of theories as {\bf Split models} and performed an
investigation of their classical and quantum integrability in \cite{cm}.
A scheme of some of the different types of Toda field theories, including the families of
models studied here was presented in the previous chapter. 

In the light of the preceding
discussion, it is also clear that the HSG- and SSSG-models are just particular examples of
NAAT-theories and it must be emphasised that they are  not the only ones of physical interest. 
Other types of NAAT-theories have been also object of study
in the literature (see for instance the work by J.F. Gomes et al. \cite{gomes}).

\vspace{0.25cm}
Having now established  the defining features of the families of models
we will be interested in, we shall devote the next sections to a more detailed
description of the properties of these theories which
have been investigated in the literature. The study of the HSG-models
has been carried out to a large extent over the last years. In \cite{HSG2}
their classical integrability and quantum integrability were established. In
\cite{HSGsol} soliton and multi-soliton solutions were explicitly constructed by
means of a semi-classical analysis. The latter semi-classical study has been
also exploited in \cite{HSGS} for the explicit construction of all S-matrices describing the
scattering theory of the HSG-models associated to simply-laced Lie algebras.
On the other hand, the results available for the SSSG-theories are considerably more
limited and the aim of this thesis  has been also to partially fill this gap \cite{cm}. However,
we will only report here the main conclusions of this work and focus more our discussion on
the properties of the HSG-models to which we will continuously appeal  in the next chapters. 

\subsection{The homogeneous sine-Gordon models}
\label{hsgsec}
\indent \ \
As we have explained in the previous paragraph there are many
properties of the homogeneous sine-Gordon theories which are
quite well understood for the time being and
their quantum study has been carried out along the general 
lines described in sections \ref{cftqft}, \ref{exactS}
and \ref{analit}. 

\vspace{0.25cm}

The simplest HSG-theory is associated
to $G=SU(2)$,  whose equation of motion is the complex sine-Gordon
equation~\cite{CSGold, Park}. This theory corresponds to the perturbation of the usual
$\mathbb{Z}_k$-parafermions \cite{paras} by the first thermal operator~\cite{CSGBAK}, whose exact
factorisable scattering matrix is the minimal one associated to
$A_{k-1}$~\cite{paras, Roland}. In fact, we will see later that the 
proposed scattering matrix \cite{HSGS} for the HSG-models related to simply laced Lie algebras consists
partially of $\ell$ copies of minimal $A_{k-1}$-affine Toda field theory,
whose mutual interaction is characterised by an S-matrix which violates parity. 

\vspace{0.25cm}

The  perturbing field characterising the action (\ref{Actq}) was identified in \cite{HSG2}
to be a spinless primary field having conformal dimension 
\begin{equation}
\Delta =\bar{\Delta}= \frac{h^\vee}{k+h^\vee},
\label{deldata}
\end{equation}
\noindent where $h^\vee$ is the dual Coxeter number of~$g$. In the following we will often
consider simply laced Lie algebras, so that we can  write $h^{\vee}=h$ for $h$ to be the 
Coxeter number of $g$ whose definition may be found for instance  in \cite{kac, Bou, olive}. 
Combining (\ref{deldata}) with the condition of super-renormalisability at first order
 (\ref{super}), we obtain the constraint
\begin{equation}
k>h^\vee.
\label{supf}
\end{equation}
\noindent It is in this regime where the quantum integrability of the HSG-models has been established.

\vspace{0.25cm}

Furthermore, the remaining characteristics of the unperturbed CFT are well understood \cite{Gep, GepQ, olive, DHS} and in particular, one of the most relevant ones is, as usual, the associated central charge $c$,
\begin{equation}
c_{G_k/U(1)^{\ell}}=\frac{kh-h^{\vee}}{k + h^\vee}\,\ell .
\label{cdata}
\end{equation}
\noindent Notice that (\ref{cdata}) is nothing but the usual expression for the coset 
central charge \cite{GKO} and we have used the general relation  $\dim g = \ell (h+1)$.

\subsubsection{Quantum integrability}
\indent \ \
The quantum integrability of the HSG-models was established
by C.R.~Fern\'{a}ndez-Pousa, M.V.~Gallas, T.J.~Hollowood and J.L.~Miramontes 
in \cite{HSG2} by means of the explicit construction of
$\ell$ higher spin conserved densities, $\mathcal{T}_{s}$, of spin $s=\pm 2$ and $s=\pm 3$. 
In addition, classically conserved quantities associated
to these same values of the spin were explicitly constructed
by means of the mentioned { \bf generalised  Drinfel'd-Sokolov construction} \cite{drinfel}.
Due to the specific choice of the objects $\{g, \sigma=I, \Lambda_{\pm}\}$
which characterise the  HSG-models, it was shown that infinitely many classically conserved
charges arise in these theories and that they are 
distributed in such a way that $\ell$ of them  exist associated to
each value of the spin $s=\pm 1, \pm2, ...$  At quantum level, this characteristic seems to  be
preserved, since the link between classical and quantum charges is achieved by identifying
\begin{equation}
J=J^A t^A=(\hbar k) \partial_{-} h h^{\dagger}, \,\,\,\,\,\,\hbar k =\frac{1}{\beta^2}.
\label{kbeta}
\end{equation}
\noindent Here we explicitly introduced Plank's constant in order to exhibit
the precise relationship between the semi-classical, $\hbar \rightarrow 0$,
the weak coupling, $\beta^2 \rightarrow 0$, and the $k \rightarrow \infty$ limits. 
Notice that the second equality in (\ref{kbeta}) means that the semi-classical and weak coupling
limits are equivalent. In Eq. (\ref{kbeta}) $J^A$ are 
the local currents which generate  the conformal symmetry of
the WZNW-model associated to the Lie algebra $g$\footnote{At this point
a key result due to Bais et al. \cite{baisC} has been used. They
established that the current-algebra associated to the coset CFT could be realized
as a subset of the current-algebra of the WZNW-model associated to the Lie algebra $g$. Therefore
the currents  $J^A (z)$ and $\bar{J}^A (\bar{z})$ are the ones corresponding to the usual
WZNW-model \cite{wzw}, whose properties and commutation relations are well known.},
and $A=1, \cdots, \dim g$. The modes $J_{n}^{A}$
arising in the Laurent expansion of these currents satisfy a so-called {\bf Kac
-Moody algebra} 
\begin{equation}
\left[J_{n}^{A}, J_{m}^{B}\right]= f^{ABC} J_{n+m}^{C}+ \frac{1}{2} k\,n\,\delta^{AB}\delta_{n+m,0},
\label{kacm}
\end{equation}
which also provides a representation of the Virasoro algebra (\ref{VV}) by means of
the so-called {\bf Sugawara construction} \cite{sugawara}. The Kac-Moody algebra is
characterised at quantum level, by a central extension given by the  integer $k$ 
which is known as the {\bf level} of the Kac-Moody representation \cite{wzw, olive, CFT} 
and takes always integer values.  Finally the antihermitian generators $t^A$ arising in (\ref{kbeta})
provide a compact basis for the Lie algebra $g$ and their commutation
relations $\left[t^A, t^B\right]=f^{ABC}t^C$ involve the structure constants $f^{ABC}$ 
arising in (\ref{kacm}).

The current $J$ and its anti-holomorphic counterpart are conserved in the original CFT,
\begin{equation}
\bar{\partial}{J(z)}=0, \,\,\,\,\,\,\,\,\, {\partial}{\bar {J}(\bar{z})}=0.
\label{JJ}
\end{equation}
\noindent  Since the holomorphic and anti-holomorphic components
of the energy momentum tensor can be expressed in terms of the currents $J, \bar{J}$ by 
means of  Sugawara's construction  \cite{sugawara} (see also \cite{olive, CFT}), the previous
equations imply the conservation of the energy momentum tensor in
the unperturbed CFT.

\vspace{0.25cm}
As a consequence of the previous observations, the infinitely many  locally conserved charges
of the unperturbed CFT are generated by combinations of normal ordered products
of the currents $J^A, \bar{J}^A$, some of which remain conserved after the perturbation.
Accordingly, the general form of  the spin-2 and 3 conserved densities computed in \cite{HSG2} is 
\begin{eqnarray}
\mathcal{T}_2 (z)&=&\mathcal{D}_{AB} (J^A J^B)(z), \label{density2}\\
\mathcal{T}_3 (z)&=&\mathcal {P}_{ABC}(J^A(J^B J^C))(z) +\mathcal{Q}_{AB} (J^A \partial J^B) +
\mathcal{R}_{A} (\partial^2 J^A) (z). 
\label{density3}
\end{eqnarray}
\noindent Here the parenthesis  $(\, (\,\cdots \,)\, )$ denote normal ordering. 
The concrete normal ordering prescription for two arbitrary operators we are using here
consists of selecting out the first regular term arising in their OPE. Such a prescription is used
for instance in  \cite{baisC} and the second reference in \cite{CFT}.

Clearly, the latter densities are conserved in the original CFT in
virtue of (\ref{JJ}) and
in \cite{HSG2} it was proven, by direct evaluation of Eq. (\ref{cruc}), 
that, all the classically
conserved spin-2 and 3 charges 
associated to the HSG-models give rise, at  quantum level after a suitable
renormalisation, to quantum conserved charges. These charges correspond
to particular choices of the tensors
$\mathcal{D}_{AB}$, $\mathcal{P}_{ABC}$, 
$\mathcal{Q}_{AB}$ and $\mathcal{R}_{A}$ in (\ref{density2})
and (\ref{density3}). Analogously, we may construct negative
spin densities by changing $J \rightarrow \bar{J}$ in
 (\ref{density2}) and  (\ref{density3}).
Appealing now to the results  reported in section \ref{exactS},
the existence of these conserved densities allows 
for concluding the quantum integrability 
of all HSG-theories. Recall that the counting argument has not
been exploited for this class of theories 
for the reasons summarised at the end of subsection \ref{PCFT}.
Therefore, the HSG-models are expected to admit an S-matrix description
and their exact S-matrix can be constructed by means of the bootstrap 
program \cite{Boot} described in section \ref{analit}. 

\subsubsection{The homogeneous sine-Gordon particle spectrum}
\indent \ \
The results of section \ref{analit} have shown that the exact S-matrices of 1+1-dimensional
massive integrable QFT's can be obtained as the solutions to a certain set of consistency
equations expressing the physical requirements of 
analyticity, crossing symmetry, unitarity and Hermitian analyticity together
with the Yang-Baxter and bootstrap equations, which are both highly non-trivial. The latter equations are the
only ones which actually encode information concerning the specific nature of the theory
under study. In particular, the bootstrap equations encode information about the pole structure
of the S-matrix, intimately related to the presence in the model 
of stable and/or unstable particle bound states. The fusing angles $u_{AB}^C$
can be determined once the masses of the stable particles present in the theory are known by means of  Eq. (\ref{cc}), which means the knowledge of the mass spectrum of the theory is fundamental in
order to unravel the pole structure of the S-matrix. 

\vspace{0.25cm}

Accordingly, we shall start our study by recalling the semi-classical 
mass spectrum associated to the HSG-models
related to simply-laced Lie algebras, $g$, which was determined by C.R. Fern\' andez-Pousa and
J. L. Miramontes in \cite{HSGsol}. In this paper the authors have shown that the equations of motion
of the HSG-models (\ref{zerocur}) admit classical time-dependent soliton and multi-soliton 
solutions. Multi-soliton solutions associated to HSG-models related to 
simply-laced Lie algebras can be constructed by using the so-called 
{\bf solitonic specialisation}, technique proposed  by D.I. Olive et al. in \cite{solitonic} 
and linked thereafter to the method of {\bf dressing transformations} in \cite{dressing}. 
Furthermore, it was shown also in \cite{HSGsol} that soliton and multi-soliton solutions associated
to arbitrary HSG-theories can be constructed by using the well known soliton solutions arising in the
complex sine-Gordon (CSG)  model \cite{CSGBAK} and noticing that the latter model is nothing but
the HSG-theory associated to $g=su(2)$. Thus, the construction of soliton solutions corresponding
to arbitrary HSG-theories can be performed by embedding CSG one-soliton solutions into the $su(2)$ 
subalgebras of $g$  spanned  by  the Chevalley generators $\{E_{\pm \alpha}, H_{\alpha}\}$ 
associated to each of its positive roots, $\alpha$ (see e.g. \cite{bau, cornwell, Bou}).

When considered in their rest frame, the soliton solutions 
for the HSG-models are time-periodic and they allow for
 the investigation of the quantum spectrum of particles
 in the semi-classical approximation. This can be done by means of the so-called
{\bf Bohr-Sommerfield quantization rule}, which establishes that 
\begin{equation}
S^{\text{sol}}+M^{\text{sol}}T^{\text{sol}}=2 \pi n \hbar, \,\,\,\,\,\,\,n\in \mathbb{Z},
\end{equation}
\noindent where $S^{\text{sol}}$, $M^{\text{sol}}$ and $T^{\text{sol}}$ are the action,
mass and time-period of a classical soliton solution, respectively. 

Proceeding this way, it was shown in \cite{HSGsol} that, in the simply-laced case,  there exist
$k-1$ soliton solutions ($k$ being again the level) associated to each of the positive roots
of $g$, whose masses are given by
\begin{equation}
M_{c}(\vec{\alpha})=\frac{k}{\pi}m (\vec{\alpha}) \sin \Big(\frac{\pi c}{k}\Big),\,
\,\,\,\,\,c=1, \cdots, k-1,
\label{masses}
\end{equation}
\noindent with 
\begin{equation}
m (\vec{\alpha})=2 m \sqrt {(\vec{\alpha}\cdot \Lambda_{+}) (\vec{\alpha}\cdot \Lambda_{-})}.
\label{massfun}
\end{equation} 
\noindent The  $m (\vec{\alpha})$ are the masses of the fundamental particles of the theory namely,
the particles produced by fluctuations of the field $h$ around a vacuum configuration $h_0$.
 Their masses were  identified by linearising the equations of motion in the approximation when the fluctuations are small ( i.e., $h=h_0 e^{\varphi}$, $\varphi \ll 1$ ). 
As a consequence of the existence of a global symmetry  associated
to the Lie group ${U(1)}^\ell$ (see point {\bf ii)} in the previous subsection), all soliton solutions
of the theory carry a conserved Noether charge which was shown in \cite{HSGsol} to have the form
\begin{equation}
Q_c (\vec\alpha)=c \, \vec\alpha \,\,\text{modulo}\,\,k \Lambda_{R}^*,\,\,\,\,\,\,\,\,\,c=1, \cdots, k-1,
\label{noether}
\end{equation}
\noindent where $\Lambda_{R}^{*}$ is the co-root lattice of $g$ \cite{kac, cornwell} which
is the same as the root lattice in the simply-laced case, and $k$ is the level.

The result (\ref{masses}) shows that we can label each soliton solution, say $C$, 
 by two quantum numbers
$C:=(c, \vec{\alpha})$, notation which, with small modifications, 
 turns out to be also natural for describing the quantum particle
spectrum. It can be seen  easily from (\ref{masses}) that whenever $c=a+b \,\,\text{modulo}\,\,k$ then
\begin{equation}
M_{c}(\vec{\alpha})<  M_{a}(\vec{\alpha})+ M_{b}(\vec{\alpha}), \,\,\,\,\,\,\,\,\,\,\,
Q_c (\vec\alpha)= Q_a (\vec\alpha) + Q_b (\vec\alpha),
\label{stablesol}
\end{equation}
\noindent
which supports the interpretation of the state $(c, \vec{\alpha})$ as a bound state
of  $(a, \vec{\alpha})$ and $(b, \vec{\alpha})$. 

\vspace{0.25cm}

On the other hand,  if we now consider the decomposition of any positive root $\vec{\alpha}$ in terms of simple roots $\vec{\alpha}_1, \cdots, \vec{\alpha}_{\ell}$ of the Lie algebra $g$, 
\begin{equation}
\vec{\alpha}=\sum_{i=1}^{\ell} {\kappa}_{i} \vec{\alpha}_i,
\label{roott}
\end{equation}
\noindent  we can see that
\begin{equation}
M_c (\vec\alpha) \geq \sum_{i=1}^{\ell} M_{c \cdot {\kappa}_i}(\vec{\alpha}_{i}),\,\,\,\,\,\,\,\,\,\,\,
Q_{c}(\vec\alpha)=\sum_{i=1}^{\ell} Q_{c \cdot{\kappa}_i}(\vec{\alpha}_{i}).
\label{unstablesol}
\end{equation}
\noindent Therefore, it is natural to expect the particle $(c, \vec\alpha)$ to be unstable
and decay into stable particles associated to the simple roots $\vec{\alpha}_i$,
of the Lie algebra $g$. 

\vspace{0.25cm}
Putting now together  the results (\ref{stablesol}) and (\ref{unstablesol}) we can already predict
 the stable and unstable bound state structure we might encounter at quantum level and,
consequently, we have already an anticipation of the pole structure we expect to find when constructing
the S-matrices associated to the HSG-models which enters the bootstrap equations (\ref{boots}).
At this stage, we have the following picture concerning the
quantum stable and unstable particle spectrum associated
to the HSG-models (recall that we will focus our attention  in the simply-laced case):

\vspace{0.25cm}

{\bf i)} Associated to each simple root $\vec{\alpha_{i}}$, with $i=1, \cdots, \ell$,
we have, according to (\ref{masses}), 
 a tower of $k-1$ soliton solutions which may be associated to the stable particle spectrum
at quantum level and whose masses will be assumed to be given by the same formula (\ref{masses}).
We rewrite now this formula  as
\begin{equation}
M_{a}^{i}:=M_{a}(\vec{\alpha}_i)=\frac{k}{\pi}m_{i}\sin\Big (\frac{a \pi}{k}\Big),\,\,\,\,\,\,\,
a=1,\cdots, k-1 \,\,\,\,\,\text{and} \,\,\,\,\,i=1, \cdots, \ell.
\label{impmasses}
\end{equation}
\noindent Here we changed the notation for the masses of the fundamental particles in the obvious way
$m (\vec{\alpha}_i)=m_i$. Therefore, we expect to find $\ell \times (k-1)$ stable particles
associated to each HSG-model corresponding to a coset of the form $G_k/U(1)^{\ell}$.
The masses (\ref{impmasses}) coincide for each fixed simple root $\vec{\alpha}_i$ with
the mass spectrum encountered for the minimal $A_{k-1}$-ATFT and,
in fact, we may observe that the S-matrices describing 
the interaction of particles labeled by
the same simple root will be the same found for 
 the latter  theories \cite{Roland, ATFTS, dis, uni, ALO}.
Accordingly, we may label the stable solitons (\ref{impmasses}) by two  quantum numbers
\begin{equation}
A=(a, i),\,\,\,\,\,\,\,\,\,\,\,\,\,\,a=1, \cdots, k-1 \,\,\,\,\,\,\,\,\,\,\,\,\,i=1,\cdots, \ell.
\label{stablelab}
\end{equation}
\noindent notation which we borrowed from \cite{FK}, as we will see later. 
Due to the $\mathbb{Z}_2$-symmetry of the $A_{k-1}$-Dynking diagram each particle
$A$ will have an associated antiparticle $\bar{A}$ of the same mass and quantum numbers
$\bar{A}=(k-a, i)=\overline{(a,i)}$ similar to ATFT.
The distinction between particles and anti-particles
is possible in the HSG-models due to the existence of  classically  conserved charges associated
to even  values of  the spin, which have opposite values  for a particle and its anti-particle. 
Consequently, a particle and its associated anti-particles belong to 
different mass-multiplets, which turn out to be non-degenerate. For that
reason the S-matrices determined by J.L. Miramontes and C.R. Fern\' andez-Pousa 
in \cite{HSGS} are diagonal, as outlined before.

It is also interesting  to present here the mass ratios $M_{c}^{i}/M_{c}^j$,
\begin{equation}
\frac{m_i}{m_j}=\frac{M_{c}^{i}}{M_{c}^j}=
\sqrt{ \frac{(\vec{\alpha}_i \cdot \Lambda_{+}) (\vec{\alpha}_i \cdot \Lambda_{-})}
{(\vec{\alpha}_j \cdot \Lambda_{-}) (\vec{\alpha}_j \cdot \Lambda_{+})}}.
\label{ratios}
\end{equation}
\noindent Since like in ATFT \cite{ATFTS, dis, uni}, the classical mass ratios are expected to remain
preserved in the full quantum theory, this relation provides a direct link 
between full quantum and purely classical quantities.

\vspace{0.25cm}

{\bf ii)} Additionally, every particle of mass (\ref{masses}) associated to a positive non-simple
root, $\vec{\alpha}$, may be identified as an unstable particle and expected to decay into
 stable particles
associated to the simple roots arising in the decomposition (\ref{roott}). 
In order to distinguish between stable and unstable particles we may 
use the same notation (\ref{stablelab}) for unstable particles but add to their
quantum numbers  a `tilde'. 
For instance, let us consider an unstable particle $\tilde{C}=(\tilde{c}, \tilde{k})$ associated to a root
$\vec{\alpha}$  which admits the decomposition $\vec{\alpha}=\vec\alpha_i + \vec\alpha_j$, in terms
of simple roots. Then, one would expect to encounter a scattering process of the form
\begin{eqnarray}
A + B \rightarrow &\tilde{C}& \rightarrow A+B,\nonumber \\
&&\nonumber \\
& \Updownarrow& \nonumber \\
&& \nonumber \\
(a, i) + (b, j) \rightarrow 
&(\tilde{c}, \tilde{k})& \rightarrow (a, i) + (b, j)
\label{scatuns}
\end{eqnarray}
\noindent at quantum level.
 Following  section \ref{unstableparticles}, the presence of the
unstable particle $(\tilde{c}, \tilde{k})$ in the spectrum amounts to 
the existence of a corresponding
resonance pole in the scattering amplitude $S_{AB}(\theta):=S_{ab}^{ij}(\theta)$ for
a certain value of the rapidity $\theta_{R}=\sigma_{AB}-i\bar{\sigma}_{AB}$. Notice that,
with respect to the general expression (\ref{unstableres}) in subsection \ref{unstableparticles},
we have dropped out the upper index $\tilde{C}$ in $\sigma_{AB}$ and $\bar{\sigma}_{AB}$,
 which will simplify notations, without loss of information. 
In particular, for a scattering process like (\ref{scatuns}) we will
denote the resonance parameter $\sigma_{AB}:=-\sigma_{ij}$ and the imaginary part of $\theta_{R}$, 
$\text{Im}(\theta_{R})=\bar{\sigma}_{AB}:=\bar{\sigma}_{ij}$.
The scattering matrices associated to the formation of unstable particles have
a novel feature in the HSG-models which distinguishes them from most of the diagonal 
scattering matrices associated to  other 1+1-dimensional integrable massive QFT's. 
This  feature is the parity breaking $S_{AB}(\theta) \neq S_{BA}(\theta)$ whose
origin is explained below. 

 \vspace{0.3cm}
In \cite{HSGsol, HSGS} the following semi-classical 
relationship between the masses
of the particles involved in the scattering process (\ref{scatuns}) whenever their  quantum
numbers satisfy $a=b=c=\tilde{c}$  was pointed out
\begin{equation}
(M_{\tilde{c}}^{\tilde{k}})^2 = 
(M_{c}^{i})^2+ (M_{c}^j)^2+ 2 M_{c}^i M_{c}^j \cosh \sigma_{ij}.
\label{clasBW}
\end{equation}
\noindent Here the definition (\ref{massfun}) is crucial  and the same relation (\ref{clasBW})
in fact holds for the corresponding fundamental particles. The parameters $\sigma_{ij}$ depend
upon the roots and the elements $\Lambda_{\pm}$ as follows,
\begin{equation}
\sigma_{ij}=-\sigma_{ji}=\ln\,
\sqrt{ \frac{(\vec{\alpha}_i \cdot \Lambda_{+}) (\vec{\alpha}_j \cdot \Lambda_{-})}
{(\vec{\alpha}_i \cdot \Lambda_{-}) (\vec{\alpha}_j \cdot \Lambda_{+})}}\,\,\,.
\label{sig}
\end{equation}
\noindent Since the HSG-theories are not parity invariant whenever the elements $\Lambda_{\pm}$ are linearly
independent, we have to pay attention to the sign of the parameter $\sigma_{ij}$ in (\ref{clasBW}) which 
determines the sign of the rapidity difference $\theta_{c}^i - \theta_{c}^j := \theta_{ij}$ of the incoming
particles $(c,i),  (c,j)$ at which the particle $(\tilde{c}, \tilde{k})$ is produced. Remarkably, it was
shown in \cite{HSGS} that only the positive sign  i.e. $\sigma_{ij}>0$ is compatible with
the classical integrability of the HSG-models related to simply-laced algebras. This conclusion
was drawn on the basis of the explicit computation of the spin-$s$ classically conserved charges
associated to the solitonic solutions constructed in \cite{HSGsol} (see also the last reference in \cite{nt}).
Notice that we have chosen the same name for the parameter $\sigma_{ij}$ as for the resonance
parameter arising at quantum level when unstable particles are present in the theory. The reason
is  that these two parameters are indeed the same since the resonance
poles of the S-matrix will be located at the values $\theta_{R}=-\sigma_{ij}-\frac{i \pi}{k}$.
Therefore, in the semi-classical limit $k\rightarrow \infty$, $\theta_{R}\rightarrow -\sigma_{ij}$,
which is fully consistent with (\ref{clasBW}).

\subsubsection{The homogeneous sine-Gordon S-matrices}
\indent \ \
At this stage of the investigation the existence of a well-defined Lagrangian description for
the HSG-models can give support to the conjecture that the formation of an unstable
particle $(a, i) + (b, j) \rightarrow (\tilde{c}, \tilde{k})$ really occurs at quantum level. It 
must be emphasised that, although other interesting
1+1-dimensional massive integrable theories whose S-matrices possess resonance poles 
have been investigated in the  literature, they all lack a consistent Lagrangian 
description and have been formulated only from the point of view of scattering theory. 
Examples of this are  the {\bf roaming trajectory or roaming sinh-Gordon models}
formulated by Al.B. Zamolodchikov  in \cite{staircase}, their generalisations, due
to M.J. Martins \cite{Martins} and  P. Dorey and F. Ravanini \cite{DoRav}, and 
the theories possessing infinitely many resonance poles studied in \cite{staircase2}.

\vspace{0.25cm}
The construction of the exact S-matrices associated to the HSG-models
related to simply-laced Lie algebras carried out in \cite{HSGS} provides an interesting example
of  how the combination of various ingredients like, for instance, data extractable from
different approaches,  physically-motivated requirements or assumptions ultimately
justified by the self-consistency of the results obtained, may lead to the exact calculation
of two-particle scattering amplitudes within the context of 1+1-dimensional
massive integrable models. Notice that, according to the results presented in section \ref{analit},
the exact S-matrix associated to a 1+1-dimensional massive QFT whose mass spectrum 
is non-degenerate may always be  expressed in terms of certain building blocks of the form
(\ref{blocks}), independently of the concrete theory under consideration. In the light
of the results already presented in this section it is expected that the S-matrices of the HSG-models
are diagonal. Thus, the really unknown data one needs to extract 
is the pole structure of the two-particle
scattering amplitudes, which enters the building blocks (\ref{blocks}) through the complex numbers
$x$. As we also know, we expect to encounter two sorts of simple poles associated to real and complex
values of $x$ in the physical strip.  The techniques exploited in \cite{HSGS} in order to fix
the positions of the latter poles can be summarised as follows:

\vspace{0.25cm}
{\bf i)} Concerning the stable particle spectrum of the theory, in \cite{HSGS} the assumption that
the soliton spectrum (\ref{impmasses}) computed semi-classically coincides with the exact mass spectrum of the stable particles encountered in the quantum  theory has been  fundamental. Once this
assumption is made, the corresponding fusing angles associated to the formation of stable bound
states are fixed by the equations (\ref{cc}). We already pointed out after Eq. (\ref{impmasses})
that the masses of the stable solitons coincide precisely with the mass spectrum arising in 
minimal $A_{k-1}$-ATFT \cite{Roland}. In \cite{HSGS} it was found that, in fact,
also the positions of the fusing angles computed through (\ref{cc}) are the same than in 
$A_{k-1}$-ATFT. Therefore, the natural conclusion is that the interaction between stable
particles associated to the same simple root $\alpha_{i}$ is described by the two-particle
scattering amplitudes $S_{ab}^{ii}(\theta)$ associated to minimal  $A_{k-1}$-ATFT \cite{Roland}.
There are also other ways to determine these S-matrices like imposing $\mathbb{Z}_k$ invariance
of the scattering amplitudes \cite{Roland} or even using the knowledge of the values of the
spin for which there are quantum conserved quantities in the theory. The mentioned S-matrices 
have the form,
\begin{eqnarray}
S_{ab}^{ii}(\theta ) &=&(a+b)_{\theta }\,(|a-b|)_{\theta }\prod_{n=1}^{\min
(a,b)-1}(a+b-2n)_{\theta }^{2}\,\, \label{S} \\
&=&\exp \int \frac{dt}{t}2\cosh \frac{\pi t}{k}\,\left( 2\cosh \frac{\pi t}{k%
}-I\right) _{ab}^{-1}e^{-it\theta },  \label{Sint}
\end{eqnarray}
\noindent where the building blocks,
\begin{equation}
f_{x/k}(\theta):=(x)_{\theta }=\frac{\sinh \frac{1}{2}
(\theta +i\frac{\pi x}{k})}{\sinh \frac{1}{2}(\theta -i\frac{\pi x}{k})}
\label{newblocks}
\end{equation}
\noindent have been used. Notice that Eq. (\ref{Sint}) provides an integral
 representation for the scattering amplitudes which was derived in \cite{CFKM}
 and is part of the original results presented in this thesis.
 Such a representation
is more convenient for the thermodynamic Bethe anstaz analysis we will
carry out in the next chapter  and also in the context of  form
factors, which will become clear in chapters \ref{ffs} and \ref{ffs2}.  The calculation
of (\ref{Sint}) from the initial block expression (\ref{S})
may be performed by specialising the analysis in \cite{FKS1,FKS2} to the
particular case at hand (see also \cite{ravanini}). 
Although integral representations of scattering matrices
 appeared before in the literature (see e.g. \cite{Oota, ravanini}),
the analysis carried out in \cite{FKS2} goes further, providing a proof 
valid for all ATFT's of the
equivalence between such a representation and the usual representation
in terms of hyperbolic functions employed in \cite{HSGS} for the 
original construction of the S-matrices of the HSG-models. 

Notice the occurrence in (\ref{Sint}) of the {\bf incidence matrix} $I$, which
is defined in terms of the Cartan matrix, $K$, as $I:=2-K$. 
The conclusion that the scattering amplitude (\ref{S}) indeed 
describes the interaction
between solitons labeled by the same simple root $\alpha_i$ 
was also checked for consistency
by exploiting the knowledge of the HSG-Lagrangian, which 
allows for performing perturbation
theory in the coupling constant, $\beta^2$. 
The  key observation is that  the results obtained from the tree-level calculation 
(see first reference in  \cite{ATFTS, dis})
of the scattering amplitude (\ref{S})
must be recovered from the exact expression  (\ref{S})  upon the substitution
$\beta^2=1/k$ and taking $k\rightarrow \infty$ thereafter, that is, in the semi-classical limit.
This has been explicitly confirmed in \cite{HSGS}.

\vspace{0.25cm}

{\bf ii)} What is now left is the computation of the scattering amplitudes which involve the formation
of unstable bound states. 
 Having the Lagrangian (\ref{Actq}) at hand one may perform perturbation theory in the coupling
constant $\beta^2$ in order to determine the decay width associated to the process (\ref{scatuns}) together with the two-particle scattering amplitude in the tree-level approximation. The outcome of this
calculation, whose details may be found in  \cite{HSGS}, 
allowed for checking that the process  (\ref{scatuns}) corresponds in fact
to the formation of an unstable bound state in the quantum theory and permitted the subsequent
deduction of  the position of the corresponding resonance poles arising in the scattering
amplitudes $S_{ab}^{ij}(\theta)$. The latter result was obtained by using the fact  that 
the general equation (\ref{BW11}) for an unstable particle should reduce to Eq. (\ref{clasBW})
in the $k\simeq 1/\beta^2 \rightarrow \infty$ limit and when considering a particle of very long
lifetime, namely $\Gamma_{\tilde{c}}^{\tilde{k}} \ll M_{\tilde{c}}^{\tilde{k}}$. At the same
time Eq. (\ref{BW22}) must give in the same limit the decay width computed perturbatively. 
These constraints lead the authors of \cite{HSGS} to the following result,

\begin{eqnarray}
S_{ab}^{ij}(\theta ) &=&(\eta _{ij})^{ab}\prod\limits_{n=0}^{\min
(a,b)-1}(-|a-b|-1-2n)_{\theta +\sigma _{ij}}\,,\quad \quad \quad
K_{ij}^{g}\neq 0,2  \label{ij} \\
&=&(\eta _{ij})^{ab}\exp -\int \frac{dt}{t}\,\left( 2\cosh \frac{\pi t}{k}%
-I\right) _{ab}^{-1}e^{-it(\theta +\sigma _{ij})},\,\,\, K_{ij}^{g}\neq 0,2\,,
\label{ijint}
\end{eqnarray}
with $K^{g}$ denoting the Cartan matrix of the simply laced Lie algebra $g$.
Here the $\eta _{ij}=\eta _{ji}^{*}$ are arbitrary $k$-th roots of $-1$
taken to the power $a$ times $b$ which have to be introduced in order
to guarantee that the bootstrap equations (\ref{boots}) are fulfilled by these
S-matrices. 
The shifts in the rapidity variables
are functions of the vector couplings $\sigma _{ij}$ given by Eq. (\ref{sig}). 
Due to the fact that these shifts are real, the function $S_{ab}^{ij}(\theta )$ for $%
i\neq j$ will have poles beyond the imaginary axis at the positions 
\begin{equation}
(\theta_{R})_{ij}=\sigma_{ji}-\frac{i\pi x}{k},
\end{equation}
\noindent with $x= |a-b|+1+2n$ such that the parameters $\sigma _{ji}$ characterise resonance poles
as we anticipated before. Therefore, the S-matrix (\ref{ij}) is
not parity invariant and the parity breaking takes place through the phases
$\eta_{ij}$ and the resonance parameters $\sigma_{ij}$.
It can be proven  that the latter amplitudes satisfy all the physical requirements
presented in section \ref{analit}. In particular, due to the parity breaking,
the amplitudes (\ref{ij}) are not real analytic but satisfy the condition of 
Hermitian analyticity given by Eq. (\ref{HS}) \cite{HERMAN, Mir, TW}.
The integral representation (\ref{ijint}) was  found in \cite{CFKM} in the context
of the thermodynamic Bethe ansatz we will present in the next chapter.

\vspace{0.25cm}

{\bf iii)} The remaining scattering amplitudes where shown to be $1$, namely
particles associated to quantum numbers $(a, i)$, $(b, j) $, with 
 $K_{ij}^{g}=0$  will interact freely. 

\vspace{0.25cm}
In the light of (\ref{S}) and (\ref{ij}) we can conclude that the proposed scattering
matrix \cite{HSGS} for the HSG-models related to simply laced Lie algebras consists
partially of $\ell$ copies of minimal $A_{k-1}$-ATFT \cite{Roland},
whose mutual interaction is characterised by the S-matrix (\ref{ij}), which violates parity. 

\subsubsection{How many free parameters do we have in our model?}
\label{howmany}
\indent \ \
The number of free parameters we have at our disposal will play
a very important role in the course of  the TBA- and form factor analysis 
since it determines the amount of plateaux we
may observe when computing numerically the
finite size scaling function and Zamolodchikov's $c$-function \cite{ZamC}. 
For this reason,
we want to devote a brief subsection to the computation
of the amount of free parameters we expect to find in
the HSG-models under consideration.
 
First of all, computing mass shifts from
renormalisation, we only expect to accumulate contributions
from intermediate states
having the same colour as the two scattering solitons. Thus, making use of
the well known fact that the masses of the minimal $A_{k-1}$-affine Toda
theory all renormalise with an overall factor 
(see the first two references in \cite{ATFTS, dis}), 
i.e.  for the solitons $(a, i)$ we have that $\delta M_{a}^{i}/M_{a}^{i}$ equals a constant
for fixed colour value $i$ and all possible values of the main quantum
number $a$, we acquire in principle $\ell$ different mass scales 
$m_{1},\ldots ,m_{\ell }$ in the HSG-models. 

The previous argument is also in perfect agreement with the counting
of free parameters which enter the S-matrix construction. As we have
seen in detail before,  we have  $\ell$ different mass scales
characterising the fundamental particles of the model (in one-to-one
correspondence to the simple roots of $g$). In addition we find $\ell-1$
independent resonance parameters in the theory,
$\sigma_{ij}=-\sigma _{ji}$ for each $i,j$ such that $K_{ij}^{g}\neq 0,2$,
characterising resonance poles in the scattering amplitudes which
are interpreted as the trace of the presence of unstable particles in the spectrum.
Such an interpretation will be later confirmed by our TBA- and
form factor analysis. 

This means overall we have $2\ell -1$ independent parameters in the quantum
theory. There is a precise correspondence to the free parameters which one
obtains from the classical point of view. In the latter case we have the $%
2\ell $ independent components of $\Lambda _{\pm }$ at our free disposal.
This number is reduced by $1$ as a result of the symmetry $\Lambda
_{+}\rightarrow c\Lambda _{+}$ and $\Lambda _{-}\rightarrow c^{-1}\Lambda
_{-}$ which introduces an additional dependence as may be seen from the
explicit expressions for the classical mass ratios and the classical
resonance shifts given by Eqs. (\ref{ratios}) and (\ref{sig}). The masses
of the stable particles of the model which will enter the TBA-equations
(\ref{fin}) are given by Eq. (\ref{impmasses}). 

As we have already said, the number of free parameters at our disposal, that is $2 \ell-1$, 
will determine the maximum number of plateaux we may observe when computing the 
corresponding finite size scaling function. Such a result will also provide further 
support for the physical picture anticipated in \cite{HSGsol, HSGS} for the HSG-models. 

\vspace{0.3cm}

The exact computation of the S-matrix 
presented in this subsection closes our general review of the quantum
properties of the HSG-models. As stressed before, one of the main results
presented in this thesis  is concerned with the development
of consistency checks of the latter S-matrix proposal and also with a further
development of  the full QFT. The mentioned S-matrices are
fully consistent at this point but their construction  is based, as we have seen, 
on certain assumptions. Our consistency checks have been
carried out within two different non-perturbative approaches known
as thermodynamic Bethe ansatz \cite{Yang, TBAZam1} and
form factor approach \cite{Kar, Smir}. We will see in later chapters
that all our results indeed confirm the consistency of the S-matrix
proposal \cite{HSGS}, meaning that all the relevant data characterising the
underlying CFT which are extractable in those two approaches will
be entirely consistent with our expectations based on the knowledge of the
properties of the underlying coset CFT \cite{Gep, GepQ, DHS}.

\vspace{0.25cm}

We will now describe the main results obtained
for the other family of massive integrable non-abelian Toda field theories
mentioned in this section, the SSSG-models. Some of the classical properties
as well as quantum aspects of these theories 
have been studied in \cite{cm, tesinha} and we shall report very
briefly in the next subsection our most prominent results. Although we will
not go into detail about this second type of theories, 
it is mainly our intention to emphasise the richness of  different
massive non-abelian  affine Toda field theories available.
The latter statement should be clear 
in the light of Eqs. (\ref{hsg22}) and (\ref{sssg22}) and, in particular,
the second of these equations shows that the SSSG-models represent
a really huge class of theories whose quantum study has only been carried
out to a very small extent in comparison to the HSG-theories described
above. The interest of carrying out a more extensive study of these theories
will be motivated below. 

\subsection{The symmetric space sine-Gordon models}
\label{SSSG}
\indent \ \
In the same fashion the HSG-theories are in one-to-one correspondence with
the finite, semisimple and compact Lie algebras, $g$, the classification of the SSSG-models 
exploits the existence of a correspondence between  the SSSG-theories  associated
to a Lie algebra $g= g_0 \oplus g_1$ and the 
compact symmetric spaces $G/G_0$, for $G$, $G_0$ to be the corresponding Lie
groups associated to the Lie algebras $g, g_0$,  which will be better justified later.
 The SSSG-models admit also a well-defined Lagrangian formulation  (\ref{Actq}),
being now $h(x^0, x^1)$ a $G_0$-valued field.
As we have also reported, these theories describe
perturbations of either the WZNW CFT corresponding to $G_0$ or a coset
CFT of the form $G_0/H$, where $H\simeq U(1)^p$ is a torus of $G_0$, not
necessarily maximal. The equations of motion of this kind of theories for more
general choices of the normal subgroup $H$ were originally considered in the
context of the so-called reduced two-dimensional $\sigma$-models~\cite{sssg1},
although their Lagrangian formulation was not known until much
later~\cite{sssg3}. The results of~\cite{HSG,HMQH} show that they fit quite
naturally into the class of non-abelian affine Toda theories and, what is more
important, that the condition of having a mass gap requires that $H$ is either
trivial or abelian.  The simplest SSSG theories are the ubiquitous sine-Gordon
field theory, which corresponds to $G/G_0 = SU(2)/SO(2)$, and the
complex sine-Gordon theory, which is related this time to $Sp(2)/U(2)$~\cite{HMQH}
(recall that it is also the HSG theory associated to $SU(2)$). Actually, these
two theories serve as paradigms of what can be expected in more complex
situations. Other theories already discussed in the literature that belong to
the class of SSSG theories are the integrable perturbations of the $SU(2)_k$
WZNW model and its $\widetilde{so(2)}$ reduction constructed by
V.A. Brazhnikov~\cite{brazhnikov}. Both of them are related to the symmetric space
$SU(3)/SO(3)$ and, moreover, the second is identified with the perturbation of
the usual $\mathbb{Z}_k$-parafermions by the second thermal operator.

\vspace{0.25cm}

The classification of the SSSG theories as perturbed CFT's is achieved through
the calculation of the central charge of the unperturbed CFT and
the conformal dimension of the perturbation. Since the unperturbed CFT is always
a coset CFT of the form $G_0/H$,  its central charge 
is given by the usual expression (\ref{cdata}) \cite{GKO}.
In contrast, the calculation of the conformal dimension of
the perturbation requires the knowledge of the structure of the symmetric
space. 

A {\bf symmetric space $G/G_0$} is associated with a Lie algebra
decomposition $g= g_0\oplus g_1$ that satisfies the commutation relations
\begin{equation}
[g_{0}\>,\> g_{0}]\subset g_{0}\>, \quad
[g_{0}\>,\>g_{1}]\subset g_{1}\>, \quad
[g_{1}\>,\>g_{1}]\subset g_{0}\>,
\lab{gradation}
\end{equation}
\noindent that is, precisely Eq. (\ref{n2}), which justifies in retrospect our
statement concerning the correspondence between SSSG-theories and
compact symmetric spaces.
Then, the conformal properties of the perturbation depend on the
structure of the representation of $g_0$ provided by $[g_{0},g_{1}]\subset
g_{1}$. First of all, if the perturbation is to be given by a single primary
field, then this representation has to be irreducible. This amounts to
restrict the choice of $G/G_0$ to the so-called `irreducible symmetric
spaces'~\cite{helgas}, which have been completely classified by Cartan and
are labeled by type~I and type~II. 

\vspace{0.25cm}

One of the most important results of our work \cite{cm, tesinha} 
has been  the explicit computation of the
conformal dimension of the perturbation corresponding to all the SSSG models
related to type~I symmetric spaces. This calculation  can be done by making use of the
relationship between the classification of type~I symmetric spaces and the
classification of the finite order automorphisms of complex Lie algebras and our
results are reported in tables \ref{class} and \ref{except}, for the cases when the Lie
 algebra, $g$, is classical and exceptional, respectively. It is
worth noticing that this analysis only depends on the structure of the
representation of $g_0$ given by $[g_{0},g_{1}]\subset g_{1}$. Therefore, our
results apply to any SSSG related to a type~I symmetric space,
irrespectively of the choice of the normal subgroup $H$ that determines the
coset $G_0/H$ and specifies the underlying CFT. For example, they provide the
conformal dimension of the perturbation in the SSSG models
constructed by I. Bakas,  Q-H. Park and H-J. Shin in~\cite{sssg3}, which include
the generalised sine-Gordon models related to the NAAT-equations based on $sl(2)$
embeddings constructed by T. Hollowood, J.L. Miramontes and Q-H. Park in~\cite{HMQH}.
It is also worth noticing that all the conformal dimensions
reported tables \ref{class}, \ref{except}
decrease as $k$ increases, which means that $\Delta <1$ (the perturbation is relevant)
above some minimal value of $k$ which is characteristic of each SSSG-theory. It is also
clear that all the conformal dimensions obtained vanish in the semi-classical or weak 
coupling limit $k \rightarrow \infty$ which, according to Eq. (\ref{cdata})
shows that the theory consists of $\dim g_0 -p$ massive bosons in this limit.

\vspace{0.25cm}

In \cite{cm} it was also established the classical and quantum integrability of the
subset of SSSG-theories which we named before as {\bf Split models}. Recall that these
models provide integrable perturbations of  WZNW-models related to compact Lie
groups $G_0$. It is expected that such integrability, which is ensured classically
for all NAAT-theories in virtue of Eq. (\ref{zerocur}), extends to the quantum
theory for all SSSG-models. Therefore, the latter family of models is expected
to provide a huge class of new 1+1-dimensional integrable massive QFT's which,
in virtue of the general discussion of sections \ref{cftqft}, \ref{exactS} and \ref{analit}, is
by itself a motivation for their further study.

\vspace{0.25cm}

There are however  various other reasons that justify the interest of
a more extensive study of the quantum properties of the SSSG-theories: 

\vspace{0.25cm}

{\bf i)} First of all, although  some already known
integrable massive QFT's are encountered 
as particular examples of SSSG-models, most of the SSSG-theories 
provide completely new perturbations of
WZNW-coset theories (or of WZNW-theories for the case $p=0$ in Eq. (\ref{sssg22})) which,
in view of the results obtained in \cite{HSG2, cm},  are expected to be quantum integrable. 
Therefore, all the study performed for the HSG-models could be generalised to the SSSG-theories and,
in particular, they are expected to admit an S-matrix description at quantum level. In particular,
in \cite{cm} part of the solitonic spectrum of the mentioned Split models was explicitly constructed. 
Also their classical and quantum integrability were established along the same 
lines as for the HSG-theories. In particular, the non-existence of classically
conserved quantities which are associated
to even values of the spin  was shown . Recall that the existence of these
charges allowed in \cite{HSGS}
for the distinction between particles and anti-particles and was on the origin of the diagonal
character of the S-matrix. Thus, as a novel feature, 
the S-matrices associated to  the SSSG-models are expected
to be in general non-diagonal, which will certainly make their construction more involved.

\vspace{0.25cm}
{\bf ii)} The SSSG-models are expected to present the same type of
distinguished features encountered
for the HSG-models with respect to other  integrable QFT's. In particular, their S-matrices may
break parity and  have also resonance poles associated to unstable particles. The 
presence of unstable particles in the spectrum makes these theories capture realistic properties
of QFT which are not often available in other bi-dimensional integrable models. Obviously,
there is also a well-defined Lagrangian formulation for the SSSG-models. The 
parity breaking and presence of unstable particles in the spectrum have also specific consequences
in the thermodynamic Bethe ansatz and form factor framework  which further
support  the interest of these models. In particular, we expect to encounter 
the characteristic ``staircase'' pattern that we will find later for the scaling function
of the HSG-models. 

\vspace{0.25cm}
{\bf iii)} Additionally, in \cite{cm} soliton solutions associated to a certain
subset of the SSSG-models were found. For the models studied there these
solutions have been shown to carry topological charges, similarly to the ones 
which characterise the sine-Gordon soliton and multi-soliton solutions
(see for instance \cite{raja}). This in contrast to the soliton solutions
constructed for the HSG-models \cite{HSGsol}, which carry Noether charges
associated to the global $U(1)^{\ell}$ symmetry of the Lagrangian. However, for general
SSSG-models, it was pointed out in \cite{cm} that, in fact, it is expected
to encounter soliton solutions which may carry at the same time topological
and Noether charges namely, ``Dyon-like'' solutions \cite{dyon} to the classical equations
of motion. To our knowledge, such feature has not been encountered before
in the literature for other 1+1-dimensional massive integrable QFT and provides
additional motivation for a further study of the SSSG-models.

\subsection{The $\protect{g|\tilde{g}}$-theories.}
\label{ggtilde}
\indent \ \
To close this chapter, we want to report some results concerning a family of theories
recently proposed by A. Fring and C. Korff in \cite{FK} which contains the HSG-models
and minimal ATFT's as particular examples and can be understood, in fact,
 as  a generalisation of them
(see also \cite{KK, FF}). The particles arising in these models
are labeled by two quantum numbers $(a, i)$, notation which we have borrowed in this
thesis  for labeling the HSG-solitons. Each of these quantum numbers is associated to
a  simply-laced Lie algebra in such a way that $a=1, \cdots, \ell$,  for $\ell$ to be rank of a Lie
algebra ${g}$ and $i=1,\cdots, \tilde{\ell}$, for $\tilde{\ell}$ to be the rank of  a Lie algebra
$\tilde{g}$. Consequently, the mentioned models have been named as {\bf $g| \tilde{g}$-theories}.
The corresponding two-particle scattering amplitudes will be denoted by
$S_{ab}^{ij}$, notation which we have also used in this thesis for the HSG S-matrices
and which slightly differs from the one used in the original literature \cite{HSGS}.
The construction of the $g| \tilde{g}$-theories provided in \cite{FK} has been very
recently generalised by C. Korff to the case when the $\tilde{g}$-algebra is non simply-laced \cite{KK2}.

The first quantum numbers $a, b$ govern the fusing structure and are
usually referred to as {\bf main quantum numbers} while the second quantum numbers $i, j$
will be named as {\bf colours} and  some of the scattering amplitudes
corresponding to $i\neq j$ will  break parity.

\vspace{0.25cm}

It is worth noticing that the construction  of the $g|\tilde{g}$-theories carried out in \cite{FK}
follows very different lines to what we have seen for the NAAT-models.
The starting point in \cite{FK} is not the existence of
a consistent Lagrangian  formulation but a new S-matrix proposal which
satisfies all the physical requirements presented in section \ref{analit} and therefore, is 
perfectly consistent form the scattering point of view.  Consequently, a Lagrangian
formulation is not known for the time being for a large subset of the 
$g|\tilde{g}$-theories, apart from the particular cases 
\begin{eqnarray}
A_{k-1}| \tilde{g}\equiv \,\,\,\,\,&&\tilde{G}_k\,\,\text{HSG-models},\label{gghsg}\\
g | A_1 \equiv \,\,\,\,\,\,\,\,\,\,&&\text{minimal ATFT}\label{minatft},
\end{eqnarray}

\vspace{0.25cm}
The S-matrix proposal presented in \cite{FK} takes advantage of the fact that
for many theories the structure of the two-particle scattering amplitudes is of the form
\begin{equation}
S_{AB}(\theta)=S_{AB}^{\text{min}}(\theta) S_{AB}^{\text{CDD}}(\theta, B).
\end{equation}
\noindent This is the case, for instance,
for all ATFT's related to simply-laced Lie algebras \cite{ATFTS, dis} and also
for the HSG-models. The first piece, $ S_{AB}^{\text{min}}(\theta)$ 
is the so-called {\bf minimal S-matrix},
which satisfies by itself all the physical requirements summarised in section \ref{analit}.
The second part,  $S_{AB}^{\text{CDD}}(\theta, B)$, is a so-called {\bf CDD-factor} \cite{CDD},
namely a function which satisfies trivially the conditions summarised in section \ref{analit} and
at the same time does not have poles in the physical sheet. It depends on the effective coupling
constant $B$.

\vspace{0.25cm}

The S-matrices characterising the $g|\tilde{g}$-theories were proposed in \cite{FK} to be,
\begin{eqnarray}
S_{ab}^{ij}(\theta)&=&S_{ab}^{\text{min}}(\theta), \,\,\,\,\,\text{for $i=j$,}\nonumber\\
S_{ab}^{ij}(\theta)&=&S_{ab}^{\text{CDD}}(\theta, B_{ij}),\,\,\,\,\,\,\text{for $i\neq j$}.
\label{abij}
\end{eqnarray}
\noindent where as a novel feature the mentioned substructure $A=(a, i)$  for the
particle quantum numbers was introduced. Accordingly, there are now a set of effective
coupling constants $B_{ij}$ labeled by colour quantum numbers and satisfying $B_{ii}=0$.
In \cite{FK} it was shown that the previous S-matrix proposal is perfectly consistent, meaning
that, it satisfies all the necessary physical requirements. This is guaranteed by its definition
(\ref{abij}), since both the minimal and CDD parts satisfy these requirements by themselves.
Therefore every  $g|\tilde{g}$-theory contains a set of $\ell \times \tilde{\ell}$ stable
particles and is characterised by the existence of  $\tilde{\ell}$ different mass scales.
The corresponding S-matrix  is given by (\ref{abij}) and its explicit form
can be found in \cite{FK}, where an integral representation, very
useful in many contexts, was provided.

\vspace{0.25cm}
The thermodynamic Bethe ansatz (see chapter \ref{tba}) carried out in \cite{FK}
revealed that if as usual  we view the $g |\tilde{g}$-theories as perturbed conformal
field theories, the Virasoro central charge associated to the underlying CFT will have the general form
\begin{equation}
c_{g| \tilde{g}}=\frac{\ell \tilde{\ell}\tilde{h}}{h+\tilde{h}},
\end{equation}
\noindent where $h, \tilde{h}$ are the Coxeter numbers of the Lie algebras $g, \tilde{g}$ respectively. 
The latter central charges coincide with the ones found in \cite{DHS} for a class of 
parafermionic theories which can be interpreted as generalisations of the ones proposed
in \cite{Gep}. Also at the conformal level, when $g$ is taken to be not $A_n$, 
there exist so far no Lagrangian description.

\begin{table}[h]
\begin{center}
\begin{tabular}{|c|c|c|}\hline
&&\\
$G/{G_0}$&$\ell_{G/G_0}$&$\Delta$\\
&&\\
\hline\hline
&&\\
$SU(3)/SO(3)$&2& $\frac{6}{k+2}$\\
&&\\
\hline
&&\\
$SU(4)/SO(4)$&3&$\frac{4}{k+2}$\\
&&\\
\hline
&&\\
$SU(n)/{SO(n)}$, \,\, $(n \geq 5)$&$n-1$&$\frac{n}{2(k+n-2)}$\\
&&\\
\hline
&&\\
$SU(2n)/Sp(n)$, \,\, $(n \geq 1)$&$ n-1$&$\frac{n}{k+n+1}$\\
&&\\
\hline
&&\\
${SO(n+2)}/{SO(n) \times U(1)}$, \,\, $(n \geq 5)$&
$2$&
$\frac{n-1}{2(k+n-2)}+\frac{1}{2k}$\\
&&\\
\hline
&&\\
${SO(n+3)}/{SO(n) \times SO(3)}$, \,\, $(n \geq 4)$&
3&
$\frac{n-1}{2(k+n-2)}+\frac{1}{k+2}$\\
&&\\
\hline
&&\\
${SO(n+m)}/{SO(n) \times SO(m)}$, \,\, $(n, m \geq 4)$&
min$(n,m)$&
$\frac{n-1}{2(k+n-2)}+\frac{m-1}{2(k+m-2)}$\\
&&\\
\hline
&&\\
$Sp(n+m)/{Sp(n) \times Sp(m)}$, \,\, $(n,m  \geq 1)$&
min$(n,m)$&
$\frac{1+2n}{4(k+n+1)}+\frac{1+2m}{4(k+m+1)}$\\
&&\\
\hline
&&\\
${SU(n+m)}/{SU(n)\times SU(m)\times U(1)}$,& min$(n,m)$&
$\frac{(n-1)(n+1)}{2n(k+n)}+
\frac{(m-1)(m+1)}{2m(k+m)}+\frac{n+m}{2nmk}
$\\
$(n, m \geq 2)$& &\\
&&\\
\hline
&&\\
${Sp(n)}/{SU(n) \times U(1)}$, \,\, $(n  \geq 2)$&n&
$\frac{(n-1)(n+2)}{n(k+n)}+\frac{2}{nk}$\\
&&\\
\hline
&&\\
${SO(2n)} /{SU(n) \times U(1)}$, \,\, $(n  \geq 3)$&$[n/2]$&
$\frac{n^2-n-4}{2n(k+n-1)} + \frac{2}{nk}$\\
&&\\
\hline
\end{tabular}
\end{center}
\caption{Conformal dimensions of the perturbations corresponding to the type~I 
SSSG-models associated to the classical Lie groups $G$.}
\label{class}
\end{table}

\begin{table}[h]
\begin{center}
\begin{tabular}{|c|c|c|}\hline
&&\\
$G/{G_0}$&$\ell_{G/G_0}$&$\Delta$\\
&&\\
\hline\hline
&&\\
${E_6}/{Sp(4)}$& 6&
 $\frac{6}{k+5}$\\
&&\\
\hline
&&\\
${E_6}/{F_4}$&2&$\frac{6}{k+9}$\\
&&\\
\hline
&&\\
${E_7}/{SU(8)}$&7&$\frac{9}{k+8}$\\
&&\\
\hline
&&\\
${E_8}/{SO(16)}$&8&$\frac{15}{k+14}$\\
&&\\
\hline
&&\\
${F_4}/{SO(9)}$&1&$\frac{9}{2(k+4)}$\\
&&\\
\hline
&&\\
${E_6}/{SU(6) \times SU(2)}$&4&
$\frac{21}{4(k+6)}+\frac{3}{4(k+2)}$
\\
&&\\
\hline
&&\\
${E_7}/{SO(12)\times SU(2)}$&4&$
\frac{33}{4(k+10)}+ \frac{3}{4(k+2)}$\\
&&\\
\hline
&&\\
${E_8}/{E_7 \times SU(2)}$&4&$\frac{57}{4(k+18)}+ \frac{3}{4(k+2)}$\\
&&\\
\hline
&&\\
${F_4}/{Sp(3)\times SU(2)}$&4&
$\frac{15}{4(k+4)}+\frac{3}{4(k+2)}$
\\
&&\\
\hline
&&\\
${G_2}/{SU(2)\times SU(2)}$&2
&$\frac{15}{4(3k+2)}+\frac{3}{4(k+2)}$
\\
&&\\
\hline
&&\\
${E_6}/{SO(10) \times U(1)}$& 2& $ \frac{767}{128(k+8)}+
 \frac{1}{128k}$\\
&&\\
\hline
&&\\
${E_7}/{E_6 \times U(1)}$& 3&  $ \frac{527}{72(k+12)}+
\frac{1}{72k}$\\
&&\\
\hline
\end{tabular}
\end{center}
\caption{Conformal dimensions of the perturbations corresponding to the 
type~I SSSG-models associated to the exceptional Lie groups $G$.}
\label{except}
\end{table}

\chapter{Thermodynamic Bethe Ansatz of the Homogeneous sine-Gordon models}
\label{tba}
\indent \ \
The thermodynamic Bethe ansatz (TBA)\ is established as an important method
which serves to investigate ``off-shell'' properties of 1+1 dimensional
quantum field theories (QFT's). Originally formulated in the context of the
non-relativistic Bose gas by Yang and Yang \cite{Yang}, it was extended
thereafter by Zamolodchikov \cite{TBAZam1} to relativistic quantum field
theories whose scattering matrices factorise into two-particle ones. The
latter property is always guaranteed when the QFT in
question is integrable, as explained in the previous chapter.
 Provided the S-matrix has been determined in some
way, for instance via the bootstrap program \cite{Boot, KTTW}
and/or by extrapolating semi-classical results as it is the case
 for the Homogeneous sine-Gordon (HSG) models, the TBA allows for calculating the ground state energy
of the integrable model on an infinite cylinder whose circumference is
identified as compactified time direction. When the circumference is sent
to zero the effective central charge of the conformal field theory (CFT)\
governing the short distance behaviour can be extracted. In the case in
which the massive integrable field theory is obtained from a conformal model
by adding a perturbative term which breaks the conformal symmetry, the TBA
constitutes an important consistency check for the S-matrix since it allows
for the explicit computation of  the Virasoro central charge associated to the underlying CFT.

The main purpose of this chapter  is to apply this technique to the
scattering matrices described in subsection \ref{hsgsec} of the previous chapter,
which have recently been proposed \cite{HSGS} to
describe the HSG-models  \cite{HSG, HSG2, HSGsol, Park, HMQH} related to simply laced Lie algebras. 
Within the TBA-context, we will  investigate  the physical picture
anticipated for these models in the original literature  i.e., 
we want to check for consistency the S-matrix proposal \cite{HSGS}.

Although we have already summarised in the previous chapter
 the main features of the HSG-models, it is worth
to emphasise again before entering our specific TBA-analysis, that the HSG-models
possess very remarkable properties which distinguish them  from  many
other 1+1-dimensional integrable quantum field theories studied in the literature so far, 
and that these properties have specific consequences in the TBA-analysis:
\vspace{0.25cm}

 {\bf i)} some of the two-particle S-matrices violate parity \cite
{HSGS}, namely $S_{AB}(\theta) \neq S_{BA}(\theta)$ in general. Such
parity violation is intimately related in these theories to the existence of resonance poles
in the S-matrix which, for the time being, we will assume are the only trace of the presence
of unstable particles in the spectrum of a 1+1-dimensional integrable 
QFT. This feature i.e., the parity violation, will reflect itself
 in various ways along  the TBA-analysis as, for instance, we may have to distinguish
two sets of different TBA-equations (see section \ref{resonance}) or the so-called $L$-functions
entering the TBA-analysis which are symmetric functions of the rapidity in parity 
invariant theories will not be symmetric anymore once any of the resonance parameters
$\sigma_{ij}$ characterising the mass scale of the unstable particles is  different from zero.

\vspace{0.25cm}

{\bf ii)} the mentioned existence of unstable particles in the spectrum is also a remarkable
feature by itself. Although the presence of resonance poles in the S-matrix has been observed 
also for other models like the roaming sinh-Gordon model originally introduced by 
Zamolodchikov in \cite{staircase} and generalised thereafter in \cite{Martins, DoRav}, and also in 
\cite{staircase2, new} where S-matrices containing an infinite number of resonance poles have been
proposed, the HSG-models are distinguished  essentially in two aspects. First, as we said,  they break
parity invariance and second, the HSG-models are not only consistent  from the scattering theory
point of view but also allow for a Lagrangian description \cite{HSG} which
makes possible the direct identification of some of the resonance poles with  unstable particles
 \cite{HSGS} by means of a  semi-classical analysis \cite{HSGsol, HSGS}.
 However, the existence of resonance poles
has similar effects in the outcome of the TBA-analysis for the HSG-models and for the models
studied in \cite{staircase, Martins, DoRav}. 
In particular we shall observe that  the finite size scaling function 
computed in the TBA turns out to have for the HSG-models  a  ``staircase-like" behaviour,
similar to the one originally  observed  for  the roaming sinh-Gordon model 
\cite{staircase} and for its generalisations \cite{Martins, DoRav}. However, 
for the HSG-models such a behaviour admits a very nice physical interpretation
when the resonance poles are understood
as the trace of the presence of unstable particles in the spectrum. In this case the 
scaling function gives information about the energy scale at which
the onset of every stable or unstable particle in the model takes place.
This is a consequence of the fact pointed out in the previous chapter
that the  plateaux in the scaling function are directly related
to the amount of free parameters in the model. Such a relationship 
does not hold neither for the roaming trajectory models \cite{staircase},
nor for the generalisations studied in \cite{Martins, DoRav}. For these
theories infinitely many plateaux arise in the scaling functions, although
the amount of free parameters available is finite.

\vspace{0.3cm}

The main  results presented in this chapter can be found in \cite{CFKM} 
(see also \cite{KK, MM, FF}). The chapter is organised as follows:

\vspace{0.2cm}

In section \ref{gentba} we recall the basic ideas entering the thermodynamic
Bethe ansatz approach \cite{Yang, TBAZam1, KM2}:
 in subsection \ref{BWF} we start by studying a 1+1-dimensional
integrable QFT in a circumference whose dimension $L$ will
be sent to infinity in the so-called thermodynamic limit which we consider in subsection \ref{tl}.
The constraints for the momenta of the particles arising as a consequence of the introduction of 
periodic boundary conditions  lead in the thermodynamic limit, when the
equilibrium configuration is considered, to a set of coupled non-linear integral equations 
which are known as thermodynamic Bethe ansatz equations. They are presented 
also in subsection \ref{tl} for a general  1+1-dimensional integrable QFT. 
The study of the QFT in the thermodynamic limit at finite temperature is shown to be equivalent
to  its  formulation on an  infinite cylinder whose circumference is identified
 as the compactified time dimension. In the ultraviolet limit,  
we will derive a crucial relation between the equilibrium free energy 
of the QFT at finite temperature and the Virasoro central charge of the corresponding  underlying CFT. 
The generalisation of this relationship from the critical
point  to the  ``off-critical'' 
situation leads to the definition of the finite size scaling function as a 
measure of effective light degrees of freedom in a QFT and closes section \ref{gentba}.
 In section \ref{resonance} we introduce the TBA equations for a
parity violating system and carry out the ultraviolet limit recovering the expected coset central charge. 
We also justify and present  in detail the approximations leading to the analytical calculation
of the latter coset central charge.  Section \ref{su3} is devoted to a detailed study of the TBA for the 
$SU(3)_{k}-$HSG model. In the subsequent subsections we discuss the ``staircase'' pattern of the scaling
function and illustrate how the UV-limit for the HSG-model may be viewed as
the UV-IR flow between different conformal models. We extract the
ultraviolet central charges of the HSG-models. We study separately the case
when parity is restored, derive universal TBA-equations and $Y$-systems. 
In section \ref{examples} we illustrate our analysis with several explicit examples which
confirm the preceding analytical arguments. In particular the semi-classical limit $k \rightarrow \infty$ is studied in subsection \ref{semi}. Finally, 
we state our conclusions and point out some open questions
in section \ref{con3}.

\section{The thermodynamic Bethe ansatz approach}
\label{gentba}
\indent \ \
Before we attempt to carry out our specific  TBA-analysis for the HSG-models
 we must introduce the basic ideas necessary for the understanding of 
 the thermodynamic Bethe ansatz approach. The key points  have already
 been anticipated very briefly in the introduction as well as the main motivation which
justifies the interest of the TBA-approach in the context of 1+1-dimensional
 integrable quantum field theories. We may view a 1+1-dimensional 
massive integrable QFT as a certain
 integrable perturbation of a CFT by means of a relevant operator of the original or underlying CFT.
 Taking the original theory to be conformally invariant,
there exists an infinite number of conserved quantities 
associated to it of which certain combinations
might  remain conserved in the perturbed QFT. For that reason,
the perturbed QFT may be also integrable, namely, the corresponding  S-matrix is factorisable 
in two-particle scattering matrices and there is no particle production in any scattering process. 
At this stage one might carry on and attempt to construct such  S-matrix in the standard way  i.e.,
 by means of the bootstrap program \cite{Boot,KTTW}.
This procedure usually involves several assumptions. For
example, for the models at hand, the HSG-models, the S-matrix
 proposal relies on the fundamental assumption  that the soliton  
spectrum determined semi-classically  coincides with the exact spectrum 
of stable particles at quantum level. This  assumption is 
 justified by the fact that the results obtained 
are self-consistent, namely the corresponding S-matrices \cite{HSGS} satisfy all 
the properties (\ref{HS}),
(\ref{unit}), (\ref{crossing}) and (\ref{bootseqs})
introduced in the previous chapter and are consistent with the semi-classical 
physical picture \cite{HSGsol, HSGS}. In addition, there exist usually some factors, 
the so-called CDD-factors
originally introduced in \cite{CDD}, which one can always ``add"  to  any  S-matrix proposal,
as they trivially satisfy all the requirements presented in section \ref{analit}
 and at the same time do not have poles in the
physical sheet  $0 <\text{Im} (\theta)< \pi$.  Therefore, 
provided  we have been able to construct a  S-matrix which is ``candidate"
  for the description of the scattering 
theory associated to a very particular 
1+1-dimensional  integrable QFT, it  is highly desirable to develop tools which permit to 
carry out this one-to-one correspondence. These consistency checks may 
consequently clarify finally the role of the mentioned 
ambiguities and assumptions.  In this context, the thermodynamic Bethe ansatz approach turns out
to be one of the tools we referred to above and so we intend to use it in order to check the
consistency of  the  S-matrix proposal \cite{HSGS} for the HSG-models. Let us now review 
the basic ideas required in the thermodynamic Bethe ansatz approach, following the arguments presented in
\cite{TBAZam1, KM2, kotba} (see also \cite{mussrev,BF2, KK, JJ}).

\subsection{The Bethe wave function}
\label{BWF}
\indent \ \
Let us then  start with a 1+1-dimensional integrable QFT whose S-matrix is factorisable
and moreover diagonal, which will be the case for the HSG-models. Suppose  also that 
 we compactify the space dimension on a   circumference of length $L$
 which we assume to be very large since later we will consider the thermodynamic 
limit $L \rightarrow \infty$, $N_A \rightarrow \infty$,
 $N_A$ being the number of particles of type $A$, with $N_A /L$ finite. In general, a relativistic QFT
can not be described by means of the  wave function formalism.  However, if we consider our system of
$N=\sum_{A} N_A$ particles  in  the asymptotic limit in which all these particles 
are  well separated  from each other, meaning that their mutual interaction
 is  negligible, we can associate to each of them a position 
$\{x_1, \cdots, x_N\}$. 
In that situation, the system is describable in terms of a wave function
$\Psi (x_1, \cdots, x_N)$ which is known as {\bf Bethe wave function} and was originally proposed
in \cite{Bethe}.

Suppose now that we take one of these particles of type  $A$, 
along the space direction, that is the circumference $[-L/2, L/2]$.
Whenever there is another particle of type $B$ such that their
positions $x_A$, $x_B$ are now close enough to allow a non-negligible interaction, 
the wave function description 
is not acceptable since the two particles can not be treated as  free anymore. 
Clearly, their  interaction will  be characterised by the corresponding  
two-particle S-matrix $S_{AB}(\theta_{A}-\theta_{B})$. Therefore,
the scattering theory must  provide conditions in order to 
match the wave functions in the regions where
the particle interactions are not negligible.  Essentially, whenever the
positions of two particles $A, B$ are interchanged, and interpreting then as $in$- and $out$-states,
respectively, the wave function  will pick up a factor
$S_{AB}(\theta_{A}-\theta_{B})$.  At the same time, since we have compactified the space
dimension in the circumference $[-L/2, L/2]$ the wave function 
must also satisfy periodic boundary conditions,
\begin{equation}
\Psi(x_1, \cdots, x_A=L/2,\cdots, x_N)=\Psi(x_1, \cdots, x_A=-L/2, \cdots, x_N).
\label{per}
\end{equation}
\noindent Therefore, if we consider again a  particle of type  $A$ which is initially at the position 
$x_A=-L/2$ and we take it along  the ``world line"  $[-L/2, L/2]$ so that  it finally reaches the position $x_A=L/2$ the wave functions describing the initial and final situations should be identical since,
taking (\ref{per}) into account, there is not difference between the initial and final configurations. 
However, as we mentioned before, on its trip along the space direction the particle has interacted
with all the other particles located at intermediate positions $-L/2 <x_i < L/2$ 
and therefore the initial
wave function must have picked up all the two-particle scattering amplitudes corresponding to each
interaction. Consequently, by imposing the conditions,
\begin{eqnarray}
\Psi (x_1, \cdots, x_A=-L/2, \cdots, x_N)& =& e^{i L M_A \sinh \theta_A}
\prod_{A \neq B}S_{AB}(\theta_{A}-\theta_{B}) \times \nonumber \\ 
&&  \Psi (x_1, \cdots, x_A=L/2, \cdots, x_N), 
\end{eqnarray}
\noindent  for $A=1,\cdots, n$, with $n$ being the number of particle species. 
Therefore, recalling now Eq. (\ref{per}),
we obtain the following set of constraints for the values of the rapidities 
$\theta_A$, 
\begin{equation}
e^{i L M_A \sinh \theta_A} \prod_{A \neq B }S_{AB}(\theta_{A}-\theta_{B})=1 \,\,\,\,\,\,\,\,\,\,
\textnormal{for}\,\,\,\,\,\,\,\,\,\, A=1, \cdots, n. 
\end{equation}
\noindent Here  $M_A$ is the mass of the particle species  $A$ and, as usual, we write  the momenta 
$p_A^1=M_A \sinh \theta_A$ in terms of the rapidities $\theta_A$. By taking now the logarithm of the latter
equations and introducing the phase shifts 
\begin{equation}
\delta_{AB}(\theta)= -i \ln S_{AB}(\theta),
\label{fs}
\end{equation}
\noindent  we obtain a set of coupled transcendental equations for the rapidities
\begin{equation}
L M_A \sinh \theta_A + \sum_{B \neq A} \delta_{AB} (\theta_A-\theta_B) = 2 \pi n_A, \,\,\,\,\,\,\,\,\,\,
\textnormal{for}\,\,\,\,\,\,\,\,\,\, A=1, \cdots, n, 
\label{bae}
\end{equation}
which are known as {\bf Bethe ansatz equations}. The numbers $n_A$ are integers
which can be interpreted as quantum numbers labeling 
the multiparticle state at hand. In other words, the specific
statistical  nature of the particles enters through 
the collection of values $\{ n_1, \cdots, n_n\}$. 
Notice that in Eq. (\ref{bae}) we are associating one value 
$n_A$ to each particle species $A=1, \cdots, n$, therefore,
each  quantum number $n_A$ should be understood as a vector
whose entries are the  quantum numbers characterising
each of the particles of the species $A$. 
For instance, if these particles are fermions, Pauli's exclusion principle ensures that
there can not exist two particles in the same quantum state, namely the
values $(\theta_A^{(i)}, n_A^{(i)})$ have to be different for each particle $i$.
On the other hand, if  the particles of a certain species $A$ are bosons,
there is not a limit to the number of particles carrying the
same quantum numbers. In addition, provided 
the $n_A$ entries are integers, the sum of  all the equations
(\ref{bae})  gives the expected quantization condition for the total momentum of a system 
\begin{equation}
P=\sum_{A=1}^{n} M_A \sinh \theta_A= \frac{2 \pi}{L} \hat{n} \,\,\,\,\,\,\,\,\,\,
\textnormal{for}\,\,\,\,\,\,\,\,\,\, \hat{n}=\sum_{A=1}^{n} n_A \in \mathbb{Z},
\end{equation} 
\noindent  subject to the periodic boundary conditions (\ref{per}). 

\subsection{The thermodynamic limit}
\label{tl}
\indent \ \
The Bethe ansatz equations (\ref{bae}) establish a set of constraints 
for the allowed rapidities of a system of $N$ particles subject to
periodic boundary conditions.
Let us now look at a particular particle species $A$ and suppose 
the system contains a number $N_A$ of particles of  this type. 
As mentioned before, the corresponding quantum numbers $n_A$
occurring in Eq. (\ref{bae}) will  have a certain substructure
$\{n_{A}^{(1)}, \cdots, n_A^{(N_A)}\}$, namely the 
values $n_A^{(i)}$ must be interpreted as the individual quantum numbers of a particle $i$
of the species $A$. Similarly,  we will  denote the associated rapidities by $\theta_A^{(i)}$. 
Therefore, when solving the system (\ref{bae}) there will be solutions
$\{\theta_A^{(i)}\}$ corresponding to the `allowed' values of the quantum numbers $\{ n_A^{(i)}\}$. 
These solutions are usually referred to as {\bf ``roots''}. Analogously, there will be a set of
values of the rapidities  $\{\theta_A^{(i)}\}$ which would be  `solutions' associated to
the non permitted
values of the quantum numbers $\{ n_A^{(i)}\}$. These rapidities will be called {\bf ``holes''}.

\vspace{0.2cm}

The system (\ref{bae}) can be solved for the rapidities
in the usual {\bf Bethe ansatz approach} but in our case we will be interested in 
the so-called {\bf thermodynamic or low-density limit}.  
In the thermodynamic limit both the dimension of the circumference 
$L$ where we compactified our system and the number of particles
of each species $N_A$ are taken to be infinite whereas the density of particles
of each type $N_A/L$ remains finite. In  this macroscopic picture, 
it is useful to introduce three new functions which, following 
\cite{TBAZam1, KM2, kotba, mussrev, BF2, KK, JJ} we denote by
$\rho^{(r)}_{A}(\theta)$,  $\rho^{(h)}_{A}(\theta)$ and 
$\rho_{A}(\theta)$ and we define in the following way for each particle species $A$:

\vspace{0.25cm}

{\bf  i)} we define the function $\rho^{(r)}_{A}(\theta)$  as the rapidity density of  ``roots''  per
unit length, namely the number of  particles of type $A$ possessing rapidities
 in an interval $[\theta, \theta + d \theta ]$  divided by $L \, d\theta$,
\begin{equation}
\rho^{(r)}_{A}(\theta)= \frac{1}{L}\frac{d n_A^{(r)}}{d \theta},
\label{ror}
\end{equation}
\noindent where the superscript $r$ indicates that the integers $ n_A^{(r)}$ 
are `allowed' on the r.h.s. of (\ref{bae}), that is, they correspond
to  rapidities which are ``roots'' of (\ref{bae}).

\vspace{0.25cm}

{\bf  ii)} analogously , we can define now a function which measures the rapidity 
density of ``holes'' per unit length, 
\begin{equation}
\rho^{(h)}_{A}(\theta)= \frac{1}{L}\frac{d n_A^{(h)}}{d \theta},
\label{roh}
\end{equation}
\noindent where the superscript $h$ in $ n_A^{(h)}$ indicates that
the corresponding rapidities $\theta_{A}$ are  what we called  ``holes ''.

\vspace{0.25cm}

{\bf iii)} Finally, we define the total  rapidity density of states per unit length $\rho_{A}$ as 
\begin{equation}
\rho_{A}(\theta) = \rho_{A}^{(r)}(\theta) + \rho_{A}^{(h)}(\theta) =
 \frac{1}{L} \frac{d n_A}{d \theta}.
\label{tot}
\end{equation}

\vspace{0.3cm}

\noindent In terms of  the functions (\ref{ror}), (\ref{roh}) and (\ref{tot}), we can now safely 
carry out the thermodynamic limit ($L, N_A \rightarrow \infty$, $N_A/L$ finite) of  (\ref{bae}). 
Hence, we obtain
\begin{equation}
\rho_A(\theta)= \rho_{A}^{(r)}(\theta) + \rho_{A}^{(h)}(\theta)= \frac{M_A}{2 \pi} \cosh \theta
+ \sum_{B=1}^{n} \Phi_{AB} * \rho_B^{(r)}(\theta),
\label{prov}
\end{equation}
\noindent where the functions $\Phi_{AB}(\theta)$ containing the information about the dynamical
interaction of the system are given, with the help of (\ref{fs}), by
\begin{equation}
\Phi _{AB}(\theta )=\Phi _{BA}(-\theta )\;= \frac{d \delta_{AB}(\theta)}{d\theta }=
-i\frac{d}{d\theta }\ln S_{AB}(\theta )\,.  \label{kernel}
\end{equation}
\noindent They are usually called  kernels, and the symbol
 ` \,* \,' denotes the rapidity 
convolution defined as
\begin{equation}
f * g(\theta):=\int\limits_{-\infty}^{\infty} \frac {d\theta ^{\prime }}{2\pi}
 \,f (\theta -\theta ^{\prime }) g (\theta ^{\prime}),
\label{convolution}
\end{equation}
\noindent for two arbitrary functions $f$ and $g$. 
In (\ref{prov}) the densities of  ``roots''  
 and ``holes''  $\rho_A^{(r)}$, $\rho_A^{(h)}$
are independent functions, which makes the system (\ref{prov}) still difficult to handle. However,
additional constrains can be obtained when the equilibrium configuration at a certain finite temperature
$T$ is studied. The thermodynamic equilibrium of the system is equivalent to the minimisation of the
total free energy per unit length, which we denote by
 $f(\rho, \rho^{(r)})= \sum_{A} 
f ( \rho_A, \rho_A^{(r)})$ and is given by
\begin{equation}
f( \rho, \rho^{(r)})= h( \rho^{(r)}) - T s ( \rho, \rho^{(r)}),
\label{fe}
\end{equation}
\noindent where the functions $h( \rho^{(r)})$ and $s( \rho, \rho^{(r)})$
are respectively the total energy and entropy of the system per unit length, i.e.,
\begin{eqnarray}
h(\rho^{(r)})&=&\sum_{A=1}^{n}h ( \rho_A^{(r)})=\sum_{A=1}^{n}
\int\limits_{-\infty}^{\infty} d\theta \, \rho_A^{(r)} (\theta) 
M_A \cosh \theta, \label{h} \\
s ( \rho, \rho^{(r)})&=&\sum_{A=1}^{n} s( \rho_A, \rho_A^{(r)})=
 \sum_{A=1}^{n}
\ln {\mathcal N} ( \rho_A, \rho_A^{(r)}).
\label{s}
\end{eqnarray}
\noindent Here  ${\mathcal N} ( \rho_A, \rho_A^{(r)})$ has to be understood as 
 the density of quantum states 
labeled by the integers $n_A^{(i)}$ with $A=1, \cdots, n$ and $i=1, \cdots, N_A$, 
which correspond  to the same density configuration  $\rho_A, \rho_A^{(r)}$.
 It can be computed
as follows: Let $N^{\prime}_{A}$  be the number of quantum levels associated to the particle species $A$ contained in a rapidity interval  $[\theta, \theta + d \theta ]$. Taking definition (\ref{tot}) into account,
it follows that $N^{\prime}_{A}\approx \rho_A (\theta) d\theta$. Let $n^{\prime}_{A}$  be the
number of particles of type $A$ distributed in the same interval, namely  $n^{\prime}_{A} \approx
\rho_A^{(r)}(\theta) d\theta$. Taking Pauli's exclusion principle into account one can easily obtain
the number of different possible distributions of  these $n^{\prime}_A$ particles of type $A$ between
the $N^{\prime}_{A}$ different quantum levels, 
\begin{eqnarray}
\frac{ N^{\prime}_A !}{ n^{\prime}_A ! (N^{\prime}_A -n^{\prime}_A)!}& \approx &
\frac{[\rho_A(\theta) d\theta] !}{ [\rho_A^{(r)}(\theta) d\theta]! 
[(\rho_A(\theta)-\rho_A^{(r)}(\theta) )d\theta] !}, \,\,\,\, \textnormal{fermionic case}\nonumber \\
\frac{( N^{\prime}_A + n^{\prime}_A -1) !}{ n^{\prime}_A ! (N^{\prime}_A-1)!} &\approx & 
\frac{[(\rho_A(\theta) - \rho_A^{(r)}(\theta)  - 1) d\theta]  !}
{[\rho_A^{(r)}(\theta) d\theta]  ! [(\rho_A(\theta)-1 ) d\theta] !}, \,\,\,\, \,\,\,\,\textnormal{bosonic case}
\end{eqnarray}
\noindent  Therefore, the corresponding entropies read
\begin{eqnarray}
s (\rho,\rho^{(r)})&=& 
\sum_{A=1}^{n} \int\limits_{-\infty}^{\infty} d \theta \Big(\rho_A \ln \rho_A-\rho_A^{(r)} \ln \rho_A^{(r)}- (\rho_A - \rho_A^{(r)}) \ln(\rho_A-\rho_A^{(r)})\Big), \,\,\,\\
s (\rho,\rho^{(r)} )&=& 
\sum_{A=1}^{n} \int\limits_{-\infty}^{\infty} d \theta \Big(\rho_A \ln \rho_A-\rho_A^{(r)} \ln \rho_A^{(r)}- (\rho_A + \rho_A^{(r)}) \ln(\rho_A+\rho_A^{(r)})\Big), \,\,\,\label{entropy}
\end{eqnarray}
\noindent for the species $A$ to be fermions or bosons respectively. Notice that, in the derivation
of the latter equations we have used Stirling's formula
$\ln N! \approx N (\ln N-1) \approx$$ N \ln N$ for $N\gg 1$.

\noindent  Having  now explicit expressions for the energy and entropy of the system in terms of the densities $\rho_A$ and $\rho_A^{(r)}$ 
we are in the position to express the free energy per unit length (\ref{fe}) in terms of the mentioned
densities and determine the extremum conditions corresponding to both fermionic and bosonic statistics.
For  this purpose it is convenient to define a set of new functions $\epsilon_A (\theta)$ which
 are referred  to as {\bf ``pseudo-energies''}  and  which can be expressed in terms
 of the densities of  ``roots'' and ``holes''  as follows,
\begin{eqnarray}
e^{\epsilon_A}&=& \frac{\rho_A}{\rho_A^{(r)}}-1=
\frac{ \rho_A^{(h)}}{\rho_A^{(r)}},\,\,\,\,\,\,
\textnormal{fermionic case},\nonumber \\
e^{\epsilon_A}&=& \frac{\rho_A}{\rho_A^{(r)}}+1=\frac{ \rho_A^{(h)}}{\rho_A^{(r)}}+2,\,\,\,\,\,\,
\textnormal{bosonic case}.
\end{eqnarray}
\noindent Therefore,  with the help of (\ref{prov}) and (\ref{kernel}),
the extremum conditions for the free energy per unit
length (\ref{fe}) read,
\begin{equation}
\epsilon_A(\theta)= R M_A \cosh \theta -\sum_{B} \Phi_{AB}* L_{B}(\theta), 
\label{fin}
\end{equation}
\noindent where $R= 1/T$ and we have introduced the $L$-functions 
\begin{eqnarray}
L_A (\theta) = \ln (1+e^{-\epsilon_A (\theta)}), \,\,\,\,\,\,\,\, \,\,\,\,\,\,\,\,
\textnormal{fermionic case},\nonumber \\
L_A (\theta) = -\ln (1-e^{-\epsilon_A (\theta)}), \,\,\,\,\,\,\,\, \,\,\,\,\,\,\,\,\,
\textnormal{bosonic case}.\label{Lf}
\end{eqnarray}
\noindent From (\ref{fin}) and (\ref{Lf}) one easily sees that whenever the S-matrix reduces to
$\pm 1$, namely we study a system of free fermions or bosons, the pseudo-energies are simply given by 
\begin{equation}
\epsilon_A(\theta)/R=  M_A \cosh \theta,
\label{ffb}
\end{equation}
\noindent that is,  the free ``on-shell'' energies.

The non-linear integral equations (\ref{fin}) are usually called {\bf thermodynamic
Bethe ansatz equations}. In general, omitting the systems of free fermions or bosons studied in
 \cite{KM2}, the system (\ref{fin}) can not be solved analytically for 
the pseudo-energies for generic temperature,
 and we may ultimately resort to numerical methods. Once the pseudo-energies
$\epsilon_A (\theta)$ have been computed in some way and, consequently the $L$-functions,
it is possible to obtain the free energy of the system  (\ref{fe}) subject to 
the extremum conditions (\ref{fin}) namely, the extremal free energy
\begin{equation}
 f (R) = -\frac{1}{R} \sum_{A=1}^{n} M_A \int\limits_{-\infty}^{\infty} 
\frac{d \theta}{2 \pi} L_A (\theta) \cosh \theta, 
\label{ex}
\end{equation}
\noindent  therefore, taking (\ref{ffb}) into account, the equilibrium free energy
for a system of free bosons or fermions reduces to the one of a relativistic Fermi or Bose gas at finite
temperature $T=1/R$ (see section 6 in \cite{KM2}). 

As we will show later, the extremal  free energy  (\ref{ex}) is directly related to the 
ground state energy of the system which at the same time, in the UV limit, depends on the Virasoro central charge of  the underlying CFT. These relationships justify the link between 
the infinite volume thermodynamics of a QFT, expressed in terms of its corresponding S-matrix by means of (\ref{fin}), and  some of the relevant data characterising  the original CFT.
 
\subsubsection{Quantum field theory on a torus}
\label{qftt}
\indent \ \
As stated in several occasions, the analysis performed before is based on the study of a QFT 
compactified on a  circumference of length $L$, which is taken to be infinity in
the thermodynamic limit. Let us assume now that we formulate our QFT on a torus generated by two
orthogonal circles of circumferences $R$ and $L$ and define a Cartesian (Euclidean) 
coordinates system in the way showed in Fig. \ref{torus}.  In that situation, we have at our disposal  two possible choices for the quantization axis i.e., two possible ways to develop a quantum mechanical formulation of the theory,
depending  on whether we identify  the directions $x$, $y$ with the space and time dimensions 
or vice-versa:

\begin{figure}[!h]
 \begin{center}
  \leavevmode
    \includegraphics[scale=0.65]{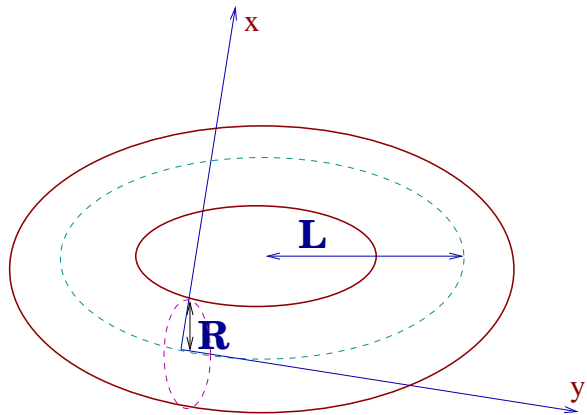}   
    \caption{Torus generated by two circles 
of circumferences $\protect{R}$ and $\protect{L}$.}
   \label{torus}
 \end{center}
\end{figure}

\vspace{0.3cm}
{\bf i)}  as a first choice, we may consider the $y$ axis as the one
 playing the role of time direction and  let
 $L \rightarrow \infty$ with $R$ remaining finite. Then, if we denote by $H_R$ the
hamiltonian describing the system in the compactified (periodic) space of dimension $R$  and by 
$\mathcal{H}_R$ the corresponding Hilbert space,  the partition function associated to the field theory $\mathcal{Z}(R, L)$ is given by,
\begin{equation} 
\mathcal{Z}(R,L) = \lim_{L \rightarrow \infty} \textnormal{Tr}_{\mathcal{H}_R} e^{-L H_R} \approx
e^{-E_0 (R) L},
\end{equation}
\noindent for $E_0 (R)$ to be the energy of the ground state, since  in the limit $L \rightarrow \infty $,
the term associated to the eigenstate of minimum energy of the Hamiltonian $H_R$ is the dominant
contribution to the partition function $\mathcal{Z}(R,L)$,  

\vspace{0.3cm}

{\bf ii)} alternatively we could exchange the roles of the axis $x$, $y$ and associate them now to
the time and space dimensions respectively. Hence, if we now perform the same limit
carried out in  {\bf i)}, namely we take $L \rightarrow \infty$ we are in a new situation in which the
theory is time-periodic, with periodicity given by $R$. Moreover, if we  identify the dimension $R$ as
the same quantity occurring in the thermodynamic Bethe ansatz equations (\ref{fin}), namely the inverse
temperature $R=1/T$, our alternative choice of axis is equivalent to a formulation of the QFT
at finite temperature. Therefore, denoting by $H_L$ the new hamiltonian and by $\mathcal{H}_L$  the corresponding Hilbert space, the partition function $\mathcal{Z}(R,L)$  is
now given by
\begin{equation} 
\mathcal{Z}(R,L) = \lim_{L \rightarrow \infty} \textnormal{Tr}_{\mathcal{H}_L} e^{-R H_L} \approx
e^{- R L f (R)},
\end{equation}
\noindent where $f(R)$ is the free energy per unit length of the system at finite temperature $T=1/R$
namely, the same function obtained in (\ref{ex}). 

\vspace{0.3cm}
{\bf  iii)} Putting now  together the results coming from   {\bf i)} and {\bf ii)} we obtain a relationship
 between the extremal free energy (\ref{ex}) and the ground state energy $E_0 (R)$ which is simply
\begin{equation}
R f(R) = E_0 (R).
\label{fegee}
\end{equation}
\noindent  We are now  able to extract information about the underlying CFT
by performing the ultraviolet limit $ R = 1/T \rightarrow 0 $. 
In this limit the partition function  $\mathcal{Z}(R,L)$ must reduce to 
 the one of  the underlying  CFT 
on a cylinder, as  the limit $L \rightarrow \infty$  is  nothing but  
the limit torus $\rightarrow$ infinite cylinder. Consequently, it is necessary at this point
to make use of some basic notions of conformal field theory on a cylinder which for this purpose have
been introduced in subsection \ref{cftbo} of the previous chapter. 
The results presented there 
guarantee that a relationship 
between the ground state energy $E_0 (R)$, related
to the equilibrium free energy by means of  (\ref{fegee}), and the energy 
of the eigenstates of the hamiltonian (\ref{hcyl}) can
be now established. 
Clearly such a relationship arises in the UV limit in which the QFT studied 
in the previous subsections reduces to its underlying CFT on a cylinder of 
infinite length $L \rightarrow \infty$ and  time dimension $R=1/T$. 
In the ultraviolet limit we obtain 
\begin{equation}
\lim_{R \rightarrow 0} R f(R) = -\frac{\pi}{6 R} c_{eff} \,\,\,\,\,\,\,\,
\textnormal{with}\,\,\,\,\,\,\,\,
c_{eff}:=c- 12(\Delta_0 +\bar\Delta_0),
\label{uvl}
\end{equation}
\noindent where $\Delta_0, \bar{\Delta}_0$ are the lowest conformal dimensions related to the two
chiral sectors. The latter relationship was originally derived in 
\cite{scaling}. The quantity $c_{eff}$ is known as 
{\bf effective central charge} and for unitary  CFT's 
coincides with the Virasoro central charge $c$, since
 the conformal dimensions  $\Delta_0, \bar{\Delta}_0$
are zero. Since the WZNW-coset models are unitary CFT's, this will be the case for the HSG-models.
Therefore, the introduction of periodic boundary conditions in a unitary CFT has the effect of generating
a non-zero value for the ground state energy in the UV-limit and in that sense the effective central charge can be interpreted as {\bf Casimir energy}. 

The relation (\ref{uvl}) will be fundamental for our TBA-analysis. Recall that the initial motivation 
for such an analysis was the search for consistency checks of the S-matrix proposal \cite{HSGS} for
the HSG-models, checks  which will  exploit the interpretation
 of these models as perturbed coset theories. 
Therefore, having solved the TBA-equations (\ref{fin}) and extracted the corresponding
$L$-functions we may introduce then in (\ref{ex}) and  perform the corresponding integrals while carrying out  the UV-limit.
Proceeding this way we may have identified the Virasoro central charge of the original CFT. 
Since the HSG-models, like many other 1+1-dimensional integrable QFT's, have been constructed as 
perturbations of a well-known CFT, the Virasoro central charge extracted from our analysis may be 
directly compared with its expected value ( see Eq. (\ref{cdata})).

However, this is not the only information we can extract from a TBA-analysis. 
In a similar fashion we defined the effective central charge
 $c_{eff}$ through (\ref{uvl}) we can define a new function
$c(R)$ which we will  refer to as {\bf finite size scaling function},
\begin{equation}
c(R) = -\frac{6 R^2}{\pi} f(R)= \frac{3 R}{\pi^2} \sum_{A=1}^{n} M_A \int\limits_{-\infty}^{\infty} 
{d \theta} L_A (\theta) \cosh \theta.
\label{fsef}
\end{equation}
\noindent In comparison to  (\ref{uvl}) the function $E_{0}(R)=-\pi c(R)/6 R$
can be interpreted as ``off-critical '' Casimir energy \cite{TBAZam1, KM2}.  
It is clear  that in the UV-limit,
\begin{equation}
\lim_{R \rightarrow 0} c(R)= c_{eff}= c-12(\Delta_0+ \bar{\Delta}_0).
\label{uvceff}
\end{equation}

Thus, the solution of the TBA-equations (\ref{fin}) does not only allow for the identification of the
Virasoro central charge of the underlying CFT but also for the computation (usually numerically) of 
the function $c(R)$ defined in (\ref{fsef}). It is also expected  that for a massive QFT
the IR-limit of the
finite size scaling function must be  zero, and in the intermediate region $0 < R < \infty$ 
it should provide information about  the onset of  the stable and unstable particles of the
 model i.e., the energy scale $R$ at which these particles can be formed. 
In particular, it will be shown later for the HSG-models
that, for finite values of $R$, the scaling function $c(R)$ exhibits a very characteristic 
``staircase'' pattern
where the number of steps is determined by the number of free parameters in the model and is also related to the
number of particles present in the theory.
Such a behaviour was observed previously for the roaming sinh-Gordon model \cite{staircase} 
and their generalisations \cite{Martins,DoRav}. This behaviour suggests that, similarly to the 
well-known Zamolodchikov's  $c$-function \cite{ZamC}, which we will  compute in the context of 
the form factor program, the scaling function admits  an interpretation as a measure of  
effective light degrees of freedom. In other words, starting in the deep UV-limit which
 corresponds to the situation in which the mass of any of the particles of the theory is 
much lower than the energy scale, and therefore they all  can be formed and 
contribute to the Virasoro central charge $c$, we will observe the consecutive ``decoupling'' 
of  the  heaviest particles of the theory as soon as an energy scale much lower than the energy
 precise for their production is reached. Finally, in the IR-limit none of the particles can be
 formed and the scaling function will vanish. It is then clear that the information provided by 
the scaling function is crucial concerning
the check of the physical picture expected for the HSG-models 
\cite{HSG, HSG2, HSGsol, Park, HMQH}. Recall that the S-matrix proposal \cite{HSGS} was based
on the extrapolation of  semi-classical results, in particular concerning the stable and unstable 
particle spectrum therefore, from that point of view, the outlined interpretation of the ``staircase'' 
 behaviour of the scaling function provides strong support for the validity of the mentioned assumptions. 
It must be emphasised that such a clear physical interpretation is not possible for the mentioned roaming sinh-Gordon
models \cite{staircase, Martins, DoRav}, amongst other reasons, 
 due to the lack of a consistent Lagrangian formulation.

We have now all the necessary ingredients to carry out our specific
application of the TBA-approach for the the HSG-models. The
main features of these theories have been already stated in subsection \ref{hsgsec} of 
the previous chapter. As a fundamental input in
the TBA-framework the knowledge of the two-particle scattering amplitudes and mass spectrum is needed.
Also, some of the most characteristic data of the associated underlying coset CFT, like 
the Virasoro central charge (\ref{cdata}) and the dimension of the perturbing field (\ref{deldata}) have been
reported in the previous chapter and we will frequently appeal to their expected values in the course
of our TBA-analysis. 

\section{TBA with parity violation and resonances}
\label{resonance}
\indent \ \
In this section we are going to determine the conformal field theory which
governs the UV-regime of the QFT associated with the
S-matrix elements (\ref{Sint}) and (\ref{ijint}). According to the defining
relation (\ref{Actq}) and the discussion of the previous section, we
expect to recover the WZNW-coset theory with effective central charge (\ref
{cdata}) in the extreme UV-limit by means of a TBA-analysis. Since up
to now such an analysis has only been carried out for parity invariant
S-matrices, a few comments are due to implement parity violation. Recall that the
starting point in the derivation of the key equations are the Bethe ansatz
equations, which are the outcome of dragging one soliton, say of type 
$A=(a, i)$, along the world line. For the time being we do not need the
distinction between the two quantum numbers. On this trip the formal
wave-function of $A$ picks up the corresponding S-matrix element as a phase
factor when meeting another soliton (see subsection \ref{BWF}). 
Due to the parity violation it matters,
whether the soliton is moved clockwise or counter-clockwise along the world
line, such that instead of the set of equations (\ref{bae}) 
we end up with two apparently different sets of Bethe Ansatz equations 
\begin{equation}
e^{iLM_{A}\sinh \theta _{A}}\prod\limits_{B\neq A}S_{AB}(\theta _{A}-\theta
_{B})=1\quad \text{and \quad }e^{-iLM_{A}\sinh \theta
_{A}}\prod\limits_{B\neq A}S_{BA}(\theta _{B}-\theta _{A})=1\,,  \label{BA}
\end{equation}
with $L$ denoting the length of the compactified space direction. These two
sets of equations are of course not entirely independent and may be obtained
from each other by complex conjugation with the help of the properties 
of unitarity and Hermitian analyticity of the S-matrix (\ref{HS}) summarised in
the previous chapter.
We may carry out the thermodynamic limit of (\ref{BA}) in the usual fashion 
\cite{TBAZam1} as described in subsection \ref{tl}
and obtain the following sets of non-linear integral
equations 
\begin{eqnarray}
\epsilon _{A}^{+}(\theta )+\sum_{B}\,\Phi _{AB}*L_{B}^{+}(\theta )
&=&r\,M_{A}\cosh \theta, \quad  \label{tbaa} \\
\epsilon _{A}^{-}(\theta )+\sum_{B}\,\Phi _{BA}*L_{B}^{-}(\theta )
&=&r\,M_{A}\cosh \theta \,\,\,.  \label{ptba}
\end{eqnarray}
Here $r=m_{1}T^{-1}$ is the
inverse temperature times the overall mass scale $m_{1}$ of the lightest
particle and we also re-defined the masses by $M_{a}^{i}\rightarrow
M_{a}^{i}/m_{1}$ keeping, however, the same notation. As pointed out before,
 it is useful in these considerations to  introduce the so-called pseudo-energies $%
\epsilon _{A}^{+}(\theta )=\epsilon _{A}^{-}(-\theta )$ and the related
functions 
\begin{equation}
L_{A}^{\pm }(\theta )=\ln (1+e^{-\epsilon _{A}^{\pm }(\theta )}).
\label{LL}
\end{equation}

Notice that (\ref{ptba}) may be obtained from (\ref{tbaa}) simply by the parity
transformation $\theta \rightarrow -\theta $ and the first equality in  (\ref{kernel}). 
The main difference of these equations in comparison with the
parity invariant case is that we have lost the usual symmetry of the
pseudo-energies as functions of the rapidities, such that we have now in
general $\epsilon _{A}^{+}(\theta )\neq \epsilon _{A}^{-}(\theta )$. This
symmetry may be recovered by restoring parity.

\vspace{0.25cm}

It is also worth mentioning  that according to (\ref{Lf}) the $L$-functions (\ref{LL}) imply 
 that we have  chosen the statistical interaction to be of fermionic type. 
This choice has been made for all the theories studied in the TBA-context until
now and relies on the fact that the S-matrix describing the interaction between
identical particles namely, particles  with the same quantum numbers and momenta is given by 
\begin{equation}
S_{aa}^{ii}(0)=-1,
\end{equation}
\noindent as follows from (\ref{S}). Therefore,
 it coincides with the S-matrix describing the interaction
between free fermions. This choice is not in contradiction with the
bosonic character of the field $h(x^0, x^1)$ entering the action (\ref{Actq}) and should
not lead to confusion. In particular it is a well-known fact that  in $1+d$-dimensions, with $d<3$
there is not an intrinsic meaning for the statistics and
any QFT described  in terms of bosonic fields admits also a description in terms of 
fermionic fields which only involves a suitable field  transformation.
Examples of such  an ``ambiguity''  are  provided by \cite{WW}.  Concerning the role of the 
statistics in the TBA-context, in \cite{BF2} the TBA-approach was formulated for
a more general choice of the statistics known as {\bf Haldane statistics} \cite{haldane}. 
It was observed 
also in \cite{BF2} that essentially any choice of the statistical interaction can be made
 leading to the same TBA-equations, 
if it is suitably compensated by a particular change in the phase of the S-matrix. 
Therefore, in the TBA-context we could extract the same information i.e., central charge, 
scaling function... even if we decided to choose a more ``exotic''  type of statistics, 
provided the S-matrix proposal was modified as suggested in \cite{BF2}. Very recently 
\cite{BB},  the TBA-approach has been generalised for {\bf Gentile statistics} \cite{gentile} 
along the same lines of \cite{BF2}. 

The scaling function, may be computed similar as in (\ref{fsef})
\begin{equation}
c(r)=\frac{3\,r}{\pi ^{2}}\sum_{A}M_{A}\int\limits_{0}^{\infty }d\theta
\,\cosh \theta \,(L_{A}^{-}(\theta )+L_{A}^{+}(\theta ))\,,  \label{scale}
\end{equation}
\noindent where we only used the property $L_{+}(\theta)=L_{-}(-\theta)$ to substitute
the integral $\int_{-\infty}^{\infty}$  in  (\ref{fsef}) by $\int_{0}^{\infty}$ in (\ref{scale}).
As we have seen in subsection \ref{qftt}, once the equations (\ref{tbaa}), (\ref{ptba}) have
been solved for the pseudo-energies $\epsilon_{A}^{\pm}$, 
of special interest is the deep UV-limit, i.e.  $r\rightarrow 0$
in which, according to (\ref{uvceff}) we expect to identify the Virasoro
central charge (\ref{cdata}) of the underlying CFT.
This assumption will turn out to be consistent with the
analytical and numerical results.

\subsection{Ultraviolet central charge for the HSG-models}
\label{UVCC}
\indent \ \
In this section we follow the usual argumentation of the TBA-analysis \cite{TBAZam1}
which leads to the effective central charge, paying, however, attention to the
parity violation. We will recover indeed the value in (\ref{cdata}) as the
central charge of the HSG-models. In fact, following  \cite{TBAZam1}
we will be able to determine the effective central charge  analytically when performing the 
limit $r\rightarrow 0$ in (\ref{scale}). The reason is that in this limit
there will be various  terms both in (\ref{scale}) and (\ref{tbaa}) which
we can  safely neglect and consequently, the evaluation of both expressions becomes simpler. 

It is useful at this point to introduce the new variable $x=\ln (r/2)$ so that the UV-limit 
corresponds now to  $x \rightarrow -\infty$. In this limit the factor $r M_A \cosh \theta$ arising both  in the integral (\ref{scale}) and in the TBA-equations (\ref{tbaa}), (\ref{ptba})  can be approximated in the following way,
\begin{equation}
rM_{A}\cosh \theta= {M_A}\Big( e^{\theta+x}+ e^{-\theta+x}\Big) \simeq {M_A} e^{\arrowvert \theta \arrowvert+x},
\label{cha}
\end{equation}
\noindent Here we find for the first time the occurrence of the term $M_A e^{x+\theta}$ in our analysis, 
which arises naturally in the  formulation the TBA-equations for massless particles. 
The concept of massless scattering has been introduced originally in \cite{triZam} 
as follows: The on-shell energy of a right and left moving particle
is given by $E_{\pm }=M/2\, e^{\pm \theta }$ which is formally obtained from
the on-shell energy of a massive particle $E=m\cosh \theta $ by the
replacement $\theta \rightarrow \theta \pm \sigma/2$ and taking the limit $%
m\rightarrow 0, \sigma \rightarrow \infty$, while keeping the expression
$M=m\,e^{\sigma/2}$ finite. We will encounter these on-shell energies in the course of
our analysis. 

The substitution of (\ref{cha}) into the expression for the scaling function (\ref{scale}) 
gives
\begin{equation}
c(r)\simeq \frac{3}{\pi ^{2}}\sum_{A}M_{A}\int\limits_{0}^{\infty
}d\theta \,e^{\theta+x} \,(L_{A}^{-}(\theta )+L_{A}^{+}(\theta ))= \frac{3}{\pi ^{2}}\sum_{A} M_{A}\int\limits_{x}^{\infty}
d\theta \,e^{\theta} \,(L_{A}^{\small{>}}(\theta )+ L_{A}^{\small{<}}(\theta )).
  \label{appc}
\end{equation}
\noindent The second equality in (\ref{appc})
 is simply obtained by shifting $\theta \rightarrow \theta+x$ and
defining the so-called ``kink''  functions
\begin{equation}
L_A^{\small{>}}(\theta):=L_A^{+}(\theta-x), \,\,\,\,\,\,\,\, \,\,\,\,\,\,\,\,  \textnormal{and}
\,\,\,\,\,\,\,\, \,\,\,\,\,\,\,\,  L_A^{\small{<}}(\theta):=L_A^{-}(\theta-x),  
\label{kink}
\end{equation}
\noindent originally introduced  in \cite{TBAZam1}.

We can now  carry out  the same limit for  the TBA-equations (\ref{tbaa}), (\ref{ptba}), 
\begin{eqnarray}
 M_A e^{\theta}&\simeq & \epsilon_A^{\small{>}}(\theta) +
\sum_{B}\,\Phi _{AB}*L_{B}^{>}(\theta ), \label{tbalimit1}\\
 M_A e^{\theta} &\simeq &\epsilon_A^{\small{<}}(\theta) +
 \sum_{B}\,\Phi _{BA}*L_{B}^{\small{<}}(\theta ), 
\label{tbalimit}
\end{eqnarray}
\noindent where the approximation (\ref{cha})  has been used again. 
Taking  now  the derivative with respect to $\theta$ of the latter equations 
and using  (\ref{kernel}) and (\ref{convolution}) we obtain
\begin{eqnarray}
M_A e^{\theta} &\simeq &\frac{d \epsilon_{A}^{>}}{d \theta} +  \sum_{B}
\int\limits_{-\infty}^{\infty} \frac{ d\theta ^{\prime}}{2 \pi} {\Phi
_{AB}( \theta -\theta ^{\prime })}\frac{d L _{B}^{>}(\theta ^{\prime })}{d\theta^{\prime }},\label{der1}\\
M_A e^{\theta} &\simeq &\frac{d \epsilon_{A}^{<}}{d \theta} +  \sum_{B}
\int\limits_{-\infty}^{\infty} \frac{ d\theta ^{\prime}}{2 \pi} {\Phi
_{AB}(- \theta +\theta ^{\prime })}\frac{d L _{B}^{<}(\theta ^{\prime })}{d\theta^{\prime }}.
\label{der2}
\end{eqnarray}
\noindent  Notice that in the convolution terms the derivative with
respect to $\theta$ has been substituted by a derivative with respect to $\theta^{\prime}$.
 This can be done using  
integration by parts and the trivial fact
\begin{equation}
\frac{d}{d \theta}\Phi_{AB}( \theta -\theta ^{\prime })= 
-\frac{d}{d \theta^{\prime}}
 \Phi_{AB}( \theta -\theta ^{\prime}).
\end{equation}

Having Eqs. (\ref{der1}) and (\ref{der2}) at hand, it is possible to substitute
the term $M_A e^\theta$ in (\ref{appc}) by the r.h.s. of these equations. By doing so, we obtain
\begin{equation}
\sum_A M_A \int\limits_{x}^{\infty} d \theta e^\theta L_A^{>}(\theta)=
 \sum_A M_A \int\limits_{x}^{\infty} d \theta L_A^{>}(\theta)  
\frac{d \epsilon_A^{>}} {d \theta} +
 \sum_{A, B} \int\limits_{x}^{\infty} L_A^{>}(\theta) \, \Phi_{AB}*
(L_{B}^{>})^{\prime} (\theta) ,
\label{cder}
\end{equation}
\noindent where $(L_A^{>})^{\prime}$ denotes the derivative of the $ L_{A}^{>}$ function.
Obviously a completely analogous equation can be obtained for the other contribution to (\ref{appc})
by using (\ref{der2}).

Let us now look separately at each of  the two contributions in (\ref{cder}). 
Using (\ref{LL}) it is trivial  to see that the first one is simply
\begin{equation}
  \sum_A M_A \int\limits_{x}^{\infty}  d \theta \frac{d \epsilon_A^{>}} {d \theta}  L_A^{>}(\theta)= \sum_A M_A \int\limits_{\epsilon_A^{>}(x)}^{\epsilon_A^{>}(\infty)} d \epsilon_A^{>} \ln ( 1 + e^{\epsilon_A^{>} (\theta)}),
\label{one}
\end{equation}
\noindent whereas for the second one we obtain
\begin{eqnarray}
 \sum_{A, B}\int\limits_{x}^{\infty} d \theta \, L_A^{>}(\theta) \,
 \Phi_{AB}* (L_{A}^{>})^{\prime} (\theta) &=&  \sum_{A,B}
\int\limits_{x}^{\infty} \frac {d\theta^{\prime}}{2 \pi} \int\limits_{-\infty}^{\infty} d \theta\,
L_{B}(\theta^{\prime}) \Phi_{AB} (\theta- \theta^{\prime}) \frac{d L_A (\theta)}{d \theta}\nonumber \\
& \simeq &
 \sum_{A, B} \int\limits_{x}^{\infty} d \theta  \frac{L_A^{>}(\theta) }{d \theta}
 \Phi_{AB}* L_{A}^{>} (\theta).
\label{mima}
\end{eqnarray}
\noindent Here we used the approximation $\int_{x}^{\infty}\simeq \int_{-\infty}^{\infty}$ to
obtain the last equality and the fact that there is a summation over the indices $A, B$,
 so that their
names can safely be interchanged and similarly for the integration variables $\theta$, $\theta^{\prime}$.
 It is also necessary to take the first property in (\ref{kernel})
into account together, of course, with the definition of the convolution (\ref{convolution}).
However, it is possible to express (\ref{mima}) still in a more suitable way 
by using now the approximated TBA-equation (\ref{tbalimit1}) to eliminate the
convolution term and the defining relation (\ref{LL}) of the $L$-functions in terms of the 
pseudo-energies
\begin{equation}
\sum_{A, B}\int\limits_{x}^{\infty} d \theta  \frac{L_A^{>}(\theta) }{d \theta}
 (\Phi_{AB}* L_{A}^{>}) (\theta) \simeq
 \sum_{A}\int\limits_{\epsilon_A^{>}(x)}^{\epsilon_A^{>}(\infty)} 
d \epsilon_A^{>} \frac{\epsilon_A^{>}}{(1+ e^{\epsilon_A^{>}})}- \sum_A \int\limits_{x}^{\infty} d \theta  M_A e^\theta L_A^{>}(\theta).
\label{two}
\end{equation}
\noindent Notice that the last term on the r.h.s., which is obtained after integration by parts, 
 is precisely our starting point (\ref{cder}). Therefore, 
substituting the contributions (\ref{one}) and (\ref{two}) into  (\ref{cder}) we obtain
\begin{equation}
 \int\limits_{x}^{\infty} d \theta  M_A e^\theta L_A^{>}(\theta)=
\frac{1}{2} \int\limits_{\epsilon_A^{>}(x)}^{\epsilon_A^{>}(\infty)} d \epsilon_A^{>}\,
 M_A \ln ( 1 + e^{\epsilon_A^{>} (\theta)}) +
\frac{1}{2}\int\limits_{\epsilon_A^{>}(x)}^{\epsilon_A^{>}(\infty)} 
d \epsilon_A^{>} \, \frac{\epsilon_A^{>}}{(1+ e^{\epsilon_A^{>}})}.
\end{equation}
\noindent A similar expression can be obtained for 
the second term in (\ref{appc}) so that, we can finally write
\begin{equation}
\lim_{r\rightarrow 0}\, c(r)\simeq \frac{3}{2\pi ^{2}}\sum\limits_{p=+,-}%
\sum_{A}\int\limits_{\epsilon _{A}^{p}(0)}^{\epsilon _{A}^{p}(\infty
)}d\epsilon _{A}^{p}\left[ \ln (1+e^{-\epsilon _{A}^{p}})+\frac{\epsilon
_{A}^{p}}{1+e^{\epsilon _{A}^{p}}}\right] \,\,\,,
\end{equation}
\noindent where we use again the standard pseudo-energies $\epsilon_A^{\pm}$ instead of the 
``kink ''  variables (\ref{kink}) and consequently substitute
\begin{equation}
\epsilon_A^{+}(0)=\epsilon_A^{>}(x) \,\,\,\,\,\,\,\,\,\,\,\,\,\,\,\textnormal{and}
\,\,\,\,\,\,\,\,\,\,\,\,\,\,\,\epsilon_A^{-}(0)=\epsilon_A^{<}(x).
\label{subs}
\end{equation}
\noindent Upon the 
substitution $y_{A}^{p}=1/(1+\exp (\epsilon _{A}^{p}))$ we obtain
the following  expression for the effective central charge 
\begin{equation}
c_{\text{eff}}=\lim_{r\rightarrow 0}\, c(r)=
\frac{6}{\pi ^{2}}\sum_{A}\mathcal{L}
\left( \frac{1}{1+e^{\epsilon _{A}^{\pm }(0)}}\right) \,\,.  \label{ceff2}
\end{equation}
Here we used the integral representation for Roger's dilogarithm function 
\begin{equation}
\mathcal{L}(x)=1/2\int_{0}^{x}dy \Bigg(\frac{\ln y}{(y-1)}-\frac{\ln (1-y)}{y}\Bigg),
\label{dlog}
\end{equation}
 and the fact
that $\epsilon _{A}^{+}(0)=\epsilon _{A}^{-}(0)$, $y_{A}^{+}(\infty)=y_{A}^{-}(\infty )=0$. 
Comparing Eq. (\ref{ceff2}) with the equivalent expression obtained in the parity invariant case 
 (see e.g. \cite{TBAZam1}), we conclude that we end up precisely with the same
situation: Having solved the TBA-equations (\ref{tbalimit1}) and (\ref{tbalimit}), 
we may compute the effective central charge in terms of Roger's
dilogarithm thereafter. Notice that in the ultraviolet limit we have
recovered the parity invariance and (\ref{ceff2}) holds for all finite values
of the resonance parameter. 
Notice also that, in fact,  in order to compute the effective central charge (\ref{ceff2})
we do not need to solve the TBA-equations (\ref{tbalimit1}), (\ref{tbalimit}) for any value of $\theta$, since only the values of the pseudo-energies at zero rapidity enter the expression (\ref{ceff2}).
Therefore, let us now consider the limit $\theta \rightarrow 0$ of Eqs. (\ref{tbalimit1}) and (\ref{tbalimit}).

When we assume that the
kernels $\Phi _{AB}(\theta )$ are strongly peaked$\footnote{%
That this assuption holds for the case at hand is most easily seen by noting
that the logarithmic derivative of a basic building block $(x)_{\theta }$ of
the S-matrix reads 
\[
-i\frac{d}{d\theta }\ln (x)_{\theta }=-\frac{\sin \left( \frac{\pi }{k}%
\,x\right) }{\cosh \theta -\cos \left( \frac{\pi }{k}\,x\right) }%
=-2\sum_{n=1}^{\infty }\sin \left( \frac{\pi n }{k}\,x\right) e^{-n|\theta
|}\,. \]
From this we can read off directly the decay properties.}$ at $\theta =0$
and develop the characteristic plateaux one observes for the scaling models,
we can take out the $L$-functions from the integral in the equations (\ref
{tbalimit1}), (\ref{tbalimit}) and obtain 

\begin{equation}
\epsilon _{A}^{\pm }(0)+\sum_{B}\,N_{AB}L_{B}^{\pm }(0)=0\qquad \quad \text{%
with \quad }N_{AB}=\frac{1}{2\pi }\int\limits_{-\infty }^{\infty }d\theta
\,\,\Phi _{AB}(\theta )\,\,,  \label{consttba}
\end{equation}
\noindent where we again appeal to (\ref{subs}).  
Note that in (\ref{consttba}) we have recovered the parity invariance.

Having the resonance parameter $\sigma $ present in our theory we may also
encounter a situation in which $\Phi _{AB}(\theta )$ is peaked at $\theta
=\pm \sigma $. This means in order for (\ref{consttba}) to be valid, we have
to assume $\epsilon _{A}^{\pm }(0)=\epsilon _{A}^{\pm }(\pm \sigma )$ in the
limit $r\rightarrow 0$ in addition, to accommodate that situation. This
assumption will be justified for particular cases from the numerical results (see e.g.
Fig. \ref{fig11}), but can also be derived from analytical considerations which make use
of Eqs. (\ref{tbalimit1}) and (\ref{tbalimit}). Recall that these equations
have been obtained as the UV-limit
of the original TBA-equations (\ref{tbaa}), (\ref{ptba}). Obviously,  
since $x \rightarrow -\infty$, 
the condition $|\sigma| \ll x$ is always guaranteed in the UV-regime.
At the same time, footnote 1 shows that
 the kernels $\Phi_{AB}(\theta )$ are strongly peaked either at $\theta=0$ or $\theta= 
\pm \sigma$. Therefore, one can safely approximate
 the convolution term in (\ref{tbalimit1}) at $\theta=x$ by
\begin{equation}
\Phi _{AB}*L_{B}^{>}(x)\simeq  \Big(\int_{-\infty}^{\infty} \frac{d y}{2 \pi} 
\Phi_{AB}(y)\Big) L_B^{>}(x)= N_{AB} L_B^{>}(x), 
\label{luis}
\end{equation}
\noindent and analogously for (\ref{tbalimit}). 
The substitution of (\ref{luis}) in (\ref{tbalimit1}) and equivalently in (\ref{tbalimit})
leads to Eqs. (\ref{consttba}) with the help of  definitions  (\ref{kink}).
\vspace{0.25cm}

For the case at hand we read off from the integral representation of the
scattering matrices (\ref{Sint}) and (\ref{ijint})
\begin{equation}
N_{AB}=N_{ab}^{ij}=\delta _{ij}\delta
_{ab}-K_{ij}^{g}\,(K^{A_{k-1}})_{\,\,\,\,\,\,\,\,\,ab}^{-1}\,\,,  \label{NN}
\end{equation}
\noindent for $K^{g}$ and $K^{A_{k-1}}$ to be the Cartan matrices of $g$ and $A_{k-1}$ 
respectively.

With $N_{ab}^{ij}$ in the form (\ref{NN}) and the identifications 
\begin{equation}
Q_{a}^{i}:=\prod_{b=1}^{k-1}(1+\exp (-\epsilon _{b}^{i}(0)))^{K_{ab}^{-1}},
\end{equation}
the constant TBA-equations (\ref{consttba}) are precisely the equations which occurred before
in the context of the restricted solid-on-solid models \cite{Kirillov, Kun1, Kun2, Kun3, Kun4}. 
It was noted in there that (\ref{consttba}) may be solved elegantly in
terms of Weyl-characters and the reported effective central charge coincides
indeed with the one for the HSG-models (\ref{cdata}). This is a highly non-trivial
result which provides determinant support for the S-matrix proposal \cite{HSGS}, since
the ultraviolet central charge of all the parafermionic theories related to the cosets
$G_k/U(1)^{\ell}$ is precisely reproduced. Also,  from a
purely mathematical point of view, the fact that the quantity $c_{eff}$ given by
Eq. (\ref{ceff2}) is a rational number is quite exceptional. When this happens, 
the system (\ref{consttba}) and (\ref{ceff2}) is known under the special name of
{\bf accessible dilogarithms}. A more general set of $N$-matrices, containing 
(\ref{NN}) as a particular example, 
was obtained in the context of the $g |\tilde{g}$-theories \cite{FK}. 

It should be noted that we understand the $N$-matrix here as defined in (\ref
{consttba}) and not as the difference between the phases of the S-matrix. In
the latter case we encounter contributions from the non-trivial constant
phase factors $\eta $. Also in that case we may arrive at the same answer by
compensating them with a choice of a non-standard statistical interaction \cite{BF2}  as
outlined in section \ref{resonance}.

\vspace{0.3cm}

We would like to close this section with a comment which links our analysis
to structures which may be observed directly inside the conformal field
theory. When one carries out a saddle point analysis, see e.g. \cite{Rich, BF1},
on a Virasoro character of the general form 
\begin{equation}
\chi (q)=\sum\limits_{\vec{m}=0}^{\infty }\frac{q^{\frac{1}{2}\vec{m}(%
\mathbf{1}-N)\vec{m}^{t}+\vec{m}\cdot \vec{B}}}{(q)_{1}\ldots (q)_{(k-1)%
\text{$\ell $}}}\,\,,  \label{chi}
\end{equation}
with $(q)_{m}=\prod_{k=1}^{m}(1-q^{k})$, one recovers the set of coupled
equations as (\ref{consttba}) and the corresponding effective central charge
is expressible as a sum of Roger's dilogarithms as (\ref{ceff2}). Note that
when we choose $g\equiv A_{1}$ the HSG-model reduces to the minimal ATFT and
(\ref{chi}) reduces to the character formulae in \cite{KM}. Thus the
equations (\ref{consttba}) and (\ref{ceff2}) constitute an interface between
massive and massless theories, since they may be obtained on one hand in the
ultraviolet limit from a massive model and on the other hand from a limit
inside the conformal field theory. This means we can guess a new form of the
coset character, by substituting (\ref{NN}) into (\ref{chi}). However, since
the specific form of the vector $\vec{B}$ does not enter in this analysis
(it distinguishes the different highest weight representations) more work
needs to be done in order to make this more than a mere conjecture. 
Further results in this direction have been provided in \cite{prep}.

\section{Thermodynamic Bethe ansatz for the $\protect{SU(3)_{k}}$-HSG model}
\label{su3}
\indent \ \
We shall now focus our discussion on $G=SU(3)_{k}$. First of all we need to
establish how many free parameters we have at our disposal in this case.
According to the discussion in subsection \ref{hsgsec} we can tune the 
resonance parameter and the mass ratio 
\begin{equation}
\sigma :=\sigma _{21}=-\sigma _{12}\quad \text{and}\quad m_{1}/m_{2}\,\,.
\end{equation}

It will also be useful to exploit a symmetry present in the TBA-equations
related to $SU(3)_{k}$ by noting that the parity transformed equations (\ref
{ptba}) turn into the equations (\ref{tbaa}) when we exchange the masses of
the different types of solitons. This means the system remains invariant
under the simultaneous transformations 
\begin{equation}
\theta \rightarrow -\theta \quad \quad \text{and\qquad }m_{1}/m_{2}%
\rightarrow m_{2}/m_{1}\,\,.  \label{inv}
\end{equation}
For the special case $m_{1}/m_{2}=1$ we deduce therefore that $\epsilon
_{a}^{1}(\theta )=\epsilon _{a}^{2}(-\theta )$, meaning that a parity
transformation amounts to an interchange of the colours\footnote{Notice that, from now on, we will
use the mentioned substructure of the quantum numbers $A=(a, i)$ for the HSG-models. 
For the particular $SU(3)_k$-case we will have $a=1, \cdots, k-1$ and $i=1,\cdots,\ell$, 
where $\ell=2$ is the rank of  $su(3)$.}. Furthermore,
we see from (\ref{ptba}) and the defining relation $\sigma =\sigma
_{21}=-\sigma _{12}$ that changing the sign of the rapidity variable is
equivalent to $\sigma \rightarrow -\sigma $. Therefore, we can restrict
ourselves to the choice $\sigma \geq 0$ without loss of generality.

\subsection{Staircase behaviour of the scaling function}
\label{stairbehaviour}
\indent  \ \
We will now come to the evaluation of the scaling function (\ref{scale}) for
finite and small scale parameter $r$. To do this we have to solve first the
TBA equations (\ref{tbaa}) for the pseudo-energies, which up to now has not
been achieved analytically for systems with a non-trivial dynamical
interaction due to the non-linear nature of the integral equations.
Nonetheless, numerically this problem can be controlled relatively well.
In \cite{FKS1} a rigorous proof of the existence and uniqueness of  solutions to 
TBA-equations of the type (\ref{tbaa}), (\ref{ptba}) was provided. Before \cite{FKS1}
such properties were simply assumed to hold in the light of the consistent analytical
and numerical results obtained. However, the proof provided in \cite{FKS1} is very
relevant since if  the equations  (\ref{tbaa}), (\ref{ptba}) allowed several different solutions
for the pseudo-energies  $\epsilon_A^{\pm}(\theta)$,  
different values for the  effective central charge could
be obtained starting with the same S-matrix and statistics and, consequently,
the results of the TBA-analysis would not provide a reliable consistency check of the 
mentioned S-matrix.

In the UV-regime, for $r\ll 1$, one is in a better position and can set up
approximate TBA-equations involving, formally,
 massless particles, for which various approximation schemes have
been developed which depend on the general form of the $L$-functions. Since
those functions are not known a priori, one may justify ones assumptions in
retrospect by referring to the numerics. In section \ref{examples} we present numerical
solutions for the equations (\ref{tbaa}) for various levels $k$ showing that
the $L$-functions develop at most two (three if $m_{1}$ and $m_{2}$ are very
different) plateaux in the range $x <\theta <-x$
and then fall off rapidly (see Fig. \ref{fig11}). This type of behaviour is similar
to the one known from minimal ATFT \cite{TBAZam1,TBAKM}, and we can
therefore adopt various arguments presented in that context. The main
difficulty we have to deal with here is to find the appropriate ``massless''
TBA equations accommodating the dependence of the TBA equations on the
resonance shifts $\sigma $.

We start by giving also an integral representation for the kernel (\ref{kernel}).
This representation can be obtained from the integral representation of the two-point scattering
matrices  (\ref{Sint}), (\ref{ijint}) by using the same kind of arguments presented in \cite{FKS1}
for ATFT's related to simply laced Lie algebras.
Moreover, we can separate the kernel (\ref{kernel}) into two parts 

\begin{eqnarray}
\phi _{ab}(\theta ) &=&\Phi _{ab}^{ii}(\theta )=\int dt\,\left[ \delta
_{ab}-2\cosh \tfrac{\pi t}{k}\left( 2\cosh \tfrac{\pi t}{k}-I\right)
_{ab}^{-1}\right] \,e^{-it\theta }\;,\quad  \label{kernelint} \\
\psi _{ab}(\theta ) &=&\Phi _{ab}^{ij}(\theta +\sigma _{ji})=\int dt\,\left(
2\cosh \tfrac{\pi t}{k}-I\right) _{ab}^{-1}\,e^{-it\theta }\;\,\, ,i\neq
j,\,\,\,K_{ij}^{g}=-1.  \label{kernelint2}
\end{eqnarray}
Here $\phi _{ab}(\theta )$ is just the TBA-kernel of the $A_{k-1}$-minimal
ATFT \cite{FKS1} and in the remaining kernel $\psi _{ab}(\theta )$ we have removed the
resonance shift. Note that $\phi ,\psi $ do not depend on the colour values $
i,j$ and may therefore be dropped all together in the notation. 
They are generically valid for all values of
the level $k$. The convolution term in (\ref{tbaa}) in terms of $\phi ,\psi $
is then re-written as 
\begin{equation}
\sum_{j=1}^{\ell }\sum_{b=1}^{k-1}\Phi _{ab}^{ij}*L_{b}^{j}(\theta
)=\sum_{b=1}^{k-1}\phi _{ab}*L_{b}^{i}(\theta )+\sum\Sb j=1  \\ j\neq i 
\endSb ^{\ell }\sum_{b=1}^{k-1}(\psi _{ab}*L_{b}^{j})(\theta -\sigma _{ji})\,.
\label{conv}
\end{equation}
These equations illustrate that whenever we are in a regime in which the
second term in (\ref{conv}) is negligible we are left with $\ell $
non-interacting copies of the $A_{k-1}$-minimal ATFT. In other words, whenever
the resonance parameters $\sigma_{ij}$ become very big, the unstable particles
can not be formed anymore so that all the two-particle scattering amplitudes
satisfy Eq. (\ref{and1}) and we have  $\ell$ 
copies of minimal $A_{k-1}$-ATFT. 

We will now specialise the discussion on the $SU(3)_{k}$-case for which 
it is very useful to 
perform the shifts $\theta \rightarrow \theta \pm \sigma /2$ in the TBA-equations. 
In the UV-limit $r\rightarrow 0$ with $\sigma \gg 1$ the shifted
functions can be approximated by the solutions of the following sets of
integral equations 
\begin{eqnarray}
\varepsilon _{a}^{\pm }(\theta )+\sum_{b=1}^{k-1}\phi _{ab}*L_{b}^{\pm
}\left( \theta \right) +\sum_{b=1}^{k-1}(\psi _{ab}*L_{b}^{\mp })\left( \theta
\right) &=&r^{\prime }\,M_{a}^{\pm }\,e^{\pm \theta }\quad \,  \label{uvTba}
\\
\hat{\varepsilon}_{a}^{\pm }(\theta )+\sum_{b=1}^{k-1}(\phi _{ab}*\hat{L}%
_{b}^{\pm })\left( \theta \right) &=&r^{\prime }\,M_{a}^{\mp }\,\,e^{\pm
\theta }\,,\quad  \label{kinktba}
\end{eqnarray}
where we have introduced the parameter $r^{\prime }=r/2 \,e^{\sigma/2}= e^{x+\sigma/2}$
 familiar from the discussion of massless scattering \cite{triZam} and the masses $
M_{a}^{+/-}=M_{a}^{1 \,/ \,2}$. Notice that the indices ``+'', ``-''  in the pseudo-energies
and $L$-functions have nothing to do with the same indices used in section \ref{resonance}.
Now, these labels refer to  the two different possible colour indices $i=1,2$, since the rank
of the Lie algebra $su(3)$ is $\ell =2$ and the functions  $\varepsilon_a^{\pm}(\theta)$ are 
in fact related to the original ones $\epsilon_{a}^{i}(\theta)=$$\epsilon_A( \theta) =$$\epsilon_A^{+}(\theta)$ by means of 
a rapidity shift, in a similar fashion to the ``kink'' equations (\ref{kink})  introduced in 
subsection \ref{UVCC}. The same relationship holds for the functions 
$\hat\varepsilon_a^{\pm}$ which make sense in a different regime of values 
of the parameters $r, \sigma$. 
More  precisely, the  relationship between the solutions of the 
system (\ref{uvTba}), (\ref{kinktba})
and those of the original TBA-equations is given by 
\begin{eqnarray}
\varepsilon _{a}^{+ | -}(\theta ) &=&\epsilon _a^{1 | 2}(\theta \pm \sigma
/2)\qquad \text{for\quad }x\ll \pm \theta \ll x +\sigma
\label{e1} \\
\hat\varepsilon _{a}^{+ | -}(\theta ) &=&{\epsilon}_{a}^{1 | 2}(\theta \pm\sigma /2)\qquad 
\text{for\quad }\pm \theta \ll \min [-x,x+\sigma ].  \label{e2}
\end{eqnarray}
 \noindent  where we have assumed the equality of the mass scales i.e, $m_1=m_2$.
Similar equations may be written down for the generic case.
The approximations leading to (\ref{uvTba}), (\ref{kinktba}) follow similar lines as 
the ones involved in the derivation of Eq. (\ref{ceff2}), with the difference that
now two parameters $x, \sigma$ are involved in the discussion and their relative value
has to be taken into account in order to establish  which terms are negligible and which are
not in the mentioned limit $r \rightarrow 0, \sigma \gg 1$. 

In deriving  (\ref{e2}) we have neglected  the
convolution terms $(\psi _{ab}*L_{b}^{1})(\theta +\sigma )$ and
 $(\psi_{ab}*L_{b}^{2})(\theta -\sigma )$ which appear in the TBA-equations for 
$\epsilon _{a}^{2}(\theta )$ and $\epsilon _{a}^{1}(\theta )$,
respectively. This is justified by the following argument: For $|\theta |\gg 
$ $-x$ the free on-shell energy term is dominant in the TBA equations,
i.e.  $\epsilon _{a}^{i}(\theta )\approx r\,M_{a}^{i}\cosh \theta \simeq M_a^{i} e^{|\theta| + x}$
and the functions $L_{a}^{i}(\theta )$ are almost zero. The kernels $\psi _{ab}$ are
centered in a region around the origin\thinspace $\theta =0$ outside of
which they exponentially decrease (see footnote in subsection \ref{UVCC} for an explanation of this).
This means that the convolution terms in question can be neglected safely if 
$\theta \ll x+\sigma $ and $\theta \gg  -x-\sigma $,
respectively. Note that the massless system provides a solution for the
whole range of $\theta $ for non-vanishing $L$-function only if the ranges of
validity in (\ref{e1}) and (\ref{e2}) overlap, i.e.  if \quad $x \ll
\min [-x, x+\sigma ]$ which is always true in the limit $r\rightarrow 0$ 
when $\sigma \gg 1$. The solution is uniquely defined in the
overlapping region. These observations are confirmed by our numerical
analysis.

The resulting equations (\ref{kinktba}) are therefore decoupled and we can
determine $\hat{L}^{+}_a$ and $\hat{L}^{-}$ individually. In contrast, the
equations (\ref{uvTba}) for $L_{a}^{\pm }$ are still coupled to each other
due to the presence of the resonance shift. Formally, the on-shell energies
for massive particles have been replaced by on-shell energies for massless
particles in the sense of \cite{triZam}, such that if we interpret $r^{\prime }$ 
as an independent new scale parameter the sets of equations 
(\ref{uvTba}) and (\ref{kinktba}) could be identified formally as massless TBA-systems
in their own right.

Introducing then the scaling function associated with the system (\ref{uvTba}) as

 \begin{equation}
c_{\text{o}}(r^{\prime })=\frac{3\,r^{\prime }}{\pi ^{2}}\sum_{a=1}^{k-1}%
\int d\theta \,\,\left[ M_{a}^{+}\,e^{\theta }L_{a}^{+}(\theta
)+\,M_{a}^{-}\,e^{-\theta }L_{a}^{-}(\theta )\right],   \label{c0}
\end{equation}
and analogously the scaling function associated with (\ref{kinktba}) as 
\begin{equation}
\hat{c}_{\text{o}}(r^{\prime })=\frac{3\,r^{\prime }}{\pi ^{2}}%
\sum_{a=1}^{k-1}\int d\theta \,\left[ M_{a}^{+}\,e^{\theta }\hat{L}%
_{a}^{+}(\theta )+M_{a}^{-}\,e^{-\theta }\hat{L}_{a}^{-}(\theta )\right],
\label{ckink}
\end{equation}
we can express the scaling function (\ref{scale}) of the HSG model in the
regime $r\rightarrow 0$, $\sigma \gg 1$ approximately by 
\begin{eqnarray}
c(r,\sigma ) &=&\frac{3\,r\,e^{\frac{\sigma }{2}}}{2\pi ^{2}}%
\sum_{i=1,2}\sum_{a=1}^{k-1}M_{a}^{i}\int d\theta \,\left[ \,e^{\theta
}L_{a}^{i}(\theta -\sigma /2)+e^{-\theta }L_{a}^{i}(\theta +\sigma
/2)\right]   \nonumber \\
&\approx &c_{\text{o}}\left( r^{\prime }\right) +\hat{c}_{\text{o}}\left(
r^{\prime }\right) \,\;.  \label{uvscale}
\end{eqnarray}
Thus, we have formally decomposed the massive $SU(3)_{k}$-HSG model in the
UV regime into two massless TBA systems (\ref{uvTba}) and (\ref{kinktba}),
reducing therefore the problem of calculating the scaling function of the
HSG model in the UV-limit, $r\rightarrow 0$, to the problem of evaluating the
scaling functions (\ref{c0}) and (\ref{ckink}) for the scale parameter $%
r^{\prime }$. The latter depends on the relative size of $x$ and the
resonance shift $\sigma.$ Keeping now $\sigma \gg 1$ fixed, and letting $r
$ vary from finite values to the deep UV-regime, i.e.  $r=0$, the scale
parameter $r^{\prime }$ governing the massless TBA systems will pass
different regions. For the regime $-x<\sigma /2$ we see that the
scaling functions (\ref{c0}) and (\ref{ckink}) are evaluated at $r^{\prime
}>1$, whereas for $-x>\sigma /2$ they are taken at $r^{\prime }<1$.
Thus, when performing the UV-limit of the HSG-models the massless TBA-systems
pass formally from the ``infrared'' to the ``ultraviolet'' regime with
respect to the parameter $r^{\prime }$. We emphasise that the scaling
parameter $r^{\prime }$ has only a formal meaning and that the physical
relevant limit we consider is still the UV-limit, $r\rightarrow 0$, of the HSG-models. However,
we will find later, in the context of the form factor approach
more reasons which support the description of systems of this type, 
containing resonance parameters, as massless systems (see section \ref{rgflow} in chapter \ref{ffs}).
Proceeding this way has the advantage that we can treat the
scaling function of the HSG-models by the UV and IR central charges of the
systems (\ref{uvTba}) and (\ref{kinktba}) as functions of $r^{\prime }$ 
\begin{equation}
c(r,\sigma )\approx c_{\text{o}}\left( r^{\prime }\right) +\hat{c}_{\text{o}%
}\left( r^{\prime }\right) \approx \left\{ 
\begin{array}{ll}
\,c_{IR}+\hat{c}_{IR}\,, & 0\ll -x \ll \frac{\sigma }{2} \\ 
c_{UV}+\hat{c}_{UV}\,, & \frac{\sigma }{2}\ll -x
\end{array}
\right. \,\,.  \label{step}
\end{equation}
Here we defined the quantities $c_{IR}:=\lim_{r^{\prime }\rightarrow \infty
}c_{\text{o}}(r^{\prime })$, $c_{UV}:=\lim_{r^{\prime }\rightarrow 0}c_{%
\text{o}}(r^{\prime })$ and $\hat{c}_{IR},\hat{c}_{UV}$ analogously in terms
of $\hat{c}_{\text{o}}(r^{\prime })$.

In the case of $c_{IR}+\hat{c}_{IR}\neq c_{UV}+\hat{c}_{UV}$, we infer from (%
\ref{step}) that the scaling function develops at least two plateaux at
different heights. A similar phenomenon was previously observed for the theories
discussed in \cite{staircase, Martins, DoRav}, where infinitely many plateaux occurred which
prompted to call them ``staircase models''. As a difference, however the resonance
shifts enter the corresponding S-matrices in a very different way.

\noindent In the next
subsection we determine the central charges in (\ref{step}) by means of the
standard TBA central charge calculation, setting up the so-called constant
TBA equations.

\subsection{Central charges from constant TBA equations}
\label{ctba}
\indent \ \
In this subsection we will perform the limits $r^{\prime }\rightarrow 0$ and 
$r^{\prime }\rightarrow \infty $ for the massless systems (\ref{uvTba}) and (%
\ref{kinktba}) referring to them formally as ``UV-'' and ``IR-limit'',
respectively, keeping however in mind that both limits are still linked to
the UV-limit of the HSG model in the scale parameter $r$, as discussed in the
preceding subsection. We commence with the discussion of the ``UV-limit'' $%
r^{\prime }\rightarrow 0$ for the subsystem (\ref{uvTba}). We then have
three different rapidity regions in which the pseudo-energies are
approximately given by 
\begin{equation}
\varepsilon _{a}^{\pm }(\theta )\approx \left\{ 
\begin{array}{ll}
r^{\prime }M_{a}\,e^{\pm \theta }, & \text{for }\pm \theta \gg -\ln
r^{\prime } \\ 
-\sum_{b}\phi _{ab}*L_{b}^{\pm }(\theta )-\sum_{b}\psi _{ab}*L_{b}^{\mp
}(\theta ), & \text{for }\ln r^{\prime }\ll \theta \ll -\ln r^{\prime } \\ 
-\sum_{b}\phi _{ab}*L_{b}^{\pm }(\theta ), & \text{for }\pm \theta \ll \ln
r^{\prime }
\end{array}
\right. \;.
\end{equation}
We have only kept here the dominant terms up to exponentially small
corrections. We proceed analogously to the discussion as may be found in 
\cite{TBAZam1,TBAKM}. We see that in the first region the particles become
asymptotically free. For the other two regions the TBA equations can be
solved by assuming the $L$-functions to be constant. Exploiting once more that
the TBA kernels are centered at the origin and decay exponentially, we can
similarly as in section \ref{resonance} take the $L$-functions outside of the integrals and
end up with the sets of equations 
\begin{eqnarray}
x_{a}^{\pm } &=&\prod_{b=1}^{k-1}(1+x_{b}^{\pm })^{\hat{N}%
_{ab}}(1+x_{b}^{\mp })^{N_{ab}^{\prime }}\quad \quad \text{for }\ln
r^{\prime }\ll \theta \ll -\ln r^{\prime },  \label{ctba1} \\
\hat{x}_{a} &=&\prod_{b=1}^{k-1}(1+\hat{x}_{b})^{\hat{N}_{ab}}\qquad \qquad
\qquad \quad \,\text{for }\pm \theta \ll \ln r^{\prime },  \label{ctba2}
\end{eqnarray}
for the constants $x_{a}^{\pm }=e^{-\varepsilon _{a}^{\pm }(0)}$ and $\hat{x}%
_{a}=e^{-\varepsilon _{a}^{\pm }(\mp \infty )}$. The $N$-matrices can be read
off directly from the integral representations (\ref{kernelint}) and (\ref
{kernelint2}) 
\begin{eqnarray}
&&\hat{N}_{ab}:=\frac{1}{2\pi }\int\limits_{-\infty}^{\infty}d \theta 
\,\, \phi_{ab}(\theta) =\delta_{ab}-2(K^{A_{k-1}})^{-1}_{\,\,\,\,ab}\,\,,\\
&&N_{ab}^{\prime }:=\frac{1}{2\pi }\int\limits_{-\infty}^{\infty} d\theta \,\,
\psi_{ab}(\theta) =(K^{A_{k-1}})^{-1}_ {\,\,\,\,ab}\,\,.
\end{eqnarray}
Notice that the latter $\hat{N}, N^{\prime}$-matrices are just the particularisation of
formula (\ref{NN}) for $K_{ij}^{g}=K_{ij}^{su(3)}$. The $\hat{N}$-matrix corresponds
to the case $i=j$ and the ${N}^{\prime}$-matrix is obtained for $i \neq j$.

Since the set of equations (\ref{ctba2}) has already been stated in the
context of minimal ATFT and its solutions may be found in \cite{TBAKM}, we
only need to investigate the equations (\ref{ctba1}). These equations are
simplified by the following observation. Changing $\theta $ by  $-\theta $ the
constant $L$-functions must obey the same constant TBA equation (\ref{ctba1})
but with the role of $L_{a}^{+}$ and $L_{a}^{-}$ interchanged. The
difference in the masses $m_{1},m_{2}$ has no effect as long as $m_{1}\sim
m_{2}$ since the on-shell energies are negligible in the central region $\ln
r^{\prime }\ll \theta \ll -\ln r^{\prime }$. Thus, we deduce $%
x_{a}^{+}=x_{a}^{-}=:x_{a}$ and (\ref{ctba1}) reduces to 
\begin{equation}
x_{a}=\prod_{b=1}^{k-1}(1+x_{b})^{N_{ab}}\;\qquad \quad \text{with}\quad
N_{ab}=\delta_{ab}-(K^{A_{k-1}})^{-1}_{\,\,\,ab}\;.  \label{cTba2a}
\end{equation}
\noindent
Remarkably, also this set of equations may be found in the literature in
the context of the restricted solid-on-solid models \cite{Kun1}.
Specializing some of the general Weyl-character formulae in \cite{Kun1, Kun2} to
the $su(3)_{k}$-case a straightforward calculation leads to 
\begin{equation}
x_{a}=\frac{\sin \left( \frac{\pi \,(a+1)}{k+3}\right) \sin \left( \frac{\pi
\,(a+2)}{k+3}\right) }{\sin \left( \frac{\pi \,\,a}{k+3}\right) \sin \left( 
\frac{\pi \,(a+3)}{k+3}\right) }-1\quad \text{and\quad }\hat{x}_{a}=\frac{%
\sin ^{2}\left( \frac{\pi \,(a+1)}{k+2}\right) }{\sin \left( \frac{\pi \,\,a%
}{k+2}\right) \sin \left( \frac{\pi \,(a+2)}{k+2}\right) }-1\,\,\text{.}
\label{cccTBA}
\end{equation}
Having determined the solutions of the constant TBA equations (\ref{ctba1})
and (\ref{cTba2a}) one can proceed via the standard TBA-calculations along
the lines of \cite{TBAZam1,triZam,TBAKM} and compute the central charges
from (\ref{c0}), (\ref{ckink}), expressing them in terms of Roger's
dilogarithm functions 
\begin{eqnarray}
c_{UV} &=&\lim_{r^{\prime }\rightarrow 0}c_{\text{o}}(r^{\prime })=\frac{6}{%
\pi ^{2}}\sum_{a=1}^{k-1}\left[ 2\mathcal{L}\left( \frac{x_{a}}{1+x_{a}}%
\right) -\mathcal{L}\left( \frac{\hat{x}_{a}}{1+\hat{x}_{a}}\right) \right]
\;,  \label{cuv} \\
\hat{c}_{UV} &=&\lim_{r^{\prime }\rightarrow 0}\hat{c}_{\text{o}}(r^{\prime
})=\frac{6}{\pi ^{2}}\sum_{a=1}^{k-1}\mathcal{L}\left( \frac{\hat{x}_{a}}{1+%
\hat{x}_{a}}\right) \,\,.
\end{eqnarray}
Using the non-trivial identities 
\begin{equation}
\frac{6}{\pi ^{2}}\sum_{a=1}^{k-1}L\left( \frac{x_{a}}{1+x_{a}}\right) =3\,%
\frac{k-1}{k+3}\quad \text{and}\quad \frac{6}{\pi ^{2}}\sum_{a=1}^{k-1}L%
\left( \frac{\hat{x}_{a}}{1+\hat{x}_{a}}\right) =2\,\frac{k-1}{k+2}\;
\end{equation}
found in \cite{Log} and \cite{Kirillov}, we finally end up with 
\begin{equation}
c_{UV}=\frac{\left( k-1\right) (4k+6)}{\left( k+3\right) \left( k+2\right) }%
\qquad \text{and\qquad }\hat{c}_{UV}=2\,\frac{k-1}{k+2}\,\,.  \label{cuvv}
\end{equation}
For the reasons mentioned above, $\hat{c}_{UV}$ coincides with the effective
central charge obtained from $A_{k-1}$ minimal ATFT describing parafermions 
\cite{Witten} in the conformal limit, in other words is the central charge associated
to the WZNW-coset $SU(2)_{k}/U(1)$. Notice that $c_{UV}$ corresponds formally to
the coset $(SU(3)_{k}/U(1)^{2})/(SU(2)_{k}/U(1))$. 

The discussion of the infrared limit may be carried out completely analogous
to the one performed for the UV-limit. The only difference is that in case
of the system (\ref{uvTba}) the constant TBA equations (\ref{ctba1}) drop
out because in the central region the free energy terms becomes dominant
when $r^{\prime }\rightarrow \infty $. Thus, in the IR-regime, the
central charges of both systems coincide with $\hat{c}_{UV}$, 
\begin{equation}
c_{IR}=\lim_{r^{\prime }\rightarrow \infty }c_{\text{o}}(r^{\prime })=\hat{c}%
_{IR}=\lim_{r^{\prime }\rightarrow \infty }\hat{c}_{\text{o}}(r^{\prime
})=2\,\frac{k-1}{k+2}\;.  \label{cir}
\end{equation}

\noindent
The results (\ref{cuvv}) and (\ref{cir}) show that we can identify the massless flow 
\begin{equation}
(SU(3)_{k}/U(1)^{2})/(SU(2)_{k}/U(1)) \rightarrow SU(2)_{k}/U(1),
\label{subflow}
\end{equation}
\noindent as a subsystem inside the $SU(3)_k$-HSG model. The study of more sophisticated 
flows is left for future investigations \cite{prep}.

In summary, collecting the results (\ref{cuvv}) and (\ref{cir}), we can
express equation (\ref{step}) explicitly in terms of the level $k$, 
\begin{equation}
c(r,M_{\tilde{c}}^{\tilde{k}})\approx \left\{ 
\begin{array}{ll}
4\,\frac{k-1}{k+2}\,\,, & \qquad \text{for\quad }1\ll \frac{2}{r}\ll M_{%
\tilde{c}}^{\tilde{k}} \\ 
6\,\frac{k-1}{k+3}\,, & \qquad \text{for\quad }M_{\tilde{c}}^{\tilde{k}}\ll 
\frac{2}{r}
\end{array}
\right. \,\,.  \label{stepk}
\end{equation}
We have replaced the limits in (\ref{step}) by mass scales in order to
exhibit the underlying physical picture. Here $M_{\tilde{c}}^{\tilde{k}}$ is
the smallest mass of an unstable bound state which may be formed in the
process $(a, i)+(b,j)\rightarrow (\tilde{c},\tilde{k})$ for $K_{ij}^{g}\neq
0,2$. We also used that the Breit-Wigner formula (\ref{BW11}) implies that $
M_{\tilde{c}}^{\tilde{k}}\sim m \, e^{\sigma /2}$ for $m$ to be the mass scale of the stable
particles and  large positive $\sigma $.
First, one should note that in the deep UV-limit we obtain the same effective
central charge as in section \ref{UVCC}, albeit in a quite different manner. On the
mathematical side, this implies some non-trivial identities for Roger's
dilogarithm and, on the physical side, the relation (\ref{stepk}) exhibits a more detailed
behaviour than the analysis in section \ref{UVCC}. In the first regime the lower
limit indicates the onset of the lightest stable soliton in the two copies
of the complex sine-Gordon model. The unstable particles are on an energy scale
much larger than the temperature of the system. Thus, the dynamical
interaction between solitons of different colours is ``frozen'' and we end
up with two copies of the $SU(2)_{k}/U(1)$-coset which do not interact with
each other. As soon as the parameter $r$ reaches the energy scale of the
unstable solitons with mass $M_{\tilde{c}}^{\tilde{k}}$, the solitons of
different colours start to interact, being now enabled to form bound states.
This interaction breaks parity and forces the system to approach the $%
SU(3)_{k}/U(1)^{2}$-coset model, with central charge given by the formula in 
(\ref{cdata}) for $G=SU(3)$.

The case when $\sigma $ tends to infinity is special and one needs to pay
attention to the order in which the limits are taken, we have 
\begin{equation}
4\,\frac{k-1}{k+2}=\lim_{r\rightarrow 0}\lim_{\sigma \rightarrow \infty
}c(r,\sigma )\neq \lim_{\sigma \rightarrow \infty }\lim_{r\rightarrow
0}c(r,\sigma )=6\,\frac{k-1}{k+3}\,.
\end{equation}
\noindent Physically, one can understand this result by arguing
that if we take $\sigma \rightarrow \infty$  before we carry out the UV-limit
the formation of bound states is never possible 
since the mass of the unstable particle $M_{\tilde{c}}^{\tilde{k}}
 \sim m \, e^{\sigma/2} \rightarrow \infty$
and therefore,  is always on an energy scale much higher 
than the temperature of the system. 
Consequently,  the  theory reduces to two non-interacting 
copies of the ${SU(2)_k}/U(1)$-coset model. 
On the other hand, if the UV-limit is considered before, as long as
 we keep $\sigma$ finite, we will
obtain the same result given in the second part of  (\ref{stepk}),
 namely the central charge related to the $SU(3)_k /U(1)^2$-coset. 
Since this  value does not depend on the concrete value of $\sigma$, once the 
UV-limit has been carried out the value obtained is not modified by
 a posterior limit $\sigma \rightarrow \infty$.

One might enforce an additional step in the scaling function by exploiting
the fact that the mass ratio $m_{1}/m_{2}$ is not fixed. So, it may be chosen
to be very large or very small. This amounts to decoupling the TBA-systems for
solitons with different colour by shifting one system to the infrared with
respect to the scale parameter $r$. The plateau then has an approximate
width of $\sim \ln |m_{1}/m_{2}|$ (see figure \ref{fig12}). However, as soon as $r$
becomes small enough the picture we discussed for $m_{1}\sim m_{2}$ is
recovered.

\subsection{Restoring parity and eliminating the resonances}
\label{rper}
\indent \ \
In this subsection we are going to investigate the special limit $\sigma
\rightarrow 0$ which is equivalent to choosing the vector couplings $\Lambda
_{\pm }$ in (\ref{Actq}) parallel or anti-parallel. For the classical
theory it was pointed out in \cite{HSG} that only then the equations of
motion are parity invariant. Also the TBA-equations become parity invariant
in the absence of the resonance shifts, albeit the S-matrix still violates
it through the phase factors $\eta $. The reason is that the phases $\eta$  cancel when 
computing the kernels  (\ref{kernelint}), (\ref{kernelint2}) due to the definition (\ref{kernel}). 
Since in the UV regime a small
difference in the masses $m_{1}$ and $m_{2}$ does not effect the outcome of
the analysis, we can restrict ourselves to the special situation $m_{1}=m_{2}
$, in which case we obtain two identical copies of the system 
\begin{equation}
\epsilon _{a}(\theta )+\sum_{b=1}^{k-1}(\phi _{ab}+\psi _{ab})*L_{b}(\theta
)=r\,M_{a}\cosh \theta \;.
\end{equation}
Then we have $\epsilon _{a}(\theta )=\epsilon _{a}^{1}(\theta )=\epsilon
_{a}^{2}(\theta )$, $M_{a}=M_{a}^{1}=M_{a}^{2}$ and the scaling
function is given by the expression 
\begin{equation}
c(r,\sigma =0)=\frac{6\,r}{\pi ^{2}}\sum_{a=1}^{k-1}M_{a}\int d\theta
\,\,L_{a}(\theta )\cosh \theta \;.
\end{equation}
The factor 2 in comparison with (\ref{scale}) takes the two copies for $%
i=1,2$ into account. The discussion of the high-energy limit follows the
standard arguments similar to the one of the preceding subsection and as may
be found in \cite{TBAZam1,TBAKM}. Instead of shifting by the resonance
parameter $\sigma $, one now shifts the TBA equations by $x=\ln (r/2)$. The
constant TBA equation which determines the UV central charge then just
coincides with (\ref{ctba1}). We therefore obtain 
\begin{equation}
\lim_{r\rightarrow 0}\lim_{\sigma \rightarrow 0}c(r,\sigma )=\frac{12}{\pi
^{2}}\sum_{a=1}^{k-1}L\left( \frac{x_{a}}{1+x_{a}}\right) =6\,\frac{k-1}{k+3}%
\;.
\end{equation}
Thus, again we recover the coset central charge (\ref{cdata}) for $G=SU(3)$,
but this time without breaking parity in the TBA-equations. This is in
agreement with the results of section \ref{UVCC}, which showed that we can obtain
this limit for any finite value of $\sigma $.

\subsection{Universal TBA equations and $Y$-systems}
\label{YY}
\indent \ \
Before we turn to the discussion of specific examples, for fixed values of
the level $k$, we would like to comment that there exists an alternative
formulation of the TBA-equations (\ref{tbaa}) in terms of a single integral
kernel. This version of the TBA-equations is of particular advantage when
one wants to discuss properties of the model and keep the level $k$ generic.
The starting point towards this re-formulation of the TBA-equations is the
 computation of  their  Fourier transform. This is particularly simple for the TBA-kernels due
to the general property,
\begin{equation}
\widetilde{f*g}(t)=\int\limits_{-\infty}^{\infty} dt \,e^{-i t \theta} f*g(\theta)= \frac{1}{2\pi}
\tilde{f}(t)\tilde{g}(t), 
\label{prop}
\end{equation}
\noindent where the `tilde' denotes now the Fourier transform. The 
Fourier transform of the TBA-kernels $\phi $ and $\psi $ can be now read
 off directly from (\ref{kernelint}) and (\ref{kernelint2}),
\begin{eqnarray}
\frac{\tilde\phi_{ab}(t)}{2 \pi}&=&\delta_{ab}-2\cosh \tfrac{\pi t}{k}\left( 2\cosh \tfrac{\pi t}{k}-I\right)
_{ab}^{-1}, \label{fker1}\\
\frac{\tilde\psi_{ab} (t)}{2\pi}&=&\left(2\cosh \tfrac{\pi t}{k}-I\right) _{ab}^{-1},
\label{fker2}
\end{eqnarray}
\noindent where $I_{ab}$ is the incidence matrix of $A_{k-1}$. 
Defining now the  functions $\xi_a^{i} (\theta) = \epsilon_a^{i}(\theta) - r M_a^{i} \cosh \theta $
one can rewrite the original TBA-equations (\ref{tbaa}), (\ref{ptba}) in their Fourier transformed  version as
\begin{equation}
2 \pi \sum_{b=1}^{k-1}
\left(2\cosh \tfrac{\pi t}{k}-I\right) _{ab}(\tilde{\xi}_b^{i}(t) + \tilde{L}_b^{i}(t))= 
\Big( 2 \cosh \tfrac{\pi t}{k}- e^{i t \sigma_{ji}}\Big) \tilde{L}_a^{i}(t). 
\end{equation}
The inverse Fourier transform of the latter equations gives the set of integral
equations 
\begin{equation}
\epsilon _{a}^{i}(\theta )+\Omega _{k}*L_{a}^{j}(\theta -\sigma
_{ji})=\sum_{b=1}^{k-1}I_{ab}\,\Omega _{k}*(\epsilon
_{b}^{i}+L_{b}^{i})(\theta )\;.\quad  \label{uni}
\end{equation}
\noindent in terms of the kernel $%
\Omega _{k}$ which, with the help of (\ref{prop}), is easily found to be 
\begin{equation}
\Omega _{k}(\theta )=\frac{k/2}{\cosh (k\theta /2)}\;.
\end{equation}
\noindent 
Similarly to the analysis carried out in \cite{TBAZamun}, 
the on-shell energies have dropped out in (\ref{uni})
because of the crucial relation \cite{dis, Freeman, ALO}
\begin{equation}
\sum_{b=1}^{k-1}I_{ab}M_{b}^{i}=2\cos \tfrac{\pi }{k}\,M_{a}^{i}\;,
\end{equation}
which is a property of the mass spectrum inherited from ATFT.
 Even though the explicit dependence on the scale parameter $r$ has been
lost, it is recovered from the asymptotic condition 
\begin{equation}
\epsilon _{a}^{i}(\theta )\stackunder{\theta \rightarrow \pm \infty }{%
\longrightarrow }rM_{a}^{i}\,e^{\pm \theta }\,\,\,.
\end{equation}
The integral kernel present in (\ref{uni}) has now a very simple form and
the $k$-dependence is easily read off.

Closely related to the TBA equations in the form (\ref{uni}) are the
following functional relations also referred to as $Y$-systems. Using complex
continuation (see e.g. \cite{FKS1} for a similar computation),  together with the property
\begin{equation}
f(\theta+ \frac{i \pi}{k}) + f( \theta - \frac{i \pi}{k}) = 2 \int\limits_{-\infty}^{\infty}
 dt \, e^{-i t\theta} \cosh \frac{\pi t}{k}\,\tilde{f}(t),
\end{equation}
\noindent which holds for any function $f(\theta)$ whose Fourier transform is well defined, 
and defining the quantities $Y_{a}^{i}(\theta )=\exp (-\epsilon _{a}^{i}(\theta ))$ the
integral equations are replaced by 
\begin{equation}
Y_{a}^{i}(\theta +\tfrac{i\pi }{k})Y_{a}^{i}(\theta -\tfrac{i\pi }{k}%
)=\left[ 1+Y_{a}^{j}(\theta -\sigma _{ji})\right] \prod_{b=1}^{k-1}\left[
1+Y_{b}^{i}(\theta )^{-1}\right] ^{-I_{ab}}\,.  \label{Y}
\end{equation}
The $Y$-functions are assumed to be well defined on the whole complex rapidity
plane where they give rise to entire functions. These systems are useful in
many aspects, for instance they may be exploited in order to establish
periodicities in the $Y$-functions, which in turn can be used to provide
approximate analytical solutions of the TBA-equations.  By doing
conformal perturbation theory (CPT) around the underlying CFT 
describing the UV-limit of the massive QFT it was shown  in \cite{TBAZam1, KM2} that
the scaling function, both for unitary and non-unitary CFT's,
can be expanded in integer multiples of the period of the $Y$-functions which is directly linked
to the dimension of the perturbing operator $\Delta$. Such expansion was found to have
the general form 
\begin{equation}
c(r)=c_{eff}+ \frac{6}{\pi} B(\lambda) r^2 + \sum_{n=1}^{\infty}{C}_n  (r^{y}\lambda)^{n},
\label{cpt}
 \end{equation}
\noindent for $\lambda$ to be the coupling constant characterising the perturbing term (see Eq. (\ref{pss}) in chapter \ref{ntft}) and $y=2-\Delta$.
As explained in \cite{TBAZam1, KM2}, the coefficients $C_n$ of the $r$-expansion can be computed by doing  CPT around the underlying CFT, as well as the function $B(\lambda)$ which might be fixed by the requirement $\lim_{r\rightarrow \infty} 
c(r)=0$ in massive QFT's.
 Clearly Eq. (\ref{cpt}) also satisfies  the condition $\lim_{r \rightarrow 0}c(r) = c_{eff}$. 

Noting that the asymptotic behaviour of the $Y$-functions is 
\begin{equation}
\lim_{\theta\rightarrow \infty }Y_{a}^{i}(\theta )\sim e^{-rM_{a}^{i}\cosh \theta },
\end{equation}
\noindent we recover for $\sigma \rightarrow \infty$ the $Y$-systems of the $A_{k-1}$-minimal ATFT 
derived originally in \cite{TBAZamun}. In this case the $Y$-systems were shown to
have a period related to the dimension of the perturbing operator (see (\ref{conj})). 
We found some explicit periods for generic values of the resonance parameter $\sigma $,
as we discuss in the next section for concrete examples.

\section{Explicit examples}
\label{examples}
\indent \ \
In this section we support our analytical discussion with some numerical
results and, in particular, justify various assumptions for which we had no
rigorous analytical argument so far. We numerically iterate the
TBA-equations (\ref{tbaa}) and have to choose specific values for the level $%
k $ for this purpose. As we pointed out in chapter \ref{ntft}, quantum
integrability has only been established for the choice $k >  h$ \cite{HSG2}. 
Since, according to Eq. (\ref{deldata}),
the perturbation is relevant also for smaller values of $k$ and, in addition, the
S-matrix makes perfect sense for these values of $k$, it will be interesting
to see whether the TBA-analysis in the case of $SU(3)_{k}$ will exhibit any
qualitative differences for $k \leq 3$ and $k > 3$. From our examples for the
values $k=2,3,4$ the answer to this question is that there is no apparent
difference. Recall that the condition $k > h$ is
a constraint which emerged as a way to ensure the super-renormalisability at first order of
the models under consideration (see Eq. (\ref{supf})). 
Provided this condition is satisfied, the task of
explicitly construct the corresponding quantum conserved 
charges becomes simpler, as we have seen in subsection \ref{PCFT}.
However, there is no reason why integrability
should not hold for different values of $k$ and, as mentioned above,
this observation is supported by our TBA-results for different particular examples.
For all cases studied  we find the staircase pattern of the scaling
function predicted in the preceding section as the values of $\sigma $ and $%
x $ sweep through the different regimes. Besides presenting numerical plots
we also discuss some peculiarities of the systems at hand. We provide the
massless TBA equations (\ref{uvTba}) with their UV and IR central charges
and state the $Y$-systems together with their periodicities. Finally, we also
comment on the classical or weak coupling limit $k\rightarrow \infty $.

\subsection{The $\protect{SU(3)_{2}}$-HSG model}
\label{modeltba}
\indent \ \
This is the simplest model for the $SU(3)_{k}$-series, since it contains only
the two self-conjugate solitons (1,1) and (1,2). The formation of stable
particles via fusing is not possible and the only non-trivial S-matrix
elements are those between particles of different colour 
\begin{equation}
S_{11}^{11}=S_{11}^{22}=-1,\quad S_{11}^{12}(\theta -\sigma
)=-S_{11}^{21}(\theta +\sigma )=\tanh \frac{1}{2}\left( \theta -i\frac{\pi }{%
2}\right) \;.  \label{ZamS}
\end{equation}
Here we have chosen $\eta _{12}=-\eta _{21}=i=\sqrt{-1}$. One easily convinces oneself
that (\ref{ZamS}) satisfies indeed (\ref{HS}) and (\ref{crossing}). This
scattering matrix may be related to various matrices which occurred before
in the literature. First of all, when performing the limit $\sigma
\rightarrow \infty $, the scattering involving different colours becomes free
and the systems consists of two free fermions leading to the central charge $
c=1$. Taking instead the limit $\sigma \rightarrow 0$ the expressions in 
(\ref{ZamS}) coincide precisely with a matrix which describes the scattering
of massless ``Goldstone fermions (Goldstinos)'' discussed in \cite{triZam}.
Apart from the factor $i$, the matrix $S_{11}^{21}(\theta )|_{\sigma =0}$
was also proposed to describe the scattering of a massive particle 
\cite{anticross, CM, DelMuss1}. Having only one colour available one is not able to set up the
usual crossing and unitarity equations and in 
\cite{anticross,CM, DelMuss1} the authors
therefore resorted to the strange concept of ``anti-crossing''. As our analysis
shows this may be consistently overcome by breaking the parity invariance.
The TBA-analysis is summarized as follows 
\begin{eqnarray*}
\text{Unstable particle formation} &:&\text{\qquad \quad }c_{su(3)_{2}}=%
\frac{6}{5}=c_{UV}+\hat{c}_{UV}=\frac{7}{10}+\frac{1}{2} \\
\text{No unstable particle formation} &:&\text{\qquad \quad }%
2c_{su(2)_{2}}=1=c_{IR}+\hat{c}_{IR}=\frac{1}{2}+\frac{1}{2}\,\,\,.
\end{eqnarray*}
It is interesting to note that the flow from the tricritical Ising to the
Ising model which was originally investigated in \cite{triZam}, emerges as a subsystem
in the HSG-model in the form $c_{UV}\rightarrow c_{IR}$. This is the particularisation
of the flow (\ref{subflow}) pointed out in subsection (\ref{ctba}). This suggests that
we could alternatively also view the HSG-system as consisting out of a
massive and a massless fermion, where the former is described by (\ref{c0}), (%
\ref{uvTba}) and the latter by (\ref{ckink}), (\ref{kinktba}), respectively.

\begin{figure}
\begin{center}
\includegraphics[width=11cm,height=14cm,angle=-90]{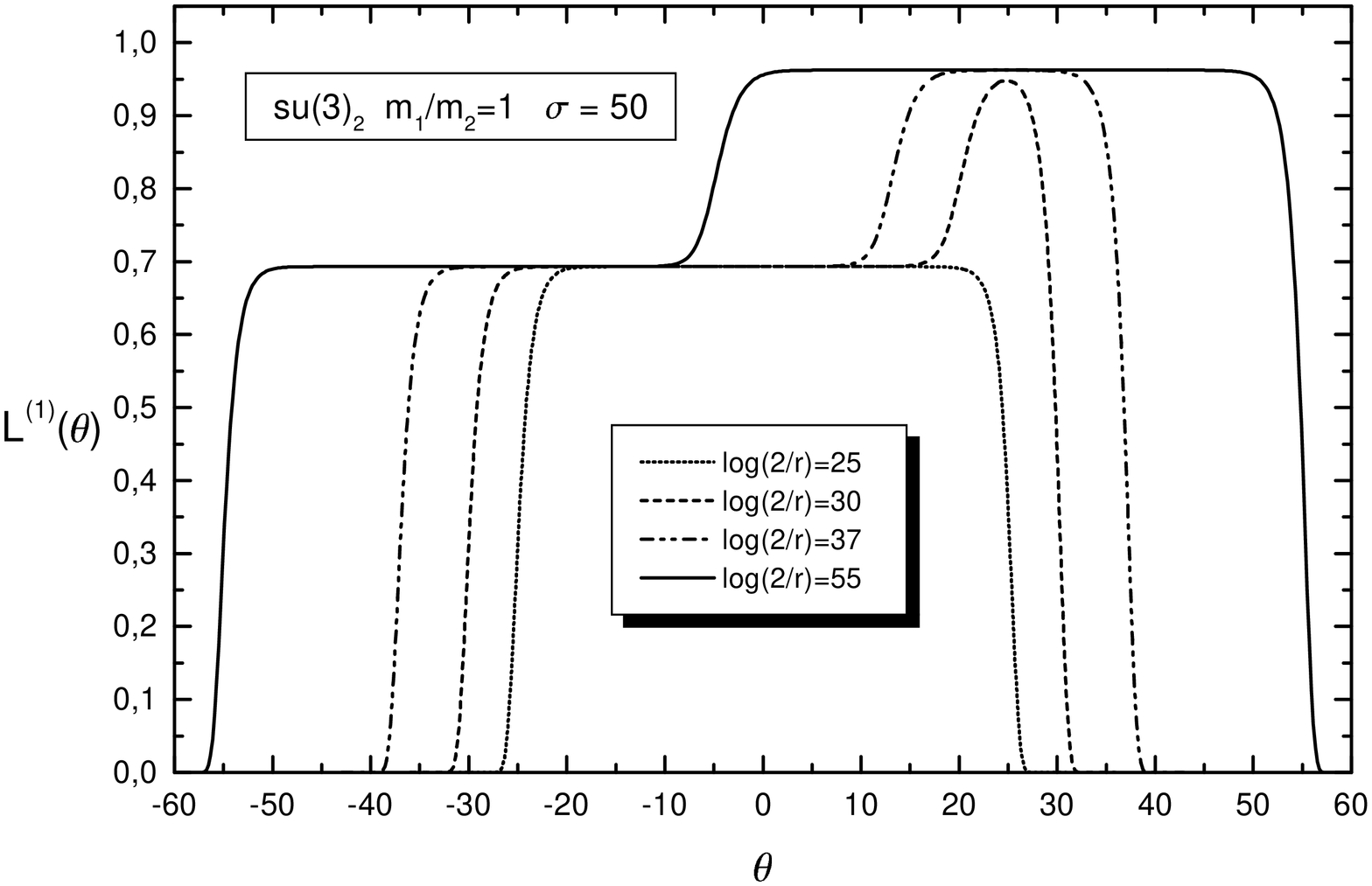}
\caption[Numerical solution for $\protect{L^{1}}(\protect\theta )$ in the
$\protect{su(3)_2}$-HSG model.]
{Numerical solution for $\protect{L^{1}(\theta )}$
of the $\protect{su(3)_{2}}$ related TBA-equations
 at different values of the scale
parameter $\protect{r}$ and fixed resonance shift and mass ratio.}
\label{fig11}
\end{center}
\end{figure}

\begin{figure}
\begin{center}
\includegraphics[width=15cm,height=20cm,angle=0]{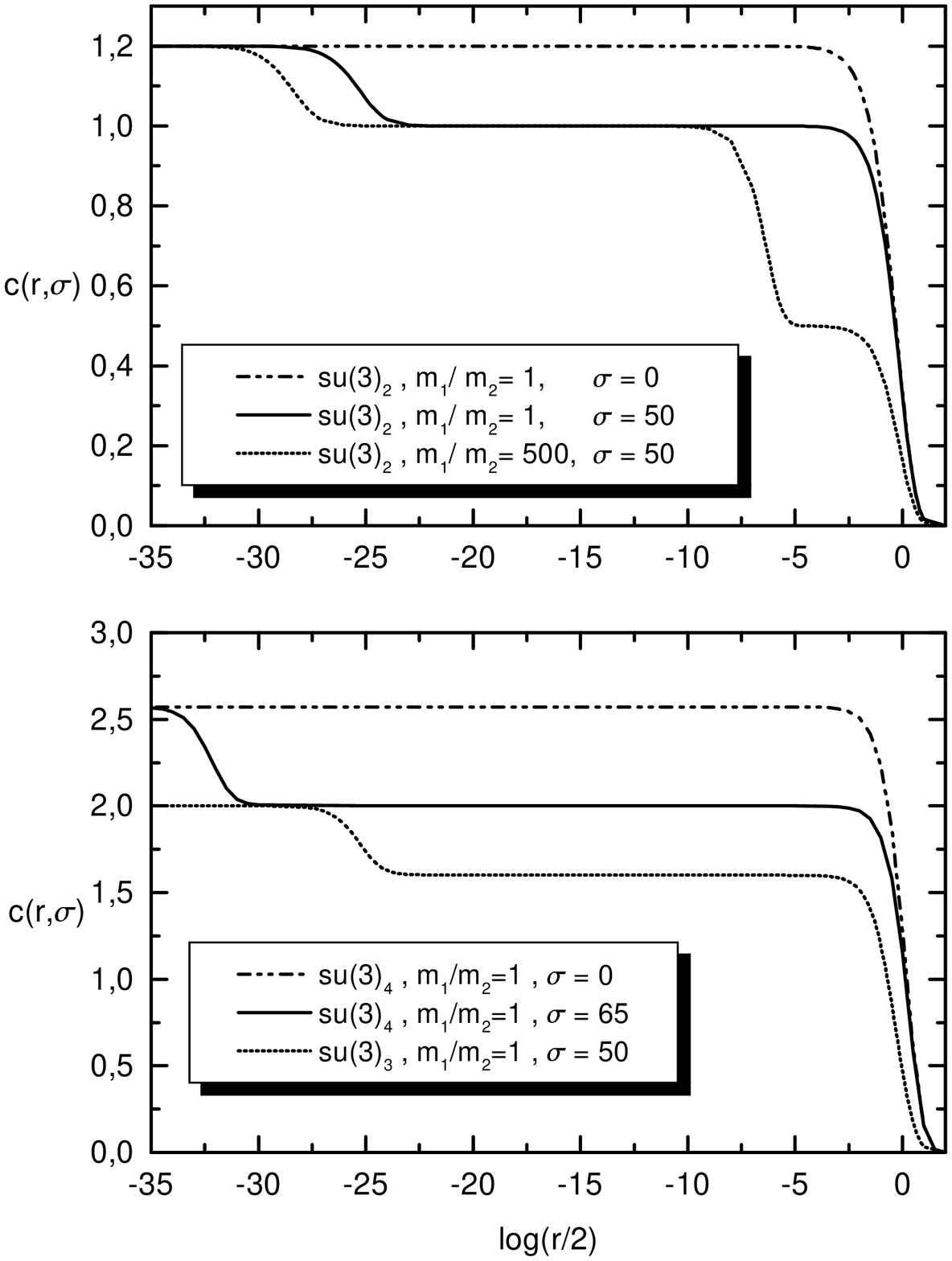}
\caption{Numerical plots of the scaling function for 
$\protect{su(3)_{k},\;k=2,3,4}$ as a function of the variable $\protect{\ln (r/2)}$ at different
values of the resonance shift and mass ratio.}
\label{fig12}
\end{center}
\end{figure}

Our numerical investigations of the model match the analytical discussion
and justifies various assumptions in retrospect. Fig. \ref{fig11} exhibits various
plots of the $L$-functions in the different regimes. We observe that for $-x=\ln
(2/r)<\sigma /2,$ $\sigma \neq 0$ the solutions are symmetric in the
rapidity variable, since the contribution of the $\psi $ kernels responsible
for parity violation is negligible. The solution displayed is just the free
fermion $L$-function, $L^{i}(\theta )=\ln (1+e^{-rM^{i}\cosh \theta })$.
Approaching more and more the ultraviolet regime, we observe that the
solutions $L^{i}$ cease to be symmetric signaling the violation of parity
invariance. The second plateau is then formed, which will extend beyond $%
\theta =0$ for the deep ultraviolet (see Fig. \ref{fig11}). The staircase pattern of
the scaling function is displayed in Fig.  \ref{fig12}  for the different cases
discussed in the previous section. We observe always the value $6/5$ in the
deep ultraviolet regime, but depending on the value of the resonance
parameter and the mass ratio, it may be reached sooner or later. The plateau
at $c=1$ corresponds to the situation when the unstable particles can not be
formed yet and we only have two copies of $SU(3)_{2}$ which do not interact.
Choosing the mass ratios in the two copies to be very different, we can also
``switch them on'' individually as the plateau at $c=1/2$ indicates. 

\vspace{0.5cm}

The $Y$-systems (\ref{Y}) for $k=2$ read 
\begin{equation}
Y_{1}^{i}\left( \theta +i\frac{\pi }{2}\right) Y_{1}^{i}\left( \theta -i%
\frac{\pi }{2}\right) =1+Y_{1}^{j}(\theta -\sigma _{ji})\quad
i,j=1,2,\;i\neq j\,\,.  \label{Y2}
\end{equation}
For $\sigma =0$ they coincide with the ones derived in \cite{triZam} for the
``massless'' subsystem. Shifting the arguments in (\ref{Y2}) appropriately,
the periodicity 
\begin{equation}
Y_{1}^{i}\left( \theta +\frac{5\pi i}{2}+\sigma _{ji}\right)
=Y_{1}^{j}(\theta )\,\,  \label{Periode}
\end{equation}
is obtained after a few manipulations. For a vanishing resonance parameter the relation
 (\ref{Periode}) coincides with the one obtained in \cite{TBAZam1,triZam}.
These periods may be exploited in a series expansion of the scaling function
in terms of the conformal dimension of the perturbing operator \cite{TBAZam1, KM2}

\vspace{0.2cm}

All the results obtained in this subsection for the $SU(3)_2$-HSG will be reproduced
in the next chapter within the context of a very different framework, the so-called form factor
approach. In particular, the relation between the presence of resonance parameters in the S-matrix
and the interpretation of the observed flows as massless flows will receive important support
in the form factor context. Notice that in the TBA-context the conformal 
dimension of the perturbation
can be related to the periodicities of the $Y$-systems, but such relationship is conjectured once the
expected value of the conformal dimension $\Delta$ is known (see (\ref{deldata})). In the
next chapter we will compute this conformal dimension in a completely different way and confirm
the result conjectured in the TBA-context. In addition, we will be able to compute conformal dimensions
of fields other than the one of the perturbing operator, which is up to now not possible in the TBA-approach.

\subsection{The $\protect{SU(3)_{3}}$-HSG model}
\label{model3}
\indent \ \
This model consists of two pairs of solitons $\overline{(1,1)}=(2,1)$ and $%
\overline{(1,2)}=(2,2)$. When the soliton $(1,  i)$ scatters with itself it
may form $(2,i)$ for $i=1,2$ as a bound state. The two-particle S-matrix
elements read 
\begin{equation}
S^{ii}(\theta )=\left( 
\begin{array}{ll}
(2)_{\theta } & -(1)_{\theta } \\ 
-(1)_{\theta } & (2)_{\theta }
\end{array}
\right) \quad \quad S^{ij}(\theta -\sigma _{ij})=\left( 
\begin{array}{ll}
\eta _{ij}\,(-1)_{\theta } & \eta _{ij}^{2}\,(-2)_{\theta } \\ 
\eta _{ij}^{2}\,(-2)_{\theta } & \eta _{ij}\,(-1)_{\theta }
\end{array}
\right) \;.
\end{equation}
Since soliton and anti-soliton of the same colour obey the same TBA
equations we can exploit charge conjugation symmetry to identify $\epsilon
^{i}(\theta ):=\epsilon _{1}^{i}(\theta )=\epsilon _{2}^{i}(\theta )$
leading to the reduced set of equations 
\begin{equation}
\epsilon ^{i}(\theta )+\varphi *L^{i}(\theta )-\varphi *L^{j}(\theta -\sigma
_{ji})=rM^{i}\cosh \theta ,\quad \varphi (\theta )=-\frac{4\sqrt{3}\cosh
\theta }{1+2\cosh 2\theta }\;.
\end{equation}
The corresponding scaling function therefore acquires a factor two, 
\begin{equation}
c(r,\sigma )=\frac{6\,r}{\pi ^{2}}\sum_{i}M^{i}\int d\theta \,\cosh \theta
\,L^{i}(\theta )\;.
\end{equation}
This system exhibits remarkable symmetry properties. We consider first the
situation $\sigma =0$ with $m_{1}=m_{2}$ and note that the system becomes
free in this case 
\begin{equation}
M^{1}=M^{2}=:M\;\Rightarrow \;\epsilon ^{1}(\theta )=\epsilon
^{2}(\theta )=rM\cosh \theta \;,
\end{equation}
meaning that the theory falls apart into four free fermions whose central
charges add up to the expected coset central charge of $2$. Also for unequal
masses $m_{1}\neq m_{2}$ the system develops towards the free fermion theory
for high energies, when the difference becomes negligible. This is also seen
numerically.

For $\sigma \neq 0$ the two copies of the minimal $A_{2}$-ATFT or
equivalently the scaling Potts model start to interact. The outcome of the
TBA-analysis in that case is summarized as 
\begin{eqnarray*}
\text{Unstable particle formation} &:&\text{\qquad \quad }%
c_{su(3)_{3}}=2=c_{UV}+\hat{c}_{UV}=\frac{6}{5}+\frac{4}{5}, \\
\text{No unstable particle formation} &:&\text{\qquad \quad }2c_{su(2)_{3}}=%
\frac{8}{5}=c_{IR}+\hat{c}_{IR}=\frac{4}{5}+\frac{4}{5}\,\,\,.
\end{eqnarray*}
As discussed in the previous case for $k=2$, the $L$-functions develop an
additional plateau after passing the point $-x=\sigma /2$. This
plateau lies at $L=\ln 2$ which is the free fermion value signaling that the
system contains a free fermion contribution in the UV-limit as soon as the
interaction between the solitons of different colours becomes relevant.
Fig. \ref{fig12} exhibits the same behaviour as the previous case.  We clearly
observe the plateau at $8/5$ corresponding to the two non-interacting copies
of the minimal $A_{2}$-ATFT. As soon as the energy scale of the unstable
particles is reached, the scaling function approaches the correct value of 
$c=2$.

The $Y$-systems (\ref{Y}) for $k=3$ read 
\begin{equation}
Y_{1,2}^{i}\left( \theta +i\frac{\pi }{3}\right) Y_{1,2}^{i}\left( \theta -i%
\frac{\pi }{3}\right) =Y_{1,2}^{i}\left( \theta \right) \frac{%
1+Y_{1,2}^{j}(\theta +\sigma _{ij})}{1+Y_{1,2}^{i}\left( \theta \right) }%
\quad i,j=1,2,\;i\neq j\,\,.  \label{Y3}
\end{equation}
Once again we may derive a periodicity 
\begin{equation}
Y_{1,2}^{i}\left( \theta +2\pi i+\sigma _{ji}\right) =Y_{1,2}^{j}(\theta ),
\end{equation}
by making the suitable shifts in (\ref{Y3}) and subsequent iteration.

\subsection{The $\protect{SU(3)_{4}}$-HSG model}
\label{model4}
\indent \ \
This model involves 6 solitons, two of which are self-conjugate $\overline{%
(2,1)}=(2,1)$, $\overline{(2,2)}=(2,2)$ and two conjugate pairs $\overline{%
(1,1)}=(3,1)$, $\overline{(1,2)}=(3,2)$. The corresponding two-particle
S-matrix elements are obtained from the general formulae (\ref{S})\ and (\ref
{ij}) 
\begin{equation}
S^{ii}(\theta )=\left( 
\begin{array}{ccc}
(2)_{\theta } & (3)_{\theta }(1)_{\theta } & -(2)_{\theta } \\ 
(3)_{\theta }(1)_{\theta } & (2)_{\theta }^{2} & (3)_{\theta }(1)_{\theta }
\\ 
-(2)_{\theta } & (3)_{\theta }(1)_{\theta } & (2)_{\theta }
\end{array}
\right),
\end{equation}
for soliton-soliton scattering with the same colour values and 
\begin{equation}
S^{ij}(\theta -\sigma _{ij})=\left( 
\begin{array}{ccc}
\eta _{ij}(-1)_{\theta } & \eta _{ij}^{2}(-2)_{\theta } & \eta
_{ij}^{3}(-3)_{\theta } \\ 
\eta _{ij}^{2}(-2)_{\theta } & -(-3)_{\theta }(-1)_{\theta } & \eta
_{ij}^{2}(-2)_{\theta } \\ 
\eta _{ij}^{3}(-3)_{\theta } & \eta _{ij}^{2}(-2)_{\theta } & \eta
_{ij}(-1)_{\theta }
\end{array}
\right),
\end{equation}
for the scattering of solitons of different colours with $\eta _{12}=e^{i%
\frac{\pi }{4}}$. In this case the numerical analysis becomes more involved, but for
the special case $m_{1}=m_{2}$ one can reduce the set of six coupled
integral equations to only two by exploiting the symmetry $%
L_{a}^{1}(\theta )=L_{a}^{2}(-\theta )$ and using charge conjugation
symmetry, $L_{1}^{i}(\theta )=L_{3}^{i}(\theta )$. The numerical outcomes,
shown in Fig. \ref{fig12} again match with the analytic expectations (\ref{stepk})
and yield for $-x=\ln (2/r)>\sigma /2$ the coset central charge of $18/7$. In
summary we obtain: 
\begin{eqnarray*}
\text{Unstable particle formation} &:&\text{\qquad \quad }c_{su(3)_{4}}=%
\frac{18}{7}=c_{UV}+\hat{c}_{UV}=\frac{11}{7}+1 \\
\text{No unstable particle formation} &:&\text{\qquad \quad }%
2c_{su(2)_{4}}=2=c_{IR}+\hat{c}_{IR}=1+1\,\,\,,
\end{eqnarray*}
which matches precisely the numerical outcome in Fig.\ref{fig12}, with the same
physical interpretation as already provided in the previous two subsections.

\subsection{The semi-classical limit $\protect{k\rightarrow \infty}$}
\label{semi}
\indent \ \
As last example we carry out the limit $k\rightarrow \infty $, which is of
special physical interest since it may be identified with the weak coupling
or equivalently the classical limit, as is seen from the relation $\hbar
\beta ^{2}=1/k+O(1/k^{2}).$ To illustrate this equivalence we have
temporarily re-introduced Planck's constant. It is clear from the
TBA-equations that this limit may not be taken in a straightforward manner.
However, we can take it in two steps, first for the on-shell energies and
the kernels and finally for the sum over all particle contributions. The
on-shell energies are easily computed by noting that the mass spectrum
becomes equally spaced for $k\rightarrow \infty $%
\begin{equation}
M_{a}^{i}=M_{k-a}^{i}=\frac{k}{\pi } m_{i} \sin \frac{\pi \,a}{k}%
\approx a\,m_{i}\;\qquad ,\quad a<\frac{k}{2}\;.  \label{MM}
\end{equation}
For the TBA-kernels the limit may also be taken easily from their integral
representations 
\begin{eqnarray}
&&\phi _{ab}(\theta )\stackunder{k\rightarrow \infty }{\longrightarrow }2\pi
\,\delta (\theta )\,\left( \delta _{ab}-2\left( K^{A_{k-1}}\right)_{\,\,\,ab}^{-1}\right)\nonumber \\
 &&\psi _{ab}(\theta )\stackunder{%
k\rightarrow \infty }{\longrightarrow }2\pi \,\delta (\theta )\,\left(
K^{A_{k-1}}\right)_{\,\,\,ab}^{-1},\;
\end{eqnarray}
when employing the usual integral representation of the delta-function.
Inserting these quantities into the TBA-equations yields 
\begin{equation}
\epsilon _{a}^{i}(\theta )\approx r\,a\,m_{i}\cosh \theta
-\sum_{b=1}^{k-1}\left( \delta _{ab}-2\left( K^{A_{k-1}}\right)_{\,\,\,ab}^{-1}
\right) L_{b}^{i}(\theta )-\sum_{b=1}^{k-1}\left( K^{A_{k-1}}\right)_{\,\,\,ab}^{-1}
L_{b}^{j}(\theta -\sigma )\;.  \label{TTT}
\end{equation}
We now have to solve these equations for the pseudo-energies. In principle
we could proceed in the same way as in the case for finite $k$ by doing the
appropriate shifts in the rapidity. However, we will be content here to
discuss the cases $\sigma \rightarrow 0$ and $\sigma \rightarrow \infty $,
which, as follows from our previous discussion, correspond to the situation of
restored parity invariance and two non-interacting copies of the minimal
ATFT, respectively. The related solutions to the constant TBA-equations (\ref{ctba2}) and (%
\ref{cTba2a}) become 
\begin{equation}
\sigma \rightarrow \infty :\;\hat{x}_{a}\stackunder{k\rightarrow \infty }{%
\longrightarrow }\frac{(a+1)^{2}}{a(a+2)}-1\quad \text{and\quad }\sigma
\rightarrow 0:\;x_{a}\stackunder{k\rightarrow \infty }{\longrightarrow }%
\frac{(a+1)(a+2)}{a(a+3)}-1\;.  \label{l1}
\end{equation}
The other information we may exploit about the solutions of (\ref{TTT}) is
that for large rapidities they tend asymptotically to the free solution,
meaning that 
\begin{equation}
\sigma \rightarrow 0,\infty \;:\;\;\;\;\;L_{a}^{i}(\theta )\stackunder{%
\theta \rightarrow \pm \infty }{\longrightarrow }\ln (1+e^{-r\,a\,m_{i}\cosh
\theta }).  \label{l2}
\end{equation}
We are left with the task to seek functions which interpolate between the
properties (\ref{l1}) and (\ref{l2}). Inspired by the analysis in \cite
{Fowler} we take these functions to be 
\begin{eqnarray}
\sigma &\rightarrow &\infty \;:\;\;\;\;\;L_{a}^{i}(\theta )=\ln \left[ \frac{%
\sinh ^{2}\left( \frac{a+1}{2}\,rm_{i}\cosh \theta \right) }{\sinh \left( 
\frac{a\,}{2}\,rm_{i}\cosh \theta \right) \sinh \left( \frac{a+2}{2}%
\,rm_{i}\cosh \theta \right) }\right],\,\,\,  \label{p1} \\
\sigma &\rightarrow &0\;:\;\;\;\;\;L_{a}^{i}(\theta )=\ln \left[ \frac{\sinh
\left( \frac{a+1}{2}\,rm_{i}\cosh \theta \right) \sinh \left( \frac{a+2}{2}%
\,rm_{i}\cosh \theta \right) }{\sinh \left( \frac{a}{2}\,rm_{i}\cosh \theta
\right) \sinh \left( \frac{a+3}{2}\,rm_{i}\cosh \theta \right) }\right].\,\,\,
\label{p2}
\end{eqnarray}
The expression (\ref{p1}) coincides with the expressions discussed in the
context of the breather spectrum of the sine-Gordon model \cite{Fowler} and (\ref{p2})
 is constructed in analogy. We are now equipped to compute the
scaling function in the limit $k\rightarrow \infty $%
\begin{equation}
c(r,\sigma )=\lim_{k\rightarrow \infty }\frac{3\,r}{\pi ^{2}}%
\sum_{i=1}^{2}\int d\theta \,\cosh \theta
\sum_{a=1}^{k-1}M_{a}^{i}L_{a}^{i}(\theta )\,\,.
\end{equation}
Using (\ref{MM}), (\ref{p1}) and (\ref{p2}) the sum over the main quantum
number may be computed directly by expanding the logarithm. We obtain for $%
k\rightarrow \infty $%
\begin{eqnarray}
\lim_{\sigma \rightarrow \infty}c(r)&=&-\dfrac{6r}{\pi ^{2}}\,\,\sum_{i=1}^{2}\!%
\!\int \,\,d\theta \,m_{i}\cosh \theta \ln \left( 1-e^{-r\,m_{i}\cosh \theta
}\right),  \label{in1} \\
\lim_{\sigma \rightarrow 0}c(r)&=&-\dfrac{6\,r}{\pi ^{2}}\,\,\sum_{i=1}^{2}\!\!%
\int \!\!d\theta \,\,m_{i}\cosh \theta [\ln \left( 1-e^{-r\,m_{i}\cosh
\theta }\right) +\nonumber \\
 &+& \,\,\,\,\,\,\,\,\ln (1-e^{-r\,2m_{i}\cosh \theta })].\,
\label{in2}
\end{eqnarray}
Here we have acquired an additional factor of 2, resulting from the
identification of particles and anti-particles which is needed when one
linearizes the masses in (\ref{MM}). Taking now the limit $r\rightarrow 0$
we obtain 
\begin{eqnarray}
\text{no unstable particle formation} &:&\text{\quad \ }2\,c_{su(2)_{\infty
}}=4\;  \label{in11} \\
\text{unstable particle formation} &:&\text{\quad \ }c_{su(3)_{\infty
}}=6\,\,.  \label{in22}
\end{eqnarray}

The results (\ref{in1}), (\ref{in11}) and (\ref{in2}), (\ref{in22}) allow a
nice physical interpretation. We notice that for the case $\sigma
\rightarrow \infty $ we obtain four times the scaling function of a free
boson. This means in the classical limit we obtain twice the contribution of
the non-interacting copies of $SU(2)_{\infty }/U(1)$, whose particle content
reduces to two free bosons each of them contributing $1$ to the effective
central charge which is in agreement with (\ref{cdata}). For the case $%
\sigma \rightarrow 0$ we obtain the same contribution, but in addition the
one from the unstable particles, which are two free bosons of mass $2m_{i}$.
This is also in agreement with (\ref{cdata}).

Finally, it is interesting to observe that, when taking the resonance poles to
be $\theta _{R}=\sigma -i\pi /k$, the semi-classical limit taken in the
Breit-Wigner formula (\ref{BW11}) leads to $m_{\tilde{k}%
}^{2}=(m_{i}+m_{j})^{2}$. On the other hand (\ref{in2}) seems to suggest
that $m_{\tilde{k}}=2m_{i}$, which implies that the mass scales should be
the same. However, since our analysis is mainly based on exploiting the
asymptotics we have to be cautious about this conclusion.

\section{Summary of results and open points}
\label{con3}
\indent \ \
Our main conclusion is that the TBA-analysis indeed confirms the consistency
of the scattering matrix proposed in \cite{HSGS}. In the deep ultraviolet
limit we recover the $G_{k}/U(1)^{\ell }$-coset central charge for any value
of the $2\ell -1$ free parameters entering the S-matrix, including the
choice when the resonance parameters vanish and parity invariance is
restored on the level of the TBA-equations. This is in contrast to the
properties of the S-matrix, which is still not parity invariant due to the
occurrence of the phase factors $\eta $, which are required to close the
bootstrap equations \cite{HSGS}. However, they do not contribute to the
TBA-analysis, which means that so far we can not make any definite statement
concerning the necessity of the parity breaking, since the same value for
the central charge is recovered irrespective of the value of the $\sigma $'s.
The underlying physical behaviour is, however, quite different as our
numerical analysis demonstrates:

\vspace{0.25cm}
\begin{itemize}

\item For vanishing resonance parameter $\sigma=0$ and taking the energy scale of the 
stable particles to be of the same order  $m_1 \simeq m_2$,  the deep
ultraviolet coset central charge is reached straight away. From the physical point of view, 
this is the expected behaviour since, according to (\ref{BW22}) whenever
the resonance parameter is  vanishing the same happens to the  decay width of the unstable particles.
Therefore, the unstable particles become ``virtual states'' characterised by
poles in the imaginary axis, beyond the physical sheet. 
They are on an energy scale of the same order as
the one of the stable particles. Being the energy scale corresponding to the onset of all stable and unstable particles of the same order,  the scaling function takes the value corresponding to the
ultraviolet coset central charge of the underlying CFT as soon as the mentioned scale is reached.
As shown in section \ref{rper}, for $\sigma=0$ and $m_1=m_2$ parity is restored at the level of the TBA-equations and consequently the corresponding $L$-functions must be symmetric in the rapidity variable.
Although no numerical results are presented which confirm this statement it is clear from Fig. \ref{fig11}
that the $L$-functions cease to be symmetric as soon as parity invariance is violated. 

\vspace{0.25cm}

\item On the other hand, for non-trivial resonance parameter, the scaling function 
passes the different regions in the
energy scale. It develops  then a ``staircase'' 
behaviour where the number and size of  the plateaux
  is determined by the relative mass scales between 
the stable and unstable particles and the stable 
particles themselves. Therefore, different choices 
of the $2 \ell -1$ free parameters at hand lead  
to a theory with a different physical content, but still possessing the same central
charge. This feature is also consistent with the physical picture anticipated for the HSG-models,
since in the deep ultraviolet limit, as long as the resonance parameter is finite, the energy
scale is much higher than the energy scales necessary 
for the production of all the stable and unstable particles. 
Therefore all the particle content of the 
model contributes to the scaling function which,
interpreted as a measure of  effective light degrees 
of freedom, will reach its maximum value, namely
the Virasoro central charge of the unperturbed CFT. 
Being parity broken through the resonance shift
$\sigma$, the $L$-functions cease to be symmetric as
 soon as the energy scale of the unstable particles
is reached and develop also plateaux  as shown in Fig. \ref{fig11}.

\vspace{0.25cm}

It must be emphasised that the sort of flows observed correspond to a system of  TBA-equations
which formally, after the introduction of the auxiliary
 parameter $r^{\prime}=r/2 \, e^{\sigma/2}$, can be re-interpreted 
as the TBA-equations corresponding to two  massless systems, in the spirit of \cite{triZam}.
 As we have mentioned, the connection between flows related to the presence of unstable particles in the spectrum and massless flows should be understood as formal at this point, since the parameter $r^{\prime}$ was only introduced aiming towards a simplification of the analytical and numerical analysis. However, in the two subsequent chapters we will collect additional arguments which support the 
belief that the observed flows should be in fact understood as effective massless flows.
 
\end{itemize}

\vspace{0.25cm}

The similar ``staircase''  behaviour observed both for the HSG-models and for the models studied in 
\cite{staircase, Martins, DoRav} has been emphasised in several occasions along this chapter. Therefore,
it is appropriate at this point to clarify our understanding about the origin of this  similarity. 
The presence of resonance poles in the spectrum is characteristic of both sorts of theories  but, 
as we will justify now, the  ``staircase''  behaviour  of the scaling function does not admit the
 same clear physical interpretation for the models studied in 
\cite{staircase, Martins, DoRav} than for the HSG-models. 

Whereas for the HSG-models the resonance parameter enters the S-matrix as a shift in the rapidity variable, 
in the models studied in \cite{staircase, Martins, DoRav}, the resonance
parameter arises as a consequence of the analytical continuation to the complex plane 
of the effective coupling  constant $B$,  which characterises the Lagrangian and S-matrix 
of  the sinh-Gordon model \cite{SSG} and, 
 in fact, of all affine Toda field theories related to simply-laced Lie algebras
\cite{ATFTS, dis, mussrev}.
The mentioned  complexification takes place in the following way
\begin{equation}
B \rightarrow 1 \pm \frac{2 i \sigma}{\pi}.  
\label{shift}
\end{equation} It is interesting to notice that the particular form of  (\ref{shift}) 
is not casual. In particular,  the real part of $B$ has necessarily to be one
 so that the consequent  
transformation of the sinh-Gordon S-matrix via  (\ref{shift})  generates a new but 
still consistent  S-matrix, in the sense described in section \ref{analit} of the preceding chapter. The 
consistency of the new S-matrix in guaranteed by the fact that, for all affine Toda field theories 
\cite{ATFTS, dis}, the coupling constant $B$ occurs always in the 
combination $B(2-B)$, which stays real under (\ref{shift}). $B=1$ is the so-called {\bf self-dual point}
 since in that case
$B=2-B$.

The introduction of the resonance parameter $\sigma$ by means of  (\ref{shift})
makes the S-matrix exhibit a resonance pole in the imaginary axis
$\theta_R = - \sigma - \frac{i \pi}{2}$
similarly to the $SU(3)_2$-HSG model. As usual this pole could be 
 understood as the trace of an unstable particle. 
However, even though only one resonance parameter has been introduced,  
the TBA-analysis carried out in \cite{staircase}
 for the roaming sinh-Gordon model shows that the corresponding scaling function
develops an infinite number of plateaux. Therefore, the results in \cite{staircase} can not be interpreted 
physically by using the same sort of arguments employed in the TBA-analysis of the HSG-models.
Equivalently, the infinite number of plateaux observed
in \cite{staircase} can not be related to the number of free parameters in the model. The same can be
said with respect to the models studied in \cite{Martins, DoRav} whose construction follows the same
lines, i.e. performing the transformation (\ref{shift})
introduced for the roaming sinh-Gordon model, with the difference that they take as an input the
S-matrices of other simply laced affine Toda field theories instead of the sinh-Gordon model  
($A_1^{(1)}$-ATFT), which is the simplest of their class.

\vspace{0.25cm}

Although all the results of this chapter confirm the consistency of the S-matrix proposal
\cite{HSGS} there are important data concerning the underlying CFT which have not been reproduced
in the TBA-context. For instance, it would  be highly desirable to
carry out the series expansion (\ref{cpt}) of the scaling function in $r$ and determine
the dimension $\Delta$ of the perturbing operator.  It will be useful for this to know the periodicities of 
the $Y$-functions. In the light of the results found in section \ref{examples} and the expected value of 
$\Delta$ given by  (\ref{deldata}), we conjecture that they will be
\begin{equation}
Y_{a}^{i}\left( \theta +i\pi (1-\Delta )^{-1}+\sigma _{ji}\right) =Y_{\bar{a}%
}^{j}(\theta ).  \label{conj}
\end{equation}
\noindent
For vanishing resonance parameter and the choice $g=su(2)$, this behaviour coincides with the one
obtained in \cite{TBAZamun}. This suggests  the form in (\ref{conj}) is of a
very universal nature beyond the models discussed here and supports our conjecture. It would be highly desirable
to have a model-independent explanation for this behaviour starting from first principles. 

\vspace{0.25cm}

Thus,  concerning the issue of the identification of the conformal dimension of
the perturbing operator we have to conclude that, although the preceding arguments
support the conjecture (\ref{conj}), more work and/or different tools are needed 
in order to make a definite statement. 
However, one should also keep in mind that the perturbing operator, although playing a distinguished role
in the construction of the massive QFT, does not of course fulfill all the operator content of the underlying CFT. The latter operator content is well known for the WZNW-coset theories \cite{Gep, DHS}. In particular, the corresponding conformal dimensions can be obtained easily by means of a general formula
which might be found also in  \cite{Gep, DHS}. Unfortunately,  the question of how to identify the whole operator content of the underlying CFT is left unanswered in the TBA-framework and we must appeal to a different method if we intend to really fulfill all the relevant data, apart from (\ref{deldata}) 
and (\ref{cdata}), characterising the ultraviolet  CFT. A different approach which allows to find an
answer to the latter question and provides at the same time all the information extracted from the preceding TBA-analysis shall be presented in the two succeeding chapters. The mentioned approach
is the so-called {\bf form factor program} originally pioneered by the members of
the Berlin group M. Karowski and P. Weisz in  \cite{Kar}. Apart from providing a consistency check,
in comparison with the TBA-analysis, this approach also serves to develop the theory further towards
a full-fledged QFT.

\vspace{0.25cm}

We also observe from our $su(N)$-example that the two regions, $
k > h$ for which quantum integrability was explicitly shown in \cite{HSG},
and $k \leq h$, for which quantum integrability has not been established up to now,
 do not show up in our analysis. As we mentioned at the beginning
of section \ref{examples}, this might be due to the fact that the constraint $k > h$ arises
when we select out those models which are super-renormalisable at first order (see definition
in subsection \ref{PCFT}). For these models the task of explicitly constructing quantum conserved charges
gets simplified as shown in \cite{Pertcft} (see also subsection \ref{PCFT}). However, there
is no reason why the HSG-models which are no super-renormalisable at first other should not
be integrable. Therefore, the results emerging from our TBA-analysis are not contradictory and
could be understood as an indication of the fact that possibly all HSG-models, irrespectively
of the value of the level $k$, are quantum integrable. 
In order to check the previous conjecture, it would be desirable to explicitly 
construct quantum conserved charges associated to the HSG-models
corresponding also to $k<h$, or prove their integrability by other means like, for instance, the
`counting-argument' reported in subsection \ref{PCFT}.

\vspace{0.25cm}

It would be very interesting to extend the case-by-case analysis of section
\ref{examples} to other algebras. The first challenge in these cases is to incorporate
the different resonance parameters, namely increase the rank of the Lie algebra.  However,
it is clear from our analysis that the number of TBA-equations to be solved increases precisely 
with the rank of the Lie algebra. This means that a TBA-analysis will become very complicate
as soon as $\ell >2$, both from the analytical and numerical  point of view. 
Concerning this problem, although having also its particular inconveniences for high rank,
the form factor program mentioned above is more advantageous  (see chapter \ref{ffs2}).

\chapter{Form Factors of the Homogeneous sine-Gordon models.}
\label{ffs}
\indent \ \
A form factor in QFT is a  matrix element of a local operator between the vacuum state
and an $n$-particle $in$-state. Despite the fact that we will focus our
 attention on the properties and
applications of form factors within the context of integrable massive 1+1-dimensional
QFT's, they are objects which have been analysed  in the framework of 1+3-dimensional QFT's
 over the last 40 years. In the latter context, they are frequently introduced 
as functions characterising the scattering amplitude associated to the interaction of
a charged particle with an external electromagnetic field or the electromagnetic 
field of another particle (see e.g. \cite{barton, Wein}). However, the form factor
approach was not exploited in the context of 1+1-dimensional QFT's until  1977,
when the pioneering work due to P. Weisz and M. Karowski \cite{Kar} was published. 
In these seminal papers,
the fundamental properties of form factors in 1+1-dimensional theories were established. It was
 found that, similarly to the construction procedure of S-matrices for integrable massive 1+1-dimensional
QFT's described in chapter \ref{ntft}, the form factors associated to a certain operator
can be obtained as the solutions to a set of consistency equations 
whose origin is based on physically-motivated requirements. It is also in 1+1-dimensions when
the form factor approach turns out to be most powerful for various reasons. First, 
the solution to the mentioned consistency equations allows in principle for computing all
$n$-particle form factors associated to any local field of the massive QFT. Also,
once the form factors associated to certain  local operators of the QFT are known, 
they might be used for many interesting applications 
like computing correlation functions, determining the operator content of the perturbed CFT or explicitly
compute other relevant quantities which characterise the underlying CFT. 

After the original papers \cite{Kar}, the development of the form factor approach in 
the context of 1+1-dimensional integrable QFT's has been carried out to a large extent 
by F.A. Smirnov et al. \cite{Smirnov, Smir2, Smir}. However, the interest of this approach
within the context of 1+1-dimensional QFT's constructed as perturbed CFT's received
renewed interest  with the work of J.L. Cardy and G. Mussardo \cite{Cardybast2, CardyBast}.
After that, the form factor program has been carried out for different models, under several
different aspects and by many authors. Some of these works are
 \cite{Zamocorr,YZam, BFKZ,CardyBast,deter1,deter2,DM,DSC, various}. 
The latter list is not meant to be complete.  

Concerning the present status of the form factor approach, there are still a lot of open problems some of
which we itemize below, mentioning also the particular contributions of our work to their understanding.

\subsubsection{Reviewing the present status of the form factor approach}
\indent \ \

{\bf Generic building blocks}

\vspace{0.25cm}

\noindent Although the form factors are obtained as the solutions to a certain set of
consistency equations which we will see in detail later, it is not known by
now whether there exist or not general `building blocks' in terms of which any form factor of a 1+1-dimensional 
integrable and massive QFT can be expressed, similarly to the situation
encountered in the construction of exact S-matrices along the lines of subsection \ref{analit}. 
There are various places in the literature 
\cite{deter1, deter2, Zamocorr} were determinant expressions in terms of
elementary symmetric polynomials depending upon the particle rapidities have been found for
the form factors associated to certain operators. 
This is also the case for a large class of local operators of the $SU(N)_2$-HSG models we
have studied here.  As we will see later, these determinant expressions can
 be equivalently written in terms of contour
integrals and various types of integral representations 
can be found in the literature whose precise inter-relation still needs to be clarified. 
Once the building blocks are known one may pose the question 
how are they combined by means of Lie algebraic quantities, analogously to the S-matrix construction
(see \cite{uni, Oota,FKS2}).

\vspace{0.25cm}
{\bf Closed solutions}
\vspace{0.25cm}

\noindent However, the open questions stated in the previous paragraph
 go already beyond the present
status of understanding of the form factor approach for many models, meaning that, the questions pointed out above,
 may be posed once all $n$-particle form factors associated to a certain operator of the QFT
have been constructed. This is not the usual case and, from that point of view,
 the situation we will encounter
in the course of our precise analysis is quite extraordinary, since we will find close formulae for all
$n$-particle form factors related to a large class of operators. Therefore, it is still an open problem how
to solve in general all the consistency equations any form factor of a 1+1-dimensional QFT has to satisfy. For
many models only the form factors associated to the lowest values of $n$ and/or to specific local operators
of the theory have been constructed. 

\vspace{0.25cm}
{\bf Identification of the operator content}
\vspace{0.25cm} 

\noindent Similarly again to the situation arising in the construction of S-matrices,
where specific information about the theory under consideration only enters at quite a late stage of
the construction, several of the consistency equations satisfied by form factors do not require any
knowledge about the precise nature of the local operator at hand. Therefore, once a solution is found
one still needs to identify the precise operator it corresponds to. There are various constraints which
can be used in order to match each form factor solution with a concrete operator of the massive QFT but
still, the techniques available need to be refined, since there is no systematic way to identify
all local operators of the QFT, as we may find in the context of our precise analysis.

\vspace{0.25cm}
{\bf Ultraviolet limit}
\vspace{0.25cm} 

\noindent Concerning the identification of operators mentioned in the previous point, one of the
open problems which is worth mentioning here is the fact that such an identification is ultimately
performed by assuming a one-to-one correspondence between the operator contents of the perturbed and
unperturbed CFT, for the primary field content. 
Provided the latter is known (which is not always the case) one can use various techniques
to identify the ultraviolet conformal dimensions of the local fields of the massive QFT by exploiting the
knowledge of their form factors. However, these techniques still need also refinement, since they
do not allow for unraveling the possible degeneracy of the operators of the underlying CFT and 
sometimes they can not be used for all operators, or they do not allow for a clear-cut identification
of the ultraviolet conformal dimension.  This means
 that having a priori a relatively good guess for its value 
turns out to be very important.

\vspace{0.25cm}
{\bf Two-point correlation functions}
\vspace{0.25cm}

As mentioned above, one of the most important objects one is
in principle in the position to compute once the form factors of certain
operators  are known is any two-point correlation function involving these
operators. It can be proven  that these two-point functions admit an expression
in terms of an infinite series where the $n$-th term is an $n$-dimensional
integral in the rapidities whose integrand depends on the $n$-particle form
factors of the two operators arising in the correlation function. Concerning
the computation of correlation functions in the form factor context i.e., the evaluation of the
aforementioned series, there are aspects which need further investigation. First, it is not known
up to now any rigorous proof of the convergence of the series mentioned above, which is generally assumed
in the light of the behaviour observed for several models. Second, the evaluation
of the multi-dimensional integrals arising in this series requires a lot of
computer time already for $n \geq 6$, which means it would be very interesting
to investigate whether it is possible to sum the mentioned series analytically.  

\vspace{0.25cm}
{\bf Momentum space cluster property}
\vspace{0.25cm}

Finally, we would like to briefly mention that some of the general
properties of form factors still lack a  rigorous  proof  for the time being. 
One of them is the momentum space cluster property, whose
investigation for the $SU(3)_2$-HSG model will provide, in our opinion, a 
valuable contribution to the  present status of understanding of this
characteristic of form factors, which has been observed for several 
concrete models in the literature  \cite{clust,Zamocorr,deter2,MS}.

\vspace{0.25cm}
{\bf Locality}
\vspace{0.25cm}

Another property which needs further
investigation is the locality of the
operators arising in the form factor definition, which is a fundamental  
requirement one should be able to prove in the form factor framework in order
to definitely confirm that the objects we want to construct characterise a
properly defined QFT. We will mention later that some attempts for such
a proof may be found in the literature but they all hold only for particular
types of QFT's and operators.

\subsubsection{Main purposes of the form factor analysis}
\indent \ \
In the light of the previous historical review and summary of open questions we can
already anticipate the main purposes our form factor analysis  will serve for:

\vspace{0.25cm}

{\bf i)} Develop further the QFT associated to the HSG-models.

\vspace{0.25cm}

{\bf ii)} After  the physical  picture emerging from \cite{HSG2, HSGsol} and  \cite{HSGS} 
has been confirmed by the thermodynamic Bethe ansatz analysis of the preceding 
chapter  \cite{Yang, TBAZam1}, the form factor analysis may be exploited as  an 
alternative approach which allows for  double-check and even go beyond the TBA-results, 
providing in this way also a more exhaustive check  for  the consistency of the S-matrix 
proposal \cite{HSGS}, i.e.  more information about the underlying CFT.

\vspace{0.25cm}

{\bf iii)} Apart from the two preceding points, which explicitly point out 
the utilisation of the form factor approach as a means for extracting information
characteristic from  the specific models at hand, our concrete form factor analysis
will also provide a valuable contribution to the understanding of various properties and
applications of form factors which, in the light of the previous historical 
introduction, need further clarification. We will specify in more detail later the
concrete contributions of our work  in this direction.

\vspace{0.25cm}

Recall that the main outcome of our TBA-analysis has been 
the identification of the Virasoro
central charge of the underlying CFT and  the numerical computation 
of the finite size scaling
function \cite{TBAZam1, KM2} for which a ``staircase'' pattern intimately 
related to the presence of unstable particles in the spectrum 
\cite{BW, ELOP, HSGS} 
and indicating the different energy scales of unstable and stable particles, has been observed.
 Also the conformal dimension of the perturbing operator emerged in the context of the TBA by conjecturing 
a relationship between the latter quantity and the periodicities of the 
so-called $Y$-systems \cite{TBAZamun}. 

However,  at present  we know that the identification of  the operator content  of 
a 1+1-dimensional QFT starting from its scattering theory is 
not achievable in the TBA-context (apart from the possible identification of the 
conformal dimension of the perturbing operator outlined above) and in fact, for 
general quantum field theories is still an outstanding issue  whose investigation 
has great interest in its own right. 

It is worth mentioning that recently a link 
between scattering theory and local interacting fields in terms of polarisation-free generators
has been developed \cite{Bert, BBS}. Unfortunately, they involve subtle domain
properties and are therefore objects which concretely can only be handled
with great difficulties.
On the other hand, the central concepts of relativistic quantum field theory, like Einstein
causality and Poincar\'{e} covariance, are captured in local field equations
and commutation relations. As a matter of fact local quantum physics
(algebraic quantum field theory) \cite{Haag} takes the collection of all
operators localised in a particular region, which generate a von Neumann
algebra, as its very starting point (for recent reviews see e.g. \cite{Buch, Haag2}).
Ignoring subtleties of non-asymptotic states, it is essentially
possible to obtain the latter picture from the former, namely the  ``particle picture'',  
by means of the LSZ-reduction formalism \cite{LSZ}.

Fortunately, in the context of 1+1-dimensional QFT's,  the form factor approach \cite{Kar} turns out to be 
very convenient for  the purposes summarised two paragraphs ago. As was said before,
form factors are matrix elements (see Eq. (\ref{ffff})) of a certain
local operator associated to a 1+1-dimensional QFT between a multi-particle 
{\em in}-state and the vacuum. These matrix elements can be obtained as
 the solutions to a certain set of 
consistency equations  \cite{Kar,Smir,Zamocorr, YZam,BFKZ} which have their 
origin on very general principles of quantum field theory  and therefore do 
not rely on the specific nature of the corresponding  local operator. Notice 
that, this is very similar to the situation arising in the construction of the
 two-particle S-matrices related to integrable massive bi-dimensional QFT's
 described in chapter \ref{ntft}, in the
sense that the bootstrap program \cite{Boot} also allows for
 the exact construction of these amplitudes by viewing then 
as the solutions to a set of consistency equations based on 
physical requirements such as unitarity,
crossing symmetry, Hermitian analyticity or Lorentz invariance.

Once a solution for these consistency equations is found, 
it is left the task of relating it to a particular local operator. 
First of all, one shall make the assumption 
dating back to the initial papers \cite{Kar}, that each solution to the form factor
consistency equations \cite{Kar,Smir,Zamocorr, YZam, BFKZ} corresponds to a
particular local operator. Based on this, numerous authors \cite
{Kar,Smir,Zamocorr,BFKZ,CardyBast,deter1,deter2,Smir2,DM,DSC} have used various
ways to identify and constrain the specific nature of the operator, e.g. by
looking at asymptotic behaviours, performing perturbation theory, taking
symmetries into account, formulating quantum equations of motion, etc. Our
analysis will  especially exploit the conjecture that each local
operator related to a primary field  has a counterpart in the ultraviolet conformal field theory.

Having identified the operator content (or at least, part of it) by means of any of the methods
just summarised, the knowledge of the form factors associated to certain local operators allows
immediately for the computation of correlation functions involving these operators, at least
up to a certain approximation. Such a relationship can be exploited 
for various applications like, for instance:

\vspace{0.3cm}

{\bf i)}  the calculation of the two-point function of the trace of
the energy momentum tensor $\Theta$, an operator whose existence in guaranteed for any QFT.
Once the two-point function of the trace of energy momentum tensor is known
it is possible to evaluate, usually numerically, Zamolodchikov's  $c$-function \cite{ZamC}. The
$c$-function contains the same physical information as the finite size scaling function 
of the TBA, namely,  in the UV-limit  reduces to the Virasoro central charge of the underlying 
CFT whence following its renormalisation group (RG)  flow we will  observe the familiar ``staircase''
  behaviour  as a sign of  the different mass scales of unstable and stable particles.  The latter
 behaviour supports the interpretation of Zamolodchikov's $c$-function as a measure of massless 
effective degrees of freedom in the Hilbert space. Also, by looking at the asymptotic UV-behaviour 
of the two-point function, it is  possible to extract the conformal dimension of the perturbing
 operator by means of the well-known proportionality between  the perturbing field and the trace 
of the energy momentum tensor established in \cite{Cardypert} (see also \cite{cardy}). Therefore, 
having determined the form factors of the trace of the energy momentum tensor, one is in principle 
in the position to obtain essentially all the information provided by the preceding TBA-analysis.

\vspace{0.3cm}

{\bf ii)} the calculation of the two-point function of any other  local operator for which the 
form factors are known. The latter application of form factors, allows for the possibility to  study 
the ultraviolet behaviour of these two-point functions and determine thereafter the 
ultraviolet conformal dimension of the operator, namely the conformal dimension of the 
operator of the underlying CFT which acts as counterpart of the one under inquiry in the UV-limit. 
In some cases, depending on the particular internal symmetries of the model at
hand, the ultraviolet conformal dimension can be also computed  by means of the so-called $\Delta$-sum
rule derived by G. Delfino, P. Simonetti and J.L. Cardy in \cite{DSC},  
provided  the correlation function of 
the operator at hand with the trace of the energy momentum tensor is known. One might also perform a
renormalisation group analysis for the conformal dimensions obtained through the $\Delta$-sum rule
 \cite{DSC} and observe the flow of the operator content  from a CFT to another as the RG-parameter varies.

\vspace{0.4cm}
The concrete analysis we present in this chapter collects the results reported in
 \cite{CFK, CF1, CF2} (see also \cite{FF}),  and  can be summarised as follows:

\vspace{0.2cm}
In this chapter we study a concrete model, the $SU(3)_2$-HSG model, within the form factor 
context.  As mentioned above, form factors are solutions to a set of general consistency equations  
\cite{Kar,Smir,Zamocorr,YZam, BFKZ}. These equations are reported in section \ref{gen},
 together with a summary of the key steps involved in their solution (subsection \ref{key}) and 
several applications of form factors: the evaluation 
of correlation functions (subsection \ref{corre}) and
the numerical methods employed for this purpose (subsection \ref{numeros}),
the computation
of the Virasoro central charge of the underlying CFT (subsection \ref{cff}), the 
re-construction of the operator content of the underlying CFT (subsection \ref{uvffs})
or the development of a renormalisation group analysis which might confirm and 
go beyond the physical picture emerging from the TBA-analysis (subsections \ref{cdflows} and \ref{betalike}).
After the general
equations to be solved have been introduced, we come in section \ref{modelff}
to the description of the main features of the $SU(3)_2$-HSG model, emphasising especially   
those characteristics  which may be more relevant in the form factor context. 
In section \ref{mini} we present a general ansatz for the solutions to the 
consistency equations summarised in section \ref{gen}. 

This ansatz depends
upon certain functions which are known in the literature as minimal form factors. We
recall their main properties and derive their explicit form for the model at hand. In section
\ref{solution} we systematically construct all  $n$-particle form factors associated to a large
class of local operators of the $SU(3)_2$-HSG model. These solutions are given in terms of building
blocks consisting out of determinants of matrices whose entries are elementary symmetric polynomials
on the rapidities. They also admit an alternative representation in terms of contour integrals which
we also present in the same section. However, at this stage of our investigation we do not provide a
rigorous proof  of the proposed solutions which are based on the extrapolation of the results obtained
up to a certain value of $n$. In section \ref{ising} we illustrate the results of the previous section by
particularising the form factor solutions for three concrete local operators of the theory. In section
\ref{proof} we carry out a rigorous proof,  based on very simple properties of determinants, of the solutions proposed in the preceding sections.
We demonstrate how these general solutions are interrelated by the 
so-called momentum space cluster property in section \ref{mscp}. In
particular we show that the cluster property serves also as a construction
principle, in the sense that from one solution to the consistency equations
we may obtain a large class, almost all, of new solutions. Having now a large set of solutions
for different operators available we might exploit the knowledge of the form factors for the various
applications reported in section \ref{gen}.
 First of all, in section \ref{cff2} we compute the Virasoro central charge of the underlying
CFT by means of  Zamolodchikov's $c$-theorem \cite{ZamC}.
Taking furthermore into account that the $SU(3)_{2}$-HSG model, like numerous other
1+1 dimensional integrable models, may be viewed as a perturbed CFT
whose entire field content is well classified and assuming now
a one-to-one correspondence between operators in the conformal and in the perturbed theory,
we carry out an identification on this level  that is, we identify each solution of the form
factor consistency equations with a local operator  which is labeled according to the ultraviolet
CFT. We present this analysis  in section \ref{identifying} where we determine
the ultraviolet conformal dimensions of all the operators whose form factors were constructed before. 
In particular, due to the outlined proportionality between the perturbing field and the trace
of the energy momentum tensor \cite{Cardypert}, the latter  
analysis makes it  possible to identify the conformal dimension of the perturbation. In section
\ref{rgflow} we compute numerically the RG-flow of  Zamolodchikov's
$c$-function \cite{ZamC} and also  extend our RG-analysis to the conformal dimensions
of those local operators for which the $\Delta$-sum rule \cite{DSC} holds. We verify that the
RG-analysis leads to a physical picture which is entirely  consistent with the TBA-results and
goes beyond them at the same time. Finally we review the main outcome of our analysis 
and point out some open problems in section \ref{sum4}. 

\section{Generalities on form factors}
\label{gen}
\indent \ \
Before the analysis of the specific results obtained for the 
$SU(3)_2$-HSG model is presented, we must recall the
definition and general  properties of  form factors. These properties,
can be rigorously justified in most cases in terms of general
principles of quantum field theory and analytic properties in 
the complex plane. Here we only pretend to provide 
a general overview of all of them in order to make comprehensible our specific study.
For a more detailed derivation we refer the reader to the seminal papers \cite{Kar}
and specially to the book \cite{Smir}. A fairly detailed review of these properties may be
also found in the papers \cite{Zamocorr,YZam,BFKZ}, where the form factor approach
is exploited for the study of concrete models.
We also make use of this section  to  set up the general framework and 
notation we shall continuously appeal to in  subsequent sections. 

\subsection{The form factor consistency equations}
\label{properties}
\indent \ \
Form factors are tensor valued functions, representing matrix elements of
some local operator $\mathcal{O}({x})$ located at the origin between a
multi-particle {\em{in}}-state and the vacuum

\begin{equation}
F_{n}^{\mathcal{O}|\mu _{1}\ldots \mu _{n}}(\theta _{1},\ldots ,\theta
_{n}):=\left\langle 0|\mathcal{O}(0)|V_{\mu _{1}}(\theta _{1})V_{\mu
_{2}}(\theta _{2})\ldots V_{\mu _{n}}(\theta _{n})\right\rangle _{\text{in}%
}\,\,.
\label{ffff}
\end{equation} 
\noindent
Recall that the vertex operators $V_{\mu_i}(\theta_i)$ have been
introduced in subsection \ref{Zamalgebra} as a means for representing
the asymptotic particle states of the QFT. Form factors, have the following general
 properties\footnote{Here we will restrict ourselves to QFT's possessing diagonal S-matrices.}:

\vspace{0.3cm}

{\bf Property 1: CPT invariance}
 
\vspace{0.6cm}

\noindent  
As a consequence of CPT-invariance or the braiding of two operators 
$V_{\mu_i}(\theta_i ), V_{\mu_{i+1}}(\theta_{i+1})$ given by  (\ref{Zalg1}),
one obtains 

\begin{equation}
F_{n}^{\mathcal{O}|\ldots \mu _{i}\mu _{i+1}\ldots }(\ldots ,\theta
_{i},\theta _{i+1},\ldots )=F_{n}^{\mathcal{O}|\ldots \mu _{i+1}\mu
_{i}\ldots }(\ldots ,\theta _{i+1},\theta _{i},\ldots )S_{\mu _{i}\mu
_{i+1}}(\theta _{i,i+1})\,.  \label{W1}
\end{equation}

\vspace{0.3cm}

{\bf Property 2: monodromy}
 
\vspace{0.6cm}

\noindent  
The analytic continuation in the complex $\theta $-plane at the cuts when 
$\theta =2\pi i$ together with the property of crossing of the S-matrix (see Eq. (\ref{crossing}) 
in chapter \ref{ntft}) lead to 
\begin{eqnarray}
F_{n}^{\mathcal{O}|\mu _{1}\ldots \mu _{n}}(\theta _{1}+2\pi i,\ldots
,\theta _{n})&=&\omega\,F_{n}^{\mathcal{O}|\mu _{2}\ldots \mu _{n}\mu _{1}}(\theta_ 
{2},\ldots ,\theta_{n},\theta_{1})=
\nonumber \\
&=& \omega\,{\prod_{i=2}^n}
 S_{\mu_i \mu_1}(\theta_{i1})\,
 F_{n}^{\mathcal{O}|\mu _{1}\ldots \mu _{n}}(\theta
_{1},\ldots ,\theta _{n}),  \label{W2}
\end{eqnarray}
\noindent where $\omega$ is the so-called {\bf factor of local commutativity} originally
introduced in \cite{YZam}. The meaning of such parameter as well as the motivation for
its introduction will be given a bit later, when introducing the so-called `kinematical
residue equation'.
\vspace{0.3cm}

\noindent 
The former two properties are usually referred to as {\bf Watson's equations}
 \cite{Watson}. 

\vspace{0.4cm}
The {\bf pole structure} of the form factors is encoded in the two following properties (3 and 4) which have the form
of recursive equations relating form factors associated to different numbers of particles. 
For reasons which become clear below, such pole structure is fundamental in order to find explicit solutions to the form factor
consistency equations. 

\vspace{0.3cm}

{\bf Property 3: kinematical residue equation}
 
\vspace{0.3cm}

\begin{figure}[!h]
 \begin{center}
  \leavevmode
    \includegraphics[scale=0.55]{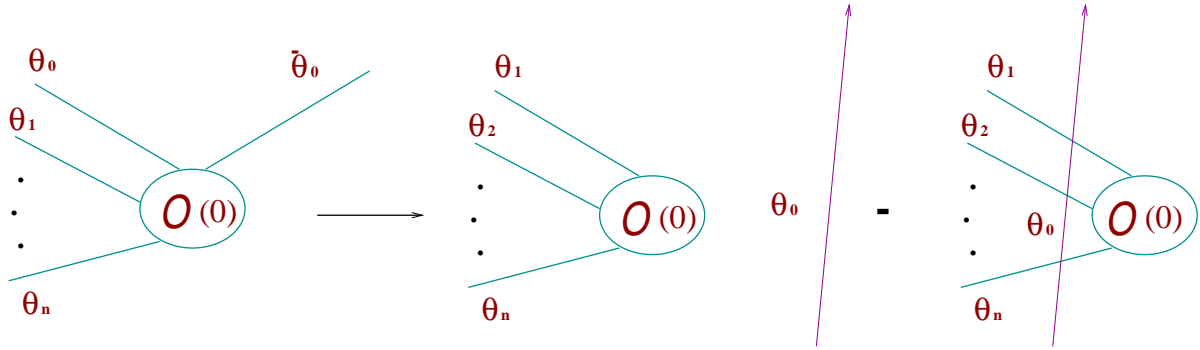}   
    \caption{Kinematical pole residue equation for form factors:
   $\protect{F_{n+2}^{\mathcal{O}}\rightarrow 
   F_{n}^{\mathcal{O}}}$.}
 \label{prop3}
 \end{center}
\end{figure}

\noindent 
The first type of simple poles arise 
for a form factor whose first two particles are conjugate to each other. In
that case we
have a kinematical pole at $\theta=i\pi$, $\theta$ being the rapidity difference between these two particles. The existence of this simple pole leads to a recursive equation relating the $(n+2)$- and the $n$-particle form factor 
\begin{equation}
\stackunder{{\small \bar{\theta}}_{0}\rightarrow {\small \theta }_{0}}{\text{%
Res}}{\small F}_{n+2}^{\mathcal{O}|\bar{\mu}\mu \mu _{1}\ldots \mu _{n}}%
{\small (\bar{\theta}}_{0}{\small +i\pi ,\theta }_{0}{\small ,\theta }_{1}%
{\small ,\ldots ,\theta }_{n}{\small )}=i(1-\omega \prod_{l=1}^{n}S_{\mu \mu
_{l}}(\theta _{0l})){\small F}_{n}^{\mathcal{O}|\mu _{1}\ldots \mu _{n}}%
{\small (\theta }_{1}{\small ,\ldots ,\theta }_{n}{\small ),}  \label{kin}
\end{equation}
with $\omega $ being the so-called 
{\bf factor of local commutativity}  \footnote{The factor of local commutativity was 
originally introduced in \cite{YZam} and  interpreted as a factor which takes care of the  possible
semilocality of the vertex operator  $V_{\mu}(\theta )$ with respect to 
the field $\mathcal{O}(0)$, namely 
\begin{equation}
V_{\mu}(\theta) \mathcal{O}(0)= \omega \, \mathcal{O}(0) V_{\mu}(\theta ). 
\nonumber
\end{equation}
However, this interpretation and the subsequent introduction of $\omega$ in Eq. (\ref{kin})
is not rigorously argued at the level of the form factor consistency equations namely,
the occurrence of the factor of local commutativity in (\ref{kin}) still needs further
clarification. Nonetheless, the need for introducing such a factor in (\ref{kin}) is
supported by the specific analysis carried out in the original paper \cite{YZam}, which
showed that only when  introducing a factor of local commutativity, in principle different
for each of the local operators of the theory, it was possible to find solutions to
the form factor consistency equations which are in correspondence with the primary field content
of the underlying CFT. The concrete model analysed in \cite{YZam} is the
 thermal perturbation of the  Ising model for which the operator content of the underlying CFT
is well known and consists of the energy density operator, $\varepsilon$, together
with the order, $\Sigma$, and disorder, $\mu$, operators (see e.g \cite{CFT, cardy}).
More concretely, the authors of \cite{YZam} realised that,
being the corresponding  S-matrix equal to -1,  
the form factor solutions for the disorder operator  $\mu$ can only 
be obtained from (\ref{kin}) once  
a factor of  local commutativity  $\omega (\mu)=-1$  has been introduced. 
This observation  is 
crucial also  for our particular study, since the S-matrices of  
the models we will be interested in
reduce to the one of the thermal perturbation of the Ising model 
when considering the interaction between particles of the same type. This is also
the ultimate reason why we have been forced to introduce the factor of  local commutativity
in the course of our concrete analysis.} and $\bar{\mu}$ the anti-particle of $\mu$ 
(see Fig. \ref{prop3}).

\newpage

\vspace{0.3cm}

{\bf Property 4: bound state residue equation}
  
\vspace{0.3cm}

\noindent 
The second type of simple poles we referred to occur only when bound states
are present in the model. 
When this happens, the corresponding form factors
have poles located at the values of the rapidity in the physical strip
 $0 \leq \text{Im}(\theta)\leq\pi$
corresponding to fusing angles of the S-matrix (see section \ref{analit} in chapter \ref{ntft}).
This gives rise to another set of recursive equations relating now
the $(n+1)$- and the $n$-particle form factor.

\begin{figure}[!h]
 \begin{center}
  \leavevmode
    \includegraphics[scale=0.55]{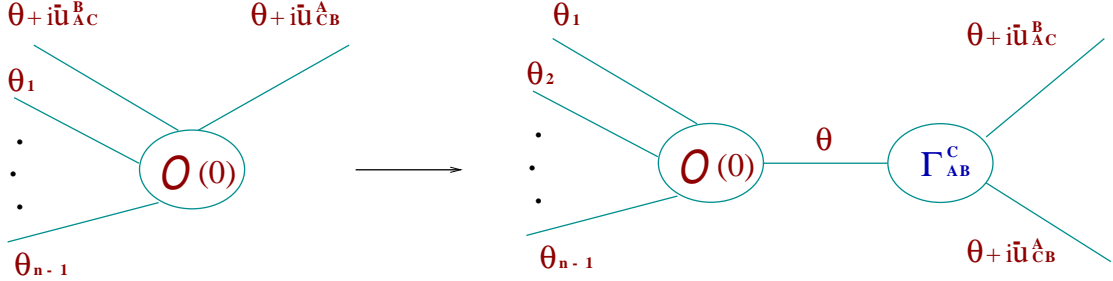}   
    \caption{Bound state residue equation for form factors:
   $\protect{F_{n+1}^{\mathcal{O}}\rightarrow 
   F_{n}^{\mathcal{O}}}$.}
  \label{prop4}
 \end{center}
\end{figure}

Recall that if  two  particles $A,B$ interact to form a stable bound state ${C}$ i.e, 
$ A + B \rightarrow {C}$, there is a pole in the corresponding
two particle amplitude at $\theta_{AB}=i u_{AB}^C$ of the form 

\begin{equation}
\stackunder{{\small \theta} \rightarrow {\small i(\pi- u_{AB}^C)}}{\text{%
Res}} S_{AB}(\theta)= i (\Gamma_{AB}^C)^2,
\end{equation}
\noindent where $\Gamma_{AB}^C$ is the three-particle vertex on mass-shell and
we are always assuming the S-matrix to be diagonal\footnote{Notice that we are not
referring now to the decay width of an unstable particle for which we used a very
similar notation in chapter \ref{ntft}.}.
The corresponding residue equation for the form factors is given by \cite{LSZ2}

\begin{eqnarray}
&&\lim_{\epsilon \rightarrow 0} \epsilon \,F_{n+1}^{\mathcal{O}|A\, B \mu_1 \ldots \mu_{n-1}}
(\theta + i \bar{u}_{AC}^B - \epsilon,\theta - i \bar{u}_{BC}^A + \epsilon, \theta_1,\ldots,\theta_{n-1})=
\nonumber \\
&&\,\,\,\,\,\,\,\,\,\,\,\,\,\,\,\,\,\,\,\,\,\,\,\,\,\,\,\,\,\,\,\,\,\,\,\,\,\,\,\,\,\,\,\,\,\,\,\,
\Gamma_{AB}^C \, F_{n}^{C \, \mu_1 \cdots \mu_{n-1}}(\theta, \theta_1,\cdots,\theta_{n-1}).
\label{bstate}
\end{eqnarray}
\noindent As usual, $ \bar{u}_{AC}^B=\pi-u_{AC}^B $ and $\bar{u}_{BC}^A=\pi-u_{BC}^A$. Eq. (\ref{bstate}) is known in the literature as {\bf bound state residue equation}
and establishes a set of recursive relations involving the $(n+1)$- and
$n$-particle form factors (see Fig.  \ref{prop4}).

\vspace{0.4cm}

\noindent
A very important feature which distinguishes the former  properties 1-4  from
the ones we are going to recall now is the fact that the first ones  are
based on very general principles and therefore do not require any  information
about the particular nature of the operator  $\mathcal{O}$ under consideration.
However, it was already pointed out in the introduction of this chapter that,
once the assumption that each different solution to the form factor consistency 
equations corresponds to a different local operator is made \cite{Kar}, 
there are different ways to identify and constrain the nature of such an operator. 
One of the most fruitful and useful ones is to look at the asymptotic behaviour
of the corresponding form factors. The following two properties establish 
important constraints for this asymptotic behaviour. 

\vspace{0.3cm}

{\bf Property 5: relativistic invariance}
 
\vspace{0.3cm}

\noindent  
Since we are describing relativistically invariant theories we expect for an
operator $\mathcal{O}$ with spin $s$%
\begin{equation}
F_{n}^{\mathcal{O}|\mu _{1}\ldots \mu _{n}}(\theta _{1}+\lambda ,\ldots
,\theta _{n}+\lambda )=e^{s \lambda }F_{n}^{\mathcal{O}|\mu _{1}\ldots \mu
_{n}}(\theta _{1},\ldots ,\theta _{n})\,\,,  \label{rel}
\end{equation}
\noindent  with $\lambda$ being an arbitrary real number. 
This equation establishes a first constraint on the asymptotic behaviour of the form 
factors.
\vspace{0.3cm}

{\bf Property 6: asymptotic bounds}
 
\vspace{0.3cm}

\noindent 
To be able to associate a solution of the equations 
(\ref{W1})-(\ref{bstate}) to a particular operator, 
the following upper bound on the asymptotic behaviour 
which was derived in \cite{DM}
\begin{equation}
\left[ F_{n}^{\mathcal{O}|\mu _{1}\ldots \mu _{n}}(\theta _{1},\ldots
,\theta _{n})\right] _{i}\leq \,\Delta^{\mathcal{O}}  \label{bound}
\end{equation}
turns out to be very useful. Here $\Delta^{\mathcal{O}}$ denotes the conformal dimension
of the operator $\mathcal{O}$ in the conformal limit. For convenience we
introduced the short hand notation,
\begin{equation}
\lim_{\theta _{i}\rightarrow \infty
}f(\theta _{1},\ldots ,\theta _{n})= \textnormal{const.} e^ {[f(\theta _{1},\ldots
,\theta _{n})]_{i}\theta _{i}},
\label{not}
\end{equation}
\noindent which later will turn out to be very useful. 
\vspace{0.4cm}

\noindent 
The next important property of form factors we want to state is known as momentum space
{\bf cluster property} 
 and has been analysed explicitly for several concrete 1+1-dimensional
QFT's \cite{clust,Zamocorr,deter2,MS}. It  also admits a perturbative interpretation
by means of Weinberg's power counting argument \cite{Kar, BK, Wein}. 
With respect to the former properties, the momentum
space cluster property differs in two basic aspects: First, it is not known 
for the time being a proof which can be considered at the same footing as 
the proofs existing for the former properties 1-6. Second, whereas the properties
1-6 do not require any knowledge about the specific nature of the local operator 
$\mathcal{O}$, the cluster property captures at least  
part of the operator nature of the theory.

\vspace{0.3cm}

{\bf Property 7: momentum space cluster property}
 
\vspace{0.3cm}

\noindent
Cluster properties in space, i.e. the observation that far separated
operators do not interact, are quite familiar in quantum field theories \cite
{Wich} for a long time. In 1+1 dimensions a similar property has also been
noted in momentum space. For the purely bosonic case this behaviour can
be explained perturbatively by means of  Weinberg's power counting theorem, see
e.g.  \cite{Kar, BK}\footnote{
There exists a heuristic argument which provides some form of intuitive
picture of this behaviour \cite{DSC} by appealing to the ultraviolet conformal field theory. 
However, the argument is based on
various assumptions, which need further clarification. For instance it
remains to be proven rigorously that the particle creation operator $V_{\mu}(\theta )$ 
tends to a conformal Zamolodchikov operator for $\theta
\rightarrow \infty $ and that the local field factorises equally into two
chiral fields in that situation. The restriction in there that $\lim_{\theta
\rightarrow \infty }S_{ij}(\theta )=1$, for $i$ being a particle which has
been shifted and $j$ one which has not, excludes a huge class of interesting
models, in particular the one at hand.}. As mentioned in
section \ref{properties}, the cluster property states for an $n$-particle form factor associated
to an operator $\mathcal{O}$ that

\begin{equation}
\mathcal{T}_{1,\kappa }^{\lambda }F_{n}^{\mathcal{O}}(\theta _{1},\ldots
,\theta _{n})\sim F_{\kappa }^{\mathcal{O}^{\prime }}(\theta _{1},\ldots
,\theta _{\kappa })F_{n-\kappa }^{\mathcal{O}^{\prime \prime }}(\theta
_{\kappa +1},\ldots ,\theta _{n})\,\,,  \label{cluster}
\end{equation}
\noindent where,  for convenience we have introduced  the operator 
\begin{equation}
\mathcal{T}_{a,b}^{\lambda }:=\lim_{\lambda \rightarrow \infty
}\prod_{p=a}^{b}T_{p}^{\lambda }\,
\end{equation}
which will allow for concise notations. It is composed of the translation
operator $T_{a}^{\lambda }$ which acts on a function of $n$ variables as 
\begin{equation}
T_{a}^{\lambda }\,f(\theta _{1},\ldots ,\theta _{a},\ldots ,\theta
_{n})\,\mapsto \,f(\theta _{1},\ldots ,\theta _{a}+\lambda ,\ldots ,\theta
_{n})\,\,\,,
\end{equation}
\noindent therefore the operator $\mathcal{T}_{1,\kappa}^{\lambda}$ has the effect of shifting to infinity the first $\kappa$ rapities entering in  $F_{n}^{\mathcal{O}}(\theta _{1},\ldots,\theta _{n})$.

Whilst Watson's equations  and the residue equations stated above,
 are operator independent features of form factors,
the cluster property captures part of the operator nature of the theory. The
cluster property (\ref{cluster}) does not only constrain the solution, but
eventually also serves as a construction principle in the sense that when
given $F_{n}^{\mathcal{O}}$ we may employ (\ref{cluster}) and construct form
factors related to $\mathcal{O}^{\prime }$ and $\mathcal{O}^{\prime \prime }$.

Despite the fact that the cluster property has been analysed for several
concrete models in  \cite{clust,Zamocorr,deter2,MS}, and that the
possibility of having form factors associated to different local fields of
the massive QFT on the r.h.s. and l.h.s. of (\ref{cluster}) was originally 
pointed out in \cite{deter2}, examples of such a situation were up to now absent in
the literature. In other words, only self-clustering  was encountered
for all the specific local operators of the various QFT's investigated in 
\cite{clust,Zamocorr,deter2,MS}. This fact makes the outcome of the cluster-analysis 
carried out in this thesis for the $SU(3)_2$-HSG very remarkable, since it provides
the first explicit example of a model where the form factors of different local
operators can be obtained from each other via clustering.
Therefore, our analysis gives definite support to the assertion that
(\ref{cluster}) constitutes a closed mathematical structure, which
relates various different solutions and whose abstract nature still needs to be
unraveled.  

\vspace{0.25cm}

{\bf Property 8: locality}

\vspace{0.25cm}
\noindent Finally, we would like to mention  a very  important  property of form factors for which, 
similarly to the momentum space cluster property, it does not exist such a well established proof as
for the mentioned consistency properties 1-6, but which is really the confirmation that the objects
we want to construct characterise a well defined QFT. This property is the {\bf locality} of the
operators $\mathcal{O}$ arising in (\ref{ffff}) which is expressed, in the bosonic case,
 by means of  the usual condition (see
for instance \cite{Wein}):
\begin{equation}
[\mathcal{O}_1({x}), \mathcal{O}_2({y})]=0, 
\end{equation}
\noindent for ${x}$ and ${y}$ to be causally disconnected points in Minkowski's 
space and $\mathcal{O}_1, \mathcal{O}_2$ two local operators of the QFT.
As mentioned above, properties 1-4 do not require any information about the particular
nature of the operator $\mathcal{O}$. Therefore, once a solution to these consistency equations is found,
it is of great interest to check that the locality of the operator involved is guaranteed 
namely, properties 1-4 and locality are self-consistent requirements.
A proof of the locality property involves the computation of  correlation functions of the form
\begin{equation} 
\langle 0 |[\mathcal{O}_1 ({x}), \mathcal{O}_2 ({y})]|0 \rangle,
\label{locality}
\end{equation}
\noindent which is possible once the form factors associated to the operator $\mathcal{O}_1, 
\mathcal{O}_2$ are known, as
we explain in more detail in the next subsection. Therefore, proving that such a
correlation function vanishes whenever the points ${x}, {y}$ are not causally connected is
equivalent to demonstrate the locality of any operator whose form factors have been obtained as solutions
to the previous consistency equations. In this direction, the work
carried out in \cite{thomas} provided a proof of  locality for bosonic QFT's whose particle
spectrum does not contain stable bound states, so that property 4 does not enter the form factor analysis.
An analysis concerning the locality property within the form 
factor approach may be also found in appendix B of \cite{Smir}, 
although the arguments exhibited there still lack the rigour of the proofs of properties 1-6 and
are concerned with the specific case of bosonic QFT's which do not contain
stable bound states. Furthermore, the analysis performed in \cite{Smir} holds for  a particular
choice of one of the operators $\mathcal{O}_1$, $\mathcal{O}_2$ so that Eq. (\ref{locality}) is
not proven for the generality of local operators of the QFT's under consideration.

\subsection{Summarising the key steps of the solution procedure}
\label{key}
\indent \ \ 
Before we enter the different applications of form factors, we want to anticipate
very generically and briefly the main steps of a possible solution procedure leading to
the construction of form factors:

\vspace{0.25cm}
{\bf Solving Watson's equations}
\vspace{0.25cm}

\noindent The starting point in the solution procedure consists of finding an ansatz for the 
solution to the two first properties of form factors, the so-called Watson's equations.
A formal ansatz for such a solution was already provided in the original papers \cite{Kar}, 
\begin{equation}
F_n^{\mathcal{O}|\mu_1 \cdots \mu_n}=
Q_n^{\mathcal{O}| \mu_1 \cdots \mu_n}(\theta_1, \cdots, \theta_n)
\, K_n^{\mu_1 \cdots \mu_n}(\theta_1, \cdots, \theta_n).
\label{step1}
\end{equation}
\noindent where $Q_n^{\mathcal{O}| \mu_1 \cdots \mu_n}(\theta_1, \cdots, \theta_n)$ is
assumed to be a function of the rapidities which neither contains zeros nor poles in the physical sheet and
$K_n^{ \mu_1 \cdots \mu_n}(\theta_1, \cdots, \theta_n)$ encodes the pole structure
of the form factors revealed by  Eqs. (\ref{kin}) and (\ref{bstate}) and does not
depend on the particular nature of the operator at hand. Additional properties
of these two functions, more concretely, the symmetries of the 
former and the pole structure of the latter, 
can be imposed thereafter by appealing to the particular features of the model under consideration.
In particular, it is common to assume the following structure for the $K$-function
\begin{equation}
K_n^{ \mu_1 \cdots \mu_n}(\theta_1, \cdots, \theta_n)=
\prod_{i<j}\frac{ F_{\text{min}}^{ij}(\theta_{ij})}{P_{ij}(\theta_{ij})},
\end{equation}
\noindent where the functions $F_{\text{min}}^{ij} (\theta_{ij})$ are the so-called {\bf minimal form factors},
whose properties will be introduced in subsection \ref{mini},
 and $P_{ij}(\theta_{ij})$ are functions which capture the pole
structure of the form factors. In this fashion, the ansatz (\ref{step1}) can 
 be chosen in such a way that Watson's equations are automatically satisfied.  

\vspace{0.25cm}
{\bf Solving the kinematical and bound state residue equations}
\vspace{0.25cm}

\noindent The next step consists of  ``plugging'' the previous ansatz into Eqs. (\ref{kin}) and
(\ref{bstate}). This gives rise to two recursive Eqs. relating the $(n+2)$- and $n$-particle form factors
and the $(n+1)$- and $n$-particle form factors, respectively. The finding of general solutions to 
these equations is one of the hardest parts, and so far least systematic,
 of the whole analysis and, as mentioned before, in many
cases one is only able to find solutions for the first lowest values of $n$ and/or for 
particular local operators. There is not an established procedure to solve in total generality
such sort of recursive equations and furthermore, this solution becomes much more involved when both equations
are present at the same time, namely when the model possesses stable bound states and property
4 enters the analysis. In order to get a first glimpse at the the general behaviour it is usually
advantageous to analyse at first models which do not contain bound states, such that one only
has to solve Eq. (\ref{kin}). We will also proceed this way for the HSG-models and find that, 
even in the situation when Eq. (\ref{bstate}) does not arise, the finding of general close formulae for
all $n$-particle form factors associated to any local operator of the QFT is highly non-trivial.

\vspace{0.25cm}
{\bf Identifying and constraining the nature of the operator}
\vspace{0.25cm}

\noindent Having found a solution to the recursive Eqs.(\ref{kin}),  and possibly (\ref{bstate})
in case stable bound states are present, there is left the task to identify the concrete
operator of the QFT to which these solutions correspond to. This is not known a priori,
 since properties 1-4 do not involve any anticipation
of the nature of such an operator. Numerous authors \cite
{Kar,Smir,Zamocorr,BFKZ,CardyBast,deter1,deter2,Smir2,DM,DSC} have used various
ways to identify and constrain the specific nature of the operator: by
looking at asymptotic behaviours, performing perturbation theory, taking
symmetries into account, formulating quantum equations of motion, etc. Our
analysis will  especially exploit the conjecture that each local
operator of the massive QFT  has a counterpart in the ultraviolet CFT
whose primary operator content is, for the models at hand, well classified \cite{Gep, GepQ, DHS}.

\subsection{Correlation functions from form factors}
\label{corre}
\indent \ \
 Once the $n$-particle form factors corresponding to two particular
operators $\mathcal{O},\mathcal{O}^{\prime}$ are known one is in principle 
in the position to compute the correlation function
 $\langle \mathcal{O}(r) \mathcal{O}^{\prime}(0) \rangle$. First of all,
 this correlation function can be rewritten in terms of form factors as follows,

\begin{eqnarray}
\langle \mathcal{O}(r) \mathcal{O}^{\prime}(0) \rangle &=&
\sum_{n=1}^{\infty} \sum_{\mu_1,\cdots,\mu_n=1}^{N}
\int{ \frac{{d\theta_1}\cdots{d\theta_n}}{{n!}(2\pi)^n}
\langle 0 | \mathcal{O}(r) |V_{\mu_1}(\theta_1)\cdots 
V_{\mu_n}(\theta_n) \rangle}
\nonumber \\
&& \,\,\,\,\,\,\,\,\,\,\,\,\,\,\,\,\,\,\,\,
\,\,\,\,\,\,\,\,\,\,\times\, \langle V_{\mu_1}(\theta_1)\cdots 
V_{\mu_n}(\theta_n) | \mathcal{O}(0)|0 \rangle
\label{correfunc}
\end{eqnarray}
\noindent where we have introduced the following orthogonal projector $P$,

\begin{equation}
P = \sum_{n=0}^{\infty} \sum_{\mu_1,\cdots,\mu_n=1}^{N}
\int{\frac{{d\theta_1}\cdots{d\theta_n}}{{n!}(2\pi)^n}
| V_{\mu_1}(\theta_1)\cdots 
V_{\mu_n}(\theta_n) \rangle \langle V_{\mu_1}(\theta_1)\cdots 
V_{\mu_n}(\theta_n)|}
\label{prox}
\end{equation}

\noindent and  the first sum runs over all the $n$-particle states while 
 the second one  runs over all the possible particle types in the theory,
 which we denote generically by $N$.
It is easy to prove that $P$ is a projector namely,  $P^2=I$ and $P^{\dagger}=P$, provided  the states are normalised as

\begin{equation}
\langle V_{\mu_1}(\theta_1)\cdots V_{\mu_n}(\theta_n)|V_{\mu_1^{\prime}}(\theta_1^{\prime})\cdots 
V_{\mu_n^\prime}(\theta_n^\prime)\rangle =\prod_{i =1}^{n}2\pi \delta_{\mu_i \mu_i^{\prime}}\delta(\theta_i -\theta_i^{\prime}),
\label{proxnorm}
\end{equation}
\noindent which can be derived from Eq.  (\ref{Zalg3}).

\noindent In order to rewrite (\ref{correfunc}) in terms of  form factors,
we only need to be able to shift the first matrix element appearing in
(\ref{correfunc}) to a matrix element located at the origin. For this purpose
we must consider the action of a translation of the Poincar\'e group $U_{T_{{x}}}$
on the operator $\mathcal{O}({x})$  and the vertex operators $V_{\mu_i}(\theta_i)$,

\begin{equation} 
U_{T_{x}} \mathcal{O}(0) U_{T_{x}}^{-1}=\mathcal{O}({x}),\,\,\,\,
\,\,\,\,\textnormal{and}\,\,\,\,\,\,\,\,
 U_{T_{x}} V_{\mu_i}(\theta_i) U_{T_{x}}^{-1}=e^{i p^{\nu}(\theta_i)x_{\nu}} 
V_{\mu_i}(\theta_i).
\end{equation}

\noindent Therefore, we obtain

\begin{equation}
\langle 0 | \mathcal{O}(r) |V_{\mu_1}(\theta_1)\cdots 
V_{\mu_n}(\theta_n) \rangle  =\exp \left( -r\sum_{i=1}^{n}m_{\mu _{i}}\cosh
\theta _{i}\right) 
\langle 0 | \mathcal{O}(0) |V_{\mu_1}(\theta_1)\cdots 
V_{\mu_n}(\theta_n) \rangle 
\end{equation}
\noindent for $p_0 (\theta_i)=m_{\mu_i} \cosh(\theta_i)$, 
for $m_{\mu_i}$ to be the mass of the particle
$\mu_i$. Here we have taken $x^{\nu}=(-i r, 0)$ in order to guarantee that $x^2=-r^2 <0$ and,
consequently, that the operators involved in the correlation function (\ref{corr}) are located at
causally connected space positions. 

\vspace{0.2cm}

Consequently, the correlation function (\ref{correfunc}) can finally be expressed  in terms of $n$-particle 
form factors of  the local operators $\mathcal{O}$ and $\mathcal{O}^{\prime}$ as
 
\begin{eqnarray}
\left\langle \mathcal{O}(r)\mathcal{O}^{\prime }(0)\right\rangle
&=&\sum_{n=1}^{\infty }\sum_{\mu _{1}\ldots \mu _{n}}\int\limits_{-\infty
}^{\infty }\ldots \int\limits_{-\infty }^{\infty }\frac{d\theta _{1}\ldots
d\theta _{n}}{n!(2\pi )^{n}}\exp \left( -r\sum_{i=1}^{n}m_{\mu _{i}}\cosh
\theta _{i}\right)  \nonumber \\
&&\times \,\,F_{n}^{\mathcal{O}|\mu _{1}\ldots \mu _{n}}(\theta _{1},\ldots
,\theta _{n})\,\left( F_{n}^{\mathcal{O}^{\prime }|\mu _{1}\ldots \mu
_{n}}(\theta _{1},\ldots ,\theta _{n})\,\right) ^{*}.  \label{corr}
\end{eqnarray}

The possibility to compute correlation functions from form factors may
be exploited to explicitly evaluate various quantities, in particular those
provided before by the TBA-analysis. In particular, the Virasoro central charge
of the underlying CFT by means of the so-called Zamolodchikov's $c$-theorem and
the identification of the operator content of the QFT. This is explained in more
detail in the next subsections.

\subsection{Numerical methods}
\label{numeros}
\indent\ \ 
Before we enter the analysis of the specific quantities we may be able to compute
once the form factors associated to a certain local operator are known, we want 
to comment very briefly on the numerical methods which will be employed later for
the explicit evaluation of correlation functions through Eq. (\ref{corr}).
In this thesis, we have concentrated on the emphasis of the physics rather than
indulging too much into numerical technicalities. However, since this part of the work
also required considerable effort, we want to give at least a flavour of what is involved.

Since many quantities we are interested in are 
at the moment  not accessible in an analytic way, the numerical part is rather essential. 
Without this numerical part the outcome of our study
will get considerably reduced since, ultimately, only by explicitly computing the 
Virasoro central charge, 
the conformal dimensions of various local operators of the theory,
 Zamolodchikov's $c$-function etc...
we will be in the position to  claim that our form factor solutions are perfectly
consistent with the physical picture anticipated for the models under consideration.
Therefore, the expressions obtained will not be merely
 interesting from a mathematical point of view but
also will enter the numerical evaluation of physical quantities.

As mentioned above, the main difficulty of the numerical analysis is the evaluation of the 
multi-dimensional integrals arising in the expansion (\ref{corr}) of the correlation functions.
In general, the two-particle contribution can be evaluated even analytically once
a suitable change of variables is performed (we will see explicit examples later). Therefore,
at this stage, no sophisticated numerical tools are really needed. However, although the two-particle 
contribution is in most cases the leading one\footnote{Due to symmetry reasons many correlation
functions will receive only contributions corresponding to an even number of particles. Consequently,
for these cases the two-particle contribution is the first and leading 
term arising in the expansion (\ref{corr}).}, we will see in the course
of our specific analysis that the results obtained at this order are still very far
from the expected ones, and that one has to sum at least up to 4-particle contribution to obtain
reasonably accurate values. The computation of the 3- and 4-particle contributions can usually
 be performed by using the MATHEMATICA program.
However, the evaluation of 5-, 6-, 7- or 8-particle contributions requires a lot of computer time
and can not be carried out in the same way. Due to the latter reason, we have resorted to a FORTRAN code 
which makes use of the widely used numerical recipe routine VEGAS \cite{NUM}
 for the evaluation of the multi-dimensional integrals.
In this fashion we have been able to obtain different correlation functions in the 8-particle
approximation in computing times always lower than one hour. We believe that more elaborated programs
and/or the use of faster computers will allow for reducing considerably the computing time and 
eventually evaluate the correlation functions up to a much higher order. However, 
in many cases we have been content with the results obtained at this order and also
we never intended to focus our analysis on the investigation and 
perfection of the numerical methods themselves,
but to compute physical quantities with an accuracy and time cost which always justify the
computational effort.

\subsection{Virasoro central charge from form factors}
\label{cff}
\indent \ \
One of the main purposes  of our form factor analysis is to provide an additional check
of  the consistency of the S-matrices proposed in \cite{HSGS} to describe the HSG-models at quantum level.  
The results arising from the TBA-analysis carried out in chapter \ref{tba} gave rise to  a physical picture 
which was in complete consistency with the mentioned S-matrix proposal and in particular allowed for the 
identification of  the Virasoro central charge of
the underlying CFT, a WZNW-coset theory associated in our case to  cosets of the form
$G_k/U(1)^{\ell}$, whose Virasoro central charge is given by Eq. (\ref{cdata}).
 At the same time, as indicated  in section \ref{corre}, once the $n$-particle form factors are 
known one might be able to compute correlation functions by means of (\ref{corr}) and consequently 
obtain also the ultraviolet Virasoro central charge of
the underlying CFT by means of  (\ref{Cth2}), the so-called $c$-theorem of 
Zamolodchikov \cite{ZamC}. Leaving for section \ref{rgflow} the study of the 
RG-flow of  Zamolodchikov's  $c$-function, we want to present now the key results 
which might lead us to extract  the Virasoro central charge in the form factor 
framework and therefore double-check one of the most important data provided by the TBA, surpassing
at the same time the results of the latter analysis.

\vspace{0.3cm}

Let us start  by summarising the information contained in the $c$-theorem
of Zamolodchikov \cite{ZamC}: For any 1+1-dimensional renormalisable and  unitary 
QFT there exists a function $c(g_1, \cdots, g_i, \cdots)$ of the coupling constants ${g_i}$
of the theory\footnote{From now on, we will arrange these coupling constants into  a vector 
${\bf g}:=({g_1},{g_2}, \cdots)$. }, having the following properties:

\vspace{0.25cm}

{\bf i)}  it is non-increasing along  renormalisation group
trajectories, namely 
\begin{equation}
\frac{d c({\bf g})}{dt} \leq 0,
\label{mono}
\end{equation}
\noindent for $t$ to be the renormalisation group parameter,

\vspace{0.25cm}

 {\bf ii)} it  is stationary at critical fixed points, usually denoted by ${\bf g}={\bf g}^{*}$, 
at which the 1+1-dimensional QFT acquires an infinite conformal symmetry generated as usual 
 by a Virasoro algebra (see subsection \ref{cftbo} in chapter \ref{ntft}).
If we recall the definition of the $\beta_i  ({\bf g})$-functions 
as the ``velocities'' of change of the renormalised coupling constants along the renormalisation
group flow \cite{CS}, 
\begin{equation}
\beta_i ({\bf g})=-\frac{d g_i}{dt},
\label{CS}
\end{equation}
\noindent property  {\bf ii)} is equivalently expressed by saying that these $\beta_i$-functions are
vanishing at critical fixed points,
\begin{equation}
\beta_i ({\bf g}^{*})=0 \,\,\,\,\,\Leftrightarrow \,\,\,\,\,
{\left[\frac{\partial  c({\bf g})}{\partial g_i}\right] }_{{\bf g}={\bf g}^*}=0, 
\end{equation}
\noindent therefore $c({\bf g}^{*})$ is stationary at renormalisation group critical fixed points,

\vspace{0.25cm}

 {\bf iii)}  in fact,  the $c$-theorem also establishes that the function $c({\bf g})$
 reduces at these fixed points to the Virasoro central charge of the corresponding conformal field theory.
It is the latter property which we will exploit in this and the next chapter
 in order to obtain the central charge 
corresponding to the $SU(N)_2/U(1)^{N-1}$-coset models.  
Having this aim in mind, the fundamental result we wish to employ
 was also established  in  \cite{ZamC} and can be summarised as follows 
\cite{cardy, Cardypert}: 
Let us consider the local operators $\Theta$, $T$ and $\bar{T}$ corresponding to the 
components of the energy momentum tensor of spin 2, 0 and  -2 respectively and define the following functions
\begin{eqnarray}
F(z\bar{z})&:=& z^4 \langle  T(z, \bar{z}) T(0,0) \rangle, \label{f}\\
G(z\bar{z})&:=&z^3 \bar{z}\langle T(z, \bar{z})\Theta(0,0) \rangle, \label{gg}\\
H(z\bar{z})&:=& z^2 \bar{z}^2 \langle  \Theta (z, \bar{z})\Theta(0,0) \rangle,\label{hhh}
\label{fgh}
\end{eqnarray}
\noindent in  terms of the usual complex coordinates $z=x^0+i x^1, \bar{z}=x^0-i x^1$.
The various components of the energy momentum tensor are interrelated  by means of  its  conservation law
\begin{equation}
\partial_{\bar{z}} T + \frac{1}{4} \partial{\Theta}=0.
\label{law}
\end{equation}
\noindent The  correlation function of the previous equation with $T(0,0)$ and $\Theta(0,0)$, 
with the help of the definitions   (\ref{f})-(\ref{hhh}) gives rise to
 the following  identity,
\begin{equation}
\frac{d c}{dt} = -\frac{3}{4} H,
\label{ourc}
\end{equation}
\noindent for $t=\ln (m^2 z \bar{z})= 2 \ln (m r) $, $m$ being a fixed
mass scale,  $r$ the radial distance and
\begin{equation}
c(t)=c(mr):= 2 F-G-\frac{3}{8}H.
\label{cardyc}
\end{equation}
Notice that, this function $c$ is always non-increasing, 
since the definition (\ref{hhh}) ensures that $H \geq 0$.
It is  also stationary at critical fixed points where $\Theta =0$ 
and therefore $H$ vanishes too. In
fact, at critical fixed points also the function $G$ vanishes  
for the same reason and (\ref{cardyc})
reduces to $F=c/2$, relation  which in a conformal field theory defines $c$ as its
Virasoro central charge \cite{BPZ, CFT} (see also subsection \ref{cftbo}
 of chapter \ref{ntft}). 
Consequently, the function (\ref{cardyc})  fulfills properties {\bf i)},
{\bf ii)} and  {\bf iii)} and can
be identified as the same function these properties referred to at 
the beginning of this subsection. 

\vspace{0.25cm}

By integrating now Eq.   (\ref{ourc}) we obtain for the difference between the ultraviolet and
infrared Virasoro central charges, 
\begin{equation}
c_{uv}-c_{ir }= \Delta c= \frac{3}{4} \int\limits_{-\infty}^{\infty} H(t) dt =\frac{3}{2} 
 {\int_0^\infty} dr r^3 \langle \Theta(r) \Theta(0) \rangle,
\label{Cth2}
\end{equation}
\noindent where we used  the definition 
of $H$ in terms of the trace of the energy momentum tensor (\ref{hhh}) and $t=2 \ln(mr)$. 
Recall that, due 
to the conservation of the energy
momentum tensor, it has been  possible to express the $c$-function, 
initially given in terms of several
 correlators  involving all the components of the energy momentum tensor 
(\ref{cardyc})  in terms of
 only  the correlation function of  its trace. 
Taking into account  that the infrared central
charge is zero for purely massive theories, Eq. (\ref{Cth2}) gives the ultraviolet central
charge of the corresponding underlying CFT.  Very remarkably, whereas on the l.h.s. we have 
a quantity 
characterising the ultraviolet conformal field theory, on the r.h.s.  we have the trace of the
energy momentum tensor, a local operator of the perturbed conformal field theory  or, in other words,  
a quantity  defined away from the critical fixed point.

Being formula (\ref{Cth2}) available we only need now to use (\ref{corr}) to express the 
correlation function of  $\Theta$ in terms of form factors of the same operator.
 By doing so, and
performing thereafter the $r$-integration we get the
 following expression for the central charge,

\begin{equation}
\Delta c= 9  \sum_{n=1}^{\infty }\sum_{\mu _{1}\ldots \mu _{n}}
\int\limits_{-\infty }^{\infty }\ldots \int\limits_{-\infty }^{\infty }%
\frac{d\theta _{1}\ldots d\theta _{n}}{ n!(2\pi)^{n} \left( \sum_{i=1}^{n}m_{\mu
_{i}}\cosh \theta _{i}\right) ^{4}}\left| F_{n}^{\mathcal{O}|\mu _{1}\ldots
\mu _{n}}(\theta _{1},\ldots ,\theta _{n})\right| ^{2}\,\,,  \label{cth}
\end{equation}
\noindent which can now in principle be computed provided all $n$-particle form
factors of the energy momentum tensor are known. 

\subsection{Ultraviolet conformal dimensions from form factors}
\label{uvffs}
\indent \ \ 
As indicated  at the beginning of this chapter, 
there are two assumptions which turn out to be crucial
in the form factor analysis: We will 
assume the existence of a one-to-one correspondence 
between the operator content of the underlying  
CFT and the massive QFT and between the solutions to
 the form factor consistency equations of subsection
\ref{properties} and the local operators of the massive QFT. 
Provided these two assumptions are
made, it is clear that the re-construction or identification 
of the operator content of the massive
QFT can be performed by directly identifying for each local operator of the massive QFT 
its corresponding counterpart in the UV-limit. Since the operator content of the underlying
CFT is well classified for all the HSG-models \cite{Gep, GepQ, DHS},
one is left with the task of matching
our form factor solutions with the operators of the underlying CFT. In our case, we will
carry out such an identification by extracting the ultraviolet conformal dimensions of those local
operators of the QFT for which we have previously computed all $n$-particle form factors
(see section \ref{solution}). 

There are various ways to determine the ultraviolet conformal dimensions of
the operators of the massive QFT. We may use the fact that
the values of  $\Delta^{\mathcal{O}}$ are
obtainable when we exploit our knowledge about the underlying CFT more deeply.
Considering an operator which in the conformal
limit corresponds to a primary field we can of course compute the conformal
dimension by appealing to the UV-limit of the two-point correlation
function 
\begin{equation}
\left\langle \mathcal{O}_{i}(r)\mathcal{O}_{j}(0)\right\rangle
=\sum_{k}C_{ijk}\,r^{2\Delta _{k}-2\Delta _{i}-2\Delta _{j}}\,\left\langle 
\mathcal{O}_{k}(0)\right\rangle +\ldots  \label{ultra1}
\end{equation}
The three-point couplings $C_{ijk}$ are independent of $r$. In particular
when assuming that $0$ is the smallest conformal dimension occurring in the
model (which is the case for unitary models), we have (see section \ref{cftqft}
in chapter \ref{ntft}),
\begin{equation}
\lim_{r\rightarrow 0}\left\langle \mathcal{O}(r)\mathcal{O}(0)\right\rangle
\sim r^{-4\Delta ^{\!\!\mathcal{O}}}\,\,\qquad \qquad \text{for }r\ll \left( 
\frac{C_{\Delta ^{\mathcal{O}}\Delta ^{\mathcal{O}}\,0}}{C_{\Delta ^{%
\mathcal{O}}\Delta ^{\mathcal{O}}\Delta ^{\mathcal{O}^{\prime \prime}}}\left\langle
 \mathcal{O}^{^{\prime \prime }}\right\rangle }\right)
^{1/2\Delta ^{\mathcal{O}^{\prime \prime }}}\,.  \label{ultra}
\end{equation}
Here $\mathcal{O}^{\prime \prime }$ is the operator with the second smallest
dimension for which the vacuum expectation value is non-vanishing. Using a
Lorentz transformation to shift the $\mathcal{O}(r)$ to the origin and
expanding the correlation function in terms of form factors in the usual
fashion, as presented in  (\ref{corr})
we can compute the l.h.s. of (\ref{ultra}) and extract $\Delta ^{\!\!%
\mathcal{O}}$ thereafter. The disadvantage to proceed in this way is
many-fold. First we need to compute the multidimensional integrals in (\ref
{corr}) for each value of $r$, which means to produce a proper curve
requires a lot of computational (at present computer) time. Second, 
for very small $r$ the $n$-th term within the sum is proportional to ($\log
(r))^{n}$ such that we have to include more and more terms in that region
whereas, at the same time, the expressions of the form factors and consequently, 
the integrals one has to evaluate,  become more
and more complicated (see appendix B) as $n$ increases. For that reason, 
 the identification of the conformal dimension $\Delta^{\mathcal{O}}$ turns out
to be very difficult,  unless one has already a relatively good guess for
 its value, which is our case. Finally, 
 the precise values of the lowest non-vanishing form
factors, i.e. in general vacuum expectation values or one particle form
factors are needed in order to compute the r.h.s. of (\ref{corr}).
This is due to the fact that  the constants $H^{\mathcal{O}|\tau, m }$ occurring 
in (\ref{consol}) are precisely fixed by the lowest non-vanishing form factor. 

A short remark is also due concerning solutions related to different sets of 
$\mu $'s. The sum over the particle types in (\ref{corr}) simplifies considerably when
taking into account that form factors corresponding to two sets, which
differ only by a permutation, lead to the same contribution in the sum. This
follows simply by using the first of  Watson's equations \cite{Kar,Smir,Zamocorr,YZam,BFKZ}. 
Recall that this property  states that when two particles are
interchanged we will pick up the related two particle scattering matrix as a
factor (\ref{W1}). Noting that the scattering matrix is a phase, the expression remains
unchanged.
 
\vspace{0.25cm}

Most of the disadvantages, which emerge when using (\ref{ultra}) to compute
the conformal dimensions, can be circumvented by formulating sum rules in
which the $r$-dependence has been eliminated. Such type of rule has for
instance been formulated by F. Smirnov \cite{clust} already more than a decade
ago. However, the rule stated there is slightly cumbersome in its evaluation
and we will therefore resort to one found more recently by G. Delfino,
P. Simonetti and J.L. Cardy \cite{DSC}. In close analogy to the spirit and
derivation of the $c$-theorem \cite{ZamC}, these authors derived an expression
for the difference between the ultraviolet and infrared conformal dimension
of a primary field $\mathcal{O}$%
\begin{equation}
\Delta _{uv}^{\!\!\mathcal{O}}-\Delta _{ir}^{\!\!\mathcal{O}}=-\frac{1}{%
2\left\langle \mathcal{O}\right\rangle } \int\limits_{0}^{\infty
}r\left\langle \Theta (r)\mathcal{O}(0)\right\rangle \,dr\,\,.
\label{deldel}
\end{equation}
\noindent Using the expansion of the correlation function in terms of form factors (
\ref{corr}) we may carry out the $\ r$-integration in (\ref{deldel}) and
obtain 
\begin{eqnarray}
\Delta _{uv}^{\!\!\mathcal{O}}-\Delta _{ir}^{\!\!\mathcal{O}} &=&-\frac{1}{%
2\left\langle \mathcal{O}\right\rangle }\sum_{n=1}^{\infty }\sum_{\mu
_{1}\ldots \mu _{n}}\int\limits_{-\infty }^{\infty }\ldots
\int\limits_{-\infty }^{\infty }\frac{d\theta _{1}\ldots d\theta _{n}}{%
n!(2\pi )^{n}\left( \sum_{i=1}^{n}m_{\mu _{i}}\cosh \theta _{i}\right) ^{2}}
\nonumber \\
&&\times F_{n}^{\Theta |\mu _{1}\ldots \mu _{n}}(\theta _{1},\ldots ,\theta
_{n})\,\left( F_{n}^{\mathcal{O}|\mu _{1}\ldots \mu _{n}}(\theta _{1},\ldots
,\theta _{n})\,\right) ^{*}\,\,\,.  \label{dcorr}
\end{eqnarray}
Notice also that, unlike in the evaluation of the $c$-theorem, which deals with
a monotonically increasing series, due to the fact that it only involves
absolute values of form factors, the series (\ref{deldel}) can in principle
be alternating. It is now worth to pause for a while and appreciate the advantages 
of this formula in comparison with (\ref{ultra}). First of all, since the $r$-dependence has
been integrated out we only have to evaluate the multidimensional integrals
once. Second, the evaluation of (\ref{dcorr}) does not involve any
anticipation of the value of $\Delta ^{\!\!\mathcal{O}}$. Third, due to the
fact that in the standard form factor construction  form
factors related to a local operator $\mathcal{O}$ 
are always normalised with respect to the lowest non-vanishing form factor,
when the latter is the vacuum
expectation value, the factor $\left\langle \mathcal{O}\right\rangle$ in (\ref{dcorr}) 
will cancel and its knowledge is not required at all in the analysis. Most
important, fourth, the difficulty to identify the suitable region in $r$
which is governed by the $(\log r)^{n}$ behaviour of the $n$-th term in the
sum in (\ref{corr}) and the upper bound in (\ref{ultra1}) have completely
disappeared.

There are however little drawbacks for theories with internal symmetries and
for the case when the lowest non-vanishing form factor of the operator we
are interested in is not the vacuum expectation value. The first problem
arises due to the fact that the sum rule is only applicable for primary
fields $\mathcal{O}$ whose two-point correlation function with the energy
momentum tensor is non-vanishing. We will see later that in fact this restricts quite severely
the amount of operators for which we can actually use formula (\ref{deldel}).

\vspace{0.3cm}
Notice that both (\ref{Cth2}) and (\ref{deldel}) are equations which relate quantities
characterising the ultraviolet CFT or critical,   to a correlator which 
involves operators of the massive QFT namely, ``off-critical'' objects.
 Recall also that
the identification  of conformal dimensions in the TBA-context
was only possible in principle for  the  perturbing operator (see section \ref{examples}). 
Therefore, via the $\Delta$-sum rule \cite{DSC} or the direct study of the 
UV-behaviour of the two point functions (\ref{ultra}),  the form factor analysis is expected  to provide 
more information about the underlying ultraviolet  CFT than the TBA-analysis.
In other words, by carrying out the form factor program,
we expect to confirm and surpass the TBA-outcome.

\subsection{Renormalisation group analysis: 
the  $\protect{c(t_0)}$- and $\protect{\Delta(t_0)}$-flows}
\label{cdflows}
\indent \ \
Denoting by $r$ the radial distance and by $t=2 \ln (mr)$ the renormalisation
group parameter, the functions $c(mr)$ and $\Delta (mr)$ were defined in \cite
{ZamC} and \cite{DSC} respectively, obeying the differential equations 
\begin{eqnarray}
\frac{dc(m\,r)}{dr} &=&-\frac{3}{2}\,r^3 \left\langle \Theta (r)\Theta
(0)\right\rangle  \label{dc} \\
\frac{d\Delta (m\, r)}{dr} &=&\frac{1}{2 \left\langle {\cal O}(0)\right\rangle }%
r\,\left\langle \Theta (r){\cal O}(0)\right\rangle \,\,.  \label{dd}
\end{eqnarray}
The r.h.s. of these equations involves the two-point correlation functions of
the trace of the energy-momentum tensor $\Theta $ and an operator ${\mathcal{ O}}$, 
which is a primary field in the sense of \cite{BPZ}. In general these
equations are integrated from $r=0 $ to $r=\infty $ and one
consequently compares the difference between the ultraviolet and the
infrared fixed points.
 Proceeding this way, we get Eqs. (\ref{Cth2}), (\ref{deldel}) presented before, which
we may use in order to compute the Virasoro central charge  and ultraviolet conformal dimensions.
 In order to exhibit the quantitative onset of the 
mass scale of the unstable particles we integrate these equations now
from some finite value $t_{0}$ to infinity.

Restricting our attention to
purely massive theories, we use the fact that for these theories the infrared
central charges are zero, such that we are left with the following identity
\begin{equation}
c(r_{0})=\frac{3}{2}\int\limits_{r_{0}}^{\infty }dr\,r^{3}\,\,\left\langle
\hat{\Theta} (r)\hat{\Theta} (0)\right\rangle \,\,.  \label{cr}
\end{equation}
\noindent  Since ultimately the functions $c(t)$ and $\Delta(t)$ depend upon the dimensionless
combination $r_0:=m\, r$ and the parameter $r$ in Eq. (\ref{cr}) is now just an integration variable, 
it is suitable for our purposes to perform the variable transformation $r \rightarrow r/m$ which shall
allow for eliminating any explicit dependence on the mass scale $m$. This is achieved
if, simultaneously, the trace of the energy-momentum
tensor, $\Theta$, is normalized as $\hat{\Theta}(r) = \Theta (r/m) /m^2$.
After the previous redefinitions  
both $r$ and $\hat{\Theta}(r)$  are 
dimensionless objects in Eq. (\ref{cr}). Consequently,
also the lower limit of the integral in (\ref{cr}), which we will denote by  $r_0$,
is a dimensionless renormalisation group parameter. 
Apart from the elimination of the explicit dependence in the mass scale,
the latter transformations are very useful in order to establish a direct
comparison between our results in this context and the ones obtained in
the TBA-framework. Recall that the finite size scaling function of the TBA
was defined in  Eq. (\ref{fsef}) and denoted by $c(R)$. In the TBA-framework 
$R$ is also a dimensionless variable, which is given by $R= m_1 T^{-1}$,
$T$ being the temperature of the system, namely, the energy scale, and 
$m_1$ being the mass of the lightest particle in the spectrum. In order
to compare the results obtained in both approaches (the TBA- and form 
factor approach), in particular, we want to compare
the finite size scaling function (\ref{fsef}) of the TBA with
Zamolodchikov's $c$-function given by (\ref{cr}), one can
draw now a formal analogy between the parameters $R$ and $r_0$.
It is important to notice that the normalisation of the energy momentum
tensor giving rise to the dimensionless operator $\hat{\Theta}$ does,
of course, involve an equivalent modification of the  corresponding form
factors presented in appendix B. Such modification is simply expressed as

\begin{equation}
F_{n}^{\hat{\Theta}}= \frac{F_{n}^{{\Theta}}}{m^2} \,\,\,\,\,\,\,\,\,\,\text{for}
\,\,\,\,\,\,\,\,\,\,
\hat{\Theta}=\frac{\Theta}{m^2}.
\end{equation}
\noindent  As is trivially
inferred from the expressions given in appendix B, as a consequence
of such redefinition the mass-dependence of the form factors of the
energy momentum tensor vanishes.

Although the latter equation can be trivially  
obtained from (\ref{dc}), it must be said
that such a function had not been analysed before
in the literature and constitutes the natural counterpart, in the renormalisation
group context, of the finite size scaling function (\ref{fsef}) 
computed in the TBA-approach.  
In fact our numerical results will show for several concrete models 
that these two functions carry the same physical information and 
have the same general features, 
sharing the characteristic ``staircase'' pattern already encountered for
the scaling function of the TBA. 

As presented  in subsection \ref{cff}, instead of the integral representation (\ref{cr}), the
$c$-function is equivalently expressible in terms of a sum of correlators involving also
other components of the energy momentum tensor \cite{ZamC, cardy} which can be eliminated
by means of the conservation law (\ref{law}). Recall that the HSG-models, for which
we ultimately want to analyse the behaviour of the  function (\ref{cr}),
contain unstable particles
in their spectrum, characterised by the resonance parameters $\sigma_{ij}$. 
Taking this observation
into account, we expect the flow of $c(r_{0})$ to surpass various steps: 
Starting with $r_{0}=0$ and assuming the masses
of the stable particles of the theory to be all of the same order, say 
$m_a, m_b, \cdots, m_n \sim m$, the theory
will leave its ultraviolet fixed point and at a certain definite value, say $
r_{0}=r_{u}$, one of the unstables particles will become massive with
respect to the energy scale determined by $m r_0$ 
such that $c(r_{0}>r_{u})$ can be associated to a different CFT. It
appears natural to identify this value 
as the point at which $%
c(r_{0})$ is half the difference between the two coset values of $c$ (see the specific
analysis carried out in section \ref{rgflow}).

\vspace{0.2cm}
In order to obtain an explicit expression for the masses of the unstable particles, 
it is convenient to recall at this point 
that the resonance parameters $\sigma_{ij}=-\sigma_{ij}$ enter the Breit-Wigner formula 
\cite{BW} in the following way:
In general an unstable particle of type $\tilde{c}$ is described by
complexifying the physical mass of a stable particle by adding a decay width 
$\Gamma _{\tilde{c}}$, such that it corresponds to a pole in the S-matrix as
a function of the Mandelstam variable $s$ at $s=M_{R}^{2}=(M_{\tilde{c}}{}-i\Gamma _{\tilde{c%
}}/2)^{2}$ in the non physical sheet (for a more detailed discussion see e.g. \cite{ELOP}). As
mentioned in \cite{ELOP} whenever $M_{\tilde{c}}\gg \Gamma _{\tilde{c}}$,
the quantity $M_{\tilde{c}}$ admits a clear-cut interpretation as the
physical mass. However, since this assumption is only required for
interpretational reasons we will not rely on it. Transforming as usual in
this context from $s$ to the rapidity plane and describing the scattering of
two stable particles of type $a\,$and $b$ with masses $m_{a}$ and $m_{b}$ by
an S-matrix $S_{ab}(\theta )$ as a function of the rapidity $\theta $, the
resonance pole is situated at $\theta _{R}=\sigma -i\bar{\sigma}$, $\sigma$ being the resonance parameter.
Identifying the real and imaginary parts of the pole then yields 
\begin{eqnarray}
M_{\tilde{c}}^{2}{}-\frac{\Gamma _{\tilde{c}}^{2}{}}{4}
&=&m_{a}^{2}{}+m_{b}^{2}{}+2m_{a}m_{b}\cosh \sigma \cos \bar{\sigma}
\label{BW1} \\
M_{\tilde{c}}\Gamma _{\tilde{c}} &=&2m_{a}m_{b}\sinh |\sigma |\sin \bar{%
\sigma}\,\,.  \label{BW2}
\end{eqnarray}
Eliminating the decay width from (\ref{BW1}) and (\ref{BW2}), we can express
the mass of the unstable particles $M_{\tilde{c}}$ in the model as a
function of the masses of the stable particles $m_{a},m_{b}$ and the
resonance parameter $\sigma $. In the regime
\begin{equation}
 e^{|\sigma|} >> \frac{m_a^2 + m_b^2}{m_a m_b},
\end{equation}
\noindent we obtain
\begin{equation}
M_{\tilde{c}}^{2}\sim \frac{1}{2}m_{a}m_{b}(1+\cos \bar{\sigma}%
)\,\,e^{|\sigma |}\,,  \label{Munst}
\end{equation}
\noindent which corresponds to a decay width
\begin{equation}
\Gamma \sim \frac{2 \sin \bar{\sigma}}{1+\cos{\bar{\sigma}}}.
\end{equation}
\noindent 
Notice  the occurrence in Eq. (\ref{Munst}) of the variable
 $me^{|\sigma |/2}$ for $m_a \sim m_b \sim m$ familiar from our TBA-analysis,
which was introduced originally in \cite{triZam} in order to describe massless
particles, i.e. one may perform safely the limit $m\rightarrow 0,\sigma
\rightarrow \infty $. Therefore one might be tempted to describe flows
related to (\ref{Munst}) as massless flows, interpretation which will be
also supported by our numerical results in section \ref{rgflow}.

\vspace{0.2cm}

It is also interesting to notice that the need for fulfilling the energetical
requirement $M_c > m_a + m_b$ leads to the following threshold for the allowed values
of the resonance parameter $\sigma$:
\begin{equation}
e^{|\sigma|} > 2\,\frac{(m_a + m_b)^2}{m_a m_b\, (1+ \cos \bar{\sigma})},
\label{boundd}
\end{equation}
\noindent in particular, for $m_a=m_b$, and $\bar{\sigma}=\pi/2$ which is
the case for all the $SU(N)_2$-HSG models, we obtain the constraint $\sigma > \ln 8$.
Such requirement is in agreement with the fact that very different results are obtained,
 both in the TBA- and form factor context, for small and large values of $\sigma$. In
particular, the development of plateaux in the scaling functions is a feature which only occurs for
values of the resonance parameter large enough (see section \ref{rgflow}). 
However, from the scattering matrix point of
view, the HSG S-matrices make perfect sense for any value of $\sigma$ and it
is an open problem to investigate whether or not evidence for the threshold
(\ref{boundd}) can be found in the context of the scattering theory. 
Recently, this problem has been
investigated in the context of the construction of S-matrices
containing infinitely many resonance poles \cite{new}.

\vspace{0.2cm}
As a consequence of (\ref{Munst}) we may relate the value of $r_u$ 
for different choices of the resonance parameter. 
This provides also a confirmation of the fact
that the renormalisation group flow
is indeed achieved by $m\rightarrow r_{0}\, m$. Increasing $r_{0}$ further,
the energy scale of the stable particles will eventually be reached at, say
at $r_{0}=r_{a},r_{b},\ldots ,r_{n}$. Depending on the relative mass scales
between the stable particles these points may coincide. Finally the flow
will reach its infrared fixed point $c(r_{0}=r_{ir})=0$.

\vspace{0.3cm}

Likewise we can integrate Eq. (\ref{dd}) 
\begin{equation}
\Delta (r_{0})=-\frac{1}{2\left\langle {\cal O}(0)\right\rangle }%
\int\limits_{r_{0}}^{\infty }dr\,r\,\,\left\langle \hat{\Theta} (r){\cal O}%
(0)\right\rangle \,\,,  \label{dr}
\end{equation}
which allows to keep track of the manner the operator contents of the
various conformal field theories are mapped into each other. We used that
all conformal dimensions vanish in the infrared limit. 
Here the same comments after Eq. (\ref{cr}) apply, so that $r_0$ is
a dimensionless RG-parameter. Fortunately, we have that $
\left\langle \hat{\Theta} (r){\cal O}(0)\right\rangle$ is proportional to 
$\left\langle {\cal O}(0)\right\rangle $ in many applications such that the vacuum expectation
value $\left\langle {\cal O}(0)\right\rangle $ cancels often when evaluating (\ref{dr}). 
One should note, however, that (\ref{dr}) is only applicable to those operators for
which its two-point correlator with the trace of the energy momentum tensor
is non-vanishing, such that one may not be in the position to investigate
the flow of the entire operator content by means of (\ref{dr}), as happened 
for the $SU(3)_2$-HSG model.

In order to evaluate (\ref{cr}) and (\ref{dr}) for a concrete model we have to compute the
two-point correlation functions in some way. Obviously, in our case we will make use of the results
of  the form factor analysis to be presented in subsequent sections for the $SU(3)_2$-HSG model
and in the next chapter for all the $SU(N)_2$-HSG models.

As we know, the two-point correlation functions occurring in (\ref{cr}) and (\ref{dr}), can
be expanded in terms of form factors of the corresponding operators (\ref{corr}).
Using this expansion we replace the correlation functions in (\ref{cr}) and (\ref{dr})
and perform the $r$-integrations thereafter. Thus we obtain 

\begin{eqnarray}
&&c(r_{0})=3\sum_{n=1}^{\infty}\sum_{\mu _{1}\ldots \mu
_{n}}\int\limits_{-\infty }^{\infty }\frac{d\theta _{1}
\ldots d\theta _{n}}{n!(2\pi )^{n}} 
\frac{(6+6r_{0}E+3r_{0}^{2}E^{2}+r_{0}^{3}E^{3})}{2 E^{4}} e^{-r_{0}\,E}  \nonumber \\ 
&& \,\,\,\,\,\, \,\,\,\,\,\,\,\,\,\,\,
\, \,\,\,\,\,\,\,\,\,\,\,\, \,\,\,\,\,\,\,\,\,\,\,\, \,\,\,\,\,\,
 \times \left| F_{n}^{\hat{\Theta} |\mu _{1}\ldots \mu _{n}}(\theta _{1},\ldots,\theta _{n})\right|^{2}  
\label{cr0}
\end{eqnarray}
and 
\begin{eqnarray}
&&\Delta (r_{0})=-\frac{1}{2\left\langle {\cal O}(0)\right\rangle }
\sum_{n=1}^{\infty }\sum_{\mu _{1}\ldots \mu
_{n}}\int\limits_{-\infty }^{\infty }\frac{d\theta _{1}\ldots d\theta _{n}}{%
n!(2\pi )^{n}}\frac{(1+r_{0}E)e^{-r_{0}\,E}}{2E^{2}}  \nonumber \\
&&\,\,\,\times F_{n}^{\hat{\Theta} |\mu _{1}\ldots \mu _{n}}(\theta _{1},\ldots
,\theta _{n})\left( F_{n}^{{\cal O}|\mu _{1}\ldots \mu _{n}}(\theta
_{1},\ldots ,\theta _{n})\,\right) ^{\ast }\,.  \label{dr0}
\end{eqnarray}
\noindent  with $E=\sum_{i=1}^{n}
 \hat{m}_{\mu_i} \cosh{\theta_i}$, where the masses 
$\hat{m}_{\mu_i}=m_{\mu_i}/m$ have been also normalised in terms of the overall
mass scale $m$ in order to make the quantity $E$ dimensionless, achieving
consistency with the dimensionless character of $r_0$.

\subsection{Renormalisation group flow of ${\protect\beta}$-like functions}
\label{betalike}
\indent \ \ 
As reported in section \ref{cff} (see Eq. (\ref{CS}), 
in case there is only one coupling constant $g$ in the model, 
the $\beta$-function should obey the defining equation 
\begin{equation}
\beta(g)=\frac{d g}{d{t_0}},
\label{bb}
\end{equation}
\noindent for  $t_0=2 \ln r_0$ to be the 
RG-flow parameter, so that we can rewrite the latter equation as
\begin{equation}
\frac{r_0}{2}\frac{d }{d{r_0}}=\beta(g) \frac{d}{d g}.
\label{bbb}
\end{equation}
\noindent  Here we applied Eq. (\ref{bb}) to the coupling constant $g$ in order to extract the 
$\beta$-function on the r.h.s. of  (\ref{bbb}).

It is now  our interest  to construct a function which allows a more clear identification
of the critical fixed points surpassed by the $c$-function along its RG-flow. It is clear
that this purpose might be achieved as soon as we construct any function which is proportional to 
${d c(t_0)}/{d {t_0}}$. Therefore, following an idea originally used by
Zamolodchikov in \cite{staircase}, let us define a  ``coupling constant'' 
 $g:=c_{\text{uv}}-c(t_0)$  normalized in such a way that it vanishes at the ultraviolet fixed point.
 If we now apply Eq. (\ref{bbb}) to this coupling constant we obtain the equation
\begin{equation}
\frac{r_0}{2}\frac{d c(r_0)}{d{r_0}}=\beta(g).
\label{fix}
\end{equation}
\noindent Clearly from the above definition,  whenever we find 
$\beta (\tilde{g})=0$, we can identify $\tilde{c}=c_{\text{uv}}-\tilde{g}$
as the Virasoro central charge of the corresponding CFT. In fact this is true
irrespectively of the presence of the factor $1/2$ on the l.h.s., thus we will
drop it out in what follows\footnote{There is also an additional reason for dropping out the
1/2 -factor. Although we do not present this analysis in this thesis one might be tempted
to compare the $\beta$-type functions one can construct from the finite size scaling function
of the TBA and from Zamolodchikov's $c$-function by means of (\ref{fix}) .
By doing so, it is possible to check that clearly the physical information carried by these two
functions is the same, namely the different CFT's associated to their plateaux have the same
Virasoro central charges. In the TBA-context one can define a similar  ``$\beta$-like '' 
function through Eq. (\ref{fix}) by simply substituting $t_0 \rightarrow  2 t= 2 \ln(r/2)$
and $c(r_0) \rightarrow c(r, \sigma)$ where $r$ is the inverse temperature variable used in the TBA and $c(r, \sigma)$ the scaling function (see chapter \ref{tba}). Such a substitution is suggested by direct comparison of  Fig. \ref{fig12} and \ref{f41}. Therefore, if we drop out the factor  1/2  in (\ref{fix}) we are redefining $t_0:=2 t_0$ which seems to be more natural in order to compare with the TBA-results.}.

Hence, taking the data obtained from (\ref{cr0}), we compute $\beta$ as a
function of $g$ by means of (\ref{fix}). Analogously, we can define also a $\beta$-like function
associated to the $\Delta(t_0)$-function (\ref{dr0}) presented above. The numerical computation
of these $\beta$-functions will be presented in the next chapter for the $SU(N=4)_2$- and 
$SU(N=5)_2$-HSG models.

\section{The $\protect{SU(3)_{2}}$-HSG model }
\label{modelff}
\indent \ \
As stated already in subsection \ref{modeltba}, when performing the TBA-analysis,
 the $SU(3)_{2}$-HSG model
contains only two self-conjugate solitons
(1,1) and (1,2). Aiming towards a more compact notation  these
two solitons are more conveniently denoted by ``+'' and ``-''. 
Therefore, the corresponding non-trivial S-matrix elements \cite{HSGS}
as functions of the rapidity $\theta $ read now 

\begin{equation}
S_{\pm \pm }=-1\quad \quad \text{and}\quad \quad S_{\pm \mp }(\theta )=\pm
\tanh \frac{1}{2}\left( \theta \pm \sigma -i\frac{\pi }{2}\right) \,\,.\;
\label{ZamSS}
\end{equation}

\noindent

Here  $\sigma$ is the resonance parameter, whose physical meaning
was already discussed both in  chapters \ref{ntft} and \ref{tba}. Eq. (\ref{ZamSS})
means the scattering of particles of the same type is simply described
by the S-matrix of the thermal perturbation of the Ising model. Also the
remaining amplitudes do not possess poles inside the physical sheet, such
that the formation of stable particles via fusing is not possible. The
latter characteristic means that the task of finding  solutions to the
form factor consistency equations (\ref{W1})-(\ref{bstate})
turns out to be a bit simpler, since the so-called bound-state residue equation 
 (\ref{bstate})  does not arise. 

For vanishing resonance parameter $\sigma=0$ the amplitudes $S_{\pm \mp }$
coincide formally with the ones which describe the massless flow between
the tricritical Ising and the critical Ising model as analysed in \cite{DMS}. 
However, there is an important conceptual difference since we view the
expressions (\ref{ZamSS}) as describing the scattering of massive particles.
This has important consequences on the construction of the form factors and
in fact the solutions we will present later are  different from the ones
proposed in \cite{DMS}. Furthermore we will construct all $n$-particle
form factors associated to a large class of operators whereas in \cite{DMS} only
solutions for some particular numbers of particles and
a restricted amount of operators where found .
Recall that in the HSG setting the massless flow
was recovered in the context of the TBA (see section \ref{modeltba}),
only as a subsystem in terms of specially introduced variables 
${r^\prime}=\frac{r}{2}e^{\sigma/2}$ combining the inverse
temperature $r$ and the resonance parameter $\sigma$.  The occurrence  
of the combination
of  variables $r e^{\sigma/2}$ (although for a different $r$-parameter) 
will be also encountered  
later in the context of the renormalisation group analysis.

The system (\ref{ZamSS}) is of special interest since it 
constitutes probably the simplest example of a
massive QFT involving two particles of distinct type.
Nonetheless, despite the simplicity of the scattering matrix we expect to
find a relatively involved operator content, since for finite resonance
parameter the $SU(3)_{2}$-HSG model describes a  WZNW-coset model with
central charge $c=6/5$ perturbed by an operator with conformal dimension $%
\Delta =3/5$. 
It is expected  from
the classical analysis and has also been confirmed by 
the TBA-analysis  that, whenever the resonance parameter $\sigma$ 
is finite, we  find the same ultraviolet central charge and
therefore the same operator content. For this reason, when computing 
quantities (usually numerically) in the UV-limit, it is really only
interesting to distinguish the situations  $\sigma \rightarrow \infty$
and   $\sigma$ finite (in particular, we will set  $\sigma=0$).
 One can trivially notice from (\ref{ZamSS}) that
\begin{equation}
\lim_{\sigma \rightarrow \infty} S_{\pm\mp}=1,
\label{decthe}
\end{equation}
 \noindent which  means that  the theory decouples into two 
copies of the Ising model and therefore the corresponding
Virasoro central charge is expected to be $\frac{1}{2}+\frac{1}{2}= 1$  in this limit.
The latter  behaviour will also be
 consistently recovered in the context of  form factors, 
in particular when computing the corresponding
Virasoro central charge by means of (\ref{Cth2}) and when studying the asymptotics
of  form factors  in the limit  $\sigma  \rightarrow \infty$. 

The corresponding underlying CFT,
a WZNW-$SU(3)_2/U(1)^{2}$-coset or $SU(3)_2$-para-
fermion theory \cite{Gep}, 
has recently \cite {Schou, ASchou} found an interesting application in the context 
of the construction of quantum Hall states which carry a spin and fractional charges.

\section{Recursive equations and minimal form factors}
\label{mini}
\indent \ \
Attempting now to solve the form factor consistency equations presented 
in subsection \ref{properties}, and  proceeding  as
usual in this context \cite{Kar, BKW},
 we start by  making  a factorisation ansatz which already
extracts explicitly some of the singularity structure we expect to find. For
the case at hand, property 3 tells us that 
we must have a kinematical pole at $i\pi $ when two
particles are conjugate to each other. Therefore the following parameterisation
\begin{equation}
F_{n}^{\mathcal{O}|\stackrel{l\,\times \,\pm}{\overbrace{\mu _{1}\ldots \mu
_{l}}}\stackrel{m\,\times \, \mp }{\overbrace{\mu _{l+1}\ldots \mu _{n}}}%
}(\theta _{1}\ldots \theta _{n})=H_{n}^{\mathcal{O}|\mu _{1}\ldots \mu
_{n}}Q_{n}^{\mathcal{O}|\mu _{1}\ldots \mu _{n}}(x_{1}\ldots
x_{n})\,\!\!\prod_{i<j}\!\!\frac{F_{\text{min}}^{\mu _{i}\mu _{j}}(\theta
_{ij})}{\left( x_{i}^{\mu _{i}}+x_{j}^{\mu _{j}}\right) ^{\delta _{\mu
_{i}\mu _{j}}}},  \label{fact}
\end{equation}
where we  introduced the variable $x_{i}=e^{\theta _{i}}$, is very convenient, since the
mentioned poles have already been isolated  in the denominator of (\ref{fact}).
The $H_{n}^{\mathcal{O}%
|\mu _{1}\ldots \mu _{n}}$ are normalisation constants and the functions 
$F_{\text{min}}^{\mu_i\mu_j}(\theta_{ij})$ are the 
the so-called  {\bf minimal form factors}.  They satisfy the following equations

\begin{equation}
F_{\text{min}}^{ij}(\theta )=F_{\text{min}}^{ji}(-\theta )S_{ij}(\theta )=F_{%
\text{min}}^{ji}(2\pi i-\theta )\,\, , \label{minii}
\end{equation}
and have neither zeros nor poles in the physical sheet $0<\text{Im}(\theta) <\pi$. 
Then, if we further
assume that the  $Q_{n}^{\mathcal{O}|\mu _{1}\ldots \mu _{n}}(\theta _{1},\ldots
,\theta _{n})$ are functions separately symmetric in the first $l$ and the last $m$
rapidities and in addition 2$\pi i$-periodic  in all rapidities, the
ansatz (\ref{fact}) solves Watson's equations (\ref{W1}) and (\ref{W2}) by
construction.  At the same time, being symmetric in the first $l$ and the last $m$
rapidities the $Q_{n}^{\mathcal{O}|\mu _{1}\ldots \mu _{n}}(\theta _{1},\ldots,\theta _{n})$
have to be combinations of elementary symmetric polynomials
in the first $l$ and last $m$ rapidities (see appendix A).
In particular, we have 
\begin{equation}
Q_{n}^{\mathcal{O}|\stackrel{l\,\times \,+}{\overbrace{\mu _{1}\ldots \mu
_{l}}}\stackrel{m\,\times \,-}{\overbrace{\mu _{l+1}\ldots \mu _{n}}}%
}(x_{1},\ldots ,x_{n})=Q_{n}^{\mathcal{O}|\stackrel{m\,\times \,-}{%
\overbrace{\mu _{l+1}\ldots \mu _{n}}}\stackrel{l\,\times \,+}{\overbrace{%
\mu _{1}\ldots \mu _{l}}}}(x_{l+1},\ldots x_{n},x_{1},\ldots x_{l})\,\,,
\label{sym}
\end{equation}
such that, provided a solution has been  constructed  for a particular ordering of
the $\mu $'s, for example the upper sign in (\ref{fact}), we can obtain the
solution for a permuted ordering by  using the monodromy properties. Especially, the
reversed order is obtained  by applying Eq. (\ref{sym}). Despite the fact
that we do not gain anything new, it is still instructive to verify (\ref
{kin}) as a consistency check also for the different ordering. The monodromy
properties allow some simplification in the notation and from now on we
restrict our attention without loss of generality  to the upper sign in (\ref{fact}). In addition,
we deduce from Eq. (\ref{rel}) that for a spinless operator $\mathcal{O}
$ the total degree of $Q_{n}^{\mathcal{O}}$ has to be
\begin{equation}
 \left[ Q_{n}^{\mathcal{O}} \right] = \frac{l(l-1)}{2}- \frac{m(m-1)}{2},
\label{Qasym}
\end{equation}
\noindent where the brackets $[ \,\,\,] $ have the meaning explained in (\ref{not}).

\vspace{0.2cm}
A solution for the minimal form factors i.e., of equations (\ref{minii}), is
found  to be
\begin{eqnarray}
F_{\text{min}}^{\pm \pm }(\theta ) &=&-i\sinh \frac{\theta }{2} \label{minising}\\ 
F_{\text{min}}^{\pm \mp }(\theta ) &=&\mathcal{N}^{\pm }(\theta
)\prod_{k=1}^{\infty }\tfrac{\Gamma (k+\frac{1}{4})^{2}\Gamma \left( k+\frac{%
1}{4}+\frac{i}{2\pi }(\theta \pm \sigma )\right) \Gamma \left( k-\frac{3}{4}-%
\frac{i}{2\pi }(\theta \pm \sigma )\right) }{\Gamma (k-\frac{1}{4}%
)^{2}\Gamma \left( k-\frac{1}{4}-\frac{i}{2\pi }(\theta \pm \sigma )\right)
\Gamma \left( k+\frac{3}{4}+\frac{i}{2\pi }(\theta \pm \sigma )\right) }\,\,
\label{gamma}\\
&=&  \mathcal{N}^{\pm }(\theta) \exp \left( -\int\limits_{0}^{\infty }%
\tfrac{dt}{t}\tfrac{\sin ^{2}\left( (i\pi -\theta \mp \sigma )\frac{t}{2\pi }%
\right) }{\sinh t\cosh t/2}\right) \,=e^{\pm \tfrac{\theta }{4}}\tilde{F}_{%
\text{min}}^{\pm \mp }(\theta )\,\,.  \label{14}
\end{eqnarray}

Here $F_{\text{min}}^{\pm \pm }(\theta )$ is the well-known minimal form
factor of the thermally perturbed Ising model \cite{BKW,YZam} and for the
upper choice of the signs, Eq.  (\ref{14}) coincides for vanishing $%
\sigma $ up to normalisation with the expression found in \cite{DMS}. We
introduced the normalisation function
\begin{equation}
 \mathcal{N}^{\pm }(\theta )=2^{\frac{1%
}{4}}\exp \left( \tfrac{i\pi (1\mp 1)\pm \theta }{4}-\tfrac{G}{\pi }\right) 
\end{equation}
\noindent with $G=0.91597$ being the Catalan constant.

The expression (\ref{14})  can be derived by using a result dating back to
the original literature  \cite{Kar, BKW} which states that, 
once an integral representation for the S-matrix is known in the form

\begin{equation}
S_{\mu_i \mu_j}(\theta)=\exp{\Bigg(\int_{0}^{\infty} \frac{dt}{t} f_{\mu_i \mu_j}(t)
 \sinh{\Big(\frac{ t \theta}{i \pi}\Big)}\Bigg)},
\label{intS}
\end{equation}
\noindent  where $ f_{\mu_i \mu_j}(t)$ is a certain function which depends on the particular
theory under consideration, the corresponding minimal form factors are given, up to some normalisation 
constant  $\mathcal{C}_{ij}$, by

\begin{equation}
F_{\text{mim}}^{\mu_i \mu_j}(\theta)= \mathcal{C}_{ij} \, \exp{\Bigg(\int_{0}^{\infty} \frac{dt}{t} f_{\mu_i \mu_j}(t) \frac{\sin{\Big(\frac{t (i \pi-\theta)}{2 \pi}\Big)}^2 }{\sinh(t)} \Bigg)}.
\label{intM}
\end{equation}
\noindent  For all the HSG models, such an integral  representation  (\ref{intS})
 was provided in chapter \ref{tba}, and in order to get  (\ref{14}), we only need 
to particularise it to the $SU(3)_2$-HSG model at hand.  
\vspace{0.2cm}

The minimal form factors (\ref{minising}), (\ref{14})
possess various properties which we would like to employ in the course of
our argumentation. They obey the functional identities 
\begin{eqnarray}
F_{\text{min}}^{\pm \pm }(\theta +i\pi )F_{\text{min}}^{\pm \pm }(\theta )
&=&-\frac{i}{2}\sinh \theta,  \label{he1} \\
F_{\text{min}}^{\pm \mp }(\theta +i\pi )F_{\text{min}}^{\pm \mp }(\theta )
&=&\frac{i^{\frac{2\mp 1}{2}}\,e^{\pm \frac{\theta }{2}}}{\cosh \frac{1}{2}%
\left( \theta \pm \sigma -\frac{i\,\pi }{2}\right) }\,\,,  \label{he2}
\end{eqnarray}
\noindent which are easily derived by using the representation (\ref{gamma}) in terms of 
$\Gamma$-functions of the minimal form factors. We will also exploit the asymptotic behaviours 
\begin{equation}
\lim\limits_{\sigma \rightarrow \infty }F_{\text{min}}^{\pm \mp }(\pm \theta
)\sim e^{-\frac{\sigma }{4}},\qquad \left[ F_{\text{min}}^{\pm \pm }(\theta
_{ij})\right] _{i}=\frac{1}{2},\qquad \left[ F_{\text{min}}^{\pm \mp
}(\theta _{ij})\right] _{i}=\QATOPD\{ . {0}{-1/2}\,\,,\label{behave}
\end{equation}
which, together with the factorisation ansatz (\ref{fact}) and (\ref{Qasym})  lead us
immediately to the relations 
\begin{eqnarray}
\left[ F_{n}^{\mathcal{O}|l,m}\right] _{i} &=&\left[ Q_{n}^{\mathcal{O}%
|l,m}\right] _{i}+\frac{1-l}{2}\qquad \qquad \,\,\,\text{for\quad }1\leq
i\leq l  \label{as1} \\
\left[ F_{n}^{\mathcal{O}|l,m}\right] _{i} &=&\left[ Q_{n}^{\mathcal{O}%
|l,m}\right] _{i}+\frac{m-l-1}{2}\,\,\quad \quad \text{for \quad }l<i\leq
n\,,  \label{as2}
\end{eqnarray}
\noindent where we recalled again (\ref{not}). These relations 
are very useful in the identification process of a particular solution with
a specific operator. Since we may restrict our attention to one particular
ordering only, we abbreviate the r.h.s. of (\ref{fact}) from now on as $%
F_{n}^{\mathcal{O}|l,m}$ and similar for the $Q$'s.

By substituting now  the ansatz (\ref{fact}) into the kinematic residue Eq. (%
\ref{kin})  the whole
problem of determining the form factors associated to a particular local operator $\mathcal{O}$
reduces, with the help of (\ref{he1}) and (\ref{he2}), to  solving  the following recursive equations 

\begin{eqnarray}
&&Q^{\mathcal{O}|l+2,m}(-x,x,\ldots ,x_{n}) =D_{\vartheta
}^{l,m}(x,x_{1},\ldots ,x_{n})Q^{\mathcal{O}|l,m}(x_{1},\ldots ,x_{n})
\label{Qrec} \\
&&D_{\vartheta }^{l,m}(x,x_{1},\ldots ,x_{n}) =\frac{1}{2}(-ix)^{l+1}\sigma
_{l}^{+}\sum_{k=0}^{m}(-i e^{\sigma} x)^{-k}(1-(-1)^{l+k+\vartheta }) {\sigma}%
_{k}^{-}\,\,.
\label{d1}
\end{eqnarray}
\noindent where $\sigma_l^{+}$ and ${\sigma}_k^{-}$ denote elementary
symmetric polynomials of degrees $l$ and $k$ in the variables  $x_i$ with  $i=1,\cdots,l$ and
$i=l+1, \cdots, l+m$ respectively. More information about the properties and definition
of these polynomials may be found in appendix A and in \cite{Don}. In particular the use of the 
generating equation (\ref{fund}) is fundamental in order to obtain (\ref{d1}).

If we now set  $l=2s+\tau$ and $m=2t+\tau^{\prime}$ then Eq. (\ref{d1}) can be rewritten
as 
 \begin{equation}
D_{\zeta }^{2s+\tau ,2t+\tau ^{\prime }}(x,x_{1},\ldots
,x_{n})=(-i)^{2s+\tau +1}\sigma _{2s+\tau }^{+}\sum_{p=0}^{t}x^{2s-2p+\tau
+1-\zeta }\hat{\sigma}_{2p+\zeta }^{-}\,\,.  \label{Dpar}
\end{equation}
In (\ref{d1}) $\vartheta $ is related to the factor of local commutativity $\omega
=(-1)^{\vartheta }=\pm 1$ introduced in (\ref{kin}). We introduced also the function $\zeta $ which is 
$0$ or $1$ for the sum $\vartheta +\tau $ being odd or even, respectively.
We shall use various notations for elementary symmetric polynomials.
We employ the symbol $\sigma _{k}$
when the polynomials depend on the variables $x_{i}$, the symbol $\bar{\sigma%
}_{k}$ when they depend on the inverse variables $x_{i}^{-1}$, the symbol $%
\hat{\sigma}_{k}$ when they depend on the variables $\hat{x}_i=x_{i}e^{-\sigma +i\pi
/2}$ and $\tilde{\sigma}_{k}$ when we set the first two variables to $%
x_{1}=-x,$ $x_{2}=x$. The number of variables the polynomials depend upon is
defined always in an unambiguous way through the l.h.s. of our equations,
where we assume the first $l$ variables to be associated with $\mu =+$ and
the last $m$ variables with $\mu =-$. In case no superscript is attached to
the symbol the polynomials depend on all $m+l$ variables, and as indicated in the previous 
paragraph ,  in case of a ``$+$%
'' they depend on the first $l$ variables and in case of a ``$-$'' on the
last  $m$ variables.

The recursive equations for the constants turn out to be 
\begin{equation}
H_{n+2}^{\mathcal{O}|l+2,m}=i^{m}2^{2l-m+1}e^{\sigma m/2}H_{n}^{\mathcal{O}%
|l,m}\,\,.  \label{Hrec}
\end{equation}
Fixing one of the lowest constants, the solutions to these equations read 
\begin{equation}
H^{\mathcal{O}|2s+\tau ,m}=i^{sm}2^{s(2s-m-1+2\tau )}e^{sm\sigma /2}H^{%
\mathcal{O}|\tau ,m},\qquad \tau =0,1\,.  \label{consol}
\end{equation}
Note that at this point an unknown constant, that is $H^{\mathcal{O}|\tau,m}$, enters into the procedure. This quantity is not constrained by the
form factor consistency equations and has to be obtained from elsewhere. 
Notice that there is a certain ambiguity
contained in the equations (\ref{Hrec}), i.e. we can multiply $H_{n}^{%
\mathcal{O}|l,m}$ by $i^{2l}$, $i^{2l^{2}}$ or $(-1)^{l}$ and produce a new
solution. However, since in practical applications we are usually dealing
with the absolute values of the form factors, these ambiguities will turn
out to be irrelevant.

\section{The solution procedure}
\label{solution}
\indent \ \
Solving recursive equations of the type (\ref{Qrec}) in complete generality
is still an entirely open problem. Ideally, one would like to reach a
situation similar to the one in the bootstrap construction procedure of the
scattering matrices, where one can state general building blocks, e.g.
particular combinations of hyperbolic functions whenever backscattering is
absent \cite{Mitra}, infinite products of gamma functions when
backscattering occurs or elliptic functions when infinite resonances are
present. Unfortunately, such a general analytical structure has not
been encountered  by now in the context of the form factor analysis. In fact, 
for most models,  only a few solutions of  (\ref{Qrec}) corresponding to the  
first smaller values of $n$  and/or  to particular  local operators have been constructed.
Consequently, the results we present now for the $SU(3)_2$-HSG model and generalise
thereafter  for  all the $SU(N)_2$-HSG models are
very remarkable as the goal of constructing systematically all $n$-particle form
factors associated to a large class of operators of the model has been achieved.
Therefore, the form factor study we have performed  for the $SU(3)_2$- and 
$SU(N)_2$-HSG models is not only relevant
as a consistency check through  (\ref{Cth2}) and (\ref{deldel})
of the S-matrix proposal  \cite{HSGS}
and the results of the TBA-analysis,
but also serves as an important contribution to the general understanding of the problem of finding general solutions to recursive problems of the type (\ref{Qrec}).

It will turn out that all solutions to the recursive
equations (\ref{Qrec}) may be constructed from some general building blocks
consisting out of determinants of matrices whose entries are elementary
symmetric polynomials in some particular set of variables. Let us therefore
define the ($t+s$)$\times $($t+s$)-matrix 
\begin{equation}
\mathcal{\,}\left( \mathcal{A}_{l,m}^{\mu ,\nu }(s,t)\right)
_{ij}:=\QATOPD\{ . {\sigma _{2(j-i)+\mu }^{+}\qquad \,\,\,\,\,\,\,\,\text{%
for\quad }1\leq i\leq t}{\hat{\sigma}_{2(j-i)+2t+\nu }^{-}\,\,\,\quad \quad
\,\,\,\,\,\,\,\text{for\quad }t<i\leq s+t}\,\,.  \label{Acomp}
\end{equation}

\noindent The superscripts $\mu ,\nu $ may take the values $0$ and $1$ and
the subscripts $l,m$ characterise the number of different variables related
to the particle species ``$+$'', ``$-$'', respectively. More explicitly the
matrix $\mathcal{A}$ reads 
\begin{equation}
\mathcal{A}_{l,m}^{\mu ,\nu }=\left( 
\begin{array}{rrrrrr}
\sigma _{\mu }^{+} & \sigma _{\mu +2}^{+} & \sigma _{\mu +4}^{+} & \sigma
_{\mu +6}^{+} & \cdots & 0 \\ 
0 & \sigma _{\mu }^{+} & \sigma _{\mu +2}^{+} & \sigma _{\mu +4}^{+} & \cdots
& 0 \\ 
\vdots & \vdots & \vdots & \vdots & \ddots & \vdots \\ 
0 & 0 & 0 & 0 & \cdots & \sigma _{2s+\mu }^{+} \\ 
\hat{\sigma}_{\nu }^{-} & \hat{\sigma}_{\nu +2}^{-} & \hat{\sigma}_{\nu
+4}^{-} & \hat{\sigma}_{\nu +6}^{-} & \cdots & 0 \\ 
0 & \hat{\sigma}_{\nu }^{-} & \hat{\sigma}_{\nu +2}^{-} & \hat{\sigma}_{\nu
+4}^{-} & \cdots & 0 \\ 
\vdots & \vdots & \vdots & \vdots & \ddots & \vdots \\ 
0 & 0 & 0 & 0 & \cdots & \hat{\sigma}_{2t+\nu }^{-}
\end{array}
\right) \,\,.  \label{sss}
\end{equation}

\noindent The different combinations of the integers $\mu ,\nu ,l,m$ will
correspond to different kinds of local operators $\mathcal{O}$. These sort of determinant
formulae, albeit with entirely different entries of symmetric polynomials, 
 have occurred before in various places in the literature \cite{deter1, deter2, Zamocorr}. These type
of expressions allow on one hand for systematic proofs, as we will demonstrate later, and on the other hand,
they are claimed to be useful in the construction of correlation functions as suggested in \cite{deter}.
In addition, the form factors will involve a function depending on two further indices $%
\bar{\mu}$ and $\bar{\nu}$%
\begin{equation}
g_{l,m}^{\bar{\mu},\bar{\nu}}:=(\sigma _{l}^{+})^{\frac{l-m+\bar{\mu}}{2}%
}(\sigma _{m}^{-})^{\frac{\bar{\nu}-m}{2}}\,\,.  \label{g}
\end{equation}
Here the $\bar{\mu},\bar{\nu}$ are integers whose range, unlike the one for $%
\mu ,\nu $, is in principle not restricted. However, it will turn out that
due to the existence of certain constraining relations, to be specified in
detail below, it is sufficient to characterise a particular operator by the
four integers $\mu ,\nu ,l,m$ only. Then, as we shall demonstrate, all $Q$%
-polynomials acquire the general form 
\begin{equation}
Q^{\mathcal{O}|l,m}=Q_{l,m}^{\mu ,\nu }=Q_{2s+\tau ,2t+\tau ^{\prime }}^{\mu
,\nu }=i^{s\nu }(-1)^{s(\tau +t+1)}g_{2s+\tau ,2t+\tau ^{\prime }}^{\bar{\mu}%
,\bar{\nu}}\,\,\det \mathcal{A}_{2s+\tau ,2t+\tau ^{\prime }}^{\mu ,\nu }\,.
\label{qpar}
\end{equation}

\noindent We used here already a parameterisation for $l,m$ which will turn
out to be most convenient. The subscripts in $g$ and $\mathcal{A}$ are only
needed in formal considerations, but in most cases the number of particles
of species ``$+$'' and ``$-$''  are unambiguously defined through the l.h.s.
of our equations. This is in the same spirit in which we refer to the number
of variables in the elementary symmetric polynomials. We will therefore drop
them in these cases, which leads to simpler, but still precise, notations.

\noindent 
It is also interesting to mention at this point that along with 
the representation of the elementary symmetric polynomials in terms of contour integrals given by Eq. 
(\ref{symint}), the  determinant (\ref{d1}) admits an equivalent integral representation of the form, 

\begin{eqnarray}
&&\det \mathcal{A}_{2s+\tau,2t+\tau^\prime}^{\mu ,\nu } =(-1)^{s t}\oint du_{1}\ldots
\oint du_{t}\oint dv_{1}\ldots \oint
dv_{s}\prod\limits_{j=1}^{t} \Bigg(  \frac{{ \prod\limits_{i=1}^{2s+\tau}}({x_i }+{u_j} )}
{{u_j}^{2s+\tau-1-\mu+2j}} \Bigg)\,\nonumber \\
&&
\prod\limits_{j=1}^{s} 
\Bigg(\frac{{ \prod\limits_{i=1}^{2t+\tau^{\prime}}}({\hat{x}_i }+{v_j} )}
{{v_j}^{2t+\tau^{\prime}-1-\nu+2j}} \Bigg) 
\prod\limits_{1 \leq i<j \leq s}(v_j^2 - v_i^2) 
\prod\limits_{1 \leq i<j \leq t}(u_j^2 - u_i^2) 
\prod\limits_{i=1}^{2s+\tau}  \prod\limits_{j=1}^{2t+\tau^{\prime}}
 (u_j^2 - v_i^2),
\label{integral}
\end{eqnarray}
\noindent  where the contour integrals are taken in the $x=e^\theta$-plane and
we abbreviate $\oint=(2 \pi i)^{-1}\oint$.
Such an integral representation provides another type of universal structure for  our
solutions (\ref{qpar}) and may be used later for different purposes. 
For instance, when providing a general  proof of the validity of the
solutions (\ref{qpar})  for any  $n$,  when studying the behaviour of such
solutions under the cluster property (\ref{cluster}) or when taking  the limit
$\sigma \rightarrow \infty$. There exist other types of universal integral expressions involving
also contour integrals (often in the $\theta$-plane)  like for example the integral 
representation used in \cite{BFKZ}. However, the precise link between all these
different representations is still an open question. 
Unfortunately, all these type of integral expressions are often only of a very formal nature
since their utilization in practise, for higher $n$-particle form factors 
requires still a lot of computational effort.
Hence, it will turn out that the determinant representation (\ref{sss}) is more 
useful for our present purposes. 

\vspace{0.25cm}

Before we present a proof of the general solutions (\ref{qpar})
we  will illustrate our present  results with some particular examples of
special relevance. The identification of such a  general
structure as (\ref{qpar}) for the solutions to  (\ref{Qrec}) is a highly non-trivial task
and our initial aim in \cite{CFK} was a bit less ambitious. As a starting point we tried    
to identify  at least  those solutions corresponding
to operators which had a direct counterpart in the thermally perturbed Ising model, namely
the trace of the energy momentum tensor and  the so-called ``order'' and ``disorder'' operators
(see e.g. \cite{cardy, CFT}).
Recall that, since the interaction amongst particles of the same type
is described by the S-matrix of the thermally perturbed Ising model,
 our solutions to the form factor consistency equations must also reduce to the
ones found for this model in \cite{BKW,YZam} whenever only  one particle specie
is considered. Furthermore, at an initial stage of our investigation \cite{CFK}
 the general solutions found were not rigorously proven but stated on the basis of
the generalisation of the form factors computed up to a certain order in $n$.
The next section provides a more detailed description and interpretation of our results.

\section{The solutions for the operators ${\protect\Theta}$, ${\protect\Sigma}$ 
and ${\protect\mu}$.}
\label{ising}
\indent \  \
As stated when summarising  the properties of the $SU(3)_2$-HSG model, 
the S-matrices (\ref{ZamSS})  
describing the interaction between particles of the same type are precisely
the ones of the thermal perturbation of the Ising model $S_{\pm\pm}=-1$. In this light,  
the form factors of the $SU(3)_2$-HSG model must reduce to the
ones of the thermally perturbed Ising model whenever the number of particles of
type ``$+$'' or ``$-$'' is zero.
On the other hand, the primary field content of the thermally perturbed Ising model  is  well known  
and consists of the trace of the energy density operator $\varepsilon$ and the so-called ``order''
 and ``disorder''
operators $\Sigma$ and $\mu$ whose counterparts in the UV-limit have
conformal dimensions $1/2$, $1/16$ and $1/16$ respectively (see e.g. \cite{cardy, CFT}).  
All the $n$-particle 
form factors associated to these three operators have been computed in \cite{BKW,YZam}, although
in the first reference only the form factors associated to the energy momentum tensor
and order operator were obtained. The solutions for these operators together with
 the disorder operator where found in \cite{YZam} after the introduction of the 
factor of local commutativity, $\omega$
(see footnote after Eq. (\ref{kin})).  The corresponding 
correlation functions can be found for instance in \cite{DSC}. Therefore, we can use the Ising solutions
$Q_n^{0,m}$  or $Q_n^{l,0}$  as a seed or  initial condition for the recursive problem (\ref{Qrec}), 
that is,  a way to fix the lowest non-vanishing form factor. Analogously we can use the Ising
model to fix the  unknown constant  $H^{\mathcal{O}|\tau,m}$ in (\ref{consol}).
Notice that for many models this seed for the recursive equations is not known a priori
and  the task of finding an initial condition to the recursive system (\ref{Qrec}), (\ref{consol})
becomes quite hard as, for instance, one might  need  to know vacuum expectation values.

\subsection{The trace of the energy momentum tensor, $\protect\Theta$}
\label{emt}
\indent \ \
The only non-vanishing form factor of the energy momentum tensor in the
thermally perturbed Ising model is well known to be 
\begin{equation}
F_{2}^{\Theta | \pm\pm}(\theta )=-2\pi im_{\pm}^{2}\sinh (\theta /2)\,\,.  
\label{2par}
\end{equation}
From this equation we deduce immediately that $[F_{n}^{\Theta |l,2}]_{i}=1/2$%
, which serves on the other hand to fix $[Q_{n}^{\Theta |l,2}]_{i}$ with the
help of (\ref{as1}) and (\ref{as2}). Recalling that the energy momentum
tensor is proportional to the perturbing field \cite{cardy} and the fact
that the conformal dimension of the latter is $\Delta =3/5$ for the $%
SU(2)_{3}$-HSG model, the value $[F_{n}^{\Theta |l,2}]_{i}=1/2$ is
compatible with the bound (\ref{bound}). As a further consequence of (\ref{2par}), we deduce 
\begin{equation}
H^{\Theta |0,2}=2\pi m_{-}^{2},\,\,\,\,\,\,\textnormal{and}  \,\,\,\,\,\, H^{\Theta |2,0}=2\pi m_{+}^{2},
\end{equation}
as the initial value for the computation of all higher constants in (\ref
{consol}). The distinction between $m_{-}$ and $m_{+}$ indicates that in
principle the mass scales could be very different as discussed in chapter \ref{tba}. However , in the following we will mostly assume  the masses of the particles  $m_{+}=m_{-}=m$. Notice that $H^{\Theta |0,0}$ is reached only formally, since the
kinematic residue equation does not connect to the vacuum expectation value. From (\ref{2par}) we
can easily identify the  initial values for the recursive equations (\ref{Qrec}). They are,
\begin{equation}
Q_{2}^{\Theta |0,2}=x_{1}^{-1}+x_{2}^{-1}\quad \quad \text{and\quad \quad }%
Q_{2t}^{\Theta |0,2t}=0\quad \text{for }t\geq 2\,\,.
\end{equation}
Taking now in (\ref{d1})  $\vartheta =0$, the solutions to (\ref{Qrec}), with the same
asymptotic behaviour as the energy momentum tensor in the thermally
perturbed Ising model, are computed to 
\begin{equation}
Q^{\Theta |2s+2,2t+2}=i^{s(2t+3)}e^{-(t+1)\sigma }\sigma _{1}\bar{\sigma}%
_{1}\,g^{0,2}\,\det \mathcal{A}^{1,1}\mathcal{\,\,},  \label{solu1} 
\end{equation}
\noindent with the notation introduced in the previous section. Here $\mathcal{A}^{1,1}$
is a ($t+s$)$\times $($t+s$)-matrix.
Notice that in comparison with (\ref{qpar}) the factor of proportionality in
(\ref{solu1}) includes  the term $\sigma _{1}\bar{\sigma}_{1}$.
 Terms of this type may
always be added since they satisfy the consistency equations 1-5 trivially, although they
modify the asymptotic behaviour of the solutions.
For $\Theta $ we are forced  to introduce the
factor $\sigma _{1}\bar{\sigma}_{1}$ in order to recover the solution of the
thermally perturbed Ising model for $2s+2=0$. Note that for $\Theta $ the
values $s=-1$, $t=-1$  formally make sense as they correspond  to 
number of particles of 
type ``$+$'' or ``$-$'' equal to zero respectively i.e., the Ising solutions.

\subsubsection{The limit $\protect{\sigma \rightarrow \infty}$}
\indent  \ \
When the resonance parameter tends to infinity 
the S-matrix  (\ref{ZamSS}) satisfies  
$\lim_{\sigma \rightarrow \infty} S_{\pm\mp}=1$,
therefore the system 
decouples into two non-interacting free fermions. Accordingly, 
if  we now collect the leading order behaviours from our
general solution 
\noindent where we made use of the relations  (\ref{behave}), we finally get
\begin{equation}
\lim_{\sigma \rightarrow \infty }F_{2s+2t}^{\Theta |2s,2t}\sim
e^{-(t+s-1)\sigma }\,\,.
\end{equation}
Hence, the only non-vanishing form factors in this limit are $F_{2}^{\Theta
|0,2}$ and $F_{2}^{\Theta |2,0}$, and we are left with the corresponding form
factors of the energy momentum tensor for  two copies of the thermally perturbed Ising model.

\subsection{The order operator, ${\protect\Sigma}$}
\label{sigma}
\indent \ \
For the other sectors we may proceed similarly, i.e. viewing always the
thermally perturbed Ising model as a benchmark. Taking now $\vartheta =0$, we
recall the solution for the order operator 
\begin{equation}
F_{2s+1}^{\Sigma }(\theta _{1},\ldots ,\theta _{2s+1})=i^{s}F_{1}^{\Sigma
}\prod_{i<j}\tanh \tfrac{\theta _{ij}}{2}=i^{s}(2i)^{2s^{2}+s}F_{1}^{\Sigma
}\left( \sigma _{2s+1}\right) ^{s}\prod_{i<j}\tfrac{F_{\text{min}}^{\pm \pm
}(\theta _{ij})}{x_{i}+x_{j}}.  \label{solord}
\end{equation}
With this information we may fix the initial values of the recursive
equations (\ref{Qrec}) and (\ref{Hrec}) at once to 
\begin{equation}
Q_{2t+1}^{\Sigma |0,2t+1}=\left( \sigma_{2t+1}\right)^{-t}=\left( 
\bar{\sigma}_{2t+1}\right)^{t}\qquad \text{and\qquad }
H^{\Sigma|0,1}=F_{1}^{\Sigma }\,\,.
\end{equation}
Furthermore, we deduce from Eq. (\ref{solord}) that $[F_{n}^{\Sigma |2s,2t+1}]_{i}=0$. Respecting these constraints we find as explicit
solutions 
\begin{equation}
Q^{\Sigma |2s,2t+1} =i^{s(2t+3)}\,(\sigma_1)^{1/2}({\sigma_1^{-}})^{-1/2}\,g^{-1,1}\,\det \mathcal{A}^{0,1}\mathcal{%
\,}.  
\end{equation}
\noindent Analogously to the case of the energy momentum tensor, we have also a proportionality 
 factor $(\sigma_1)^{1/2}({\sigma_1^{-}})^{-1/2} $ in comparison to (\ref{qpar}). As pointed out
before, factors of this type satisfy trivially all the consistency equations for the form factors and for
this reason, although they were originally introduced in \cite{CFK} we can now safely drop then out.
Additional  reasons for this modification will be provided below. Therefore, we can take 
\begin{equation}
Q^{\Sigma |2s,2t+1} =i^{s(2t+3)}\,g^{-1,1}\,\det \mathcal{A}^{0,1}, \label{solu2} 
\end{equation}
\noindent as the final solution for the form factors of the order operator $\Sigma$. As usual,
$ \mathcal{A}^{0,1}$ is a $(t+s)$$\times$$(t+s)$-matrix whose explicit form is obtained
from (\ref{sss}) by setting $\mu=0, \nu=1$.

\subsubsection{The limit $\protect{\sigma \rightarrow   \infty}$}
\indent \ \
When the resonance parameter tends to infinity  we obtain the following
asymptotic behaviour 
\begin{equation}
\quad \lim_{\sigma \rightarrow \infty }Q_{2s+2t+1}^{\Sigma|2s,2t+1} 
\sim e^{-s\sigma } 
\end{equation}

\begin{equation}
\lim_{\sigma \rightarrow \infty }H_{2s+2t+1}^{\Sigma |2s,2t+1}\prod_{i<j}F_{%
\text{min}}^{\mu _{i}\mu _{j}}(\theta _{ij}) =\text{const\negthinspace
\negthinspace \negthinspace \negthinspace }\prod_{1\leq i<j\leq 2s}\text{%
\negthinspace \negthinspace \negthinspace \negthinspace }F_{\text{min}%
}^{++}(\theta _{ij})\text{\negthinspace \negthinspace \negthinspace
\negthinspace }\prod_{2s<i<j\leq 2s+2t+1}\text{\negthinspace \negthinspace
\negthinspace \negthinspace }F_{\text{min}}^{--}(\theta _{ij})\,,
\end{equation}
\noindent where again  we made use of (\ref{behave}).
This means unless $s=0$, that is a reduction to the thermally perturbed
Ising model, the form factors will vanish in this limit.

\subsection{The disorder operator, ${\protect\mu}$}
\label{mu}
\indent \ \
For the disorder operator we have $\vartheta =1$ in (\ref{d1})  and the solution acquires the
same form as in the previous case 
\begin{equation}
F_{2s}^{\mu }(\theta _{1},\ldots ,\theta _{2s})=i^{s}F_{0}^{\mu
}\prod_{i<j}\tanh \frac{\theta _{ij}}{2}\,\,.
\end{equation}
Similar as for the order variable we can fix the initial values of the
recursive equations (\ref{Qrec}) and (\ref{Hrec}) to 
\begin{equation}
Q_{2t}^{\mu |0,2t}=\left( \sigma _{2t}\right) ^{1/2-t}=\left( \bar{\sigma}%
_{2t}\right) ^{t-1/2}\qquad \text{and\qquad }H^{\mu |0,0}=F_{0}^{\mu }\,\,.
\end{equation}
Furthermore, we deduce $[F_{n}^{\mu |2s,2t}]_{i}=0$. Respecting these
constraints we find as a general solution 
\begin{equation}
Q^{\mu |2s,2t} =i^{2s(t+1)}\,g^{-1,1}\,\det \mathcal{A}^{0,0}\mathcal{\,}, 
\label{solu3}
\end{equation}
\noindent where once again $\mathcal{A}^{0,0}$ is the  $(s+t)$$\times$$(s+t)$-matrix given
by (\ref{sss}) when  $\mu, \nu=0$.

\subsubsection{The limit $\protect{\sigma \rightarrow \infty}$}
\indent \ \
In this case, when  the resonance parameter tends to infinity we observe, with the help of (\ref{behave}), the following
asymptotic behaviour 
\begin{eqnarray}
\quad \lim_{\sigma \rightarrow \infty }Q_{2s+2t}^{\mu |2s,2t}
&\sim &Q_{2s}^{\mu |2s,0}Q_{2t}^{\mu |0,2t}\quad \\
\lim_{\sigma \rightarrow \infty }H_{2s+2t}^{\mu |2s,2t}\prod_{i<j}F_{\text{%
min}}^{\mu _{i}\mu _{j}}(\theta _{ij}) &=&\text{const\negthinspace
\negthinspace \negthinspace }\prod_{1\leq i<j\leq 2s}\text{\negthinspace
\negthinspace \negthinspace }F_{\text{min}}^{++}(\theta _{ij})\text{%
\negthinspace \negthinspace \negthinspace }\prod_{2s<i<j\leq 2t+2s}\text{%
\negthinspace \negthinspace \negthinspace }F_{\text{min}}^{--}(\theta _{ij})
\end{eqnarray}
such that 
\begin{equation}
\lim_{\sigma \rightarrow \infty }F_{2s+2t}^{\mu |2s,2t}\sim F_{2t}^{\mu
|0,2t}F_{2s}^{\mu |2s,0}\,\,.
\end{equation}
This means also in this sector we observe the decoupling of the theory into
two free fermions, as we expected.

\section{A rigorous proof for the general solutions}
\label{proof}
\indent \ \
Let us enumerate  now  the principle steps of the general solution procedure
for the form factor consistency equations \cite{Kar,Smir,Zamocorr, YZam, BFKZ}. For
any local operator $\mathcal{O}$ one may anticipate the pole structure of
the form factors and extract it explicitly in form of an ansatz of the type (%
\ref{fact}). This might turn out to be a relatively involved matter due to
the occurrence of higher order poles in some integrable theories, e.g. \cite
{DM}, but nonetheless it is possible. Thereafter the task of finding
solutions may be reduced to the evaluation of the minimal form factors and
to solving a (or two if bound states may be formed in the model) recursive
equation of the type (\ref{Qrec}). The first task can be carried out
relatively easily, especially if the related scattering matrix is given as a
particular integral representation \cite{Kar}. Then an integral
representation of the type (\ref{14}) can be deduced immediately. The second
task is rather more complicated and the heart of the whole problem. Having a
seed for the recursive equation, that is the lowest non-vanishing form factor,
which in our case  is provided by the knowledge of the form factors of the thermally 
perturbed Ising model, 
one can in general compute from them several form factors which involve more
particles. However, the equations become relatively involved after several
steps. Aiming at the solution for all $n$-particle form factors, it is
therefore highly desirable to unravel a more generic structure which enables
one to formulate rigorous proofs. Several examples \cite
{clust,Zamocorr,deter1, deter2} have shown that often the general solution may be
cast into the form of determinants whose entries are elementary symmetric
polynomials. Presuming such a structure which, at present, may be obtained
by extrapolating from lower particle solutions to higher ones or by some
inspired guess, one can rigorously formulate proofs as we now demonstrate
for the SU(3)$_{2}$-HSG-model, for which at the beginning only  the solutions
for the trace of the energy momentum tensor and the order and disorder operators were merely stated
in \cite{CFK} without a rigorous proof.

As indicated above, we have the two universal structures 
(\ref{sss}) and (\ref{integral})
at our disposal. We could either exploit the integral representation for the determinant $\mathcal{A}$,
or exploit simple properties of determinants. For the reasons stated  in subsection \ref{solution},
here we shall pursue the latter possibility. For this purpose it is
convenient to define the operator $C_{i,j}^{x}$ $(R_{i,j}^{x})$ which acts
on the $j^{\text{th}}$ column (row) of an ($n\times n$)-matrix $\mathcal{A}$
by adding $x$ times the $i^{\text{th}}$ column (row) to it 
\begin{eqnarray}
C_{i,j}^{x}\mathcal{A} &:&\qquad \mathcal{A}_{kj}\mapsto \mathcal{A}_{kj}+x%
\mathcal{A}_{ki}\qquad \quad 1\leq i,j,k\leq n \\
R_{i,j}^{x}\mathcal{A} &:&\qquad \mathcal{A}_{jk}\mapsto \mathcal{A}_{jk}+x%
\mathcal{A}_{ik}\qquad \quad 1\leq i,j,k\leq n\,\,.
\end{eqnarray}
Naturally the determinant of $\mathcal{A}$ is left invariant under the
actions of $C_{i,j}^{x}$ and $R_{i,j}^{x}$ on $\mathcal{A},$ such that we
can use them to bring $\mathcal{A}$ into a suitable form for our purposes.
Furthermore, it is convenient to define the ordered products, i.e. operators
related to the lowest entry act first, 
\begin{equation}
\mathcal{C}_{a,b}^{x}:=\prod_{p=a}^{b}\!\!\,\,C_{p,p+1}^{x}\qquad \text{%
and\qquad }\mathcal{R}_{a,b}^{x}:=\prod_{p=a}^{b}\!\!\,\,R_{p,p-1}^{x}\,\,.
\end{equation}
It will be our strategy to use these operators in such a way that we produce
as many zeros as possible in one column or row of a matrix of interest to
us. In order to satisfy (\ref{Qrec}) we have to set now the first variables
in $\mathcal{A}$ to $x_{1}=-x,\,x_{2}=x$, which we denote as $\tilde{%
\mathcal{A}}$ thereafter and relate the matrices $\tilde{\mathcal{A}}%
_{l+2,m}^{\mu ,\nu }$ and $\mathcal{A}_{l,m}^{\mu ,\nu }$. Taking relation (%
\ref{symp}) for the elementary symmetric polynomials into account, we can
bring $\tilde{\mathcal{A}}_{l+2,m}^{\mu ,\nu }$ into the form 
\begin{equation}
\left( \!\!\mathcal{R}_{t+2,s+t+1}^{-x^{2}}\mathcal{C}_{1,s+t-1}^{x^{2}}%
\tilde{\mathcal{A}}_{l+2,m}^{\mu ,\nu }\right) _{ij}\!=\left\{ 
\begin{array}{l}
\sigma _{2(j-i)+\mu }^{+}\qquad \qquad \qquad \,\,\,\,\,\,\,\,\,\,\,\,1\leq
i\leq t \\ 
\hat{\sigma}_{2(j-i)+2t+\nu }^{-}\qquad \quad \quad \,\,\quad
\,\,\,\,\,\,\,\,\,\,t<i\leq s+t \\ 
\sum\limits_{p=1}^{j}x^{2(j-p)}\hat{\sigma}_{2(p-s-1)+\nu }^{-}\quad
\,\,\,\,\,\,\,\,\,\,\,\,i=s+t+1
\end{array}
\right. .
\end{equation}
It is now crucial to note that since the number of variables has been
reduced by two, several elementary polynomials may vanish. As a consequence,
for $2s+2+\mu >l$ and $2t+2+\nu >m$, the last column takes on the simple
form 
\begin{equation}
\left( \!\!\mathcal{R}_{t+2,s+t+1}^{-x^{2}}\mathcal{C}_{1,s+t-1}^{x^{2}}%
\tilde{\mathcal{A}}_{l+2,m}^{\mu ,\nu }\right) _{i(s+t+1)}\!=\left\{ \QATOP{%
0\qquad \,\,\,\,\,\,\,\,\,\,\,\,\,\,\,\,\,\,\,\,\,\,\,\,\,\,\,\,\,\,\,1\leq
i\leq s+t}{\sum\limits_{p=0}^{t}x^{2(t-p)}\hat{\sigma}_{2p+\nu }^{-}\quad
\,\,\,\,\,\,i=s+t+1\quad }\right. .\,  \label{19}
\end{equation}
Therefore, developing the determinant of $\tilde{\mathcal{A}}_{l+2,m}^{\mu
,\nu }$ with respect to the last column, we are able to relate the
determinants of $\tilde{\mathcal{A}}_{l+2,m}^{\mu ,\nu }$ and $\mathcal{A}%
_{l,m}^{\mu ,\nu }$ as 
\begin{equation}
\det \tilde{\mathcal{A}}_{l+2,m}^{\mu ,\nu }=\left(
\sum\limits_{p=0}^{t}x^{2(t-p)}\hat{\sigma}_{2p+\nu }^{-}\right) \,\det 
\mathcal{A}_{l,m}^{\mu ,\nu }\,\,\,\,\,\,\,\,.  \label{arec}
\end{equation}
We are left with the task to specify the behaviour of the function $g$ with
respect to the ``reduction'' of the first two variables 
\begin{equation}
\tilde{g}_{l+2,m}^{\bar{\mu},\bar{\nu}}=i^{l-m+\bar{\mu}+2}\,x^{l-m+\bar{\mu}%
+2}\,\sigma _{l}^{+}\,g_{l,m}^{\bar{\mu},\bar{\nu}}\,\,\,.  \label{grec}
\end{equation}
Assembling the two factors (\ref{arec}) and (\ref{grec}), we obtain, in
terms of the parameterization (\ref{qpar}) 
\begin{equation}
\tilde{Q}_{2s+2+\tau ,2t+\tau ^{\prime }}^{\bar{\mu},\bar{\nu},\mu ,\nu
}=(-i)^{2s+\tau +1}\sigma _{2s+\tau }^{+}\left(
\sum\limits_{p=0}^{t}x^{2(s-p+1)+\tau -\tau ^{\prime }+\bar{\mu}}\hat{\sigma}%
_{2p+\nu }^{-}\right) \,Q_{2s+\tau ,2t+\tau ^{\prime }}^{\bar{\mu},\bar{\nu}%
,\mu ,\nu }\,\,.  \label{soll}
\end{equation}
We are now in the position to compare our general construction (\ref{soll})
with the recursive equation for the $Q$-polynomials of the $SU(3)_{2}$-HSG
model (\ref{Dpar}). We read off directly the following identifications 
\begin{equation}
\nu =\zeta \qquad \text{and\qquad}\zeta =\tau ^{\prime }-\bar{\mu}-1\,\,\,.
\label{rest}
\end{equation}
A further constraint results from relativistic invariance, which implies
that the overall power in all variables $x_{i}$ of the form factors has to
be zero for a spinless operator. By using again the short hand notation 
$[F_{n}^{\mathcal{O}}]$ for the total power, we have to evaluate 
\begin{equation}
\left[ Q_{2s+\tau ,2t+\tau^{\prime }}^{\bar{\mu},\bar{\nu},\mu ,\nu }\right]
=\,\left[ g_{2s+\tau ,2t+\tau ^{\prime }}^{\bar{\mu},\bar{\nu}}\right] +\,
\left[ \det \mathcal{A}_{2s+\tau ,2t+\tau ^{\prime }}^{\mu ,\nu}\right]
\,\,.  \label{QT}
\end{equation}
Combining (\ref{rest}) and (\ref{QT}) with the explicit expressions 
\begin{eqnarray}
\left[\det \mathcal{A}_{2s+\tau ,2t+\tau^{\prime }}^{\mu ,\nu }\right]&=&s(2t+\nu )+\mu t,
\nonumber\\
\left[g_{2s+\tau, 2t+\tau^{\prime}}^{\bar{\mu},\bar{\nu}}\right]&=& {l(l-m+\bar{\mu})}/2+{m(\bar{\nu}-m)}/2,
\nonumber\\
\left[Q_{2s+\tau,2t+\tau^{\prime}}^{\bar{\mu},\bar{\nu},\mu,\nu}\right]&=&
{l(l-1)}/2-{m(m-1)}/2,
\label{pow}
\end{eqnarray}
\noindent  we find the additional constraints 
\begin{equation}
\mu =1+\tau -\bar{\nu}\qquad \text{and\qquad }\tau \nu =\tau ^{\prime }(\bar{
\nu}-1)\,\,. 
\label{rest2}
\end{equation}
Collecting now everything we conclude that different solutions to the form
factor consistency equations can be characterised by a set of four distinct
integers. Assuming that each solution corresponds to a local operator, there
might be degeneracies of course, we can label the operators by $\mu ,\nu
,\tau ,\tau ^{\prime }$, i.e. $\mathcal{O}\rightarrow \mathcal{O}_{\tau
,\tau ^{\prime }}^{\mu ,\nu }$, such that we can also write $Q_{m,l}^{\mu
,\nu }$ instead of $Q_{m,l}^{\bar{\mu},\bar{\nu},\mu ,\nu }$. Then each $Q$%
-polynomial takes on the general form 
\begin{equation}
Q^{\mathcal{O}|2s+\tau ,2t+\tau ^{\prime }}=Q^{\mathcal{O}_{\tau ,\tau
^{\prime }}^{\mu ,\nu }|2s+\tau ,2t+\tau ^{\prime }}=Q_{2s+\tau ,2t+\tau
^{\prime }}^{\mu ,\nu }\sim g_{2s+\tau ,2t+\tau ^{\prime }}^{\tau ^{\prime
}-1-\nu ,\tau +1-\mu }\,\,\det \mathcal{A}_{2s+\tau ,2t+\tau ^{\prime
}}^{\mu ,\nu }\,\,
\end{equation}
and the integers $\mu ,\nu ,\tau ,\tau ^{\prime }$ are restricted by 
\begin{equation}
\tau \nu +\tau ^{\prime }\mu =\tau \tau ^{\prime },\quad \qquad 2+\mu >\tau
,\quad \qquad 2+\nu >\tau ^{\prime }\,\,.  \label{25}
\end{equation}
We combined here (\ref{rest}) and (\ref{rest2}) to get the first relation in
(\ref{25}). The inequalities result from the requirement in the proof which
we needed to have the form (\ref{19}). We find 12 admissible solutions to (%
\ref{25}), i.e. potentially 12 different local operators, whose quantum
numbers are presented in table \ref{t1}.
\begin{table}
\begin{center}
\vspace{0.2cm} 
\begin{tabular}{|r|r|r|r|r|r|r|}
\hline
$\mu $ & $\nu $ & $\tau $ & $\tau ^{\prime }$ & $\left[ F_{\tau \tau
^{\prime }}^{\mu \nu }\right] _{+}$ & $\left[ F_{\tau \tau ^{\prime }}^{\mu
\nu }\right] _{-}$ & $\Delta $ \\ \hline\hline
0 & 0 & 0 & 0 & 0 & 0 & 1/10 \\ \hline
0 & 0 & 1 & 0 & 0 & 0 & 1/10 \\ \hline
0 & 0 & 0 & 1 & 0 & 0 & 1/10 \\ \hline
0 & 1 & 0 & 1 & -1/2 & 0 & 1/10 \\ \hline
0 & 1 & 1 & 1 & -1/2 & 0 & 1/10 \\ \hline
0 & 1 & 0 & 2 & -1/2 & 0 & 1/10 \\ \hline
1 & 0 & 1 & 1 & 0 & -1/2 & 1/10 \\ \hline
1 & 0 & 2 & 0 & 0 & -1/2 & 1/10 \\ \hline
1 & 0 & 1 & 0 & 0 & -1/2 & 1/10 \\ \hline
1 & 1 & 2 & 2 & -1/2 & -1/2 & 1/10 \\ \hline
1 & 1 & 0 & 0 & -1/2 & -1 & * \\ \hline
1 & 0 & 0 & 0 & 0 & -1/2 & * \\ \hline
\end{tabular}
\end{center}
\caption{Operator content and form factor asymptotics 
of the $\protect{SU(3)_{2}}$-HSG model.}
\label{t1}
\end{table}

\noindent Comparing with our previous results, we have according to this
notation $F^{\mathcal{O}_{0,0}^{0,0}|2s,2t}=F^{\mu |2s,2t}$, $F^{\mathcal{O}%
_{0,1}^{0,1}|2s,2t+1}=F^{\Sigma |2s,2t+1}$ and $F^{\mathcal{O}%
_{2,2}^{1,1}|2s,2t+1}\sim F^{\Theta |2s+2,2t+2}$. The last two solutions in table \ref{t1} are
only formal in the sense that they solve the constraining equations (\ref{25}%
), but the corresponding explicit expressions turn out to be zero. In the last three columns
we present information concerning the asymptotic behaviour of each solution and its
corresponding conformal dimension in the UV-limit which will become clear
in the next section.

In summary, by taking the determinant of the matrix (\ref{sss}) as the
ansatz for the general building block of the form factors, we constructed
systematically generic formulae for  all $n$-particle form factors possibly
related to 12 different operators. In the next section we will prove that if only one of the first
9 solutions is known  we could  have obtained  all the rest by
means of the so-called momentum space cluster property (\ref{cluster}).

\section{Momentum space cluster properties}
\label{mscp}
\indent \ \
We shall now systematically investigate the cluster property (\ref{cluster})
for the $SU(3)_{2}$-HSG model. Choosing w.l.g. the upper signs for the
particle types in equation (\ref{fact}), we have four different options to
shift the rapidities 
\begin{eqnarray}
\mathcal{T}_{1,\kappa \leq l}^{\pm \lambda }F_{n}^{\mathcal{O}|l\times
+,m\times -} &=&\mathcal{T}_{\kappa +1<l,n}^{\mp \lambda }F_{n}^{\mathcal{O}%
|l\times +,m\times -}  \label{s1} \\
\mathcal{T}_{1,\kappa >l}^{\pm \lambda }F_{n}^{\mathcal{O}|l\times +,m\times
-} &=&\mathcal{T}_{\kappa +1\geq l,n}^{\mp \lambda }F_{n}^{\mathcal{O}%
|l\times +,m\times -}\,  \label{s2}
\end{eqnarray}
which a priori might all lead to different factorizations on the r.h.s. of
Eq. (\ref{cluster}). The equality signs in the equations (\ref{s1}) and
(\ref{s2}) are a simple consequence of the relativistic invariance of form
factors, i.e. we may shift all rapidities by the same amount, for $\mathcal{O%
}$ being a scalar operator.

Considering now the ansatz (\ref{fact}) we may first carry out part of the
analysis for the terms which are operator independent. Noting that 
\begin{equation}
\mathcal{T}_{1,1}^{\pm \lambda }F_{\text{min}}^{++}(\theta )=\mathcal{T}%
_{1,1}^{\pm \lambda }F_{\text{min}}^{--}(\theta )\sim e^{\frac{(\lambda \pm
\theta )}{2}}\qquad \text{and\qquad }\mathcal{T}_{1,1}^{\pm \lambda }F_{%
\text{min}}^{+-}(\theta )\sim \QATOPD\{ . {\mathcal{O}(1)}{e^{\frac{(\theta
-\lambda )}{2}}}\,,  \label{eq}
\end{equation}
we obtain for the choice of the upper signs for the particle types in the
ansatz (\ref{fact}) 
\[
\mathcal{T}_{1,\kappa \leq l}^{\pm \lambda }\,\prod_{i<j}\hat{F}^{\mu
_{i}\mu _{j}}(\theta _{ij})\sim \!\!\!\!\!\!\prod_{1\leq i<j\leq \kappa
}\!\!\!\!\!\hat{F}^{++}(\theta _{ij})\!\!\!\!\!\!\!\prod_{\kappa <i<j\leq
l+m}\!\!\!\!\!\!\hat{F}^{\mu _{i}\mu _{j}}(\theta _{ij})\left\{ \QTATOP{%
\tfrac{\sigma _{\kappa }(x_{1},\ldots ,x_{\kappa })^{\frac{\kappa -l}{2}}e^{%
\frac{\lambda \kappa (1-l)}{2}}}{\sigma _{l-\kappa }(x_{\kappa +1},\ldots
,x_{l})^{\frac{\kappa }{2}}}}{\tfrac{\sigma _{\kappa }(x_{1},\ldots
,x_{\kappa })^{\frac{m-l+\kappa }{2}}e^{\frac{\lambda \kappa (l-m-1)}{2}}}{%
\sigma _{n-\kappa }(x_{\kappa +1},\ldots ,x_{n})^{\frac{\kappa }{2}}}}%
\right. 
\]
\[
\mathcal{T}_{n+1-\kappa <m,n}^{\pm \lambda }\,\prod_{i<j}\hat{F}^{\mu
_{i}\mu _{j}}(\theta _{ij})\sim \!\!\!\!\!\!\!\!\prod_{1\leq i<j\leq
n-\kappa }\!\!\!\!\!\!\!\hat{F}^{\mu _{i}\mu _{j}}(\theta
_{ij})\!\!\!\!\!\!\!\prod_{n-\kappa <i<j\leq n}\!\!\!\!\!\!\!\hat{F}%
^{--}(\theta _{ij})\left\{ \QATOP{\tfrac{\sigma _{n-\kappa }(x_{1},\ldots
,x_{n-\kappa })^{\frac{\kappa }{2}}e^{\frac{\lambda \kappa (m-l-1)}{2}}}{%
\sigma _{\kappa }(x_{n+1-\kappa },\ldots ,x_{n})^{\frac{l-m+\kappa }{2}}}}{%
\tfrac{\sigma _{m-\kappa }(x_{l+1},\ldots ,x_{n-\kappa })^{\frac{\kappa }{2}%
}e^{\frac{\lambda \kappa (1-m)}{2}}}{\sigma _{\kappa }(x_{n+1-\kappa
},\ldots ,x_{n})^{\frac{\kappa -m}{2}}}}\right. . 
\]
The remaining cases can be obtained from the equalities (\ref{s1}) and (\ref
{s2}). Turning now to the behaviour of the function $g$ as defined in (\ref
{g}) under these operations, we observe with help of the asymptotic
behaviour of the elementary symmetric polynomials (\ref{as+}) and (\ref{as-}) reported in
appendix A,

\begin{eqnarray}
\mathcal{T}_{1,\kappa \leq l}^{\pm \lambda }\,g_{l,m}^{\bar{\mu},\bar{\nu}%
}\!\!\! &=&\!\!\![e^{\pm \lambda \kappa }\sigma _{\kappa }(x_{1},\ldots
,x_{\kappa })\sigma _{l-\kappa }(x_{\kappa +1},\ldots ,x_{l})]^{\frac{l-m+%
\bar{\mu}}{2}}\,(\sigma _{m})^{\frac{\bar{\nu}-m}{2}} \\
\mathcal{T}_{n+1-\kappa <m,n}^{\pm \lambda }g_{l,m}^{\bar{\mu},\bar{\nu}%
}\!\!\! &=&\!\!\!(\sigma _{l})^{\frac{l-m+\bar{\mu}}{2}}[e^{\pm \lambda
\kappa }\sigma _{\kappa }(x_{n+1-\kappa },\ldots ,x_{n})\sigma _{m-\kappa
}(x_{l+1},\ldots ,x_{n-\kappa })]^{\frac{\bar{\nu}-m}{2}}.\,\,\,\,\,\,\,
\end{eqnarray}

\noindent In a similar fashion we compute the behaviour of the determinants 
\begin{eqnarray}
\mathcal{T}_{1,2\kappa +\xi \leq l}^{\lambda }\det \mathcal{A}_{l,m}^{\mu
,\nu } &=&e^{\lambda t(2\kappa +\xi )}(\sigma _{2\kappa +\xi })^{t}((-1)^{t}%
\hat{\sigma}_{\nu }^{-})^{\kappa +\xi (1-\mu )}\det \mathcal{A}_{l-2\kappa
-\xi ,m}^{1-\mu ,\nu } \\
\mathcal{T}_{1,2\kappa +\xi \leq l}^{-\lambda }\det \mathcal{A}_{l,m}^{\mu
,\nu } &=&(\hat{\sigma}_{2t+\nu }^{-})^{\kappa +\xi }\det \mathcal{A}%
_{l-2\kappa -\xi ,m}^{\mu ,\nu } \\
\mathcal{T}_{n+1-2\kappa -\xi <m,n}^{\lambda }\det \mathcal{A}_{l,m}^{\mu
,\nu } &=&e^{\lambda s(2\kappa +\xi )}(\sigma _{\mu }^{+})^{\kappa +\xi
(1-\nu )}(\hat{\sigma}_{2\kappa +\xi })^{s}\det \mathcal{A}_{l,m-2\kappa
-\xi }^{\mu ,1-\nu } \\
\mathcal{T}_{n+1-2\kappa -\xi <m,n}^{-\lambda }\det \mathcal{A}_{l,m}^{\mu
,\nu } &=&((-1)^{s}\sigma _{2s+\mu }^{+})^{\kappa +\xi }\det \mathcal{A}%
_{l,m-2\kappa -\xi }^{\mu ,\nu }\,\,.
\end{eqnarray}
We have to distinguish here between the odd and even case, which is the
reason for the introduction of the integer $\xi $ taking on the values $0$
or $1$. Collecting now all the factors, we extract first the leading order
behaviour in $\lambda $%
\begin{equation}
\mathcal{T}_{1,\kappa \leq l}^{\pm \lambda }F_{2s+\tau ,2t+\tau ^{\prime
}}^{\mu ,\nu }\sim e^{-\lambda \kappa \left( \pm \nu +\tau ^{\prime }\frac{%
(1\mp 1)}{2}\right) }\qquad \mathcal{T}_{n+1-\kappa <m,n}^{\pm \lambda
}F_{2s+\tau ,2t+\tau ^{\prime }}^{\mu ,\nu }\sim e^{-\lambda \kappa \left(
\pm \mu +\tau \frac{(1\mp 1)}{2}\right) }\,.  \label{lead}
\end{equation}
Notice that, if we require that all possible actions of $\mathcal{T}%
_{a,b}^{\pm \lambda }$ should lead to finite expressions on the r.h.s. of (%
\ref{cluster}), we have to impose two further restrictions, namely $\mathcal{%
\tau }^{\prime }\geq \nu $ and $\tau \geq \mu $. These restrictions would
also exclude the last two solutions from table \ref{t1}. 
We observe further that $%
F_{2,2}^{1,1}$ tends to zero under all possible shifts. Seeking now
solutions for the set $\mu ,\nu ,\tau ,\tau ^{\prime }$ of (\ref{lead})
which at least under some operations leads to finite results and in all
remaining cases tends to zero, we end up precisely 
with the first 9
solutions in table \ref{t1}.

\begin{figure}[!h]
\begin{center}
\leavevmode
\includegraphics[width=14cm,height=14cm]{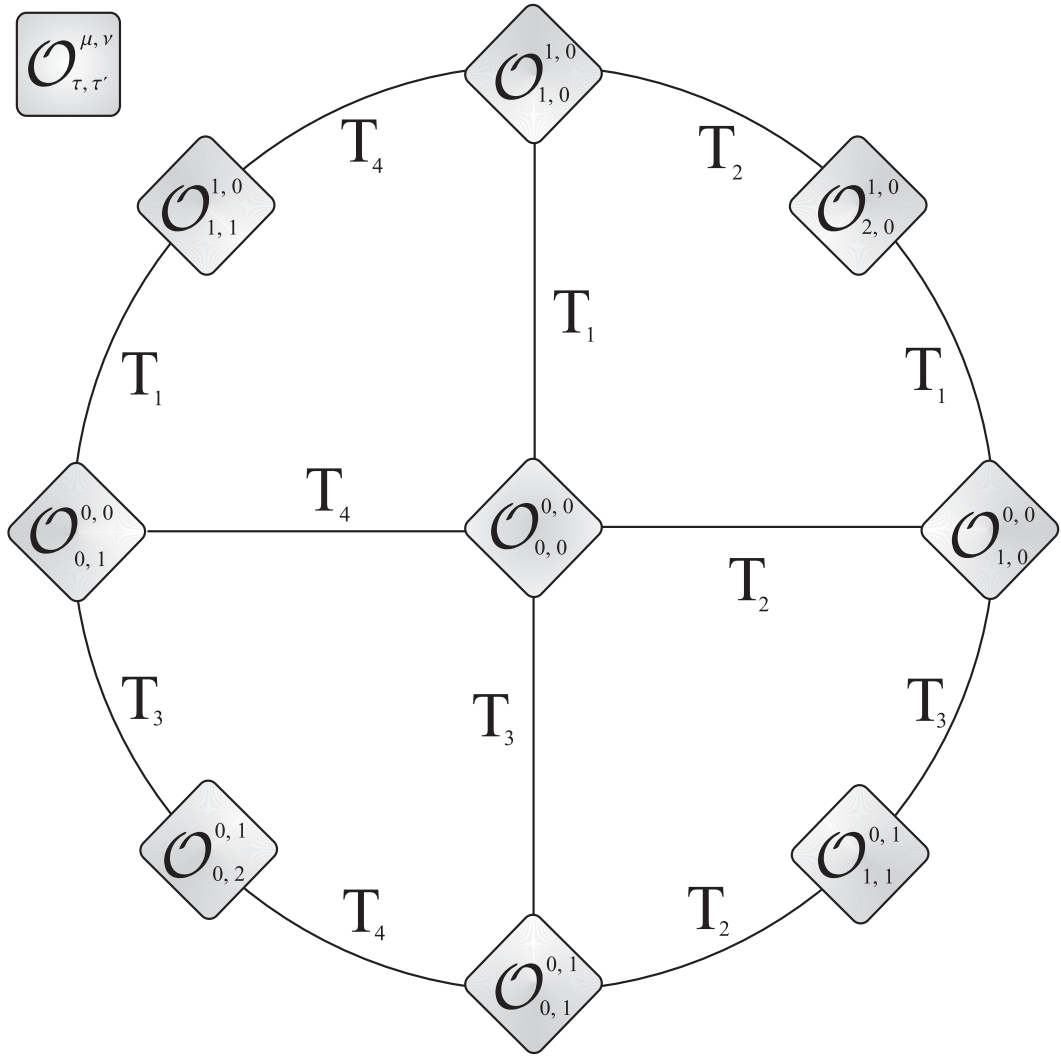}
\caption[Interrelation of various operators via
clustering.]
{Interrelation of various operators via
clustering. In this figure we use the abbreviations
 $\protect{T_{1}\equiv \mathcal{T}_{1,2\kappa +1\leq l}^{\lambda }}$, 
$\protect{T_{2}\equiv \mathcal{T}_{1,2\kappa+1\leq l}^{-\lambda }}$,
 $\protect{T_{3}\equiv \mathcal{T}_{n-2\kappa <m,n}^{\lambda }}$, 
$\protect{T_{4}\equiv \mathcal{T}_{n-2\kappa <m,n}^{-\lambda }}$. 
We also drop the 2s and 2t in the subscripts of the $\protect{\mathcal{O}}${\small 's}. 
The $\protect{T_{i}}$ on the links operate in both directions.}
\label{f21}
\end{center}
\end{figure}

Concentrating now in more detail on these latter cases which behave like $%
\mathcal{O}(1)$, we find from the previous equations the following cluster
properties 
\begin{eqnarray}
\mathcal{T}_{1,2\kappa +\xi \leq l}^{\lambda }F_{2s+\tau ,2t+\tau ^{\prime
}}^{\mu ,0} &\sim &F_{2\kappa +\xi ,0}^{0,0}F_{2s+\tau -2\kappa -\xi
,2t+\tau ^{\prime }}^{\mu +\xi (1-2\mu ),0}  \label{40} \\
\mathcal{T}_{1,2\kappa +\xi \leq l}^{-\lambda }F_{2s+\tau ,2t+\nu }^{\mu
,\nu } &\sim &F_{2\kappa +\xi ,0}^{0,0}F_{2s+\tau -2\kappa -\xi ,2t+\nu
}^{\mu ,\nu } \\
\mathcal{T}_{n+1-2\kappa -\xi <m,n}^{\lambda }F_{2s+\tau ,2t+\tau ^{\prime
}}^{0,\nu } &\sim &F_{2s+\tau ,2t+\tau ^{\prime }-2\kappa -\xi }^{0,\nu +\xi
(1-2\nu )}F_{0,2\kappa +\xi }^{0,0} \\
\mathcal{T}_{n+1-2\kappa -\xi <m,n}^{-\lambda }F_{2s+\mu ,2t+\tau ^{\prime
}}^{\mu ,\nu } &\sim &F_{2s+\mu ,2t+\tau ^{\prime }-2\kappa -\xi }^{\mu ,\nu
}F_{0,2\kappa +\xi }^{0,0}\,\,.  \label{41}
\end{eqnarray}
We may now use (\ref{40})-(\ref{41}) as a means of constructing new
solutions, i.e. we can start with one solution and use (\ref{40})-(\ref{41})
in order to obtain new ones. Fig.\ref{f21} demonstrates that when knowing just
one of the first nine operators in table \ref{t1} 
it is possible to (re)-construct
all the others in this fashion.

\vspace{0.5cm}

\subsection{The energy momentum tensor}
\label{emtclus}
\indent \ \
As we observed from our previous discussion the solution $F_{2,2}^{1,1}$ is
rather special. In fact this solution is part of the expression  (\ref{solu1}), which 
 was identified as the trace of the energy momentum tensor 
\begin{equation}
Q^{\Theta |2s+2,2t+2}=i^{s(2t+3)}e^{-(t+1)\sigma }\sigma _{1}\bar{\sigma}%
_{1}F_{2s+2,2t+2}^{1,1}\,\,.
\end{equation}
The pre-factor $\sigma _{1}\bar{\sigma}_{1}$ will, however, alter the
cluster property. The leading order behaviour reads now 
\begin{equation}
\mathcal{T}_{1,\kappa \leq 2s}^{\pm \lambda }F^{\Theta |2s,2t}\sim \mathcal{T%
}_{n+1-\kappa <2t,n}^{\pm \lambda }F^{\Theta |2s,2t}\sim e^{\lambda
(1-\kappa /2)}\,\,.
\end{equation}
We observe that still in most cases the shifted expressions tend to zero,
unless $\kappa =1$ for which it tends to infinity as a consequence of the
introduction of the $\sigma _{1}\bar{\sigma}_{1}$. There is now also the
interesting case $\kappa =2$, for which the $\lambda $-dependence drops out
completely. Considering this case in more detail we find 
\begin{eqnarray}
\mathcal{T}_{1,2}^{\pm \lambda }F^{\Theta |2s,2t} &\sim &F^{\Theta
|2,0}F^{\Theta |2s-2,2t}\times \QATOPD\{ . {\frac{\sigma
_{1}(x_{2s+1},\ldots ,x_{2s+2t})}{\sigma _{1}(x_{3},\ldots ,x_{2s+2t})}}{%
\frac{\bar{\sigma}_{1}(x_{2s+1},\ldots ,x_{2s+2t})}{\bar{\sigma}%
_{1}(x_{3},\ldots ,x_{2s+2t})}}  \label{fa1} \\
\mathcal{T}_{n+1-\kappa ,n}^{\pm \lambda }F^{\Theta |2s,2t} &\sim &F^{\Theta
|2s,2t-2}F^{\Theta |0,2}\times \QATOPD\{ . {\frac{\sigma _{1}(x_{1},\ldots
,x_{2s})}{\sigma _{1}(x_{1},\ldots ,x_{2s+2t-2})}}{\frac{\bar{\sigma}%
_{1}(x_{1},\ldots ,x_{2s})}{\bar{\sigma}_{1}(x_{1},\ldots ,x_{2s+2t-2})}}\,.
\label{fa2}
\end{eqnarray}

\noindent Note that unless $s=1$ in (\ref{fa1}) or $t=1$ in (\ref{fa2}) the
form factors do not ``purely'' factorise into known form factors, but in all
cases a parity breaking factor emerges.

\noindent We now turn to the cases $\kappa =2s$ or $\kappa =2t$ for which we
derive 
\begin{equation}
\mathcal{T}_{1,2s}^{\pm \lambda }F^{\Theta |2s,2t}\sim \mathcal{T}%
_{n+1-2t,n}^{\pm \lambda }F^{\Theta |2s,2t}\sim e^{\lambda (2-t-s)}\,\,.
\end{equation}
We observe that once again in most cases these expressions tend to zero.
However, we also encounter several situations in which the $\lambda $%
-dependence drops out altogether. It may happen whenever $t=2$, $s=0$ or $%
s=2 $, $t=0$, which simply expresses the relativistic invariance of the form
factor. The other interesting situation occurs for $t=1$, $s=1$. Choosing
temporarily (in general we assume $m_{-}=m_{+}$) $H_{2}^{\Theta |0,2}=2\pi
m_{-}^{2}$ , $m=m_{-}=m_{+}e^{2G/\pi }$, we derive in this case 
\begin{equation}
\mathcal{T}_{1,2}^{\lambda }F_{4}^{\Theta |2,2}=\frac{F_{2}^{\Theta
|2,0}F_{2}^{\Theta |0,2}}{2\pi m^{2}}\,\,.
\end{equation}
In general when shifting the first $2s$ or last $2t$ rapidities we find the
following factorization 
\begin{equation}
\mathcal{T}_{1,2s}^{\pm \lambda }F^{\Theta |2s,2t}\sim \mathcal{T}%
_{n+1-2t,n}^{\pm \lambda }F^{\Theta |2s,2t}\sim \text{ }F^{\Theta
|2s,0}F^{\Theta |0,2t}\,\,.
\end{equation}
This equation holds true when keeping in mind that the r.h.s. of this
equation vanishes once it involves a form factor with more than two
particles. Note that only in these two cases the form factors factorise
``purely'' into two form factors without the additional parity breaking
factors as in (\ref{fa1}) and (\ref{fa2}).

\section{Computing the Virasoro central charge}
\label{cff2}
\indent \ \ 
Having computed all form factors of the energy momentum tensor (see section \ref{emt}),
we are in the position to evaluate explicitly (\ref{cth}). We can now collect all the factors entering
in the form factors $F_{n}^{\Theta |\mu_1 \cdots \mu_n}$, namely the constants given by  (\ref{consol}) the $Q$'s  presented in (\ref{solu1}) and the part coming from the minimal form factors, as presented 
in the general ansatz  (\ref{fact}). The 2-particle contribution is given by  (\ref{2par})
and for the 4-particle form factor  we obtain,
\begin{equation}
F_{4}^{\Theta |++ --} =\frac{-\pi m^{2}(2+\sum_{i<j}\cosh (\theta _{ij}))}{%
2\cosh (\theta _{12}/2)\cosh (\theta _{34}/2)}\prod_{i<j}\tilde{F}_{\text{min%
}}^{\mu _{i}\mu _{j}}(\theta _{ij}).
\label{4par}
\end{equation}
\noindent Notice that the $\sigma$-dependence is ``hidden'' in the minimal form 
factors  (\ref{14}).
All the explicit expressions of  the 6-particle form factors of the energy momentum tensor for 
$\sigma=0$ can be found in appendix B.
The results obtained for the central charge are,
\begin{equation}
\Delta c^{(2)}=1,\qquad \Delta c^{(4)}=1.197...,\qquad \Delta
c^{(6)}=1.199\ldots \,,\quad \text{for }\sigma <\infty, 
\label{ce}
\end{equation}
where in the notation $\Delta c^{(n)}$, the superscript $n$ indicates the
upper limit in (\ref{cth}) namely, when carrying out the sum (\ref{cth}), 
contributions until the $n$-particle form factor have been taken into account.
 Thus, the expected value of $c=6/5=1.2$ is well
reproduced. Apart from the 2-particle contribution  $\Delta c^{(2)}$, in which case the calculation can
be performed analytically
\begin{equation}
\Delta c^{(2)}= \frac{3}{2} \int\limits_{-\infty}^{\infty} d\theta\frac{{ |F_{\text{mim}}^{++}(2\theta) |}^2}{({\cosh\theta})^4}=\frac{3}{2}\int\limits_{-\infty}^{\infty}d\theta \Bigg( \frac{{\tanh\theta}}{{\cosh\theta}}\Bigg)^2 =1, 
\label{anal}
\end{equation} 
\noindent the integrals in (\ref{cth}) have been computed
directly via a brute force Monte Carlo integration.
Typical standard deviations we achieve correspond to the order of the last
digit we quote.
Taking into account that, in the limit $\sigma \rightarrow \infty$ the only non-vanishing form factors
of the energy momentum tensor are $F_2^{\Theta | \pm \pm}$ (see section \ref{emt}), we obtain for the
central charge
\begin{equation}
\lim_ {\sigma \rightarrow \infty}\Delta c =1,
\label{siginf}
\end{equation}
\noindent as we expected, since according to (\ref{decthe}), in the limit $\sigma \rightarrow \infty$ we are left with  a system of two non-interacting  free fermions.

\vspace{0.25cm}

In short,  the main outcome of  this section is  that  the calculation of the ultraviolet Virasoro central charge associated to the $SU(3)_2$-HSG model in the context of the form factor analysis by means of  Zamolodchikov's $c$-theorem (\ref{cth}) provides a result which is in complete agreement  with the S-matrix proposal  \cite{HSGS} and with the physical picture obtained in chapter \ref{tba} 
through a TBA-analysis. 
However, we shall confirm in the next section that a form factor
study can provide more information about the underlying CFT apart from its Virasoro
central charge. In particular, we might be able to identify the ultraviolet conformal dimensions of 
those  local operators for which at least the first non-vanishing form factors are known.
 Conversely, in the context of the TBA
it remains,  an open question how to identify the operator
content. As we showed in chapter \ref{tba}, it is sometimes possible to
determine at least the dimension of the perturbing operator, despite the fact that
the reason is unclear, by
investigating periodicities in the so-called $Y$-systems \cite{TBAZamun},
 but no  information  about other local operators is  available.

\section{Identifying the operator content}
\label{identifying}
\indent \ \
Having solved Watson's and the residue equations one has still little
information about the precise nature of the operator corresponding to a
particular solution. There exist, however, various non-perturbative (in the
standard coupling constant sense) arguments which provide this additional
information and which we now wish to exploit for the model at hand.
Basically, all these arguments rely on the assumption that the superselection
sectors of the underlying conformal field theory remain separated after a
mass scale has been introduced. We will therefore first have a brief look at
the operator content of the $G_{k}/U(1)^{\ell}$-WZNW coset models and attempt
thereafter to match them with the solutions of the form factor consistency
equations. For these theories the different conformal dimensions in the
model can be parameterised by two quantities \cite{Gep, DHS}: a highest
dominant weight $\Lambda $ of level smaller or equal to $k$ and its
corresponding lower weights $\lambda $
\begin{equation}
\Delta (\Lambda ,\lambda )=\frac{(\Lambda \cdot (\Lambda +2\rho ))}{2(k+h)}-%
\frac{(\lambda \cdot \lambda )}{2k}\,\,.  \label{wei}
\end{equation}
The  lower weight $\lambda$, may be constructed in the usual fashion (see e.g. \cite{FH}): 
Consider a complete weight string $\lambda +n\alpha ,\ldots ,\lambda ,\ldots ,\lambda
-m\alpha $, that is all the weights obtained by successive additions
(subtractions) of a root $\alpha $ from the weight $\lambda $, such that $%
\lambda +(n+1)\alpha $ ($\lambda -(m+1)\alpha $) is not a weight anymore. It
is then a well known fact that the difference between the two integers $m,n$
is $m-n=\lambda \cdot \alpha $ for simply laced Lie algebras. This means
starting with the highest weight $\Lambda $, we can work our way downwards
by deciding after each subtraction of a simple root $\alpha _{i}$ whether
the new vector, say $\chi $, is a weight or not from the criterion $%
m_{i}=n_{i}+\chi \cdot \alpha _{i}>0$. With the procedure just outlined we
obtain all possible weights of the theory

In (\ref{wei})  $h$ is the Coxeter number of $G$ and $\rho $ is the Weyl vector which
is defined as (see for instance \cite{Bou})
\begin{equation}
\rho=\frac{1}{2}\sum_{\alpha>0} \alpha = \sum_{i=1}^{\ell} \lambda_i,
\end{equation}
\noindent i.e. half the sum over the positive roots or, equivalently,  the
sum over all fundamental weights, $\lambda_i$. Denoting the highest root of $G$ by $\psi $,
 the conformal dimension related to the adjoint representation $\Delta
(\psi ,0)$ is of special interest since it corresponds to the one of the
perturbing operator which leads to the massive HSG-models. 
Taking the square length
of $\psi $ to be $2$ and recalling the well known fact that the height of $%
\psi $, that is ht($\psi $), is the Coxeter number minus one, such that $%
(\psi \cdot \rho )=ht(\psi )=h-1$, it follows that 
$\mathcal{O}^{\Delta (\psi ,0)}$is an operator with 
conformal dimension $\Delta (\psi ,0)=h/(k+h)$,
 which  is precisely
the expression of the conformal dimension of the perturbing field  \cite{HSG2}.
Remarkably, the latter conformal dimension is obtained for a single combination
of weights $(\Lambda, \lambda)$.
Notice that, with the procedure described in the above paragraph it may happen
that a weight corresponds to more than one linearly independent weight vector,
such that the weight space may be more than one-dimensional. 
However,  we will  not take the multiplicities of the $\lambda$-states into account in this case and
leave a more detailed study of this issue for the next chapter (see section \ref{operator}).
 For $SU(3)_{2}$ the
expression (\ref{wei}) is easily computed and since we could not find the
explicit values in the literature we report them for reference in table \ref{t2}. 
\begin{table}
\begin{center}
\begin{tabular}{|c||c|c|c|c|c|}
\hline
$\lambda \backslash \Lambda $ & $\lambda _{1}$ & $\lambda _{2}$ & $\lambda
_{1}+\lambda _{2}$ & $2\lambda _{1}$ & $2\lambda _{2}$ \\ \hline\hline
$\Lambda $ & $1/10$ & $1/10$ & $1/10$ & $0$ & $0$ \\ \hline
$\Lambda -\alpha _{1}$ & $1/10$ & $*$ & $1/10$ & $1/2$ & $*$ \\ \hline
$\Lambda -\alpha _{2}$ & $*$ & $1/10$ & $1/10$ & $*$ & $1/2$ \\ \hline
$\Lambda -\alpha _{1}-\alpha _{2}$ & $1/10$ & $1/10$ & $3/5$ & $1/2$ & $1/2$
\\ \hline
$\Lambda -2\alpha _{1}$ & $*$ & $*$ & $*$ & $0$ & $*$ \\ \hline
$\Lambda -2\alpha _{2}$ & $*$ & $*$ & $*$ & $*$ & $0$ \\ \hline
$\Lambda -2\alpha _{1}-\alpha _{2}$ & $*$ & $*$ & $1/10$ & $1/2$ & $*$ \\ 
\hline
$\Lambda -\alpha _{1}-2\alpha _{2}$ & $*$ & $*$ & $1/10$ & $*$ & $1/2$ \\ 
\hline
$\Lambda -2\alpha _{1}-2\alpha _{2}$ & $*$ & $*$ & $1/10$ & $0$ & $0$ \\ 
\hline
\end{tabular}
\end{center}
\caption[Local operator content of the
$\protect{SU(3)_{2}}/{U(1)^{2}}$-coset model.]
{Conformal dimensions for 
$\protect{\mathcal{O}^{\Delta (\Lambda,\lambda )}}$ in
the $\protect{SU(3)_{2}}/{U(1)^{2}}$-coset model. The `\,*\,' indicate
those cases for which the corresponding weight $\protect{\lambda_i}$ 
does not arise as lower weight of $\protect\Lambda$.}
\label{t2}
\end{table}

\noindent Turning now to the massive theory and once  the spinless character of the operators 
 (\ref{rel}) has  been taken into account via  (\ref{Qasym}), a crude constraint which gives 
a first glimpse at possible solutions to the form factor consistency
equations based on their asymptotics  is provided by the bound \cite{DM}, which we presented 
at the beginning of this chapter ( \ref{bound}) and referred to also as 
property 6 of the form factors. 

It is useful to introduce  at this point some new notation in addition to the useful abbreviation
(\ref{not}) we have extensively used in preceding sections. 
In the same spirit,  we  will use now the notation $[\,\,\,]_{\pm }$ when we take the
limit in the variable $x_{i}$ related to the particle species $\mu
_{i}=``\pm "$, respectively. For the different solutions we constructed, we
reported the asymptotic behaviours  in table \ref{t1}. Notice that, assuming 
each of the 12 admissible solutions presented in table \ref{t1} has a counterpart
in the underlying conformal field theory whose conformal dimension is contained
in table \ref{t2}, 
the asymptotic behaviours presented also in table \ref{t1} are always  compatible
with the bound (\ref{bound}), irrespectively of which of the conformal dimensions in \ref{t2}
is the one associated to the  operator at hand.

Being  in a position in
which we already anticipate the conformal dimensions in table \ref{t2}, the bound (\ref{bound})
will severely restrict the possible inclusion of factors like $\sigma _{1},%
\bar{\sigma}_{1},\sigma _{1}^{-},\sigma _{1}^{+}$ and therefore the amount 
of different solutions to the recursive equations (\ref{Qrec}). Recall that it was stated  in
section \ref{ising} that factors of the type above  mentioned may always be added to any of our solutions 
of table \ref{t1},  since they trivially satisfy the consistency equations.

\subsection{$\protect\Delta$-sum rules}
\label{dsum}
\indent \ \
We may now turn to the identification of the operator content of the $SU(3)_2$-HSG
model by exploiting the second of the techniques anticipated in subsection \ref{uvffs}: the $\Delta$-sum
rule provided in \cite{DSC}.
Since in our model the $n$-particle form
factors related to the energy momentum tensor are only non-vanishing for
even particle numbers, we might only use the $\Delta$-sum
rule (\ref{deldel}) \cite{DSC} for the operators $\mathcal{O}_{0,0}^{0,0}$, 
$\mathcal{O}_{0,2}^{0,1}$, $\mathcal{O}_{2,0}^{1,0}$ and $\mathcal{O}_{2,2}^{1,1}$, 
where the latter operator is plagued by one of the problems stated in subsection \ref{uvffs}
namely, its lowest non-vanishing form factor is not the vacuum expectation value.

We will now compute the sum rule for the operators $\mathcal{O}_{0,0}^{0,0}$%
, $\mathcal{O}_{0,2}^{0,1}$, $\mathcal{O}_{2,0}^{1,0}$ up to the 6-particle
contribution. The explicit expressions or the corresponding form factors, until the 8-particle contribution have been collected in appendix B. We commence with the two particle contribution which is always
evaluated effortlessly. Noting that, as we saw in section \ref{emt} the 2-particle form factor  is given
by  (\ref{2par})  and the fact that $\Delta _{ir}^{\!\!\mathcal{O}}$ is zero in a purely
massive model, the two particle contribution acquires the  particular simple
form 
\begin{equation}
(\Delta ^{\!\!\mathcal{O}})^{(2)}=\frac{i}{4\pi \left\langle \mathcal{O}%
\right\rangle }\int\limits_{-\infty }^{\infty }d\theta \frac{\tanh \theta }{%
\cosh \theta }\left( F_{2}^{\mathcal{O}|++}(2\theta )\right) ^{*}\,\,\,.\,
\label{twop}
\end{equation}
Using now the explicit expressions for the two-particle form factors (\ref
{twopp}), we immediately find 
\begin{equation}
(\Delta ^{\!\!\mathcal{O}_{0,0}^{0,0}})^{(2)}=(\Delta ^{\!\!\mathcal{O}%
_{0,2}^{0,1}})^{(2)}=(\Delta ^{\!\!\mathcal{O}_{2,0}^{1,0}})^{(2)}=1/8\,\,\,.
\end{equation}
\noindent  Recall that, in the limit $\sigma\rightarrow \infty $ performed in section \ref{emt}
we obtained two copies of the thermally perturbed  Ising model  for which the 2-particle
form factor (\ref{2par}) is the only non-vanishing  one 
related to  the energy momentum tensor. Hence,  in this limit, the
sum over all particle types in (\ref{dcorr}) will receive only contributions from terms
involving particles of the same type or, in other words, only the 2-particle contribution will occur.
Consequently there will be two equal  contributions, namely 1/16 from $F_2^{\mathcal{O}| ++}$
and 1/16 from  $F_2^{\mathcal{O}| --}$, such that the operator $\mathcal{O}_{0,0}^{0,0}$ plays the role of the disorder operator, as we expected.

To distinguish the operators $\mathcal{O}_{0,0}^{0,0}$, $%
\mathcal{O}_{0,2}^{0,1}$, $\mathcal{O}_{2,0}^{1,0}$ from each other we have
to proceed to higher particle contributions. At present there exist no
analytical arguments for this and we therefore resort to a brute force
numerical computation.

\noindent Denoting by $(\Delta ^{\!\!\mathcal{O}})^{(n)}$ the contribution
up to the $n$-th particle form factor, our numerical Monte Carlo integration
yields 
\begin{eqnarray}
\qquad (\Delta ^{\!\!\mathcal{O}_{0,0}^{0,0}})^{(4)} &=&0.0987\qquad (\Delta
^{\!\!\mathcal{O}_{0,0}^{0,0}})^{(6)}=0.1004,  \label{98} \\
\qquad (\Delta ^{\!\!\mathcal{O}_{0,2}^{0,1}})^{(4)} &=&0.0880\qquad (\Delta
^{\!\!\mathcal{O}_{0,2}^{0,1}})^{(6)}=0.0895, \label{100} \\
\qquad (\Delta ^{\!\!\mathcal{O}_{2,0}^{1,0}})^{(4)} &=&0.0880\qquad (\Delta
^{\!\!\mathcal{O}_{2,0}^{1,0}})^{(6)}=0.0895\,\,.  \label{99}
\end{eqnarray}

\noindent We shall be content with the precision reached at this point, but
we will have a look at the overall sign of the next contribution. From the
explicit expressions of the 8-particle form factors we see that for $%
\mathcal{O}_{0,0}^{0,0}$ the next contribution will reduce the value for $%
\Delta $. For the other two operators we have several contributions with
different signs, such that the overall value is not clear a priory. In this
light, we conclude that the operators $\mathcal{O}_{0,0}^{0,0}$, $\mathcal{O}%
_{0,2}^{0,1}$, $\mathcal{O}_{2,0}^{1,0}$ all possess conformal dimension $%
1/10$ in the UV-limit. Unfortunately, the values for the latter two operators do
not allow such a clear-cut deduction as for the first one. Nonetheless,
 we base our statement on the knowledge of the operator content of the 
conformal field theory and confirm them 
also by elaborating directly on  (\ref{ultra}) and (\ref{corr}).

\begin{figure}[!h]
\begin{center}
\includegraphics[width=9.3cm,height=12.65cm,angle=-90]{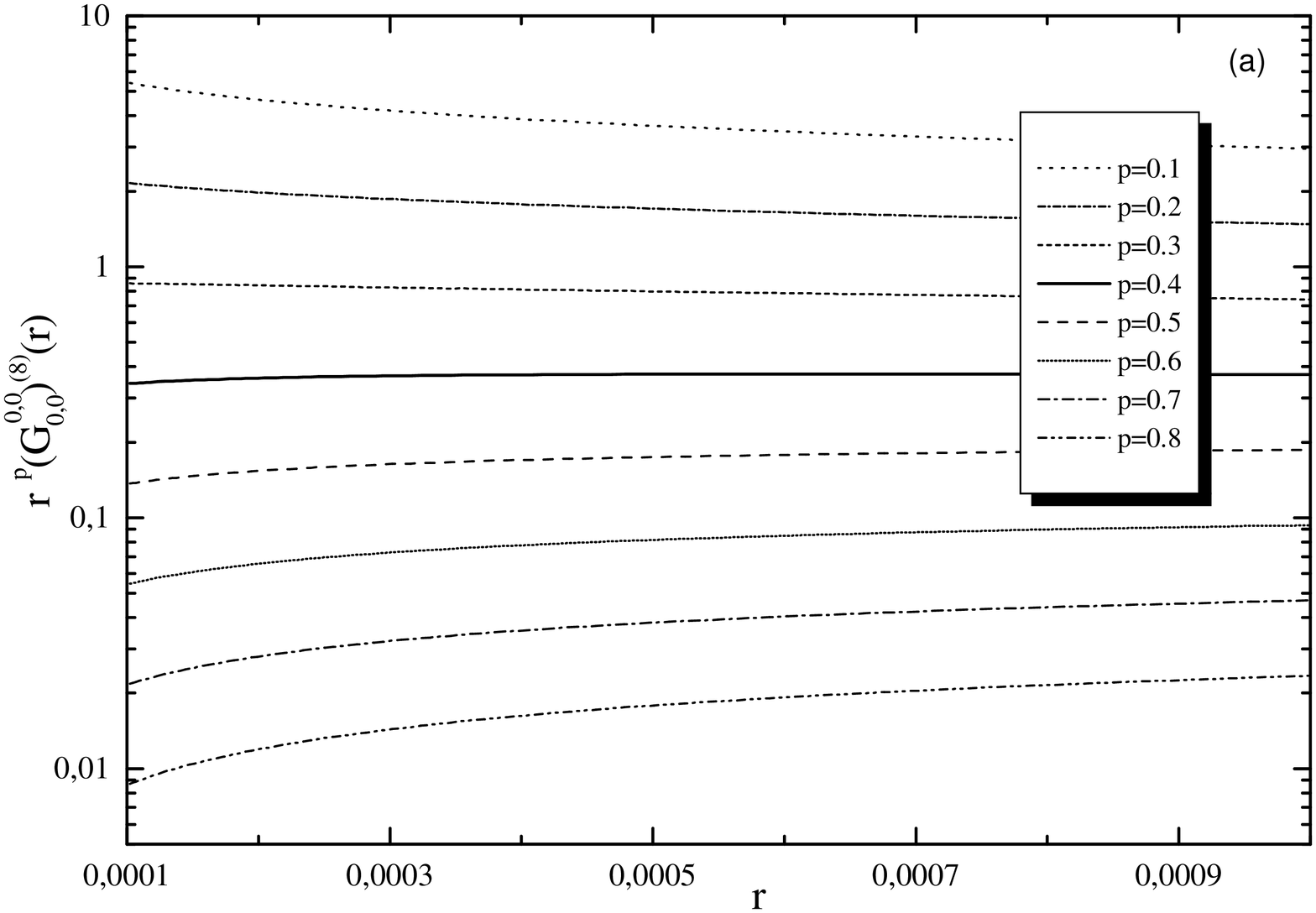}

\includegraphics[width=9.3cm,height=12.65cm,angle=-90]{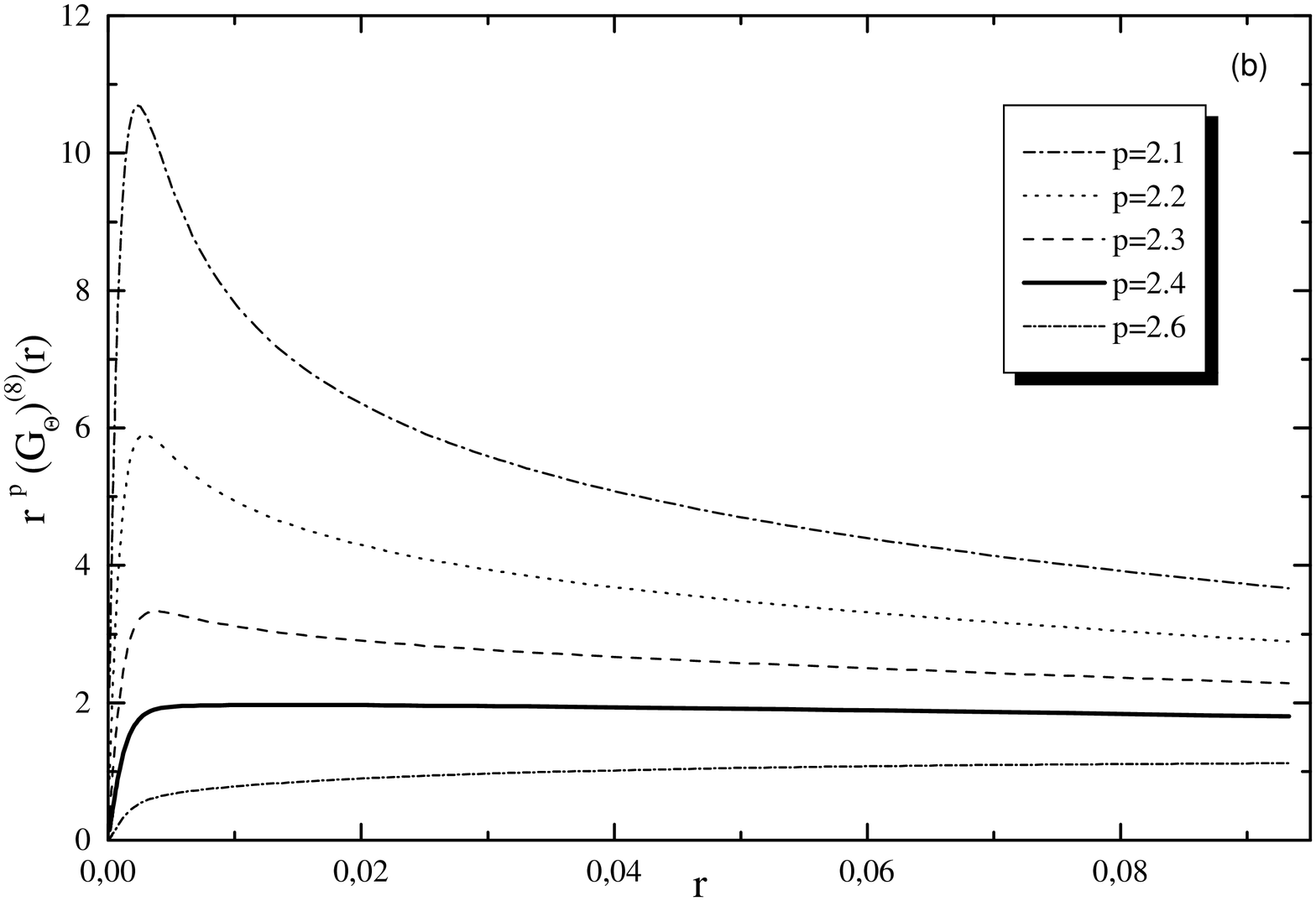}
\caption{{\bf (a)} Rescaled correlation function 
$\protect{G_{0,0}^{0,0}(r):=\left\langle \mathcal{O}_{0,0}^{0,0}(r)
\mathcal{O}_{0,0}^{0,0}(0)\right\rangle/{\langle\mathcal{O}_{0,0}^{0,0}\rangle }^2}$
summed up to the eight particle contribution as a function of $r$. 
{\bf (b)} Rescaled correlation function 
$\protect{(G_{\hat{\Theta}})^{(8)}(r):=\left\langle \hat{\Theta} (r)\hat{\Theta} (0)\right\rangle}$ summed up to
the eight particle contribution as a function of $\protect{r}$.}   
\label{f22}
\end{center}
\end{figure}

\subsection{$\protect{\Delta} $ from correlation functions}
\label{dff}
\indent \ \
In the light of the results of the previous section, it is clear
we can not use the $\Delta$-sum rule for a large class of operators of 
our model. Therefore we may resort to the study of the UV-behaviour of the
two-point functions (\ref{ultra}), as explained in subsection \ref{uvffs}.
We start by not assuming anything  about the conformal dimension of the
operator $\mathcal{O}$ and multiply its two-point correlation function (\ref
{corr}) by $r^{p}$ with $p$ being some arbitrary power. Once this
combination behaves as a constant in the vicinity of $r=0$ we take this
value as the first non-vanishing three-point coupling divided by the vacuum
expectation value of $\mathcal{O}$ and $p/4\,$as its conformal dimension.
This means even without knowing the vacuum expectation value we have a
rationale to fix $p$, but we can not determine the first term in (\ref{ultra1}
). Figure \ref{f22} (a) exhibits this analysis for the operator 
$\mathcal{O}_{0,0}^{0,0}$
up to the $8$-particle contribution and we conclude from there that its
conformal dimension is $1/10$.

The result of the same type of analysis for the energy momentum tensor is
depicted in figure \ref{f22} (b), from which we deduce the conformal dimension $3/5$.
Recalling that the energy momentum tensor is proportional to
the dimension of the perturbing field \cite{Cardypert} this is 
precisely what we expected to find. 

Furthermore, we observe that the relevant interval for $r$ differs by two
orders of magnitude, which by taking the upper bound for the validity of (%
\ref{ultra}) into account should amount to $C_{\frac{1}{10}\frac{1}{10}0}C_{%
\frac{3}{5}\frac{3}{5}\frac{1}{10}}/(C_{\frac{3}{5}\frac{3}{5}0}C_{\frac{1}{%
10}\frac{1}{10}\frac{1}{10}})\sim \mathcal{O}(10^{-2})$. Since to our
knowledge these quantities have not been computed from the conformal side,
this inequality can not be double checked at this stage.

\begin{figure}[!h]
\begin{center}
\includegraphics[width=9.4cm,height=12.65cm, angle=-90]{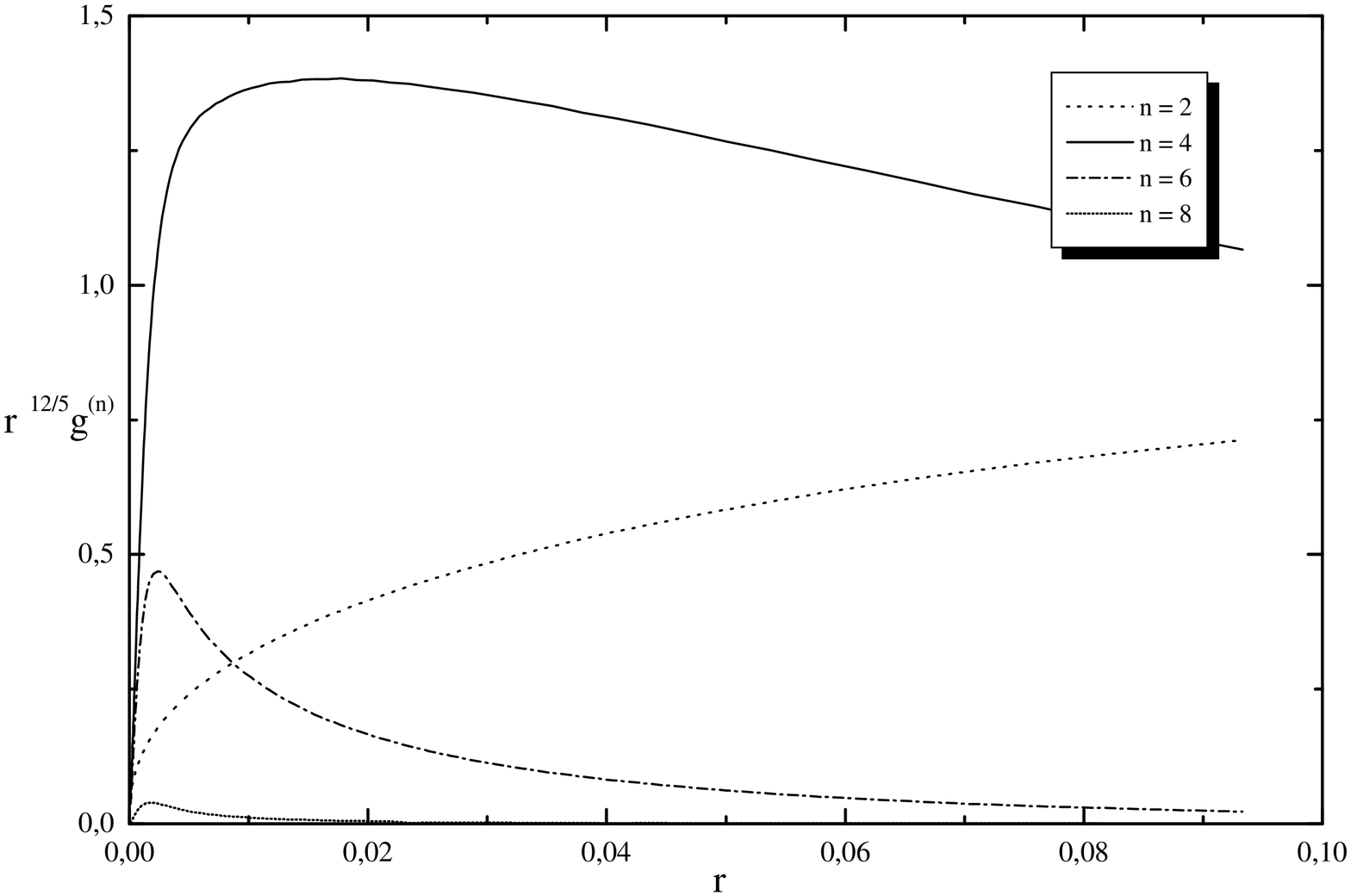}
\caption{Rescaled individual $n$-particle
contribution $\protect{g^{(n)}(r)}$ to the correlation function of  $\protect\hat{\Theta}$ 
as a function of $\protect{r}$.}
\label{f24}
\end{center}
\end{figure}

In figure \ref{f24} we also exhibit the individual $n$-particle contributions to the energy momentum tensor correlation function.
Excluding the two particle contribution, these data also confirm the
proportionality of the $n$-th term to ($\log (r))^{n}$. It is also interesting to notice
that for small $r$ the 4-, and 6-particle contributions are higher than the 2-particle
one which confirms the need  of adding more and more terms in this region and the convenience 
of a more rigorous investigation of the convergence of the series (\ref{corr}). 

We have carried out similar analysis for the other solutions we have
constructed and report our findings in table \ref{t1}. We observe that the
combination of the vacuum expectation value times the three-point coupling
for these operators differ, which is the prerequisite for unraveling the
degeneracy.

\section{RG-flow with unstable particles}
\label{rgflow}
\indent \ \
Renormalisation group methods have been developed originally \cite{GL} to
carry out qualitative analysis of regions of quantum field theories which
are not accessible by perturbation theory in the coupling constant. 
Having computed  in section \ref{cff} the ultraviolet Virasoro central charge associated to the
$SU(3)_2$-HSG model and confirmed its consistence with the result achievable  in
the thermodynamic Bethe ansatz context (see section \ref{ctba}),
we wish now to confirm and refine the physical picture emerging from 
the TBA-analysis by means of the former methods.

Recall that, in the TBA-context, 
the relative mass scales between the unstable and stable particles and the stable particles
themselves were investigated  by computing the finite size scaling function  $c(r,\sigma)$. For this
function a ``staircase''  behaviour  (see figure \ref{fig12})  was observed.  
Such behaviour allowed a very interesting
physical picture which we summarise now:

\vspace{0.25cm}
{\bf i)} \,In the UV-limit, when the temperature is very high 
all the particles of the model (for $\sigma$ finite) have masses much lower than the present energy scale
and therefore all of them contribute to the total effective central charge reproducing the expected value
$6/5$. 

\vspace{0.25cm}
 {\bf ii)}\, As soon as the  scaling parameter $r$ becomes higher or, in other words, 
 the temperature is lower we may reach
an energy scale much lower than the mass of the unstable particle so that it  can not be formed
anymore. Hence,  we are left with two non-interacting stable particles which give a contribution of  1/2
each to the scaling function. Being the mass of the unstable particle determined  by the value of the resonance parameter $\sigma$ as explained in subsection \ref{ctba} of chapter \ref{tba}, such a ``decoupling''  will take place at different energy
scales for different values of  $\sigma$. 

\vspace{0.25cm}
  {\bf iii)}\, If the masses of these two stable particles are taken to be very different,
eventually an energy scale for which only one of these particles can be produced may be reached
and the corresponding scaling function flows to the value 1/2.

\vspace{0.25cm}
  {\bf iv)}\,Finally, in the infrared limit the scaling function tends to zero
 for  purely massive QFT's.

\vspace{0.25cm}

It is interesting to notice that the former physical picture required 
in the TBA-analysis the introduction of a  parameter $r^{\prime}=\frac{r}{2}e^{\sigma/2}$ 
familiar from the discussion of the massless scattering  \cite{triZam}. 
This parameter which had only a formal meaning in
the TBA-context  arises naturally in the renormalisation group context,
supporting the interpretation of this sort of flows as massless flows.

Recall that, for the $SU(3)_2$-HSG model the resonance pole $\theta_R=\mp 
\sigma - i \pi/2$ has an imaginary part $\bar{\sigma}=\pi/2$ so that, for arbitrary resonance parameter
$\sigma$ one can deduce from (\ref{BW1}), (\ref{BW2}) that the condition
$M_{\tilde{u}}\gg \Gamma _{\tilde{u}}$ is not fulfilled.
However, as indicated in section \ref{analit}, this condition only helps for a clearer
identification of the mass parameter. For the HSG-models this condition
starts to hold when the level is large, which indicates that $M_{\tilde{u}}\gg \Gamma _{\tilde{u}}$ 
just in the semi-classical regime \cite{HSGsol, HSGS}. However, since we are not 
considering here the semi-classical regime, the interpretation of the flows observed
as the trace of the presence of unstable particles in the model is not contradictory with
the fact that the mentioned inequality does not hold for the model at hand.

\vspace{0.3cm}

\begin{figure}[!h]
\begin{center}
\includegraphics[width=10.7cm,height=8.5cm]{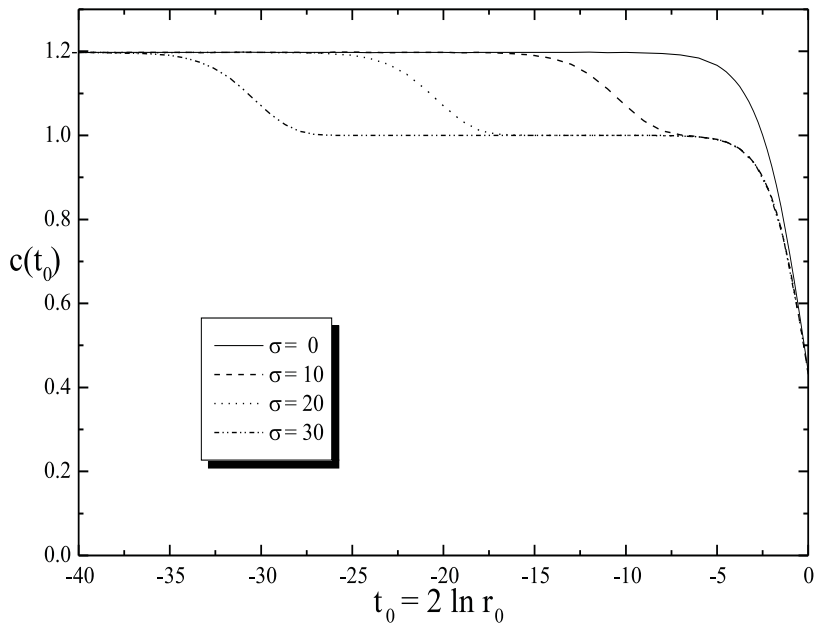}
\caption{Renormalisation group flow for the Virasoro
central charge $\protect{c(r_{0})}$ for various values of the resonance parameter $\protect\sigma$.}
\label{f41}
\end{center}
\end{figure}

Let us now analyse (\ref{cr0}), (\ref{dr0}) and (\ref{Munst}) for the $SU(3)_{2}$-HSG model.
As usual, we carry out the integrals by means of a Monte Carlo
computation. For $c(r_{0})$ we take contributions up to the 4-particle form
factor into account, namely the form factors (\ref{2par}) and  (\ref{4par}),
and display our results in figure \ref{f41}.

Following the renormalisation group flow from the ultraviolet
to the infrared, figure \ref{f41} illustrates the flow from the $SU(3)_{2}/U(1)^{2}$%
- to the $SU(2)_{2}/U(1)\otimes SU(2)_{2}/U(1)$-coset, or in other words, from the values 
$c=6/5$ to the value $c=1/2+1/2=1$, when the unstable
particle becomes massive. This confirms qualitatively the previous
observation of the TBA analysis which we recalled at the beginning of this section and
admits an entirely analogue physical explanation. Consequently, figures \ref{f41} and  \ref{fig12}
are hardly distinguishable, despite the fact that both functions are different. 

Although  for the $SU(3)_2$-HSG model it is not possible to evaluate analytically 
either the finite size scaling function nor the Zamolodchikov $c$-function, one can 
realise that these two functions are different by direct comparison of figures \ref{f41} 
and \ref{fig12}. However, one could think this direct comparison is not quite reliable in 
our case since, for example, when evaluating (\ref{cr0}) we are only taking terms up to the 
4-particle contribution. In order to convince oneself that these two functions are indeed different one can 
look at a simpler model  like  the  free fermion, for which both the finite size scaling function \cite{KM2}
and the $c$-function \cite{DSC} have been obtained analytically and admit expressions
in terms of Bessel functions.
These analytical expressions, found in \cite{KM2} and \cite{DSC} respectively,
show that the finite size scaling function
and $c$-function are clearly different.

\vspace{0.3cm}
As we said, we also wish to analyse  (\ref{Munst}) for the $SU(3)_{2}$-HSG model 
in order to give a physical interpretation to our numerical results depicted in Fig. \ref{f41}.
Taking  the mass scales of
the stable particles to be the same, i.e. $m_{+}=m_{-}=m$ we want to compute now,
for different choices of the resonance parameter $\sigma$,
the values of the RG-parameter, $t_u$, at which the unstable particle
becomes effectively massive (see paragraph before Eq. (\ref{dr})). 
As we can see in Fig. \ref{f41}, the flow between the two cosets is smooth and takes place over some
range of $t_{0}$. For this reason we have to select one particular point $t_{u}$ in the mentioned
range. As already indicated in general in subsection \ref{cdflows}, 
it is convenient to identify $t_u$ as the
point at which $c(t_{0})$ is half the difference between the two coset values of $c$. 
Indeed we find 
\begin{equation}
(t_u, \sigma) \approx (-30.8, 30), (-20.8, 20), ( -10.8,  10).
\label{umass}
\end{equation}
 It is clear from Fig. \ref{f41} that, since the overall shape of
the curves between two values of $c$ is identical for different values of $%
\sigma $, any other value in the interval would lead to the same results in
comparative considerations. Analogously to situation encountered for the stable particles, for the unstable
particles the RG-flow is indeed achieved by $M_{\tilde{c}}\rightarrow r_u M_{\tilde{c}}$, 
where $t_u=2 \ln r_u$,  $M_{\tilde{c}}$ is the mass of the unstable particle given by Eq. (\ref{Munst}) and
the combination $M_{\tilde{c}}(t_u, \sigma):=r_u M_{\tilde{c}}$ must be interpreted as the `effective' mass
of the unstable particle with respect to the RG-group energy scale. 

Actually, according to Eq. (\ref{Munst}),
and taking into account that, the values
of $t_u$ given in (\ref{umass}) should be determined by the condition,
\begin{equation}
M_u (t_u, \sigma)= r_u M_{\tilde{c}}\approx m \Rightarrow \frac{1}{\sqrt{2}}\,e^{(|\sigma |+t_u)/2} \approx 1,
\label{lu2}
\end{equation}
\noindent which just expresses the outlined assertion that the unstable particle
starts `contributing' to the value of the $c$-function or  equivalently,  can be excited,
when its associated mass $M_{\tilde{c}}$ is of order $m/r_u$, that is, the energy scale fixed
by the RG-parameter. The outcome of the  evaluation of (\ref{lu2}) for the values
(\ref{umass}) is however $0.47$ instead of 1. Nonetheless, 
the latter result should not be taken too literally since the point $t_{u}$
is only chosen because it can be easily fixed. 

\vspace{0.25cm}

\begin{figure}[h!]
\begin{center}
\includegraphics[width=10.7cm,height=8.5cm]{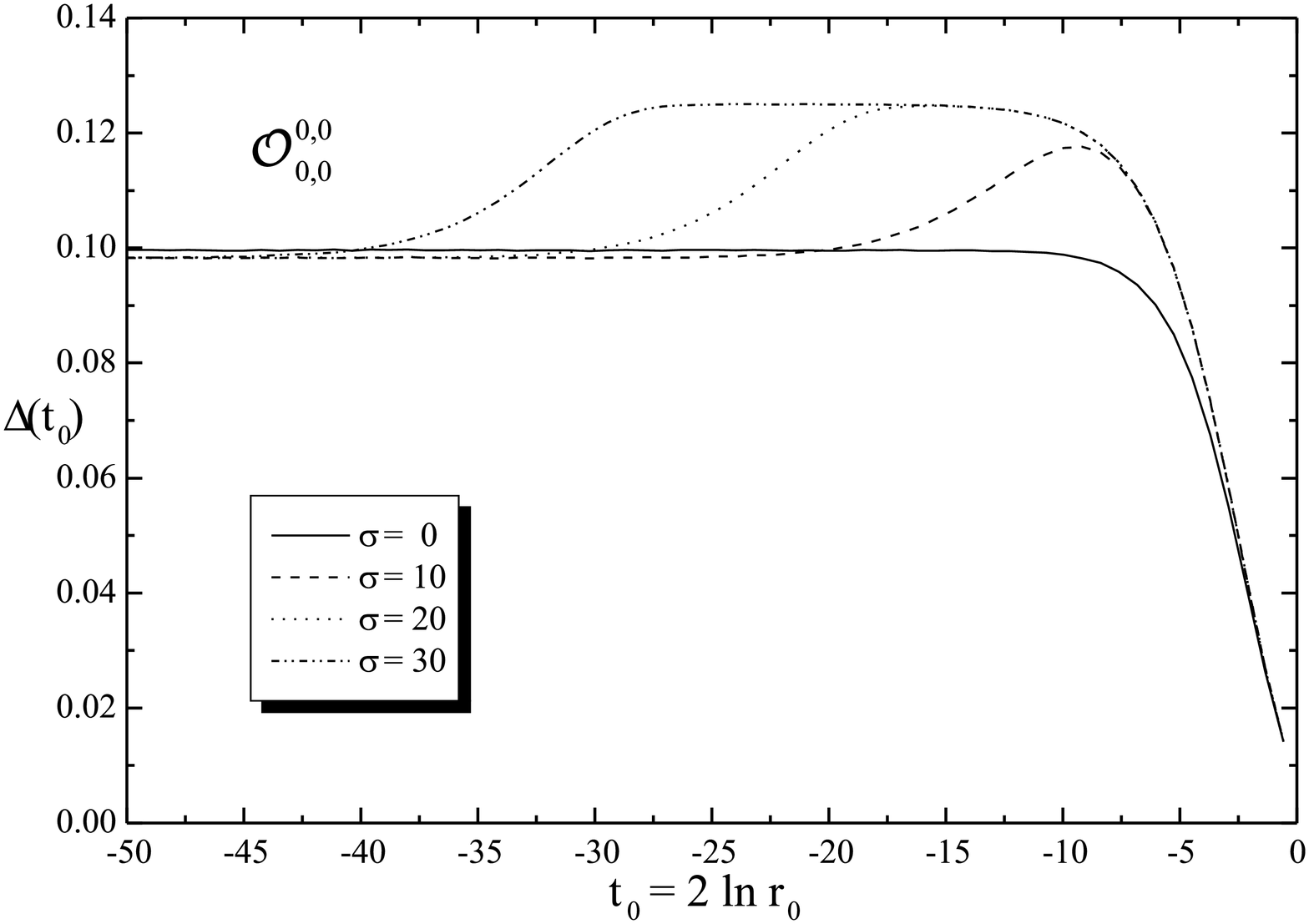}
\caption{Renormalisation group flow for the conformal
dimension $\protect{\Delta (r_{0})}$ of the operator 
$\protect{{\cal O}_{0,0}^{0,0}}$ for
various values of the resonance parameter $\protect\sigma$.}
\label{f42}
\end{center}
\end{figure}

\begin{figure}[!h]
\begin{center}
\includegraphics[width=10.7cm,height=8.5cm]{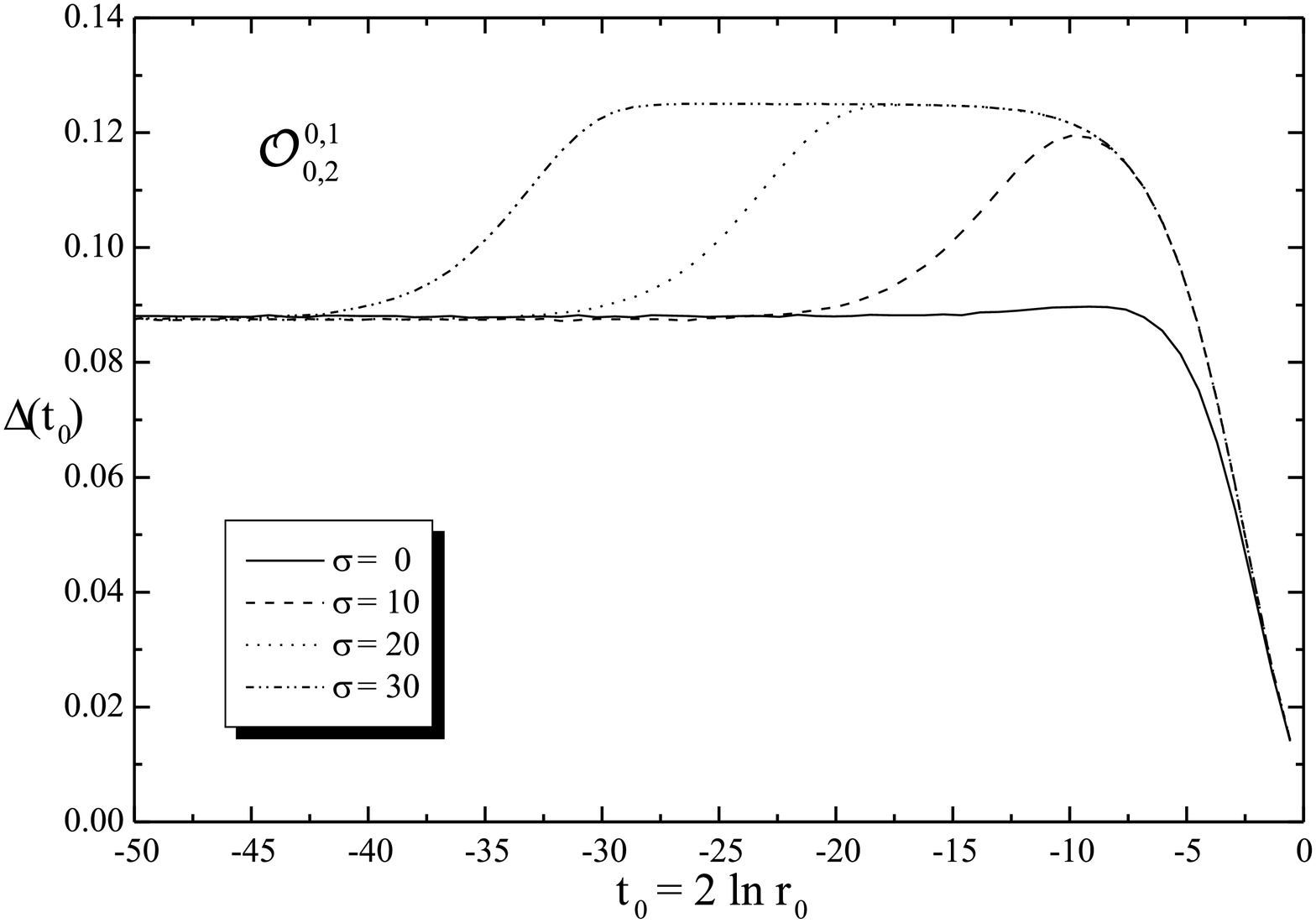}
\caption{Renormalisation group flow for the conformal
dimension $\protect{\Delta (r_{0})}$ of the operator 
$\protect{{\cal O}_{0,2}^{0,1}}$ for various values of the resonance parameter $\protect\sigma$.}
\label{f43}
\end{center}
\end{figure}

For the evaluation of the scaled conformal dimension (\ref{dr0}) we proceed
similarly. For the solutions corresponding to the operators ${\cal O}%
_{0,0}^{0,0}$, ${\cal O}_{0,2}^{0,1}${\small \ }and ${\cal O}_{2,0}^{1,0}$,
whose conformal dimension in the UV-limit was identified in section \ref{dsum}
to be $1/10$, we take up to the 6-particle form factors into account
 (see appendix B for their explicit expressions). 
For the former two operators our results are presented in figures \ref{f42} and \ref{f43}.

We observe that the conformal dimension of the operator 
${\cal O}_{0,0}^{0,0}$ flows to the value $1/8$, which is twice the conformal
dimension of the disorder operator $\mu $ in the Ising model. The factor $2$
is expected from the mentioned coset structure, i.e. we find two copies of $%
SU(2)_{2}/U(1)$. The nature of the operator is also anticipated, since by
construction the form factors $F_{n}^{{\cal O}_{0,0}^{0,0}}$ of the $SU(3)_{2}$%
-HSG model coincide precisely with $F_{n}^{\mu }$ of the thermally
perturbed Ising model when the number of particles of type ``+ '' or `` - '' is zero. It is also
clear that we could alternatively obtain (\ref{umass}) from the analysis of $%
\Delta (r_{0})$.

Despite the fact that the explicit expressions for  the form factors of $%
{\cal O}_{0,2}^{0,1}${\small \ }and ${\cal O}_{2,0}^{1,0}${\small \ }differ (see appendix B)
the values of $\Delta (r_{0})$ are hardly distinguishable and we therefore
omit the plots for the latter case. We also note the previously observed
fact, that the higher particle contributions for the latter
operators are more important than for ${\cal O}_{0,0}^{0,0}$, which explains
the fact that the starting point at the ultraviolet fixed point is not quite 
$0.1$ but $0.0880$ as observed in (\ref{99}). The operators also flow to the value $1/8$, such that the degeneracy of the $SU(3)_{2}$-HSG model disappears surjectively when the unstable
particle becomes massive.

\vspace{0.3cm}
In conclusion, the RG-flow
of Zamolodchikov's $c$-function  and of the conformal dimensions of various operators, 
allows a clear physical interpretation which is in complete agreement
with the interpretation of the ``staircase'' pattern observed for the scaling function computed in the TBA.

\section{Summary of results and open problems}
\label{sum4}
\indent \ \
Concerning the main objective we addressed in this chapter, we draw the overall conclusion
that the outcome of  the form factor analysis for the $SU(3)_2$-HSG model confirms the physical
picture obtained by means of the thermodynamic Bethe ansatz in chapter \ref{tba}.  The 
Virasoro central charge of the underlying CFT $(c=6/5)$  has been obtained 
by means of  Zamolodchikov's $c$-theorem \cite{ZamC} and qualitatively
 the same ``staircase''  behaviour \cite{staircase} observed for the finite 
size scaling function (\ref{scale}) as a function of the energy scale 
in chapter \ref{tba} has been found  for the renormalisation group flow of  
Zamolodchikov's $c$-function as a function of the renormalisation group parameter $r_0$. In this context, 
the variable $m e^{|\sigma|/2}$, originally introduced in \cite{triZam}
to describe the massless scattering, arises naturally through formula (\ref{Munst}),
which supports the idea, also present in the TBA-context, that the sort of flows observed, related to the presence of unstable particles in the spectrum, can be understood as massless flows. 

\vspace{0.3cm}

In addition to the TBA-results, our form factor analysis allowed for the identification of, at least an 
important part of  the operator content of the $SU(3)_2$-HSG model via the calculation
 of the corresponding  ultraviolet conformal dimension of those operators for which all
 the $n$-particle form factors had been previously computed. In particular, the conformal
 dimension of the perturbing operator ($\Delta=3/5$) was obtained by studying the ultraviolet
 behaviour of the two-point function of the energy momentum tensor. In this light, it can be stated 
 that solutions of the form factor consistency
equations can be identified with operators in the underlying ultraviolet
conformal field theory. In this sense one can give meaning to the operator
content of the integrable massive model. Being the mentioned identification uniquely based on 
the values of the ultraviolet conformal dimensions there is the problem that
once the conformal field theory is degenerate in
this quantity, as it is the case for the model we investigated, the
identification can not be carried out in a one-to-one fashion and therefore
the procedure has to be refined. 
In principle this would be possible by
including the knowledge of the three-point coupling of the conformal field
theory and the vacuum expectation value into the analysis. The former
quantities are in principle accessible by working out explicitly the
conformal fusion structure, whereas the computation of the latter still
remains an open challenge. In fact what one would like to achieve ultimately
is the identification of the conformal fusion structure within the massive
models.  Considering the total number of operators present in the conformal
field theory (see table \ref{t2}), one still expects to find additional solutions,
in particular the identification of the fields possessing conformal
dimension $1/2$ is an outstanding problem.

\vspace{0.3cm}

With respect to the explicit calculation of conformal dimensions, 
technically we have confirmed that the sum rule (\ref{deldel}) is clearly
superior to the direct analysis of the correlation functions. However, it has also the inconvenient that
in theories with internal symmetries,  as the one at hand, it only applies for certain operators. It would
therefore be highly desirable to develop arguments which also apply for
theories with internal symmetries and possibly to resolve the mentioned
degeneracies in the conformal dimensions.

\vspace{0.3cm}

As mentioned in several occasions throughout this chapter,
 the question of how to identify the operator content
of a massive integrable quantum field theory is left unanswered in the TBA-context, apart
from the identification of the conformal dimension of  the  perturbing field, sometimes related to the  periodicity of the so-called $Y$-systems  \cite{TBAZamun}. From that point of view,
one can say that the form factor program provides more information
about the underlying CFT than the TBA and therefore a more rigorous
check of the S-matrix proposal \cite{HSGS}. In addition, our analysis
contributes to the further development of the QFT associated to the HSG-models.

\vspace{0.3cm}

At the mathematical level, a closed formula (\ref{qpar}) for all $n$-particle 
 form factors associated to a large class of operators has been found.  This 
formula is given in terms of building blocks which can be expressed both as 
determinants whose entries are elementary symmetric polynomials (\ref{sss})
 or by means of an integral representation (\ref{integral}).
However, it remains  an open question, whether the general solution procedure
presented in section \ref{solution} can be generalised to the degree that 
determinants of the type (\ref{sss}) will serve as generic building blocks of form
factors. It remains also for us to be understood  how an integral
 representation  of the type (\ref{integral}) might
be used in practise, for instance,  to formulate rigorous  proofs of
 the type presented in section \ref{proof}.
In general, it  would be very interesting to exploit this integral representation for the same
purposes we have used the determinant formula (\ref{sss}) namely, providing general proofs or analysing
the cluster property. Another interesting open problem  is to find out the 
precise relationship between this integral representation, which
 involves contour integral in the $x=e^{\theta}$-plane, 
and the integral representation used for instance in \cite{BFKZ} which involved 
contour integrals in the $\theta$-plane.

It would be also desirable to put further constraints on the solutions
to the form factor consistency equations by means of other arguments,
that is exploiting the symmetries of the model,
formulating quantum equations of motion, possibly performing perturbation
theory etc...

\vspace{0.3cm}

Concerning the momentum space cluster property, our analysis also provides remarkable results.  
It has been proven in section \ref{mscp} that it
does not only constrain the solutions to the form factor consistency
equations but also serves as a construction principle to obtain new solutions. In particular, for the operators
itemized in table \ref{t1} we have verified  that starting with the solutions for  any of the first 9 operators one could re-construct all the others, obtained initially by solving the recursive equations  (\ref{Qrec}).
Clearly, it would be very interesting to develop arguments which allowed us to reach a level
of understanding  of this property similar to the one we have for the rest of the form factor consistency equations \cite{Kar, Smir,Zamocorr,YZam,BFKZ}. 

\vspace{0.3cm}

Regarding the RG-analysis carried out in section \ref{rgflow}, we have recovered the
TBA-picture extracted from the computation of the finite size scaling function. We find
that the $c$-function defined by means of Zamolodchikov's $c$-theorem encodes the
same physical information than the scaling function or Casimir energy computed in the
TBA. Therefore, it would be extremely desirable to elaborate on the precise relationship
 between these two functions which providing  the same physical information are however different.
Being the finite size scaling function and  Zamolodchikov's $c$-function defined  in
  such, at first sight, disconnected  contexts, the nature of such relationship is not clear a priori. 
Also the relation to the
intriguing proposal in \cite{Foda} of a renormalisation group flow between
Virasoro characters remains unclarified. 
It is also relevant the fact that, although in the 
``excited'' TBA-framework \cite{marfen, zamdota}
the conformal dimensions of certain fields of
the underlying CFT different from the perturbing field can be identified, 
 the analogue  of  $\Delta (r_{0})$ still needs to be found in the 
TBA as well as in the context of \cite{Foda}.

\vspace{0.3cm}

To finish this chapter, it is worth emphasising  that,
apart from providing a double-check of the results of our
TBA-analysis, the form factor program developed for the $SU(3)_2$-HSG model  has also contributed
to the general understanding of the form factor program itself in various aspects like, for instance, the
finding of general solutions to recursive problems of the type (\ref{Qrec}), the identification of the operator content of a 1+1-dimensional massive integrable QFT or, the understanding of the momentum space cluster property (\ref{cluster}) as a construction principle of new solutions to the form factor consistency equations. Concerning the physical picture
presented  for the $SU(3)_2$-homogeneous sine-Gordon model in \cite{HSG2, HSGsol, HSGS},
one can surely claim that it  rests now on quite firm ground and 
it is now a challenge to extend the results
of this chapter, to higher rank Lie groups. As a first step in this direction, 
a generalisation of the form factor construction
and the renormalisation group analysis for all  the $SU(N)_2$-HSG 
models shall be presented in the next chapter.

\chapter{Generalizing the form factor program to the $ \protect{SU(N)_{2}}$-HSG models.}
\label{ffs2}
\indent \ \
For most integrable quantum field theories in 1+1 space-time dimensions it
remains an open challenge to complete the entire bootstrap program, i.e. to
compute the exact on-shell S-matrix, closed formulae for the $n$-particle
form factors, identify the entire local operator content and in particular
thereafter to compute the related correlation functions. In the previous chapter we
investigated  a particular model, the $SU(3)_{2}/U(1)^{2}$-homogeneous 
Sine-Gordon model \cite{HSG, HSG2, HSGsol, Park, HMQH} (HSG), 
for which this task was completed to a large extend. In particular, general formulae for
the $n$-particle form factors related to a large class of local operators were provided. In
order to understand the generic group theoretical structure of the $n$-particle form factor 
expressions and to provide further consistency checks of the S-matrix proposal 
\cite{HSGS} it is highly desirable to extend our analysis to higher rank  Lie algebras 
as well as to higher level.

The extension of the form factor program to higher level  is 
not straightforward since once  $k>2$, stable bound states
 might be formed (see \cite{HSGS}). As a consequence,  
the corresponding form factors possess not only kinematical
 simple poles but also bound state poles, as indicates 
 the so-called  ``bound state residue equation'' \cite{LSZ2} (see also section \ref{gen}
 of the previous chapter) and therefore, the mentioned equation has to be solved in addition
 to all the other form factor consistency equations \cite{Kar, Smir,Zamocorr,YZam, BFKZ}.  
As we have seen, the bound state residue equation establishes a recursive structure between the $(n+1)$- and $n$-particle form factor which we should solve in addition to the familiar recursive equations relating the $(n+2)$- and $n$-particle form factor, whose solution was already very intricate. On this light, 
we expect the form factor program to become much more complicated for models where stable
 bound states are present. Some of the simplest examples of  models containing stable bound states are
the Yang-Lee model \cite{Zamocorr} and the Bullough-Dodd model \cite{BD}.
 The spectrum of the latter model contains a single particle state, say $A$, 
 which can be formed as bound state of itself in a scattering 
process of the type $A \times A \rightarrow A \rightarrow A \times A$.
A closed formula for all $n$-particle form factors of the
elementary field of the Bullough-Dodd model, and for particular values of the effective
coupling constant was found in \cite{deter1}.
 
Consequently, we shall start the generalisation of the analysis performed for the $SU(3)_2$-HSG model
by increasing the rank of the Lie algebra, that is, we intend to investigate 
the $SU(N)_{2}$-HSG models within  the form factor context.

\vspace{0.3cm}
The results presented in this chapter might be found in \cite{CF3}. 
Our analysis is organised as follows:

\vspace{0.2cm}

In section \ref{themodel} we report the main 
characteristics of the $SU(N)_2$-HSG model. In particular the basic data characterising the corresponding underlying CFT, the scattering matrix and the stable and unstable particle content. Thereafter, in section \ref{ffss} we start  the form factor construction in the usual way,  by providing a general ansatz for the form factor solutions. We derive the corresponding recursive equations and  determine the minimal form factors.  In section \ref{solution2} we systematically construct all $n$-particle solutions to the form factor consistency equations. These solutions correspond to a large class of local operators, although they do not fulfill the entire  operator content of the model. It turns out that these solutions involve the same sort of building blocks i.e, determinants whose entries are elementary symmetric polynomials, which we found in chapter \ref{ffs}.  It is also possible to formulate a general proof for the $SU(N)_2$-solutions along the same lines pointed out in section \ref{proof}. In section \ref{rg2} we investigate the RG-flow of Zamolodchikov's $c$-function \cite{ZamC}, reproducing the physical picture expected for the HSG-models, namely the decoupling of the theory into two non-interacting systems  whenever one of the resonance parameters becomes much larger than the energy scale considered. The same decoupling is
also observed for  the corresponding Virasoro central charge of the underlying CFT in the deep UV-limit.
 In section \ref{operator} we report  the operator content in terms of primary fields of the underlying CFT of the $SU(4)_2$-HSG model. We also compute the conformal dimension of some operators present
in any $SU(N)_2/U(1)^{N-1}$-WZNW coset theory  as, for instance, the perturbing operator. In both
cases we make use of the general formula derived  in \cite{Gep} for the conformal dimensions
of  primary fields of  the  $G_k$-parafermion theories.
 In subsection \ref{rgoperator} we investigate the RG-flow of  the conformal dimension of a certain local
operator  by means of the $\Delta$-sum rule \cite{DSC} and identify its ultraviolet central charge
in the limit  when the renormalisation group parameter $r_0 \rightarrow 0$. 
In subsection \ref{pp} we identify the conformal
dimension of the perturbing operator in the $SU(4)_2$- and $SU(5)_2$-cases by investigating 
the UV-behaviour of the two-point function of the trace of the energy momentum tensor $\Theta$.
Finally we report our main conclusions and point out some open questions in section \ref{con5}.

\section{The $\protect{SU(N)_2}$-homogeneous sine-Gordon models.}
\label{themodel}
\indent \ \
The particularisation of  the results presented in chapter \ref{ntft} for the generality of the 
HSG models allows for viewing  the $SU(N)_2$-HSG model
as the perturbation of a  gauged  WZNW-coset  $SU(N)_2/U(1)^{N-1}$ whose associated 
Virasoro central charge is given by
\begin{equation}
c_{SU(N)_{2}/U(1)^{N-1}}=\frac{N(N-1)}{(N+2)},
\label{char}
\end{equation}
by means of an operator of conformal dimension 
\begin{equation}
\Delta =\frac{N}{(N+2)}.
\label{perN}
\end{equation}
\noindent The theory
possesses already a fairly rich particle content, namely $N-1$
asymptotically stable particles characterised by their mass scales $m_{i}$, $i=1,\cdots, N-1$ 
 and $N-2$ unstable particles whose energy scale is characterised by the resonance
parameters $\sigma _{ij}=-\sigma_{ji}$, $i,j=1, \cdots, N-1$. As found in \cite{HSGsol, HSGS}, 
the stable particles can be related in a one-to-one fashion to the vertices of the $SU(N)$-Dynkin
diagram and  one can associate to the link between vertex $i$ and $j$ the $N-2$
linearly independent resonance parameters $\sigma _{ij}$. 

The physical picture provided in \cite{HSG, HSG2, HSGsol} and \cite{HSGS} for
the homogeneous sine-Gordon models and checked  in chapters \ref{tba} and \ref{ffs} for the
$SU(3)_2$-HSG model indicates that,  once an  unstable particle becomes extremely heavy the original
coset decouples into a direct product of two cosets different from the
original one in the following way

\begin{equation}
\lim\limits_{\sigma _{i,i+1}\rightarrow \infty }SU(N)_{2}/U(1)^{N-1}\equiv SU(i+1)_{2}/U(1)^{i}\otimes SU(N-i)_{2}/U(1)^{N-i-1}\,.  
\label{flow}
\end{equation}

The latter equation means that, when investigating the renormalisation group flow
of  Zamolodchikov's $c$-function, in a similar fashion we did for the $SU(3)_2$-HSG
model, we expect to observe starting in the ultraviolet limit, 
a  flow between  the value (\ref{char}) of the Virasoro central charge and
the value obtained  from the sum of the Virasoro central charges associated to the cosets arising on
the r.h.s. of Eq. (\ref{flow}). Equivalently, when studying the RG-flow 
of the conformal dimension of any local operator of the $SU(N)_2$-HSG
model computed by means of the $\Delta$-sum  rule \cite{DSC}, we expect to identify
the flow between the operator contents of the HSG models constructed as 
perturbations of  the  WZNW-cosets arising on the l.h.s. and r.h.s. of Eq. (\ref{flow}).

In other words,  we may summarize the flow along the renormalisation group
trajectory with increasing RG-parameter $r_{0}$ to cutting the related
Dynkin diagrams at decreasing values of the $\sigma $'s. For instance taking 
$\sigma _{i,i+1}$ to be the largest resonance parameter at some energy scale
the following cut takes place:

\begin{center}
\unitlength=1.0cm 
\begin{picture}(8.20,1.0)(-0.5,0.00)
\put(0.00,0.00){\circle*{0.2}}
\put(-0.10,-0.50){$\alpha_1$}
\put(0.00,-0.01){\line(1,0){1.}}
\put(1.00,0.00){\circle*{0.2}}
\put(0.90,-0.50){$\alpha_2$}
\put(0.20,0.30){$\sigma_{12}$}
\put(1.00,-0.01){\line(1,0){0.5}}
\put(1.6,0.00){$\ldots$}
\put(2.70,0.30){$\sigma_{i,i+1}$}
\put(2.2,-0.01){\line(1,0){0.5}}
\put(2.70,0.00){\circle*{0.2}}
\put(2.60,-0.50){$\alpha_i$}
\put(2.70,-0.01){\line(1,0){1.0}}
\put(3.70,0.00){\circle*{0.2}}
\put(3.10,-1.20){$\Downarrow$}
\put(3.60,-0.50){$\alpha_{i+1}$}
\put(3.70,-0.01){\line(1,0){0.5}}
\put(4.3,0.00){$\ldots$}
\put(4.90,-0.01){\line(1,0){0.5}}
\put(5.40,0.00){\circle*{0.2}}
\put(5.30,0.30){$\sigma_{N-2,N-1}$}
\put(5.40,-0.01){\line(1,0){1.0}}
\put(6.40,0.00){\circle*{0.2}}
\put(6.30,-0.50){$\alpha_{N-1}$}
\put(5.20,-0.50){$\alpha_{N-2}$}
\end{picture}

\unitlength=1.0cm 
\begin{picture}(8.20,1.5)(-0.5,0.00)
\put(0.00,0.00){\circle*{0.2}}
\put(-0.10,-0.50){$\alpha_1$}
\put(0.00,-0.01){\line(1,0){1.}}
\put(1.00,0.00){\circle*{0.2}}
\put(0.90,-0.50){$\alpha_2$}
\put(1.00,-0.01){\line(1,0){0.5}}
\put(1.6,0.00){$\ldots$}
\put(2.2,-0.01){\line(1,0){0.5}}
\put(2.70,0.00){\circle*{0.2}}
\put(2.60,-0.50){$\alpha_i$}
\put(3.70,0.00){\circle*{0.2}}
\put(3.60,-0.50){$\alpha_{1}$}
\put(3.70,-0.01){\line(1,0){0.5}}
\put(4.3,0.00){$\ldots$}
\put(4.90,-0.01){\line(1,0){0.5}}
\put(5.40,0.00){\circle*{0.2}}
\put(5.40,-0.01){\line(1,0){1.0}}
\put(6.40,0.00){\circle*{0.2}}
\put(6.20,-0.50){$\alpha_{N-i-1}$}
\end{picture}
\end{center}
\vspace*{1.2cm}

Resorting to  the usual expressions for the coset central charge \cite{GKO} of the WZNW-cosets (\ref{flow}), the decoupled system has the central charge

\begin{eqnarray}
\lim\limits_{\sigma _{i,i+1}\rightarrow \infty
}\!\!\!\!c_{SU(N)_{2}/U(1)^{N-1}}&=&c_{SU(i+1)_{2}/U(1)^{i} \otimes SU(N-i)_{2}/U(1)^{N-i-1}}=
\nonumber \\
c_{SU(i+1)_{2}/U(1)^{i}}+c_{ SU(N-i)_{2}/U(1)^{N-i-1}}&=&N-5+\frac{6(N+5)}{(N+2-i)(3+i)}\,.
\label{cdec}
\end{eqnarray}

As an input necessary  for the computation of form factors and correlation
functions thereafter we must introduce now the exact scattering matrix proposed in \cite{HSGS}
to describe the scattering theory of the $SU(N)_2$-HSG models. The
two-particle S-matrix describing the scattering of two stable particles of
type $i$ and $j$, with $1\leq i,j\leq N-1$ related to this model,
as a function of the rapidity $\theta $, may be written as 

\begin{equation}
S_{ij}(\theta )=(-1)^{\delta _{ij}}\left[ c_{i}\tanh \frac{1}{2}\left(
\theta +\sigma _{ij}-i\frac{\pi }{2}\right) \right] ^{I_{ij}}\,.
\label{ZamN}
\end{equation}
\noindent  Here, to make notation more compact, 
 we have  introduced  the incidence
matrix of the $SU(N)$-Dynkin diagram,  denoted by $I$.  
This  matrix is defined in terms of the Cartan matrix $K$ as $I=2-K$. 
 In particular, for the $su(N)$ Lie algebra at hand we have 
\begin{equation}
I_{ij}=\delta_{i, j-1}+ \delta_{i, j+1}.
\label{Inc}
\end{equation}
\noindent Both in (\ref{Inc}) and in (\ref{ZamN}) 
we used the usual Kronecker delta defined as,
\begin{displaymath}
\delta_{ij}=\left\{
\begin{array}{ll}
0 & \text{for}\,\, i \neq j,\\
1 & \text{for}\,\,i=j
\end{array}\right.
\end{displaymath}
The parity breaking which is characteristic for the HSG models and manifests
itself by the fact that $S_{ij}\neq S_{ji}$, takes place through the
resonance parameters $\sigma _{ij}=-\sigma _{ji}$ and the phases encoded in the 
colour value $c_{i} $. The latter quantity arises from a partition of the $SU(N)$-Dynkin diagram
into two disjoint sets, which we refer to as ``$+$'' and ``$-$''. We then
associate the values $c_{i}=\pm 1$ to the vertices $i$ of the Dynkin diagram
of $SU(N)$, in such a way that no two vertices related to the same set are
linked together. These colour values $c_i$ provide a natural generalisation of the indices 
``$+$'' and ``$-$'' we used in the previous chapter to label the two stable particles associated to
 the $SU(3)_2$-HSG model. They turn out to be also very useful  aiming towards a generalisation of the
form factor analysis performed for the latter model. Likewise we 
could simply divide the particles into odd and even, however, such a division would be specific to $SU(N)$ and the
bi-colouration just outlined admits a generalization to other groups as
well. The resonance poles in $S_{ij}(\theta )$ at $(\theta
_{R})_{ij}=-\sigma _{ij}-i\pi /2$ may be understood in the usual Breit-Wigner 
fashion \cite{BW} as the trace of $N-2$ unstable particles as explained for instance in \cite
{ELOP,HSGS} and chapter \ref{ntft}. Recall also that the mass of the
unstable particle $M_{\tilde{c}}\,\ $formed in the scattering between the
stable particles $i$ and $j$ behaves as $M_{\tilde{c}}\sim m  e^{\left| \sigma
_{ij}\right| /2}$, for $m$ to be the mass scale of the stable particles. 
This fact was observed both in the TBA- and the form factor 
analysis carried out in the previous chapters as well as in the original literature \cite{HSGS}.
There are no poles present in the physical region of the imaginary axis, which
indicates that no stable bound states may be formed and therefore the form factor construction
does not require the use of the mentioned bound state residue equation. Hence, we shall
follow the lines of the analysis performed in chapter \ref{ffs}.

Notice that the two-particle S-matrix  (\ref{ZamN}) reduces to the S-matrix of the thermally
perturbed Ising model whenever the interaction of particles of the same type is considered
\begin{equation}
S_{ii}(\theta)=-1.
\label{is2}
\end{equation}
\noindent
This property will be widely exploited in the course of
our form factor analysis for the same reasons argued before, namely
it serves as a ``seed''  for the recursive problem relating the $(n+2)$- and
$n$-particle form factor we  have to solve in order to determine explicitly the form factors
associated to any local operator of the $SU(N)_2$-HSG model.

It is also clear from the expression of the scattering matrix (\ref{ZamN}), that
whenever a resonance parameter $\sigma _{ij}$ with $I_{ij}\neq 0$
 (which always corresponds to  $i\neq j$) 
goes to infinity, we observe for the corresponding S-matrix that 
\begin{equation}
\lim_{\sigma_{ij}\rightarrow \infty} S_{ij}(\theta)= 1 
\label{trivial}
\end{equation}
and consequently,  we may view the whole system as consisting out of two sets of
particles which only interact freely amongst each other. 
A physical consequence is that  the unstable particle, whose mass behaved as
 $M_{\tilde{c}}\sim m  e^{\left| \sigma_{ij}\right| /2}$, for $m$ to be the mass scale of
the stable particles, and  was created in interaction process between these two
theories before taking the limit, becomes so heavy that it can not be formed anymore at any energy scale.
It can be easily deduced from (\ref{trivial}) and (\ref{is2}) that in the case when all the resonance parameters tend to infinity or all the unstable particle become extremely heavy, we will deal with $N-1$ non-interacting copies of the thermally perturbed Ising model.  Equivalently, when studying
the renormalisation group flow of Zamoldchikov's $c$-function we may reach an energy scale $r_0$, for
which  no formation of unstable particles is possible. Accordingly,  the $c$- function will flow to the value $(N-1)/2$.

\section{Recursive equations and minimal form factors}
\label{ffss}
\indent \ \
We are now in the position to compute the $n$-particle form factors related
to this model, i.e. the matrix elements of a local operator ${\cal O}(\vec{x})$
 located at the origin between a multi-particle {\em in}-state of particles
(solitons) of species $\mu_i$, created by the vertex operators  $V_{\mu_i }(\theta_i )$, and the vacuum.

We proceed in the standard way by solving the form factor
consistency equations \cite{Kar, Smir, Zamocorr, YZam, BFKZ}.
For this purpose we extract explicitly, according to standard procedure, the singularity structure.
Since no stable bound states may be formed during the scattering of two
stable particles the only poles present are the ones associated to the
kinematic residue equations, that is a first order pole for particles of the
same type whose rapidities differ by $i\pi $. Therefore, we parameterise the 
$n$-particle form factors as 
\begin{equation}
F_{n}^{{\cal O}|{\frak {M}}(l_{+},l_{-})}(\theta _{1}\ldots \theta _{n})
=H_{n}^{{\cal O}|{\frak {M}}(l_{+},l_{-})}Q_{n}^{{\cal O}|{\frak {M}}%
(l_{+},l_{-})}({\frak {X})}  
{\frak \,\!\!}\times \prod_{1\leq i<j\leq n}\frac{F_{\text{min}}^{\mu
_{i}\mu _{j}}(\theta _{ij})}{\left( x_{i}^{c_{\mu _{i}}}+x_{j}^{c_{\mu
_{j}}}\right) ^{\delta _{\mu _{i}\mu _{j}}}}\,\,\,,  
\label{fact2}
\end{equation} 
\noindent  in clear analogy to the parameterisation (\ref{fact}) presented in section \ref{solution}.
Aiming towards a universally applicable and concise
notation, it is convenient to collect the particle species $\mu _{1}\ldots
\mu _{n}$ in form of particular sets 
\begin{eqnarray}
{\frak {M}_{i}}(l_{i}) &=&\{\mu \,|\,\mu =i\} \\
{\frak {M}_{\pm }}(l_{\pm }) &=&\bigcup_{i\in \pm }{\frak {M}_{i}}(l_{i}) \\
{\frak {M}}(l_{+},l_{-}) &=&{\frak {M}_{+}}(l_{+})\cup {\frak {M}_{-}}%
(l_{-}).
\end{eqnarray}
\noindent that is, we collect in ${\frak {M}_{i}}(l_{i})$ a set of $l_i$  particles
of the same type `$i$' i.e., a set of $l_i$ particles associated to the same vertex 
$\alpha_i$ in the $SU(N)$-Dynkin diagram.
Analogously, the sets ${\frak {M}_{\pm }}(l_{\pm })$ collect the
$l_{\pm}$ particles associated to vertices of the $SU(N)$-Dynkin
diagram corresponding to $c_i=\pm 1$ respectively, according to the bi-colouration 
of the roots in the Dynkin diagram introduced in the previous section.  
 We understand here that inside the sets 
${\frak {M}_{\pm }}$
the order of the individual sets ${\frak {M}_{i}}$ is arbitrary. This simply
reflects the fact that particles of different species but identical colour
interact freely. However, ${\frak {M}}$ is an ordered set since elements of $%
{\frak {M}_{+}}$ and ${\frak {M}_{-}}$ do not interact freely and w.l.g. we
adopt the convention that particles belonging to the ``$+$''-colour set come
first. As usual, the $H_{n}$ are some overall constants and the $Q_{n}$ are polynomial
functions depending on the variables $x_{i}=\exp \theta _{i}$ which are
collected in the sets ${\frak {X},}{\frak {X}_{i},}
{\frak{X}_\pm}$ in a one-to-one fashion with respect to the particle species sets $
{\frak {M},}{\frak {M}_{i},}{\frak{M}_{\pm}}$. The functions $%
F_{\text{min}}^{\mu _{i}\mu _{j}}(\theta _{ij})$ are the  minimal
form factors introduced in section \ref{mini},
which by construction contain no singularities in the physical
sheet and solve Watson's equations \cite{Kar,Smir} for two particles. For
the $SU(N)_{2}$-HSG model they are found to be 
\begin{equation}
F_{\text{min}}^{ij}(\theta )={({\cal N}_{c_i})}^{I_{ij}}\left( \sin \frac{\theta }{2i}%
\right) ^{\delta _{ij}}e^{-I_{ij}\int\limits_{0}^{\infty }{{\frac{dt}{t}}}%
{\textstyle{\sin ^{2}\left( (i\pi -\theta \mp \sigma _{ij})\frac{t}{2\pi }\right)  \over \sinh t\cosh t/2}}%
}\,,  \label{10}
\end{equation}
where, the normalisation constants ${\cal N}_{c_i}$, are given by
\begin{equation}
{\cal N}_{c_i}=2^{\frac{1}{4}}\exp \left( i\pi \frac{(1-c_{i}(1-\theta))}{4}-\frac{G}{4}\right)
\label{constants}
\end{equation}
and depend on the Catalan constant  $G$. Notice the parallelism between (\ref{10})
and equations (\ref{minising}), (\ref{14}) in the previous chapter. Also the following
asymptotic behaviours are very close to the ones presented in section \ref{mini}
\begin{equation}
\lim\limits_{\sigma \rightarrow \infty }F_{\text{min}}^{\mu_i \mu_{i+1} }(\pm \theta
)\sim e^{-\frac{\sigma_{i, i+1} }{4}},\qquad \qquad 
 \left[ F_{\text{min}}^{\mu_i, \mu_j}(\theta _{ij})\right] _{i}=
\frac{\delta_{c_i,c_j}-\delta_{c_{i+1},0}\delta_{c_{j-1},0}}{2}.
\label{behave2}
\end{equation}
\noindent were we used the abbreviation  (\ref{not}).

\noindent  It is convenient also to introduce the notation
\begin{equation}
 \tilde{F}_{\text{min}}^{ij}(\theta )=(e^{-c_{i}\theta /4}F_{\text{min}}^{ij}(\theta))^{I_{ij}},
\label{tildemin}
\end{equation}

\noindent The minimal form factors obey the functional identities 
\begin{equation}
F_{\text{min}}^{ij}(\theta +i\pi )F_{\text{min}}^{ij}(\theta )=\left( -\frac{%
i}{2}\sinh \theta \right) ^{\delta _{ij}}\left( \frac{i^{\frac{2-c_{i}}{2}%
}\,e^{\frac{\theta }{2}c_{i}}}{\cosh \left( \frac{\theta }{2}-\frac{i\,\pi }{%
4}\right) }\right) ^{I_{ij}}.  \label{hhe2}
\end{equation}

\noindent Substituting the ansatz (\ref{fact2}) into the kinematic residue equation ,
 we obtain with the help of (\ref{hhe2}) a recursive
equation for the overall constants for \thinspace $\mu _{i}\in {\frak {M}_{+}%
}$ 
\begin{equation}
H_{n+2}^{{\cal O}|{\frak {M}}(l_{+}+2,l_{-}{\frak )}}=i^{\bar{l}%
_{i}}2^{2l_{i}-\bar{l}_{i}+1}e^{I_{ij}\sigma _{ij}l_{j}/2}H_{n}^{{\cal O}|%
{\frak {M}(}l_{+},l_{-}{\frak )}}\,\,{\frak \,\,.}  \label{Hhrec}
\end{equation}
We introduced here the numbers $\bar{l}_{i}=\sum_{\mu _{j}\in {\frak {M}_{-}}%
}I_{ij}l_{j}$, which count the elements in the neighbouring sets of ${\frak {%
M}_{i}}$.

The $Q$-polynomials have to obey the recursive equations 
\begin{eqnarray}
Q_{n+2}^{{\cal O}|{\frak {M}}(2+l_{+},l_{-})}({\frak {X}}%
^{xx})&=& D_{\varsigma_i}^{2 s_i + \tau_i, 2 \bar{s}_i+ \bar{\tau}_i}(\frak{X}^{xx})  Q^{{\cal O}|{\frak M}(l_{+},l_{-})}{(\frak {X})\,}\,\,\,  \label{ddsun}\\
D_{\varsigma_i}^{2 s_i + \tau_i, 2 \bar{s}_i+ \bar{\tau}_i}(\frak{X}^{xx}) &=& \sum_{k=0}^{\bar{s}_{i}}x^{2s_{i}-2k+\tau _{i}+1-\varsigma
_{i}}\sigma _{2k+\varsigma _{i}}(I_{ij}{\frak {\hat{X}}}_{j})  
(-i)^{2s_{i}+\tau _{i}+1}\sigma _{2s_{i}+\tau _{i}}({\frak X}_{i}).\,\,\, 
\end{eqnarray}
For convenience we defined the sets ${\frak {X}}^{xx}:=\{-x,x\}\cup {X}$, $%
{\frak {\hat{X}}:=}ie^{\sigma _{i,i+1}}{\frak X}$ and the integers $\zeta
_{i}$ which are $0$ or $1$ depending on whether the sum $\vartheta +\tau
_{i} $ is odd or even, respectively. $\vartheta $ is related to the factor
of local commutativity \cite{YZam} $\omega =(-1)^{\vartheta }=\pm 1$. As usual $\sigma
_{k}(x_{1},\ldots ,x_{m})$ is the $k$-th elementary symmetric polynomial.
Furthermore, we used the sum convention $I_{ij}{\frak {\hat{X}}}_{j}\equiv 
\mathop{\textstyle\bigcup}
\sum_{\mu _{j}\in {\frak M}}I_{ij}{\frak {\hat{X}}}_{j}$ and parameterised $l_{i}=2s_{i}+\tau _{i}$, $\bar{l}_{i}=2\bar{s}_{i}+\bar{\tau}_{i}$ in order
to distinguish between odd and even particle numbers.

Recall that, in the $SU(3)_2$-case the total degree of the Q-polynomials was determined
by the spinless character of the local operators through Eq. (\ref{Qasym}). Likewise,
for the $SU(N)_2$-case, restricting also our study to spinless local operators, we have
\begin{equation}
\left[ Q_{n}^{\mathcal{O}} \right] = \sum_{\mu_i \in  \frak{M}_{+}}\frac{l_i (l_i -1)}{2}-
\sum_{\mu_i \in \frak{M}_{-}} \frac{l_i (l_i-1)}{2}.
\label{Qasym2}
\end{equation}

\section{The solution procedure}
\label{solution2}
\indent \ \
We will now solve the recursive equations (\ref{Hhrec}) and (\ref{ddsun})
systematically. The equations for the constants are solved by 
\begin{equation}
H_{n}^{{\cal O}|{\frak {M}(}l_{+},l_{-}{\frak )}}=\,\,\,\prod\limits_{\mu
_{i}\in {\frak M}_{+}}\,\,\,i^{s_{i}\bar{l}_{i}}\,2^{s_{i}(2s_{i}-\bar{%
l}_{i}-1-2\tau _{i})}e^{\frac{s_{i}I_{ij}\sigma _{ij}l_{j}}{2}}H^{{\cal O}|{%
\tau }_{i},\bar{l}_{i}}.
\end{equation}
The lowest non vanishing constants $H^{{\cal O}|{\tau }_{i},\bar{l}_{i}}$ are
fixed by demanding, similarly as in the $SU(3)_{2}$-case,
that any form factor which involves only one particle type should correspond
to a form factor of the thermally perturbed Ising model, as indicated by Eq. (\ref{is2}).
To achieve this we exploit the ambiguity present in (\ref{Hhrec}), that is the fact that we can
multiply it by any constant which only depends on the quantum
numbers, $l_{-}$.

Concerning the Q-polynomials, as the main building blocks for their construction
serve the determinants of the ($t+s$)$\times $($t+s$)-matrix 
\begin{equation}
{\cal \,A}_{2s+\tau ^{+},2t+\tau ^{-}}^{\nu ^{+},\nu ^{-}}({\frak {X}}_{+},%
{\frak {\hat{X}}}_{-})_{ij}=%
{\sigma _{2(j-i)+\nu ^{+}}({\frak {X}}_{+})\text{,}1\leq i\leq t \atopwithdelims\{. \sigma _{2(j-i+t)+\nu ^{-}}({\frak {\hat{X}}}_{-})\,\text{, }t<i\leq s+t}%
\end{equation}
\noindent which are generalisations of the ones introduced in the previous chapter (see Eq. (\ref{sss})).
Here  $\nu ^{\pm },\tau ^{\pm }=0,1$ like for the $SU(3)_2$-case.
 The determinant of ${\cal A}$ essentially captures the summation in (\ref{ddsun})
due to the fact that it satisfies the recursive equations 
\begin{equation}
\det {\cal A}_{2+l,2t+\tau ^{-}}^{\nu ^{+},\nu ^{-}}({\frak {X}}_{+}^{\,xx},{\frak {\hat{X}}}_{-})\,\,=\left(\sum\limits_{p=0}^{t}x^{2(t-p)}\sigma _{2p+\nu ^{-}}({\frak {\hat{X}}}%
_{-})\right) \,  \det {\cal A}_{l,2t+\tau
^{-}}^{\nu ^{+},\nu ^{-}}({\frak X}_{+},{\frak {\hat{X}}}_{-})\,  \label{xx}
\end{equation}
as was shown in chapter \ref{ffs}. Analogously to the procedure developed there,  we
can build up a simple product from elementary symmetric polynomials which
takes care of the pre-factor in the recursive Eq. (\ref{ddsun}). Proceeding this way the 
solution for the Q-polynomials is
\begin{eqnarray}
&&Q_{n}^{\mathcal{O}|{\frak {M}}(l_{+},l_{-})}({\frak {X}}_{+},{\frak X}%
_{-})=\!\!\!\!\prod\limits_{\mu _{i/k}\in {\frak M}_{+/-}}\!\!\!\!\!\!\!\!\!%
\det {\cal A}_{2s_{i}+\tau _{i},\bar{l}_{i}}^{\nu _{i},\varsigma _{i}}(%
{\frak {X}}_{i},I_{ij}{\frak {\hat{X}}}_{j})  \label{sol2} \\
&&\times \sigma _{2s_{i}+\tau _{i}}({\frak {X}}_{i})^{\frac{2s_{i}+\tau _{i}-2%
\bar{s}_{i}-1-\varsigma _{i}}{2}}\sigma _{\bar{l}_{i}}(I_{ij}{\frak {\hat{X}}}%
_{j})^{\frac{\bar{\nu}_{i}-1}{2}}\sigma _{l_{k}}({\frak {\hat{X}}}_{k}))^{%
\frac{1-l_{k}}{2}}.  \nonumber
\end{eqnarray}
\noindent
It is interesting to notice that the product of elementary symmetric polynomials occurring in 
(\ref{sol2}) is not a simple generalisation of the $g$-polynomials defined  in (\ref{g}). 
In fact, one could naively try  to obtain the solution to the recursive equations (\ref{ddsun})
as a product of the type
\begin{equation}
Q_{n}^{\mathcal{O}|{\frak {M}}(l_{+},l_{-})}({\frak {X}}_{+},{\frak X}_{-})=
\prod_{\mu_i \in \frak{M}_{+}} Q_{2s_i +\tau_i, 2\bar{s}_i+\bar{\tau}_i}^{\nu_i,\varsigma_i}
({\frak {X}}_{i},I_{ij}{\frak {\hat{X}}}_{j}),
\label{naive}
\end{equation}
\noindent for $ Q_{2s_i +\tau_i, 2\bar{s}_i+\bar{\tau}_i}^{\nu_i,\varsigma_i}
({\frak {X}}_{i},I_{ij}{\frak {\hat{X}}}_{j})$ to be the solutions computed for the $SU(3)_2$-case
in Eq. (\ref{qpar}). However, if doing so we shall observe  that such  solution proposal contradicts two essential requirements:
 it does not respect the constraint (\ref{Qasym2}), which takes care of the spinless nature of the corresponding operator $\mathcal{O}$ and it also does not reduce to the thermally perturbed Ising model solution whenever only particles associated to one vertex of the $SU(N)$-Dynkin diagram are involved. Therefore, the solution proposal  (\ref{naive}) must be modified
by multiplying it with the necessary elementary symmetric polynomials so that the two constraints above
mentioned are respected. By doing so we will obtain the solution (\ref{sol2}) presented before.

It  follows immediately with the help of property (\ref{xx}) that the solutions (\ref{sol2}) obey
the recursive equations 
\begin{eqnarray}
&&Q_{2+n}^{\mathcal{O}|{\frak {M}}(2+l_{+},l_{-})}({\frak {X}}_{+}^{xx},{\frak {X}}%
_{-})=Q_{n}^{\mathcal{O}|{\frak {M}}(l_{+},l_{-})}({\frak {X}}_{+},{\frak {X}}_{-})\sigma
_{2s_{i}+\tau _{i}}({\frak {X}}_{i})\,  \nonumber \\
&&\qquad \qquad \qquad \qquad \times \sum\limits_{p=0}^{\bar{s}_{i}}x^{2(s_{i}-p)+%
\tau _{i}+1-\varsigma _{i}}\sigma _{2p+\varsigma _{i}}(I_{ij}{\frak {\hat{X}}}%
_{j})\,\,.  \label{an}
\end{eqnarray}
Comparing now the equations (\ref{ddsun}) and (\ref{an}) we obtain complete
agreement. Notice that the numbers $\bar{\nu}_{i}$ are not constrained at
all at this point of the construction. However, by demanding relativistic
invariance, which on the other hand means that the overall power in (\ref
{fact2}) has to be zero (namely, the Q's satisfy (\ref{Qasym2})), we obtain the additional constraints 
\begin{equation}
\nu _{i}=1+\tau _{i}-\bar{\nu}_{i}\qquad \text{and\qquad }\tau _{i}\varsigma
_{i}=\bar{\tau}_{i}(\bar{\nu}_{i}-1)\,\,\,.
\end{equation}
Taking in addition into account the constraints which are needed to derive (\ref{xx}) 
(see the derivation of Eq. (\ref{25})), this is most conveniently written as 
\begin{equation}
\tau _{i}\varsigma _{i}+\bar{\tau}_{i}\nu _{i}=\tau _{i}\bar{\tau}%
_{i},\qquad 2+\varsigma _{i}>\bar{\tau}_{i},\qquad 2+\nu _{i}>\tau
_{i}\,\,\,.  \label{consist}
\end{equation}

For each $\mu _{i}\in {\frak {M}}_{+}$ the equations (\ref{consist}) admit the
10 feasible solutions  presented in table \ref{t1} for the $SU(3)_2$-HSG model.
Therefore, in some sense, the $SU(3)_2$-solutions serve as building blocks for the
construction of at least part of  the solutions corresponding to the local operators of the
$SU(N)_2$-HSG model.  However,  the individual solutions for different values of $i$ are not all independent of each other. We would like to stress that despite the fact that (\ref{sol2})  
represents a large class of independent solutions, it does certainly not
exhaust all of them. Nonetheless, many additional solutions, like the energy
momentum tensor, may be constructed from (\ref{consist}) by simple
manipulations like the multiplication of some CDD-like ambiguity factors of the type
$\sigma_1 \bar{\sigma_1}$ like for the $SU(3)_2$-case (see section \ref{emt}) or
by setting some expressions to zero on the base of asymptotic considerations
(see chapter \ref{ffs} for more details). 

In order to study the renormalisation group flow of  Zamolodchikov's $c$-function \cite{ZamC}
as well as for  the conformal dimensions of certain operators by means of the $\Delta$-sum
rule \cite{DSC} (see section \ref{rgflow}), the explicit expressions of the form factors
 of the trace of the energy momentum tensor $\hat{\Theta}$, normalised in such a way
that it becomes a dimensionless object,  are needed. The first non-vanishing terms read

\begin{eqnarray}
F_{2}^{\hat{\Theta} |\mu _{i}\mu _{i}} &=&-2\pi i\sinh (\theta /2) \label{22par} \\
F_{4}^{\hat{\Theta} |\mu _{i}\mu _{i}\mu _{j}\mu _{j}} &=&\frac{\pi
(2+\sum\limits_{i<j}\cosh (\theta _{ij}))\prod\limits_{i<j}\tilde{F}_{%
\text{min}}^{\mu _{i}\mu _{j}}(\theta _{ij})}{-2\cosh (\theta _{12}/2)\cosh
(\theta _{34}/2)}\, \label{44par} \\
F_{6}^{\hat{\Theta} |\mu _{i}\mu _{i}\mu _{i}\mu _{i}\mu _{j}\mu _{j}} &=&\frac{%
\pi (3+\sum\limits_{i<j}\cosh (\theta _{ij}))\prod\limits_{i<j}\tilde{F}%
_{\text{min}}^{\mu _{i}\mu _{j}}(\theta _{ij})}{4\prod_{1\leq i<j\leq
4}\cosh (\theta _{ij}/2)} \label{66par} \\
F_{6}^{\hat{\Theta} |\mu _{i}\mu _{i}\mu _{k}\mu _{k}\mu _{j}\mu _{j}} &=&\frac{%
\pi (3+\sum\limits_{i<j}\cosh (\theta _{ij}))\prod\limits_{i<j}\tilde{F}%
_{\text{min}}^{\mu _{i}\mu _{j}}(\theta _{ij})}{4\cosh (\theta _{12}/2)\cosh
(\theta _{34}/2)} \label{666par}
\end{eqnarray}
for $I_{ij}\neq 0$ and $I_{kj}\neq 0$. Recall also the 
definition of the functions $\tilde{F}_{\text{min}}^{\mu_i \mu_j}$ given by (\ref{tildemin}).
Also the following limit, 
\begin{equation}
\lim_{\sigma _{i,i+1}\rightarrow \infty }F_{n}^{\hat{\Theta} |\mu _{i}\mu
_{i+1}\ldots }(\theta )=0\,\,,  \label{limF}
\end{equation}
\noindent which is a direct consequence of the asymptotic behaviour stated in the first 
equation in (\ref{behave2}), will be also  required in the next 
section in order to interpret  the physical picture
arising from the RG-analysis.

\section{RG-flow of Zamolodchikov's $\protect{c}$-function}
\label{rg2}
\indent \ \
As mentioned in section \ref{rgflow}, renormalisation group methods have been introduced 
originally \cite{GL} to carry out qualitative analysis of regions of quantum field theories which
can not be investigated  by doing perturbation theory in the coupling constants. In
particular the $\beta_i $-functions defined in section \ref{cff}
provide an insight into various possible asymptotic behaviours and especially 
allow to identify the fixed points of the theory, where they vanish according 
to property ii) in section \ref{cff}. Having this in mind, 
we now want to employ  a RG-analysis  to check the consistency of our solutions and
at the same time the physical picture advocated for the HSG-models.

For this purpose we want to investigate first of all the renormalisation
group flow of the $c$-function, in a similar spirit as for the $SU(3)_{2}$-case, by evaluating  numerically 
the c-function (\ref{cr0}).

Having determined the form factors in section \ref{solution2},
we are in principle in the position to
compute the two-point correlation function between two local operators in
the usual way, that is by expanding it in terms of $n$-particle form factors as indicated
by Eq. (\ref{corr}) in the previous chapter.

In order to evaluate (\ref{cr0})
we need to compute the two-point function  for the trace of the energy momentum tensor $\Theta $.
This can be done easily  by using (\ref{corr}) together with 
the form factor expressions (\ref{22par})-(\ref{666par}) of the previous
section. The individual $n$-particle contributions obtained from the evaluation of (\ref{Cth2})
read
\begin{eqnarray}
\Delta c_2 &=&(N-1)\cdot 0.5  \label{2p} \\
\Delta c_4 &=&(N-2)\cdot 0.197  \label{4p} \\
\Delta c_6 &=&(N-2)\cdot 0.002+(N-3)\cdot 0.0924  \label{6p} \\
\sum_{k=2}^{6}\Delta c_k &=&N\ast 0.7914-1.1752\,\,.  \label{sum}
\end{eqnarray}
Apart from the two particle contribution (\ref{2p}), which is usually quite
trivial and in this situation can even be evaluated analytically (see Eq. (\ref{anal})), we have
carried out the multidimensional integrals in (\ref{corr}) by means of a
Monte Carlo method as usual. We use this method up to a precision which is higher
than the last digit we quote. For convenience we report some explicit
numbers in table \ref{t31}.

\begin{table}[h]
\begin{center}
\noindent 
\begin{tabular}{|c||c|c|c|c|c|}
\hline
$N$ & $c=\frac{N(N-1)}{N+2}$ & $\Delta c_2$ & $\Delta c_4$ & $\Delta c_6$ & $%
\sum_{k=2}^{6}\Delta c_k$ \\ \hline\hline
$3$ & $1.2$ & $1$ & $0.197$ & $0.002$ & $1.199$ \\ \hline
$4$ & $2$ & $1.5$ & $0.394$ & $0.096$ & $1.990$ \\ \hline
$5$ & $2.857$ & $2$ & $0.591$ & $0.191$ & $2.782$ \\ \hline
$6$ & $3.75$ & $2.5$ & $0.788$ & $0.285$ & $3.573$ \\ \hline
$7$ & $4.\bar{6}$ & $3$ & $0.985$ & $0.380$ & $4.365$ \\ \hline
$8$ & $5.6$ & $3.5$ & $1.182$ & $0.474$ & $5.156$ \\ \hline
\end{tabular}
\end{center}
\caption{$\protect{n}$-particle contributions to the $\protect{c}$-theorem versus the
$\protect{SU(N)_{2}/U(1)^{N-1}}$-WZNW coset model  Virasoro central charge.}
\label{t31}
\end{table}

As can be seen in table \ref{t31}, the evaluation of (\ref{2p})-(\ref{sum}) 
illustrates that the series (\ref{corr})
 converges slower and slower for increasing values of $N$, such that
the higher $n$-particle contributions become more and more important to
achieve high accuracy. Our analysis suggests that it is not the functional
dependence of the individual form factors which is responsible for this
behaviour. Instead this effect is simply due to the fact that the symmetry
factor, that is the sum $\sum_{\mu _{1}\ldots \mu _{n}}$, resulting from
permutations of the particle species increases drastically for larger $N$.

Having confirmed the expected ultraviolet central charge, we now study the
RG-flow by varying $r_{0}$ in (\ref{cr0}). We expect to find that whenever
we reach an energy scale at which an unstable particle can be formed, the
model will flow to a different coset. This means following the flow with
increasing $r_{0}$ we will encounter a situation in which certain $\sigma
_{i,i+1}$ are considered to be large and we observe the decoupling into two
freely interacting systems in the way described in (\ref{flow}). For
instance for the situation $\sigma _{12}>\sigma _{23}>\sigma _{34}>\ldots $
we observe the following decoupling along the flow with increasing $r_{0}$:

\begin{center}
\begin{tabular}{cc}
$SU(N)_{2}/U(1)^{N-1}$ \,\,\,&\,\,\, ${N(N-1)}/{(N+2)}$\\ 
$\downarrow $\,\,\, &\,\,\,$\downarrow $  \\ 
$SU(N-1)_{2}/U(1)^{N-2}\otimes SU(2)_{2}/U(1)$ \,\,\,&\,\,\,${(N-1)(N-2)}/{(N+1)}+{1}/{2}$ \\ 
$\downarrow $\,\,\, &\,\,\,$\downarrow $ \\ 
$SU(N-2)_{2}/U(1)^{N-3} \otimes \left( SU(2)_{2}/U(1)\right) ^{2}$\,\,\,&\,\,\, ${(N-2)(N-3)}/{N}+1$ \\ 
$\downarrow $\,\,\,&\,\,\,$\downarrow $  \\ 
$\vdots $ \,\,\,&\,\,\,\vdots\\ 
$\downarrow $\,\,\, &\,\,\, $\downarrow $  \\ 
$\left( SU(2)_{2}/U(1)\right) ^{N-1}$ \,\,\,&\,\,\,${(N-1)}/{2}$\\
\end{tabular}
\end{center}

The numbers  reported  on the right are the values of the Virasoro central charge 
at the different fixed points  the function $c(r_0)$  supasses along its renormalisation group flow.
Here we also consider the masses of all the stable particles in the model to be of the same order
of magnitude. This means that, after the function $c(r_0)$ reaches the last value $(N-1)/2$, it
will directly flow to the IR-value $c(\infty)=0$. In other words,
the individual decoupling of each of the stable particles, which will successively
reduce the value of the $c$-function by a factor of $1/2$, is not observed here.

We can understand this type of behaviour in a ``semi-analytical'' way. The
precise difference between the central charges related to (\ref{flow}) is 
\begin{equation}
c_{SU(i+1)_{2}/U(1)^{i}\otimes SU(N-i)_{2}/U(1)^{N-i-1}}= 
c_{SU(N)_{2}/U(1)^{N-1}}-\frac{2i(N+5)(N-i-1)}{(N+2)(i+3)(N-i+2)}. 
\label{diffcex}
\end{equation}
Noting with (\ref{limF}) that we loose at each step all the contributions 
$F_{n}^{\Theta |\mu _{i}\mu _{i+1}\ldots }(\theta )$ to $\Delta c$, we may
collect the values (\ref{2p})-(\ref{6p}), which we have determined
numerically and find 
\begin{equation}
\lim_{\sigma _{i,i+1}\rightarrow \infty }{\small \Delta c({\sigma }_{i,i+1} ,\ldots )}=  \Delta c({\sigma }_{i,i+1}= 0,\ldots )-{0.2914I}_{i,i+1} -{0.0924I}_{j,j-1},
\label{diffc}
\end{equation}
\noindent where $I_{i,i+1}$ and $I_{j,j-1}$ are components of the incident matrix of $SU(N)$ (\ref{Inc}), 
and the last contribution ${0.0924I}_{j,j-1}$ only occurs when $j \neq 1, N-2$.  

Similarly as for the deep ultraviolet region, we find a relatively good agreement
between (\ref{diffcex}) and (\ref{diffc}) for small values of $N$. The
difference for larger values is once again due to the convergence behaviour
of the series in (\ref{corr}).

For $r_{0}=0$ qualitatively a similar kind of behaviour was previously
observed in \cite{ADM}, for the two particle contribution only, in the context of the
roaming  Sinh-Gordon model originally introduced in \cite{staircase} through
the S-matrix
\begin{equation}
S(\theta)=\frac{\sinh \theta- i \cosh \theta_0}{\sinh \theta + i \cosh \theta_0}.
\end{equation}
\noindent and generalised thereafter by other authors \cite{Martins, DoRav}.
 Nonetheless, there is a slight difference between the two situations which is 
due to the different ways the resonance parameter enters the S-matrix for
 the HSG-models and roaming  Sinh-Gordon model (see discussion in section \ref{con3}). 
Instead of a decoupling into
different cosets in this type of model the entire S-matrix takes on the
value $-1$, whenever  the only  resonance parameter, denoted in \cite{staircase} by 
$\theta_0$, goes to infinity. Consequently, the Virasoro central charge 
tends to the value $1/2$ as soon as $\theta_0$ is big enough, a behaviour which was pointed out
by the authors of  \cite{ADM}. Therefore,  the resulting effect, i.e. a depletion of $\Delta c$, 
is the same as for the HSG-models. However, we do not
comply with the interpretation put forward in \cite{ADM}, namely that such a
behaviour should constitute a ``violation of the c-theorem sum-rule''. The observed
effect is precisely what one expects from the physical point of view and the
$c$-theorem.

\begin{figure}
\begin{center}
\includegraphics[width=10.7cm,height=8.5cm]{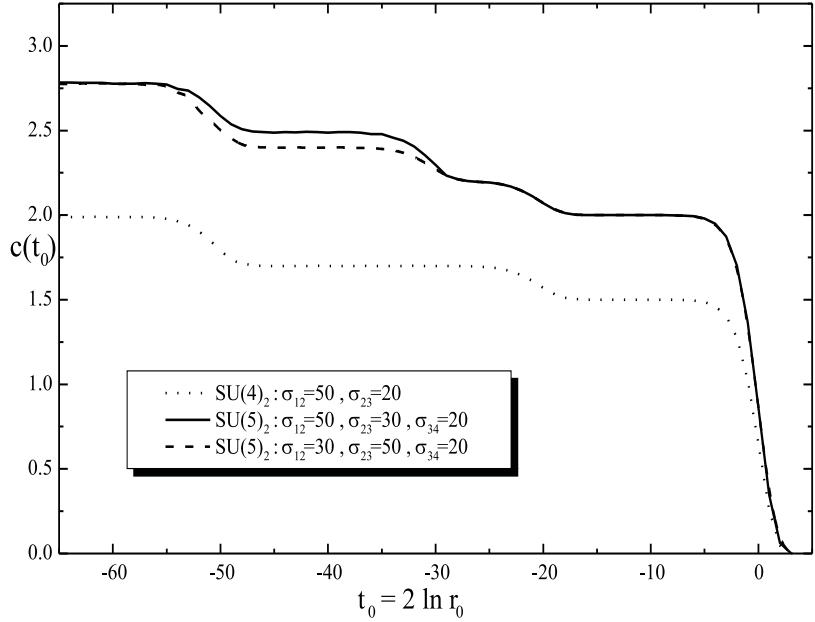}
\caption{Renormalisation group  flow for the $\protect{c}$-function.}
\label{f31}
\end{center}
\end{figure}

We present our full numerical results in Fig.  \ref{f31}, 
which confirm the outlined flow for various values of $N$ and reproduce the same
scaling behaviour for the mass of the unstable particles 
$M_{\tilde{u}}(t_u, \sigma) \sim m e^{|\sigma|+t_{\tilde{u}}}$ in terms of the RG-parameter
pointed out in section \ref{rgflow} of the previous chapter. However,
the situation in that case is more involved, since the amount of different
resonance parameters and therefore, the number of unstable particles involved, is higher. 

We observe that the $c$-function remains constant, at a value corresponding
to the new coset, in some finite interval of $r_{0}$. In particular, we
observe the non-equivalence of the flows when the relative order of
magnitude amongst the different resonance parameters is changed. For $N=5$
we confirm (we omit here the $U(1)$-factors and report the corresponding
central charges as superscripts on the last factor).

\begin{center}
\begin{tabular}{c}
\frame{$\sigma _{12}>\sigma _{23}$}$\quad SU(5)_{2}^{\frac{20}{7}}\quad 
\frame{$\sigma _{23}>\sigma _{12}$}$ \\ 
$\swarrow \qquad \qquad \searrow $ \\ 
$\quad \quad SU(4)_{2}\otimes SU(2)_{2}^{\frac{5}{2}}\quad \quad \quad \quad
\quad \qquad SU(3)_{2}\otimes SU(3)_{2}^{\frac{12}{5}}\qquad $ \\ 
$\searrow \qquad \qquad \qquad \swarrow $ \\ 
$SU(3)_{2}\otimes SU(2)_{2}\otimes SU(2)_{2}^{\frac{11}{5}}$ \\ 
$\downarrow $ \\ 
$SU(2)_{2}\otimes SU(2)_{2}\otimes SU(2)_{2}\otimes SU(2)_{2}^{2}$%
\end{tabular}
\end{center}

The precise difference in the central charges is explained with (\ref{diffc}), 
since the contribution $0.0924I_{i,i-1}$ only occurs for $i=2$.

\begin{figure}
\begin{center}
\includegraphics[width=10.7cm,height=8.5cm]{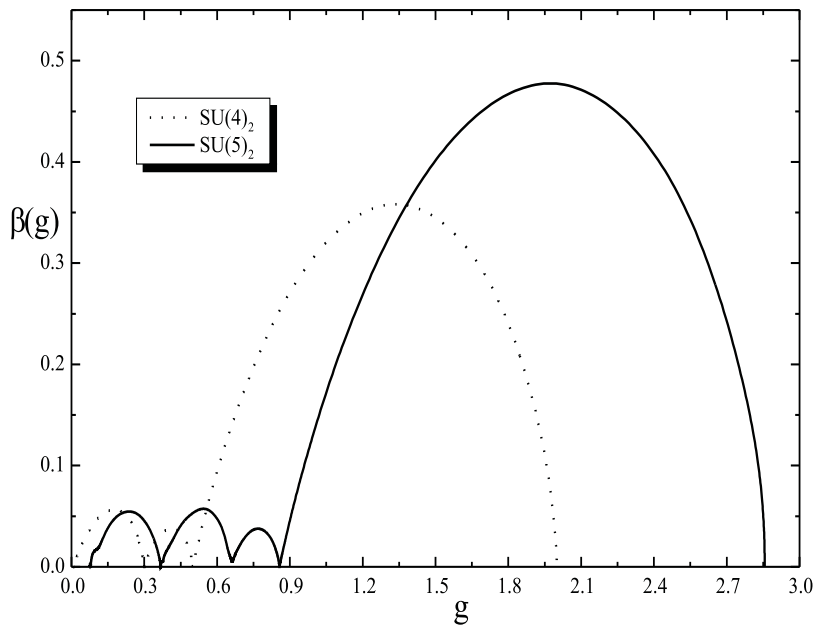}
\caption{The $\protect\beta$-function.}
\label{f32}
\end{center}
\end{figure}

To establish more clearly that the plateaux  in Fig.  \ref{f31} admit indeed an interpretation
as fixed points, namely zeros of the corresponding $\beta$-function (see section \ref{cff})
and extract the definite values of the corresponding
Virasoro central charge we can also, following \cite{ZamC, staircase}, determine a
 $\beta $-type function from $c(r)$ along the lines of subsection \ref{betalike}.
 Our results for various values of N
are depicted in Fig.  \ref{f32}, which allow a definite identification of the fixed
points corresponding to the coset models expected from the decoupling (\ref{flow}). 

For $SU(4)_{2}$ we clearly identify from Fig. \ref{f32} the four fixed points 
{\small \ }$\tilde{g}=0,0.3,0.5,2$ with high accuracy. The five fixed points 
$\tilde{g}=0,0.357,0.657,0.857,2.857$, which we expect to find for $%
SU(5)_{2} $ are all slightly shifted due to the absence of the higher order
contributions.

\section{Identifying the operator content of  $\protect{SU(N)_{2}}$-HSG model}
\label{operator}
\indent \ \
In the same fashion we did for the $SU(3)_2$-HSG model, 
we now want to identify the operator content of our theory by carrying out
the ultraviolet limit in the corresponding two-point functions or by using in some
cases the $\Delta$-sum rule \cite{DSC} and match the conformal dimension of each operator
with the one in the $SU(N)_{2}/U(1)^{N-1}$-WZNW-coset model (see section \ref{identifying}).
For this purpose we have to determine first of all the entire operator content of the
conformal field theory.

As we did in section \ref{identifying}, we shall use the following general formula for
the conformal dimensions of the parafermionic vertex operators originally derived in \cite{Gep}

\begin{equation}
\Delta (\Lambda ,\lambda )=\frac{(\Lambda \cdot (\Lambda +2\rho ))}{(4+2N)}-%
\frac{(\lambda \cdot \lambda )}{4}\,\,.  \label{dim}
\end{equation}

Here $\Lambda $ is a highest dominant weight of level smaller or equal to $k=2$
and $\rho =1/2\sum_{\alpha >0}\alpha $ is the Weyl vector, i.e. half the sum
of all positive roots. As explained in section \ref{identifying} the $\lambda$'s are
lower weights which are obtained from $\Lambda$ by subtracting  multiples of simple roots
$\alpha_i$ until the lowest weight is reached. 
Nonetheless, it may happen that a
weight corresponds to more than one linear independent weight vector, such
that the weight space may be more than one dimensional. The dimension of
each weight vector $n_{\lambda }^{\Lambda }$ is computed by means of 
\begin{equation}
n_{\lambda }^{\Lambda }=\frac{\sum_{\alpha >0}\sum_{l=1}^{\infty
}2\,n_{\lambda +l\alpha }((\lambda +l\alpha )\cdot \alpha )}{((\Lambda
+\lambda +2\rho )\cdot (\Lambda -\lambda ))}\,\,\,.  \label{mult}
\end{equation}
For consistency it is useful to compare the sum of all these multiplicities
with the dimension of the highest weight representation computed directly
from the Weyl dimensionality formula (see e.g. \cite{FH}) 
\begin{equation}
\sum_{\lambda }n_{\lambda }^{\Lambda }=\dim \Lambda =\prod_{\alpha >0}\frac{%
((\Lambda +\rho )\cdot \alpha )}{(\rho \cdot \alpha )}\,\,.  \label{weyl}
\end{equation}
To compute all the conformal dimensions $\Delta (\Lambda ,\lambda )$
according to (\ref{dim}) in general is a tedious task and therefore we
concentrate on a few distinct ones for generic $N$ and only compute the
entire content for $N=4$ which we present in table \ref{t32}.

Noting that $\,\lambda _{i}\cdot \,\lambda _{j}=K_{ij}^{-1}$, with $K$ being
the Cartan matrix, we can obtain relatively concrete formulae from (\ref{dim}). 
For instance 
\begin{equation}
\Delta (\lambda _{i},\lambda _{i})=\frac{%
4\sum_{l=1}^{N-1}K_{il}^{-1}-NK_{ii}^{-1}}{8+2N}\,\,\,.
\end{equation}

Similarly we may compute $\Delta (\lambda _{i}+\lambda _{j},\lambda
_{i}+\lambda _{j})$, etc. in terms of components of the inverse Cartan
matrix. Even more explicit formulae are obtainable when we express the
simple roots $\alpha _{i}$ and fundamental weights $\lambda _{i}$ of $SU(N)$
in terms of a concrete basis. For instance we may choose an orthonormal
basis \{$\varepsilon _{i}$\} in ${\Bbb R}^{N}$ (see e.g. \cite{Bou}), i.e. 
$\varepsilon _{i}\cdot \varepsilon _{j}=\delta _{ij}$ 
\begin{equation}
\alpha _{i}=\varepsilon _{i}-\varepsilon _{i+1},\,\,\,\lambda
_{i}=\sum\limits_{j=1}^{i}\varepsilon _{j}-\frac{i}{N}\sum\limits_{j=1}^{N}
\varepsilon _{j},\,\,i=1,\ldots N-1\,. 
\end{equation}

Noting further that the set of positive roots is given by $\{\varepsilon
_{i}-\varepsilon _{j}:1\leq i<j\leq N\}$, we can evaluate (\ref{dim}), (\ref
{mult}) and (\ref{weyl}) explicitly. This way we obtain for instance 
\begin{equation}
\Delta (\lambda _{i},\lambda _{i})=\frac{i(N-i)}{8+4N}\,\quad \,\text{%
and\quad }\,\,\Delta (2\lambda _{i},2\lambda _{i})=0.
\label{someope}
\end{equation}
Of special physical interest is the dimension of the perturbing operator. As
was already argued in the previous chapter, 
it corresponds to $\Delta (\psi ,0)$, with 
$\psi $ being the highest root and, therefore, arises for a single choice
of the weights $(\Lambda, \lambda)$.
Noting that for $SU(N)$ we have 
$\psi =\lambda _{1}+\lambda _{N-1}$, we confirm once more 
\begin{equation}
\Delta (\psi =\lambda _{1}+\lambda _{N-1},0)=\frac{N}{N+2}\,\,,
\label{pertur}
\end{equation}
\noindent being the expected value for the conformal dimension of the perturbing field which
we will determine later for the $SU(N=4)_2$- and $SU(N=5)_2$-HSG models by studying
the ultraviolet behaviour of the two-point function of the trace of the energy momentum
tensor $\Theta$. Other dimensions may be computed similarly.

As mentioned before we present now  the result of the computation of the entire
operator content in table \ref{t32} for the particular coset  $SU(4)/U(1)^{3}$.
 In case the multiplicity of a weight vector is
bigger than one, we indicate this by a superscript on the conformal
dimension.

\begin{table}[h]
\begin{center}
\noindent {\small 
\begin{tabular}{|c||c|c|c|c|c|c|}
\hline
$\lambda \backslash \Lambda $ & $\lambda _{1}$ & $\lambda _{2}$ & $2\lambda
_{1}$ & $2\lambda _{2}$ & $\lambda _{1}+\lambda _{2}$ & $\lambda
_{1}+\lambda _{3}$ \\ \hline\hline
dim $\Lambda $ & $4$ & $6$ & $10$ & $20$ & $20$ & $15$ \\ \hline
$\Lambda $ & $1/8$ & $1/6$ & $0$ & $0$ & $1/8$ & $1/6$ \\ \hline
$\Lambda -\alpha _{1}$ & $1/8$ &  & $1/2$ &  & $1/8$ & $1/6$ \\ \hline
$\Lambda -\alpha _{2}$ &  & $1/6$ &  & $1/2$ & $1/8$ &  \\ \hline
$\Lambda -\alpha _{3}$ &  &  &  &  &  & $1/6$ \\ \hline
$\Lambda -\alpha _{1}-\alpha _{2}$ & $1/8$ & $1/6$ & $1/2$ & $1/2$ & $%
5/8^{2} $ & $1/6$ \\ \hline
$\Lambda -\alpha _{2}-\alpha _{3}$ &  & $1/6$ &  & $1/2$ & $1/8$ & $1/6$ \\ 
\hline
$\Lambda -\alpha _{1}-\alpha _{3}$ &  &  &  &  &  & $1/6$ \\ \hline
$\Lambda -2\alpha _{1}$ &  &  & $0$ &  &  &  \\ \hline
$\Lambda -2\alpha _{2}$ &  &  &  & $0$ &  &  \\ \hline
$\Lambda -2\alpha _{1}-2\alpha _{2}$ &  &  & $0$ & $0$ &  &  \\ \hline
$\Lambda -2\alpha _{2}-2\alpha _{3}$ &  &  &  & $0$ &  &  \\ \hline
$\Lambda -2\alpha _{1}-\alpha _{2}$ &  &  & $1/2$ &  & $1/8$ &  \\ \hline
$\Lambda -\alpha _{1}-2\alpha _{2}$ &  &  &  & $1/2$ & $1/8$ &  \\ \hline
$\Lambda -2\alpha _{2}-\alpha _{3}$ &  &  &  & $1/2$ &  &  \\ \hline
$\Lambda -\alpha _{1}-\alpha _{2}-\alpha _{3}$ & $1/8$ & $1/6$ & $1/2$ & $%
1/2 $ & $5/8^{2}$ & $2/3^{3}$ \\ \hline
$\Lambda -2\alpha _{1}-\alpha _{2}-\alpha _{3}$ &  &  & $1/2$ &  & $1/8$ & $%
1/6$ \\ \hline
$\Lambda -\alpha _{1}-2\alpha _{2}-\alpha _{3}$ &  & $1/6$ &  & 1$^{2}$ & $%
5/8^{2}$ & $1/6$ \\ \hline
$\Lambda -\alpha _{1}-\alpha _{2}-2\alpha _{3}$ &  &  &  &  &  & $1/6$ \\ 
\hline
$\Lambda -2\alpha _{1}-2\alpha _{2}-\alpha _{3}$ &  &  & $1/2$ & $1/2$ & $%
5/8^{2}$ & $1/6$ \\ \hline
$\Lambda -\alpha _{1}-2\alpha _{2}-2\alpha _{3}$ &  &  &  & $1/2$ & $1/8$ & $%
1/6$ \\ \hline
$\Lambda -2\alpha _{1}-2\alpha _{2}$ &  &  &  &  & $1/8$ &  \\ \hline
$\Lambda -2\alpha _{1}-2\alpha _{2}-2\alpha _{3}$ &  &  & $0$ & 0 & $1/8$ & $%
1/6$ \\ \hline
$\Lambda -\alpha _{1}-3\alpha _{2}-2\alpha _{3}$ &  &  &  & $1/2$ &  &  \\ 
\hline
$\Lambda -\alpha _{1}-3\alpha _{2}-\alpha _{3}$ &  &  &  & $1/2$ &  &  \\ 
\hline
$\Lambda -2\alpha _{1}-3\alpha _{2}-\alpha _{3}$ &  &  &  & $1/2$ & $1/8$ & 
\\ \hline
$\Lambda -2\alpha _{1}-3\alpha _{2}-2\alpha _{3}$ &  &  &  & $1/2$ & $1/8$ & 
\\ \hline
$\Lambda -2\alpha _{1}-4\alpha _{2}-2\alpha _{3}$ &  &  &  & $0$ &  &  \\ 
\hline
\end{tabular}}
\end{center}
\caption[Local operator content of the 
$\protect{SU(4)_{2}/U(1)^{3}}$-WZNW coset model.]{Conformal dimensions
 for $\protect{{\cal O}^{\Delta(\Lambda ,\lambda )}}$
in the $\protect{SU(4)_{2}/U(1)^{3}}$-WZNW coset model.}
\label{t32}
\end{table}

The remaining dominant weights of level smaller or equal to $2$, namely $%
\Lambda =\lambda _{3},2\lambda _{3},\lambda _{2}+\lambda _{3}$, including
their multiplicities may be obtained from table  \ref{t32} simply by the exchange $%
1\leftrightarrow 3$, which corresponds to the ${\Bbb Z}_{2}$-symmetry of the 
$SU(4)$-Dynkin diagram.

Summing up all the fields corresponding to different lower weights, 
we have the following operator content

\begin{equation} 
{\cal O}^{2/3},{\cal O}^{1},14\times {\cal O}^{0},8\times {\cal O}%
^{5/8},18\times {\cal O}^{1/6},24\times {\cal O}^{1/2},32\times {\cal O}%
^{1/8}, 
\end{equation}
\noindent
that is 98 fields.  Therefore, although in section \ref{solution2} a large number of solutions
to the form factor consistency equations was found we see now that the operator content also
increases quite drastically when $N$ does. 

Once all the conformal dimensions of the operators of the underlying CFT corresponding to
the $SU(4)_2$-HSG  have been explicitly computed and even some of them for generic $N$ 
in (\ref{someope}) and (\ref{pertur}) we are now in the position to obtain
al least some of these dimensions  by using the sum  rule \cite{DSC} for
the operators whose symmetry makes it possible or by analysing the
ultraviolet behaviour of the corresponding correlation functions computed via (\ref{corr}).
Similarly to the $SU(3)_2$-case, whenever we find a local operator $\mathcal{O}$ for which 
the $\Delta$-sum rule \cite{DSC} holds,  we may  investigate also the RG-flow 
of the operator content for the model at hand. We shall attempt all  these tasks in the next section for
the $SU(N)_2$-HSG models with $N=4,5$.

\subsection{RG-flow of conformal dimensions}
\label{rgoperator}
\indent \ \
We will now turn to the massive model and evaluate the flow of the conformal
dimension of a local operator $\mathcal{O}$ by means of Eq. (\ref{dr0}),
which we found in subsection \ref{cdflows}. Recall that the field
${\cal O}$ entering Eq. (\ref{dr0}) is a local operator which in the conformal limit corresponds
to a primary field in the sense of \cite{BPZ}. In particular for $r_{0}=0$,
the expression (\ref{dr0}) constitutes the $\Delta$-sum rule \cite{DSC} discussed in subsection 
\ref{uvffs}, which expresses the difference between the ultraviolet and infrared conformal
dimension of the operator ${\cal O}$.

We start by investigating the operator which in the case when all particles
are of the same type corresponds to the disorder operator 
$\mu $ in the Ising model. Using the fact that we should always be able to
reduce to that situation, we consider the solution corresponding to $\tau
_{i}=\bar{\tau}_{i}=\nu _{i}=\varsigma _{i}=0$ for all $i$. Then the $\Delta 
$-sum rule (\ref{dr0}) yields for the individual $n$-particle contributions 
\begin{eqnarray}
\Delta ^{\mu}_2 &=&(N-1)\cdot 0.0625  \label{lll1} \\
\Delta ^{\mu}_4 &=&(2-N)\cdot 0.0263  \label{lll2} \\
\Delta ^{\mu}_6 &=&(N-2)\cdot 0.0017+(3-N)\cdot 0.0113  \label{l3} \\
\sum_{k=2}^{6}\Delta ^{\mu}_k &=&0.0266+N\ast 0.0206\,\,.  \label{l4}
\end{eqnarray}

We assume that this solution has the conformal dimension $\Delta (\lambda_{1},\lambda _{1})$ 
in the ultraviolet limit. For comparison we report a few
explicit numbers in table \ref{t33}.

\begin{table}[h]
\begin{center}
\noindent 
\begin{tabular}{|c||c|c|c|c|c|}
\hline
$N$ & $\Delta (\lambda _{1},\lambda _{1})$ & $\Delta ^{\mu}_2$ & $\Delta
^{\mu}_4$ & $\Delta ^{\mu}_6$ & $\sum_{k=2}^{6}\Delta ^{\mu}_k$ \\ 
\hline\hline
$3$ & $0.1$ & $0.125$ & $-0.0263$ & $0.0017$ & $0.1004$ \\ \hline
$4$ & $0.125$ & $0.1875$ & $-0.0526$ & $-0.0079$ & $0.1270$ \\ \hline
$5$ & $0.143$ & $0.25$ & $-0.0789$ & $-0.0175$ & $0.1536$ \\ \hline
$6$ & $0.156$ & $0.3125$ & $-0.1052$ & $-0.0271$ & $0.1802$ \\ \hline
$7$ & $0.1\bar{6}$ & $0.375$ & $-0.1315$ & $-0.0367$ & $0.2068$ \\ \hline
$8$ & $0.175$ & $0.4375$ & $-0.1578$ & $-0.0463$ & $0.2334$ \\ \hline
\end{tabular}
\end{center}
\caption{$\protect{n}$-particle contributions to the $\protect\Delta$
-sum rule versus conformal dimensions in the $\protect{SU(N)_{2}/U(1)^{N-1}}$-WZNW
coset model.}
\label{t33}
\end{table}

As we already observed for the $c$-theorem, the series converges slower for
larger values of $N$. The reason for this behaviour is the same, namely the
increasing symmetry factor. Note also that the next contribution is negative.

\begin{figure}
\begin{center}
\includegraphics[width=9cm,height=11.5cm, angle=-90]{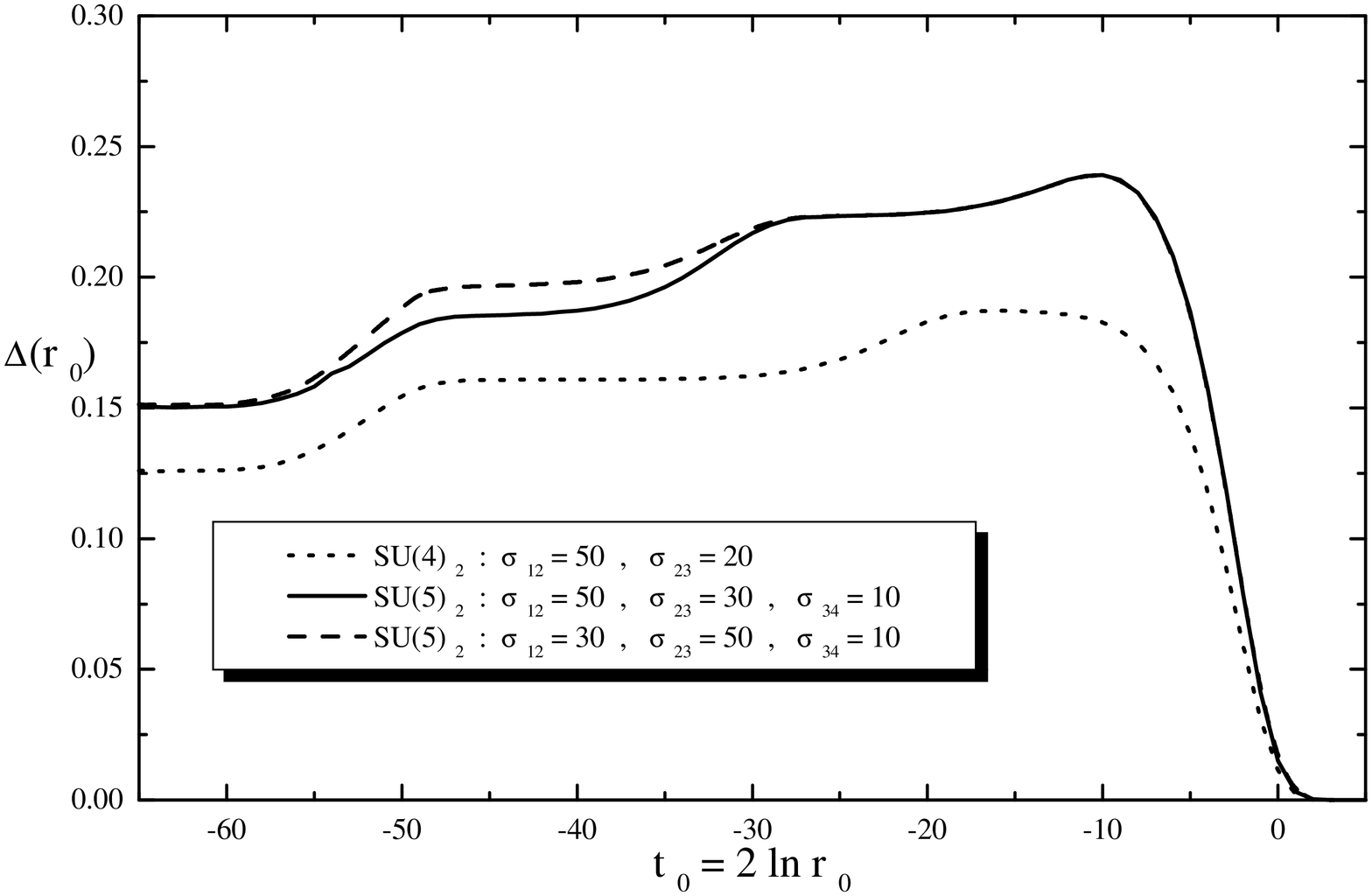}
\caption{RG-flow for the conformal dimension of $\protect\mu$.}
\label{f33}
\end{center}
\end{figure}

Following now the renormalisation group flow for the conformal dimension (\ref{dr0})
 by varying $r_{0}$, we assume that the $\Delta (\lambda _{1},\lambda _{1})$-field flows
to the $\Delta (\lambda _{1},\lambda _{1})$-field in the corresponding new
cosets. Similar as for the Virasoro central charge we may compare the exact
expression 
\begin{eqnarray}
&&\Delta (\lambda _{1},\lambda _{1})_{SU(i+1)_{2}/U(1)^{i}\otimes
SU(N-i)_{2}/U(1)^{N-i-1}}= \nonumber \\
&&\Delta (\lambda _{1},\lambda _{1})_{SU(N)_{2}/U(1)^{N-1}}+\frac{%
i(N+5)(N-i-1)}{4(N+2)(i+3)(N-i+2)},  
\end{eqnarray}
with the numerical results. The contributions (\ref{l1})-(\ref{l3}) yield 
\begin{equation}
\lim_{\sigma _{i,i+1}\rightarrow \infty }{\small \Delta }^{\mu }{\small %
(\sigma }_{i,i+1}{\small ,\ldots )}= \\
{\small \Delta }^{\mu }{\small (\sigma }_{i,i+1}={\small 0,\ldots )+0.0359I%
}_{i,i+1}+{\small 0.0113I}_{j,j-1}, \nonumber
\end{equation}
\noindent where again the contribution $ {0.0113I}_{j,j-1}$ only occurs for $j\neq 1, N-2$.
Once again we find good agreement between the two computations for small
values of $N$. Our complete numerical results are presented in Fig.  \ref{f33},
which confirm the outlined flow for various values of N.

\begin{figure}
\begin{center}
\includegraphics[width=9cm,height=11.5cm, angle=-90]{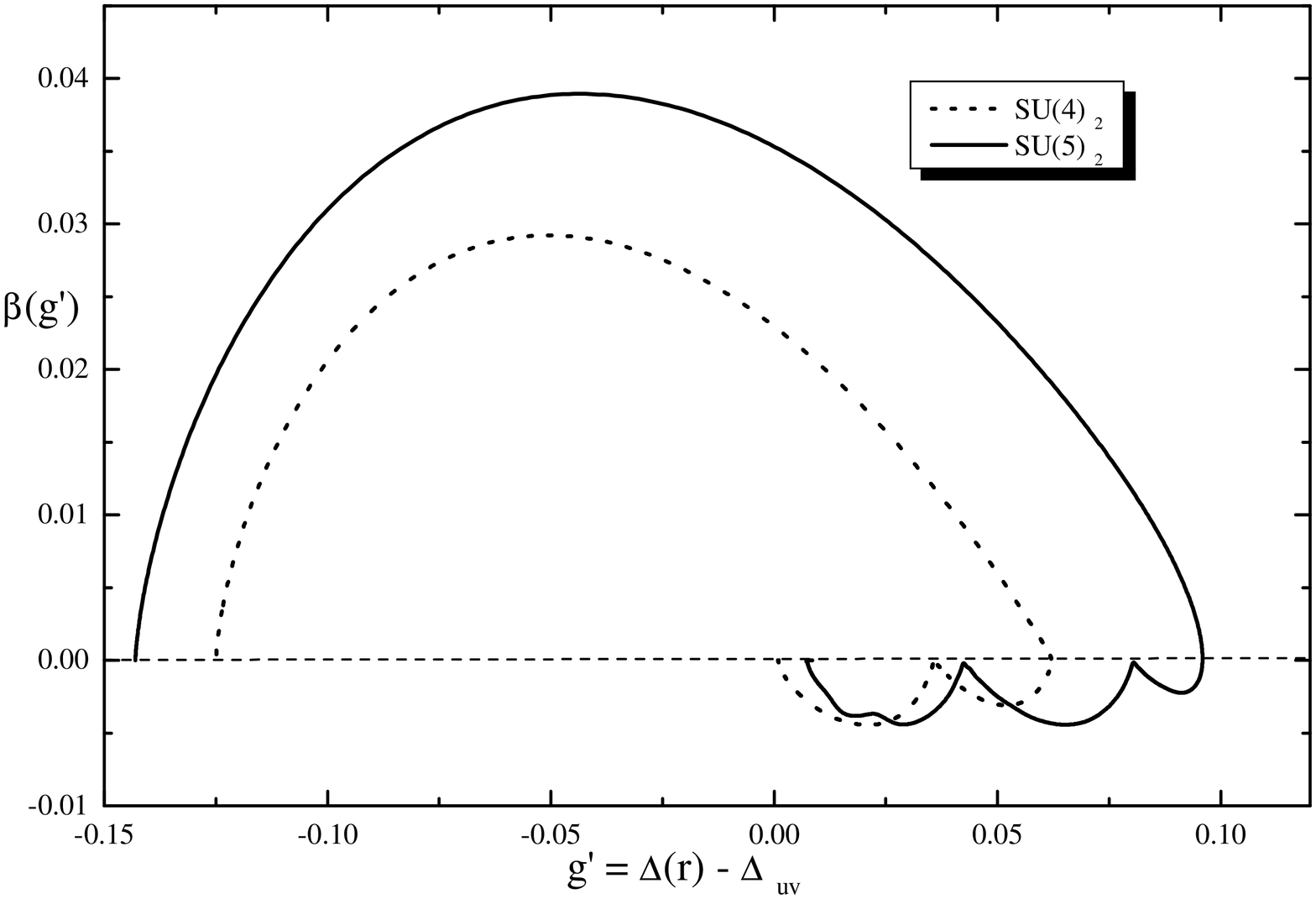}
\caption{The $\protect\beta^{\protect\prime}$-function .}
\label{f34}
\end{center}
\end{figure}

Notice by comparing  Fig.  \ref{f33} and \ref{f31}, that, as we expect, the transition
from one value for $\Delta $ to the one in the decoupled system occurs at
the same energy scale $t_{0}$ at which the value of the Virasoro central
charge flows to the new one which makes our picture consistent.

In analogy to (\ref{fix}) we may now define a function ``$\beta ^{\prime }$''
and demand that it obeys the Callan-Symanzik equation \cite{CS}
\begin{equation}
r_0 \frac{d}{d{r_0}}g^{\prime }=\beta ^{\prime }(g^{\prime })\,\,.  \label{betade}
\end{equation}
\noindent
The ``coupling constant'' related to $\beta ^{\prime }$ is normalized
in such a way that it vanishes at the ultraviolet fixed point, i.e. $
g^{\prime }:=\Delta (r_0)-\Delta _{\text{uv}}$, such that whenever we find $%
\beta ^{\prime }(\tilde{g}^{\prime })=0$, we can identify $\hat{\Delta}=%
\tilde{g}^{\prime }-\Delta _{\text{uv}}$ as the conformal dimension of the
operator under consideration of the corresponding conformal field theory.
From our analysis of (\ref{dr0}) we may determine $\beta ^{\prime }$ as a
function of $g^{\prime }$ by means of (\ref{betade}). Our results are
presented in Fig.  \ref{f34}.

Once again, for $SU(4)_{2}$ the accuracy is very high and we clearly read
off from Fig.  \ref{f34} the expected fixed points 
{\small \ }$\tilde{g}^{\prime}=-0.125,0,0.0375,0.0625$. 
The $SU(5)_{2}$-fixed points $\tilde{g}^{\prime}=
-0.1429,0,0.0446,0.0821,0.1071$, are once again slightly shifted.

Notice that the behaviour of the function $\beta^{\prime}$, although
it appears to be  a bit peculiar  is just what one expects, taking into account that the function
$\Delta(r_0)$ displayed in Fig.  \ref{f33} is not monotonically increasing nor decreasing but it
is non-decreasing in some finite interval of values of $r_0$ and decreases thereafter until the infrared value 
$\Delta_{ir}=0$ is reached. Consequently its derivative, $\beta^{\prime}$ must change sign 
once the energy scale for which the $\Delta(r_0)$-function starts decreasing is reached.

\subsection{Identifying the conformal dimension of the perturbing operator}
\label{pp}
\begin{figure}
\begin{center}
\includegraphics[width=10.7cm,height=8.5cm]{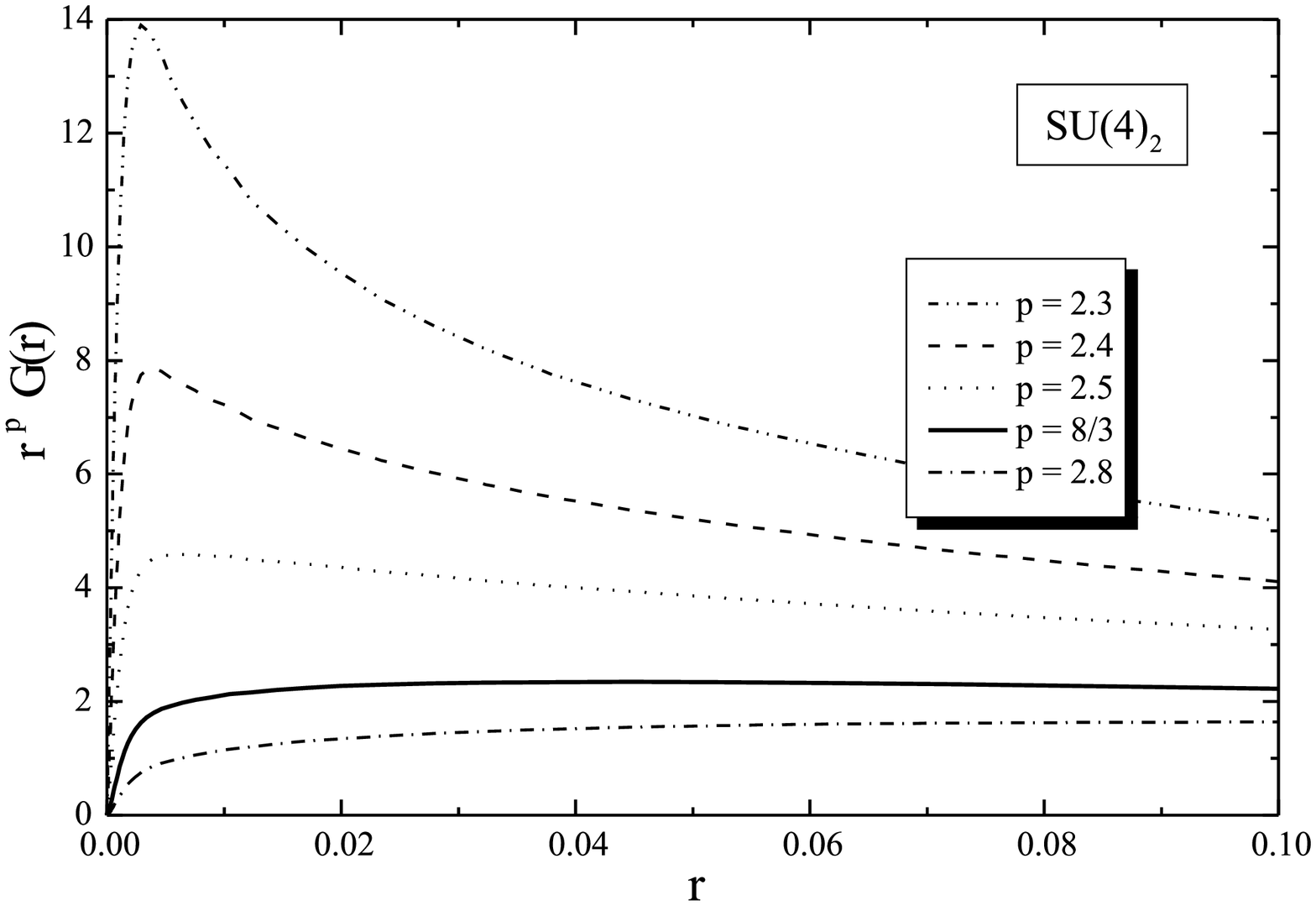}
\caption{Rescaled correlation function $\protect{G(r)=\left\langle
\hat{\Theta} (r)\hat{\Theta} (0)\right\rangle}$ as a function of $\protect{r }$ for the
 $\protect{SU(4)_2}$-HSG model.}
\label{f35}
\end{center}
\end{figure}

\begin{figure}
\begin{center}
\includegraphics[width=10.7cm,height=8.5cm]{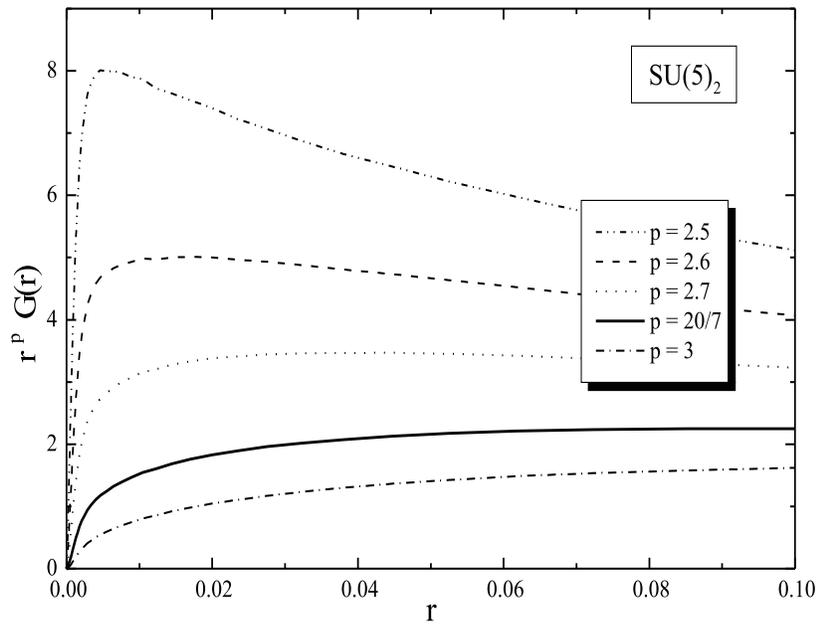}
\caption{Rescaled correlation function $\protect{G(r)=\left\langle
\hat{\Theta} (r)\hat{\Theta} (0)\right\rangle \,}$ as a function of $\protect{r}$ for the 
$\protect{SU(5)_2}$-HSG model.}
\label{f36}
\end{center}
\end{figure}

\indent \ \
As mentioned in several occasions, whenever the correlation function between $\mathcal{O}$
and $\Theta$ is vanishing or we consider an operator which does not flow to a primary field, 
we can not employ (\ref{dr0}) with $r_0=0$  to identify the ultraviolet conformal
dimension and we must resort to the well known relation reported in section \ref{cftqft} of
chapter \ref{ntft} and in Eq. (\ref{ultra}) 
\begin{equation}
\lim_{r\rightarrow 0}\left\langle {\cal O}(r){\cal O}(0)\right\rangle \sim
r^{-4\Delta ^{\!\!{\cal O}}}.  \label{ultra2}
\end{equation}
near the critical point in order to determine the conformal dimension. To
achieve consistency with the proposed physical picture we would like  to identify now 
in particular the conformal dimension of the perturbing operator which we identified in section \ref{operator} to be  $N/N+2$. Recalling that the trace of the energy momentum tensor is proportional to the perturbing field \cite{cardy} we analyse $\left\langle \hat{\Theta} (r)\hat{\Theta} (0)\right\rangle $ 
for this purpose and present our results in Figs.  \ref{f35} and \ref{f36}.

According to (\ref{ultra2}), we deduce from Figs. \ref{f35} and \ref{f36} that $\Delta =2/3,5/7$ for $%
N=4,5$, respectively, which coincides with the expected values.

\section{Summary of results and open problems}
\label{con5}
\indent \ \
One of the main deductions from our analysis is that the scattering matrix
proposed in \cite{HSGS} may certainly be associated to the perturbed gauged
WZNW-coset. This is based on the fact that we reproduce all the predicted
features of this picture, namely the expected ultraviolet Virasoro central charge, by means
of Zamolodchikov's $c$-theorem, various conformal dimensions of local operators by means
of the $\Delta$-sum rule \cite{DSC} or the direct investigation of the UV-limit
of the two-point correlation functions, and the
characteristics of the unstable particle spectrum, that is the decoupling (\ref{flow}) which 
generates the sort of flows observed for the $c(r_0)$- and $\Delta(r_0)$-functions.

\vspace{0.3cm}

In the previous chapter we presented general $n$-particle solutions
to the form factor consistency equations. Likewise, 
our construction of general solutions to the form factor consistency
equations involves the same type of  determinant building blocks found for the $SU(3)_2$-case.
Therefore, our results support the belief that such type of determinants might constitute the
basis of a generic group theoretical structure which is ``hidden'' in the form factor  consistency
equations of section \ref{gen}. Concerning this issue, our results
certainly constitute a further important step towards a generic
group theoretical understanding of the $n$-particle form factor expressions.
The next natural step is to extend our investigation towards higher level
algebras which involves  the complication outlined in the introduction.

\vspace{0.3cm}

By performing a renormalisation group analysis both of the $c$-function \cite{ZamC}
and of the conformal dimensions \cite{DSC} we found that,
concerning the computation of correlation functions, our results 
indicate that the fast convergence of the series expansion of (\ref{corr}) 
observed for models involving only one stable particle, or 
even two, like in the $SU(3)_2$-case, is spoilt when the
number of particles involved is large. Our concrete analysis for
the $SU(N)_2$-HSG models seems to suggest that such a behaviour 
is more a consequence of the increasing importance of the symmetry 
factor entering the expansion (\ref{corr}) for
increasing values of the particle number, than of the particular
functional dependence of the corresponding form factors. 
Concerning the study of the convergence behaviour of (\ref{corr}),
qualitative arguments providing an intuitive understanding of
the suppression of the higher particle form factors in the series
(\ref{corr}) have been presented in \cite{Giu}. It would be interesting
to generalise the latter arguments to the models under investigation and 
highly desirable to have  more concrete quantitative criteria at hand.

\vspace{0.3cm}

Despite the fact of having identified part of the operator content, it
remains a challenge to perform a definite one-to-one identification between
the solutions to the form factor consistency equations and the local
operators. It is clear that we require new additional technical tools to do
this, since the $\Delta $-sum rule (\ref{dr0}) may not be applied in all
situations and (\ref{ultra2}) does not allow a clear cut deduction of $\Delta$ unless one has
already a good guess for the expected value.

\vspace{0.3cm}

In comparison with other methods to achieve the same goal, we should note
that in principle we could obtain, apart from conformal dimensions different
from the one of the perturbing operator, the same qualitative picture from a
TBA-analysis similar to the one performed in chapter \ref{tba}. For instance, the
scaling functions obtained in this chapter exhibit qualitatively the same kind of
staircase pattern.
However, in the TBA-approach the number of
coupled non-linear integral equations to be solved increases with $N$, which
means the system becomes extremely complex and cumbersome to solve even
numerically. Computing the scaling function with the help of form factors
only adds more terms to each $n$-particle contribution, but is technically
not more involved.  Therefore, again the form factor  program seems to have more advantages than
a TBA-analysis,  the price we pay in this setting is, however, the slow convergence of (\ref{corr}).

\vspace{0.3cm}

We conjecture that the ``cutting rule'' (\ref{flow}) 
which describes the renormalisation group flow also holds for other groups different from $SU(N)$.
This is supported by the general structure of the HSG-scattering matrix \cite{HSGS}.

\vspace{0.3cm}
It would be desirable to put further constraints on the solutions to the form factor consistency
equations by using the symmetries of the HSG-models, formulating quantum equations of motion, doing
perturbation theory or by other means.

\chapter{Conclusions and open questions}
\label{conclu}
\indent \ \
Although we have already provided each of chapters \ref{tba},
 \ref{ffs} and \ref{ffs2} 
of this thesis with a summary of the main results and open points, it is now 
our aim to provide the reader with a more general and concise
review of all the original results presented in this thesis and
also,  to match the outcome of the different approaches we
have exploited along the previous chapters in a way which was not
possible when closing each independent chapter, since all the results had not
been presented yet. 

\vspace{0.4cm}
One of the main purposes of the work presented in this thesis has been the
development of different non-perturbative consistency checks for the S-matrices proposed
in \cite{HSGS} to describe the scattering theory corresponding
to the  {\bf Homogeneous sine-Gordon (HSG) models} \cite{HSG2}
associated to simply-laced Lie algebras. The HSG-models are particular
examples of a large family of theories which are known in the literature
as {\bf Non-abelian affine Toda field theories} \cite{nt}. These are 1+1-dimensional 
theories  possessing very interesting properties which have motivated their study
over the last years. For instance, they are all  classically integrable and, at quantum
level, they give rise to two families of unitary, massive and integrable quantum field
theories which have been named as symmetric space and homogeneous 
sine-Gordon models \cite{HSG}. As we have mentioned, our work has been mainly
concerned with the second class of models, although we also provided some results 
arising in the study of the quantum properties of the  {\bf symmetric space sine-Gordon (SSSG)}  models. 

\vspace{0.25cm}

With regard to the main objective of this work recalled above we draw the fundamental
conclusion that all the results presented in this thesis confirm the consistency of the S-matrix
proposal \cite{HSGS}. In particular, not only the concrete S-matrix proposal but also the physical
picture advocated for the HSG-models in \cite{HSG2, HSGsol, HSGS} has been confirmed by
our analysis. The latter analysis has been carried out by making use of two different approaches:
The {\bf thermodynamic Bethe ansatz} (TBA) \cite{Yang, TBAZam1}
and the {\bf form factor} approach  \cite{Kar, Smir}. 

\vspace{0.25cm}
However, it must be emphasised that the interest of the present work is not reduced  to
the single (although remarkable) fact that, after different extensive tests,
the S-matrices proposed in \cite{HSGS} have proved to be entirely consistent.
An important part of our motivation has been also 
to contribute to the improvement of the present understanding
of  the thermodynamic Bethe ansatz and form factor approaches themselves. In this direction,
the work presented in this thesis represents a valuable contribution, since the latter approaches
had never been exploited to such an extent in the context of the study of models possessing all the 
novel features introduced by the  HSG-models. These features are first,  the {\bf parity breaking}
occurring  both at the level of the Lagrangian and the two-particle S-matrices and second,
the presence of {\bf resonance poles} in the scattering amplitudes. The last characteristic is 
not a novel feature by itself, since other models containing resonance poles have
been treated in the literature \cite{staircase, Martins, DoRav, staircase2, new}. 
However, the novelty is that these poles  admit
a physical interpretation as the trace of {\bf unstable particles} in the spectrum
and at the same time,  a well-defined Lagrangian description of the
theory is available. This is not the case for many other  1+1-dimensional 
integrable quantum field theories (QFT's), which are only consistent 
from the scattering theory viewpoint. In addition, the HSG-models are characterised by
the existence of several independent mass scales, $m_i$,  and resonance parameters, $\sigma_{ij}$, 
associated to resonance poles in the scattering amplitudes. This feature is also remarkable, since
these models turn out to be integrable for arbitrary values of the mentioned free parameters, 
and the freedom for choosing  their  relative values gives rise to 
a very interesting physical picture whose interpretation 
is one of the most important results of the present work.

As many other 1+1-dimensional massive integrable QFT's, the HSG-models can be understood as
perturbed conformal field theories (CFT's). The corresponding underlying CFT's are 
{\bf Wess-Zumino-Novikov-Witten} (WZNW) coset models \cite{wzw, wzw2} related to cosets of the form
$G_k/U(1)^{\ell}$, for $\ell$ to be the rank of the Lie algebra $g$ associated
to the Lie group $G$ and $k$ to be an integer called the level. Such CFT's are also
known under the name of {\bf $G_k$-parafermion theories} \cite{Gep, GepQ, DHS}.
It is 
worth mentioning that even  the simplest examples of HSG-models associated to the lowest
values of the rank of the Lie algebra, $\ell$, and the level, $k$, give rise to theories whose operator 
content is quite involved. This is related to the fact that
the  underlying CFT's corresponding to the HSG-models have always Virasoro central charge
$c>1$, which implies the existence of infinitely many conformal primary fields in
the unperturbed theory. This aspect is responsible of some of the most interesting results obtained
within the form factor framework: the understanding of the {\bf momentum space cluster property}
not as a mere property observed for the form factors of some particular theories but also 
as a construction principle of solutions to the form factor consistency equations 
\cite{Kar, Smir, Zamocorr, YZam, BFKZ}, and the re-construction of a large part of the operator
content of a quantum field theory by exploiting the knowledge of the operator content of
its underlying CFT and of the correlation functions involving the mentioned operators, available
in the form factor context. 

\vspace{0.25cm}
Recall that, the thermodynamic Bethe ansatz and form factor approach
are non-perturbative methods which allow, amongst other applications, 
for reproducing  the most characteristic data of the underlying CFT arising in 
the UV-limit of a certain 1+1-dimensional QFT. The starting point for carrying out both, a 
TBA- or a form factor analysis, is  the knowledge of the exact S-matrix characterising
the latter QFT and consequently its mass spectrum. Provided this initial input, which
is available for many 1+1-dimensional integrable massive QFT's due to their distinguished
properties, the following quantities characterising the unperturbed CFT 
and massive QFT are in principle extractable:

\vspace{0.25cm}
{\bf i)} the {\bf Virasoro central charge} of the underlying CFT,

\vspace{0.25cm}
{\bf ii)} the flow of the Virasoro central charge of the ultraviolet CFT 
  from its critical value to the infrared limit, surpassing the different energy scales fixed
by the coupling constants of the model. 
It is possible to construct different  functions reproducing qualitatively the same type of flow namely,
carrying the same physical information. In the TBA- and form factor approach these functions are
the so-called {\bf finite size scaling function} \cite{scaling} and {\bf Zamolodchikov's c-function} \cite{ZamC}, respectively. Both functions admit an interpretation as a measure of effective light
degrees of freedom in a QFT. They reproduce the successive decoupling of massive degrees of freedom of the QFT as soon as the energy  necessary for the onset of a certain particle in the spectrum is much higher that the temperature of the system or energy scale considered, 

\vspace{0.25cm} 

{\bf iii)}  the {\bf conformal dimension} of the {\bf perturbing field} which
takes the underlying CFT away from its renormalisation group fixed point in
the construction of   the massive QFT \cite{Pertcft},

\vspace{0.25cm} 

{\bf iv)} the conformal dimensions of other local operators arising in the underlying
CFT different from the perturbing field. In other words, one may  be able to reconstruct,
at least partially,  the {\bf operator content} of the underlying CFT (at least for the
primary fields) by exploiting 
data characteristic from the massive QFT and assuming a one-to-one correspondence
between the operator content of the unperturbed CFT and the massive QFT, 

\vspace{0.25cm}
{\bf v)} the renormalisation group flow of the operator content of the underlying CFT from
the ultraviolet to the infrared limit by
explicit evaluation of the conformal dimensions of certain local operators of the unperturbed
CFT as functions of the renormalisation group parameter,

\vspace{0.25cm}

{\bf vi)} the correlation functions involving certain local operators of the massive QFT, whose
UV-behaviour  is governed by the ultraviolet conformal dimension of the 
operator under consideration. 

\vspace{0.25cm}
We will now report in more detail our original results, 
going through  each of  the previous points and explaining  how
these quantities may be available within each of the approaches considered. As
stressed along the different chapters, the results presented in this thesis
may be found in \cite{cm, CFKM, CFK, CF1, CF2, CF3}:

\vspace{0.25cm}

Concerning the evaluation of the Virasoro central charge of the underlying CFT in the TBA-framework,
in the deep ultraviolet limit we recover the $G_{k}/U(1)^{\ell }$-coset central charge for any value
of the $2\ell -1$ free parameters entering the S-matrix, including the
choice when the resonance parameters vanish and parity invariance is
restored on the level of the TBA-equations. This is in contrast to the
properties of the S-matrix, which is still not parity invariant due to the
occurrence of the phase factors $\eta $, which are required to close the
bootstrap equations \cite{HSGS}. The same observations hold  in the form
factor context. However, within the latter framework, the Virasoro central charge of 
the underlying CFT  has been explicitly evaluated only for some concrete models by 
means of  Zamolodchikov's $c$-theorem \cite{ZamC}. Therefore, 
in what concerns the evaluation of the UV central charge of the underlying CFT,
both the concrete models studied and the nature of the results obtained
are quite different in the TBA- and form factor analysis.

Whereas in the TBA-context the Virasoro central
charge can be exactly determined due to the fact that the TBA-equations admit an
analytical solution in the deep UV-limit for any HSG-model, in the form factor context
the central charge can be evaluated numerically with a certain accuracy which is very high 
when the rank of the Lie algebra is small but becomes worse as soon as the rank is increased.
Recall that the number of stable particles in the model is given by $\ell \times (k-1)$
which means increasing (decreasing) the rank of the Lie algebra is equivalent
to increasing (decreasing) the number of particles in the spectrum. 
 Moreover, the evaluation of the Virasoro central charge is the
form factor context has been carried out only for some of the $SU(N)_2$-HSG
models, which is in contrast with the generality of the results available in the TBA-context.
Technically, the computation of the Virasoro central charge by means of 
Zamolodchikov's $c$-theorem \cite{ZamC} requires the  
numerical computation of a series of multi-dimensional integrals 
arising in the  expansion of the two-point function of the trace
of the energy momentum tensor in terms of its associated $n$-particle 
form factors. Since the mentioned series contains in general an infinite
number of terms and it is not known for the time being how to sum exactly 
all these terms, one has to evaluate separately each individual contribution to the expansion. 
Furthermore, the dimension of the integrals increases with
the number of particles and with the order of the term in the series, which means for
a reasonable computer time, 
the numerical methods available at present only permit to evaluate
the first contributions to  the expansion. Our particular analysis has shown that
for the $SU(3)_2$-HSG model the value of the Virasoro central charge
can be obtained with high accuracy by summing terms up to the 6-particle
contribution. This is still true for the $SU(4)_2$- and $SU(5)_2$-HSG
models but for higher values of the rank of the Lie algebra the precision
reached gets worse and one might be forced to take into account contributions
from higher order terms. 

\vspace{0.25cm}

 As we have mentioned, the same value for the central charge is obtained 
for every finite value of the resonance parameters, 
which reflects the fact that in the ultraviolet limit 
parity invariance is restored both at the level of the
TBA-equations and also when evaluating the central charge
by means of Zamolodchikov's $c$-theorem in the form factor
context. The underlying physical behaviour is, however, quite different for
different values of the resonance parameters as our
numerical analysis has demonstrated. The numerics 
turn out to be entirely consistent with the physical picture anticipated in 
\cite{HSG2, HSGsol, HSGS} for the HSG-models in the two approaches
we have analysed. In the TBA-framework we have evaluated numerically 
the finite size scaling function \cite{scaling}
whereas its counterpart when using form factors 
has been Zamolodchikov's $c$-function \cite{ZamC}.
Both functions turned out to have an entirely analogous qualitative behaviour. 
They exhibit a very characteristic ``staircase''  pattern 
in their flow from the UV- to the IR-regime similar to the one found in  \cite{staircase} for the
roaming trajectory models. However, in the latter case the scaling function is
characterised by  infinitely many plateaux, whereas for the HSG-models only a finite
number of plateaux, in correspondence to the amount of free parameters available in the quantum theory,
was found. Also, for the HSG-models, the interpretation of the scaling functions
as a measure of light degrees of freedom in the QFT allows for a clear physical
interpretation of the results obtained:

\begin{itemize}

\item For vanishing resonance parameter $\sigma=0$ and taking the energy scale of the 
stable particles to be of the same order  $m_1 \simeq m_2$,  the deep
ultraviolet coset central charge is reached straight away. From the physical point of view 
this is the expected behaviour since, according to Breit-Wigner formula (\ref{BW11}), (\ref{BW22}), whenever
the resonance parameter is  vanishing, the same happens to the  decay width of the unstable particles.
Therefore, the unstable particles become `virtual states' associated
to poles on the imaginary axis beyond the physical $\theta$-sheet 
and they are on an energy scale of the same order as
the one of the stable particles. Being the energy scale corresponding to 
the onset of all stable and unstable particles of the same order
  the scaling function takes the value corresponding to the
ultraviolet coset central charge of the underlying CFT as soon as the mentioned scale is reached.

\item 
On the other hand, for non-vanishing resonance parameter the scaling 
function and Zamolodchikov's $c$-function surpass different regions in the
energy scale developing a ``staircase'' behaviour where the number and size
 of  the plateaux  is determined by the relative mass scales between the
 stable and unstable particles and the stable particles themselves. 
Therefore, different choices of the $2 \ell-1$ free parameters at hand lead
  to a theory with a different physical content, but still possessing the same central
charge. This feature is  consistent with the physical picture anticipated for the HSG-models,
since in the deep ultraviolet limit, as long as the resonance parameter is finite, the energy
scale is much higher than the energy scales necessary for the production of all the stable 
and unstable particles. Therefore all the particle content
 of the model contributes to the scaling function which,
interpreted as a measure of  effective light degrees of freedom, will reach its maximum value, namely
the Virasoro central charge of the unperturbed CFT. 

\end{itemize}

\vspace{0.25cm}
The concrete models for which the scaling function and the $c$-function have been evaluated are
different. Whereas in the TBA-context we focused our study in the $SU(3)_k$-HSG models and
evaluate the scaling function for $k=2, 3$ and $4$, in the context of form factors, we studied
the opposite case namely, we fixed the level $k=2$ and tuned the rank of the Lie algebra. This 
shows that the two approaches exploited here are somehow complementary, since the solution
of the TBA-equations, even numerically, gets very involved when the rank increases whereas in
the form factor context the increment of the level means the existence of asymptotic bound states
in the theory which make the so-called bound state residue equations enter the analysis, complicating
the evaluation of the corresponding form factors. In particular the $SU(3)_2$-HSG model is the
only one which has been studied in both approaches, being the results obtained entirely consistent.

\vspace{0.25cm}
Concerning the computation of the scaling function and the $c$-function, the evaluation of the
latter in the form factor context provides very interesting results
 when the rank is increased which 
have not a counterpart in our TBA-analysis. In particular the {\bf decoupling} of the $SU(N)_2$-HSG
model into the $SU(i+1)_2$- and $SU(N-i)_2$-HSG models whenever 
the resonance parameter $\sigma_{i,i+1}\rightarrow \infty$,  provides a further consistency
check for the corresponding S-matrices. Although, as we have said, such
a decoupling has not been studied in the TBA-context apart from the $N=3$ case,
what has been exploited in the latter approach is
the freedom for considering the mass scales of the stable particles to be very different, which
generates an additional plateau in the scaling function for the $SU(3)_k$-HSG models. In the form
factor context the mass scales of the stable particles have been always taken to be of the same order 
and only the freedom for  varying the masses of the unstable particles which depend on the resonance
parameters has been used  in order to reproduce the outlined decoupling. 

\vspace{0.25cm}
Another interesting result concerning the evaluation of the 
scaling and $c$-function is that the sort of flows described by the first of these functions
are associated  to a system of  TBA-equations
which, after the introduction of the auxiliary
 parameter $r^{\prime}=r/2 e^{\sigma/2}$, can be re-interpreted 
as the TBA-equations corresponding to two  massless systems, in the spirit of \cite{triZam}.
However, the connection between flows related to the presence of unstable particles in the spectrum and massless flows was observed only formally in the TBA-context, since the parameter $r^{\prime}$ was introduced aiming towards a simplification of the analytical and numerical analysis. Our later 
renormalisation group analysis has confirmed that the parameter $r^\prime$ has in fact
a deeper physical meaning and arises naturally when studying different sorts of 
renormalisation group flows, supporting the idea that the 
observed flows should be in fact understood as {\bf massless flows}.
 
\vspace{0.25cm}

The similar ``staircase''  behaviour observed both for the HSG-models and for the models studied in 
\cite{staircase, Martins, DoRav} has been emphasised many times along this thesis. However, the
features and interpretation of this staircase pattern are very different for the models studied
in \cite{staircase, Martins, DoRav} and for the ones investigated in this thesis. 
First of all, although both for the former and the latter models resonance poles are present
in the scattering amplitudes, only for the HSG-models these poles admit a physical interpretation
as the trace of the presence of unstable particles which is supported by all the results
found in this manuscript. Second, and closely related to the first observation, the amount
of plateaux present in the scaling functions of the HSG-models is intimately related to the
amount of particles, both stable and unstable, in the spectrum. Consequently, the number of
plateaux observed is always finite, which is in contrast to the roaming trajectory models \cite{staircase}
and their generalisations \cite{Martins, DoRav}, whose scaling functions were shown to develop
infinitely many plateaux. 

Whereas for the HSG-models the resonance parameter enters the S-matrix as a shift in the rapidity variable,
 in the models studied in \cite{staircase, Martins, DoRav} the resonance
parameter arises as a consequence of the analytical continuation to the complex plane 
of the effective coupling  constant $B$,  which characterises the Lagrangian and S-matrix 
of  the sinh-Gordon model \cite{SSG} and,  in fact, of all affine Toda field theories 
\cite{ATFTS, dis, uni}.
The mentioned  complexification takes place in the following way
\begin{equation}
B \rightarrow 1 \pm \frac{2 i \sigma}{\pi}.  
\label{shiftt}
\end{equation} It is interesting to notice that the particular form of  (\ref{shiftt}) 
has an explanation. The real part of $B$ has necessarily to be one so that the
consequent  transformation of the sinh-Gordon S-matrix via  (\ref{shiftt})  generates a new but 
still consistent  S-matrix. The 
consistency of the new S-matrix in guaranteed by the fact that for all affine Toda field theories 
\cite{ATFTS, dis, mussrev} the coupling constant $B$ occurs always in the 
combination $B(2-B)$ which stays real under (\ref{shiftt}).

The introduction of the resonance parameter $\sigma$ by means of  (\ref{shiftt})
makes the S-matrix exhibit a resonance pole in the imaginary axis $\theta_R = - \sigma - \frac{i \pi}{2}$
similarly to the $SU(3)_2$-HSG model. As usual one could try to understand such pole
as the trace of an unstable particle. However, even though only one resonance parameter has been introduced, 
the TBA-analysis carried out in \cite{staircase} for the roaming sinh-Gordon model shows
that the corresponding scaling function develops an infinite number of plateaux. 
Therefore, the results in \cite{staircase} can not be interpreted 
physically by using the same sort of arguments employed in the TBA-analysis of the HSG-models.
Equivalently, the infinite number of plateaux observed
in \cite{staircase} can not be related to the number of free parameters in the model. The same can be
said with respect to the models studied in \cite{Martins, DoRav} whose construction follows the same
lines summarised for the roaming sinh-Gordon model, with the difference that they take as an input the
S-matrices of other simply laced affine Toda field theories instead of the sinh-Gordon model  ($A_1^{(1)}$-ATFT), which is the simplest of their class.

\vspace{0.25cm}
An interesting open problem is the investigation of the precise relationship between 
the scaling function arising in the TBA-context and Zamolodchikov's $c$-function.
As we have repeatedly mentioned, both functions 
provide the same physical information and are qualitatively very similar.
However, by reviewing their definitions and the contexts in which they arise, 
we find that the nature of their possible relationship is not clear a priori. 
Also the relation to the intriguing proposal in \cite{Foda} of a renormalisation group flow between
Virasoro characters remains unclarified. 

\vspace{0.25cm}

Regarding now point {\bf iii)}, the evaluation of the conformal dimension of the
perturbing operator, $\Delta$, has been performed both in the TBA- and form factor  
approach. Both methods have supplied entirely consistent results. 

In the TBA-context, the computation of the conformal dimension of the 
perturbing field is possible by exploiting its relation to the periodicities
of the so-called $Y$-systems, noticed originally in \cite{TBAZamun}, despite the fact
that the reason why this relationship should occur is not known for the time being. The mentioned
conformal dimension 
also arises in the series expansion of the finite size scaling function in
terms of the inverse temperature of the system.  In this thesis we have
exploited the former relationship and identified the conformal dimension
of the perturbing field by determining the periodicities of the $Y$-systems
corresponding to various concrete examples of $SU(3)_k$-HSG models corresponding to $k=2, 3, 4$. 
However, we have not been able to provide a general formula expressing
the dependence  between these periodicities and the conformal dimension of the perturbation. 
Such a relationship has been conjectured in the light of the particular results
obtained for concrete models and more work is needed in order to make
a definite statement. Nevertheless, for vanishing resonance parameter 
and the choice $g=su(2)$, the behaviour observed coincides with the one
obtained in \cite{TBAZamun}, which  suggests  the conjectured dependence
of the periodicities of the $Y$-systems in $\Delta$  is of a
very universal nature, beyond the models discussed here. In order to clearly
determine the conformal dimension of the perturbation it would be
very interesting to carry out the mentioned series expansion of the scaling
function in terms of the inverse temperature, $r$.

Within the form factor framework, the identification of the conformal dimension
of the perturbing operator has been done for the particular  $SU(N)_{2}$-HSG models,
with $N=3, 4, 5$. In that case three fundamental results have been exploited: first of all,
the proportionality between the perturbing field and the trace of the
energy momentum tensor of the massive QFT is well known \cite{Cardypert}. Second, once the form factors of the
trace of the energy momentum tensor are known its two-point function may be
computed up to a certain approximation, as explained for  the Virasoro
central charge. Third, in the UV-limit the two-point function of the trace
of the energy momentum tensor must reduce to the two-point function of its
counterpart in  the underlying CFT. That is a quasi-primary field of the CFT
whose two-point function is forced to diverge in the UV-limit as $r^{-4\Delta}$ in terms of
the conformal dimension of the perturbation. 

\vspace{0.25cm}

Since the perturbing operator plays a distinguished role
in the construction of the massive QFT, the identification
of its conformal dimension constitutes a fundamental check for the
consistency of the proposed S-matrices. However, once the perturbing field has been identified, one
may pose the question of how to identify the remaining operator content of the underlying CFT.
With regard to point {\bf iv)},  the  re-construction of the operator content of the unperturbed CFT from the one of the massive QFT, apart from providing a further consistency check of
the S-matrices under consideration,  is an issue which has great interest in its own right.
Fortunately, the  operator content is well known for the WZNW-coset theories \cite{Gep, DHS},
a fact of which our investigation has taken advantage. Another relevant aspect is that it
becomes soon fairly involved even for low rank Lie algebras. The latter feature 
was an additional motivation for our analysis, since to our knowledge, no other QFT's possessing such an involved operator content had been studied before to such an extent in the form factor context, at least for the purpose
outlined.

Nonetheless,  the question of how to identify the whole operator content of the underlying
 CFT is left unanswered in the TBA-framework. In fact, one of the most important results of 
this thesis has been the identification of the form factor approach as a means  for re-constructing
 at least an important part of the operator content of the unperturbed CFT. Such  a re-construction makes use
of the fundamental assumptions of the existence of a one-to-one correspondence between the operator
 content of the massive QFT and its associated underlying CFT and between the solutions to the 
form factor consistency equations and the local operators of the massive QFT. The mentioned
 identification of part of the operator content has been performed to a large extent for the $SU(3)_2$-HSG
model and for one of the local operators of the $SU(N)_2$-HSG models. For all the
$SU(N)_2$-HSG models, a large subset of solutions to the form factor consistency equations have
been constructed which permitted the former investigation. 

\vspace{0.25cm}

As we have said in the previous paragraph, the identification of an important part of 
the operator content of the massive QFT within 
the form factor approach has been mainly performed for the $SU(3)_2$-HSG model via the calculation of the corresponding  ultraviolet conformal dimension of those operators for which all the $n$-particle form factors had been previously computed. In particular, the conformal dimension of the perturbing operator ($\Delta=3/5$) 
and of several other operators of conformal dimension $\Delta= 1/10$ were obtained by studying 
the ultraviolet behaviour of their two-point functions, which is constrained by the conformal symmetry
of the underlying CFT. For some operators we have determined the
conformal dimensions by means of the $\Delta$-sum rule derived in \cite{DSC}, which involves the
knowledge of the correlation functions of a certain local operator and the trace of the energy momentum
tensor and whose derivation is close in spirit to Zamolodchikov's $c$-theorem. 

In this light, it can be stated  that solutions of the form factor consistency
equations can be identified with operators in the underlying CFT.
In this sense one can give meaning to the operator
content of the integrable massive model. Being the mentioned identification uniquely based on 
the values of the ultraviolet conformal dimensions there is the problem that
once the conformal field theory is degenerate in this quantity, 
as it is the case for the models we investigate,
the identification can not be carried out in a one-to-one fashion and therefore
the procedure has to be refined.  In principle this would be possible by
including the knowledge of the three-point coupling of the conformal field
theory and the vacuum expectation value into the analysis. The former
quantities are in principle accessible by working out explicitly the
conformal fusion structure, whereas the computation of the latter still
remains an open challenge. In fact, what one would like to achieve ultimately
is the identification of the conformal fusion structure within the massive
models.  Considering the total number of operators present in the conformal
field theory (a $SU(3)_2/U(1)^2$-WZNW coset theory for the $SU(3)_2$-HSG model )
one still expects to find additional solutions, in particular the identification of the fields possessing conformal dimension $\Delta=1/2$ is outstanding.

Despite the fact of having identified some part of the operator content, it
remains a challenge to perform a definite one-to-one identification between
the solutions to the form factor consistency equations and the local
operators, at least for the primary field part. In this direction our analysis also proves that  
new additional technical tools are required to do
this, since the $\Delta $-sum rule can not be applied in all
situations, and the study of the ultraviolet behaviour of the two-point functions
does not allow a clear-cut deduction of $\Delta $ unless one has
already a priori a good guess for the expected value.

With respect to the explicit calculation of conformal dimensions, 
technically we have confirmed that the mentioned sum rule  is clearly
superior to the direct analysis of the UV-limit of 
correlation functions. However, it has also the drawback that
the existence of internal symmetries in the massive QFT
may force the two-point function of a large number of local operators with
the trace of the energy momentum tensor to vanish and so, it only applies for part of the
operator content. It would therefore be highly desirable to develop arguments which also apply for
theories with internal symmetries and possibly to resolve the mentioned
degeneracies in the conformal dimensions.

\vspace{0.25cm}
Concerning point {\bf v)}, we have also exploited the mentioned $\Delta$-sum rule as
as a tool which allows for determining the flow of the operator content of a certain CFT
from the UV- to the IR-limit, in a similar spirit to the scaling and $c$-function described
above. For this purpose, we have modified the original sum rule derived in \cite{DSC} by
the introduction of the renormalisation 
group parameter, $r_0$, whose variation from the deep UV-limit, in which we expect to 
recover the conformal dimension of an operator of the underlying CFT, to the IR-regime reveals the same 
 physical picture observed when studying the RG-flow of Zamolodchikov's $c$-function. 
Accordingly, the corresponding  $\Delta(r_0)$-function
develops a series of plateaux whose form and size depend upon
the relative mass scales of the stable and unstable particles. This function has been numerically determined
for the $SU(N)_2$-HSG models corresponding to $N=3, 4, 5$ and for different values of the resonance
parameters, reproducing again the decoupling of massive degrees of freedom
whenever one of the resonance parameters is very large in comparison to the scale
fixed by the RG-parameter $r_0$.

\vspace{0.25cm}

As mentioned above, the physical information extractable from the scaling function, 
Zamolodchikov's $c$-function and the $\Delta(r_0)$-function associated to a certain operator
of the massive QFT is mostly the same. In particular, all these functions develop plateaux 
which are commonly identified with RG-fixed points. In order 
to have a more clear-cut identification of these fixed points it is desirable to 
have new functions at hand whose zeros are precisely in one-to-one correspondence with
the plateaux mentioned before. If we now think of Zamolodchikov's $c$-function,
the sort  of functions we are looking for have similar features to 
the $\beta$-functions entering the Callan-Symmanzik equation \cite{CS}, which are
always vanishing at RG-fixed points. However, any function proportional to the derivative
of the scaling function, Zamolodchikov's $c$-function or the $\Delta(r_0)$-functions
may have also zeros where the latter functions had plateaux. In this thesis we have
exploited this simple observation in order to construct what we called $\beta$-like functions,
originally introduced in \cite{staircase}.
This has been performed for the $SU(4)_2$- and $SU(5)_2$-HSG models and allowed for
clearly identifying the different RG-fixed points the three mentioned functions surpass along
their respective flows. 

\vspace{0.25cm}

Concerning point {\bf vi)}, we have already mentioned many  times the possibility
of computing correlation functions from form factors. In principle, any correlation
function involving local operators of the massive QFT can be computed provided
all $n$-particle form factors associated to these local operators are known. In practise,
we have seen that the correlation functions admit a series expansion in terms
of form factors of the operators involved, whose exact evaluation is outstanding
for all  models except for the thermally perturbed Ising model. 
In the thermally perturbed Ising model the two point functions
$\langle \Theta(r) \Theta(0) \rangle$, $\langle \Theta(r) \mu(0) \rangle$, 
$\langle \Theta(r) \Sigma(0) \rangle$,  involving the trace of the energy momentum
tensor and the order and disorder operators can be exactly evaluated, since
the only non-vanishing form factor of the trace of the energy momentum tensor
is the two-particle one, and the corresponding integrals can even be performed
analytically. However, this is not the case for the models we have studied here and
in our analysis we have summed at most terms up to the  8-particle contribution, which
already requires a lot of computational effort. Also, although we already pointed
it out in the context of the evaluation of the Virasoro central charge, we have
observed in this thesis that the common assumption that the mentioned series
converges rapidly does not hold  anymore when the number of particle
types present in the model increases.

\vspace{0.25cm}
Regarding the form factor program itself, our main results are the following. 
At the mathematical level, a closed formula  for all $n$-particle  form factors associated to a large class of operators has been found for all $SU(N)_2$-HSG models. 
This formula is given in terms of building blocks which can be expressed both as determinants whose entries are elementary symmetric polynomials or by means of an integral representation.
However, it remains  an open question, whether the general solution procedure
presented in chapters \ref{ffs} and \ref{ffs2} can be generalised to the degree that 
determinants of the type found will serve as generic building blocks of form
factors. It remains also for us to be understood 
 how the integral representation obtained for these determinants might
be used in practise, for instance,  to formulate rigorous  proofs of the type presented in the previous
two chapters. In general, it  would be very interesting  to clarify whether such a representation 
has a purely formal nature or  it is maybe more fundamental than the determinant representation  we
have exploited in this thesis.  Another interesting open problem  is to find out the
 precise relationship between this integral representation, which involves
 contour integrals in the $x=e^{\theta}$-plane, and the integral representation
 used for instance in \cite{BFKZ} which involved contour integrals in the $\theta$-plane.

It would be also desirable to put further constraints on the solutions
to the form factor consistency equations in order to 
make more clear their identification with a particular local operator of the model.
As we have mentioned, such an identification has being carried out by analysing the
UV-behaviour of the corresponding two-point functions, a procedure which has the
inconveniences already reported. The identification procedure 
could be possibly refined by means of other arguments,
that is, exploiting the symmetries of the model,
formulating quantum equations of motion, possibly performing perturbation theory etc...

\vspace{0.25cm}

Concerning the momentum space cluster property, our analysis also provides
remarkable results. We have shown for a concrete model, the $SU(3)_2$-HSG model,
that it does not only constrain the solutions to the form factor consistency
equations but also serves as a construction principle for new solutions. 
Clearly, it would be very interesting to develop arguments which allowed us to reach a level
of understanding  of this property similar to the one we have for the rest of the form factor consistency equations \cite{Kar, Smir,Zamocorr,YZam,BFKZ}.

\vspace{0.25cm}
As a way to close this chapter, we can draw the overall conclusion 
that the physical picture advocated
for the $SU(N)_2$-homogeneous sine-Gordon models in \cite{HSG2, HSGsol, HSGS},
rests now on quite firm ground and, apart from the open problems already mentioned
along this chapter, there is also the more general challenge to extend our 
analysis to the 
complete generality of the HSG-models. Also, although it has only 
been mentioned very briefly in chapter \ref{ntft}
of this thesis, there is still a lot of interesting work to be done in what concerns the
study of the other family of unitary, massive and quantum integrable NAAT-theories,
the symmetric space sine-Gordon models  \cite{HSG, cm, tesinha}, which is left for 
future investigations.

\newpage

\indent \ \ 

\vspace{1cm}

{\bf Acknowledgments}

\vspace{0.3cm}

Before I add here, in my mother tongue, 
the original acknowledgments which appear in the version
of this thesis, presented on the 31st July 2001, I would like in 
addition to express my acknowledgment to my thesis supervisor J.L.~Miramontes
Antas for introducing me to this field of research,  
and to the members of my thesis commission, P.E.~Dorey, A.~Fring, G.~Mussardo, 
J.~S\'anchez Guill\'en and G.~Sierra 
Rodero, for their participation and judgment. I am very specially indebt 
to A.~Fring, in collaboration with who a large part of this work was carried out,  
and to P.~Dorey and G.~Mussardo, for sending me various valuable comments and
suggestions. 

\vspace{0.3cm}

This work has been supported partially by DGICYT (PB96-0960), 
CICYT (AEN99-0589), the EC Commission via a TMR Grant (FMRX-CT96-0012),
Institut f\"ur Theoretische Physik,
Freie Universit\"at Berlin and  Deutsche
Forschungsgemeinschaft (Sfb288).

\vspace{1cm}

{\small

\hspace{8cm}{Berlin,  4 de xu\~no do 2001}

\vspace{1cm}

\noindent {\em{ \huge \bf Q}ueridos colegas e amigos,

\vspace{0.3cm}

Penso que a presentaci\'on da mi\~na tese doutoral \'e acontecemento
de suficiente importancia como para, alomenos intentar, facer xustiza a
todos aqueles que sodes, dun xeito ou doutro, responsables de que as p\' axinas
que seguen estean agora nos vosas mans. Agardo que saibades perdoa-la lonxitude
do que segue (no que a agradecementos se refire) mais, tiven a fortuna neste
tempo de contar co apoio persoal e/ou profesional de moitas persoas, e non 
quixera, nunha ocasi\'on coma esta, esquecer a ningu\'en.

\vspace{0.15cm}

Adicarei pois 
a mi\~na primeira menci\'on \'o Catedr\'atico desta universidade,
D. Jos\'e Luis Miramontes Antas. Gracias a \'el tiven a oportunidade
de traballar en F\'{\i}sica Te\'orica, un traballo non sempre doado, pero 
que me proporcionou, e ainda o fai, moitas satisfacci\'ons.

\noindent A Luis Miramontes d\'ebolle tam\'en a mi\~na introducci\'on \'o campo
m\' ais espec\'{\i}fico dos sistemas integrables en 1+1-dimensi\'ons, 
compartindo comigo a s\'ua ampla experiencia 
no que se refire \'as teor\'{\i}as de Toda non abelianas, tan presentes,
tanto no meu traballo de tesi\~na, coma na tese que vai a continuaci\'on. 

\hyphenation{di-fe-ren-tes}

\noindent 
Dada a enorme relevancia que a colaboraci\'on coa Freie Universit\"at Berlin, 
en particular con Andreas Fring, tivo para a realizaci\'on desta tese, 
quixera tam\'en agradecerlle a Luis Miramontes o seu apoio \' as
mi\~nas diferentes estancias en Berl\'{\i}n durante o ano 2000, impartindo
incluso parte das horas de docencia que a min me correspond\'{\i}an,  e a
s\'ua interese no traballo desenvolvido nestes per\'{\i}odos.

\noindent Finalmente, non podo deixar de agradecerlle a Luis Miramontes a
 lectura exhaustiva
deste manuscrito a cal, de seguro, ten contribuido \'a
 mellora na presentaci\'on
dos resultados obtidos, e as moitas discusi\'ons mutuas e tempo adicado
a mi\~na formaci\'on nestes anos de doutorado. 

\vspace{0.15cm}

En segundo lugar, a\'{\i}nda que, definitivamente, non menos importante
para a realizaci\'on desta tese, debo menciona-la 
colaboraci\'on con Andreas Fring,
da Freie Universit\"at Berlin.

\noindent 
Penso que non ser\'a necesario po\~ner demasiada \'enfase na importancia
de dita colaboraci\'on, a cal queda clara \'a luz da lista de artigos
nos que esta tese se basea, e que presento algunhas follas m\'as adiante.

De seguro, todo o que  aqu\'{\i} poida dicir ser\'a insuficiente, no
que respecta \'o meu agradecemento a Andreas Fring, mais farei o posible
por mencionar, alomenos, os aspectos m\'ais relevantes:

\noindent 
Profesionalmente, debo agradecerlle o ter compartido connosco a s\'{u}a
ampla experiencia na amplicaci\'on do denominado `Bethe ansatz termodin\'amico'
a diferentes teor\'{\i}as integrables en 1+1-dimensi\'ons, o cal deu
lugar \'o primeiro froito da nosa colaboraci\'on.  Para \'el vai tam\'en 
o meu agradecemento por terme introducido \'o campo do c\'alculo de factores
de forma
e \'as diferentes aplicaci\'ons de tales obxectos. Agrad\'ezolle
tam\'en o ter compartido comigo
a s\' ua experiencia no referente \'os diferentes 
m\'etodos computacionais que son imprescindibles, tanto
no campo de factores de forma, coma no contexto do
Bethe ansatz termodin\'amico.

\noindent 
De xeito m\'ais xeral, debo agradecerlle a Andreas Fring a enorme confianza
que depositou en min  dende a mi\~na primeira visita a Berl\'{\i}n, facendo
posibles outras d\'uas visitas posteriores, parcialmente financiadas pola
Freie Universit\"at Berlin, e propo\~n\'endome e apoi\'andome
posteriormente para  o posto que ocupo na actualidade. 
Tanto durante as mencionadas estancias en Berl\'{\i}n, coma durante
a s\'ua estancia na Universidade de Santiago de Compostela, en novembro
do 2000, coma na actualidade, d\'ebolle incontables descusi\'ons e
explicaci\'ons e agrad\'ezolle, dun xeito moi especial, o seu constante
aprecio e respeto pola mi\~na contribuci\'on \'o noso traballo en 
com\'un, que de seguro aumentou a mi\~na motivaci\'on e seguridade.
Tam\'en debo extende-lo meu agradecemento a un plano m\'ais persoal,
xa que tiven (e te\~no) a fortuna de contar con apoio persoal e amistade
de Andreas Fring, aspectos que foron de gran importancia no desenvolvemento 
do noso traballo en com\'un, fac\'endoo m\'ais levadeiro e fruct\'{\i}fero. 
Este apoio foi especialmente importante durante a etapa de escritura desta
tese doutoral, e tam\'en na mi\~na adaptaci\'on (a\'{\i}nda en progreso)
a unha nova vida en Berl\'{\i}n. 

\noindent Para rematar, quero agradecerlle a Andreas Fring a s\'ua lectura
coidadosa desta tese, que a mellorou sustancialmente e tam\'en o terme
dado a oportunidade de traballar cunha persoa \'a que s\'o poido 
calificar como  humana e profesionalmente excepcional.

\vspace{0.15cm}

Continuando estes longos agradecementos, non podo esquecer a Christian
Korff, persoa coa que tam\'en tiven a sorte de colaborar, en d\'uas das
publicaci\'ons que se recollen nesta tese. Agrad\'ezolle de xeito especial
a s\'ua amabilidade e enorme axuda durante a
 mi\~na primeria estancia en Berl\'{\i}n, en marzo do 2000, e o seu
talante persoal, que fixo sempre agradables e levadeiras as horas
de traballo en com\'un.

\vspace{0.15cm}

A continuaci\'on, quero adicar tam\'en unha especial menci\'on \'o Catedr\'atico
da Universidade de Santiago, D. Joaqu\'{\i}n S\'anchez Guill\'en, pola s\'ua
continua interese, tanto no meu traballo de tesi\~na coma nesta tese doutoral
e adicionalmente, por ter sido a s\'ua visita a Berl\'{\i}n, no ver\'an
do ano 1999, a que fixo posible a posterior estancia de Christian Korff
no Departamento de F\'{\i}sica de Part\'{\i}culas da Universidade de Santiago
e, polo tanto, propiciou o comenzo da colaboraci\'on xa mencionada con
anterioridade.

\vspace{0.15cm}

Quixera tam\'en agradecerlle \'o Profesor da Freie Universit\"at Berlin,
  Robert Schrader, o ter autorizado a financiaci\'on
de parte das mi\~nas estancias en Berl\'{\i}n no ano 2000 e tam\'en
a s\'ua acollida no grupo que encabeza,
tanto naqueles per\'{\i}odos, coma na actualidade.

\vspace{0.15cm}
Non podo deixar de mencionar aqu\'{\i} \'o Profesor Em\'erito da Freie Universit\"at
Berlin, Bert Schroer, pola gran interese que mostrou no
traballo desenvolvido nesta tese e os interesantes comentarios e suxerencias
que, referentes \'o seu contido, nos fixo chegar nos \'ultimos meses.

\vspace{0.15cm}

De xeito xeral quero expres\'arlle-lo meu agradecemento a t\'oda-los
membros do Departamento de F\'{\i}sica de Part\'{\i}culas da Universidade
de Santiago, polo inmellorable ambiente de traballo e compa\~neirismo
 no que tiven a oportunidade de iniciar a mi\~na formaci\'on. 
Tam\'en, quero adic\'arlle-lo meu agradecemento, a t\'odo-los membros
do grupo encabezado polo profesor Robert Schrader do
Institut f\"ur Theoretische Physik, Freie Universit\"at Berlin,
pola s\'ua amabilidade e boa acollida,
tanto durante as mi\~nas diferentes estancias do ano pasado, coma
dende a mi\~na incorporaci\'on como membro do mencionado grupo, 
en febreiro deste ano.  

\vspace{0.3cm}

Ademais das anteriores persoas, debo mencionar tam\'en a outras moitas,
a axuda das cales est\'a m\'ais ligada o plano da amistade ca \'o 
cient\'{\i}fico mais que, non obstante, foron tam\'en moi importantes 
para o desenvolvemento do traballo que se presenta aqu\'{\i} e encheron
de lembranzas agradables os meus anos de doutorado.

En primeiro lugar, pola s\'ua axuda impagable
 durante o per\'{\i}odo
de escritura desta tese, te\~no que citar a Ricardo V\'azquez, unha persoa
\'a que me une unha gran amistade persoal. Penso que non ter\'{\i}a sido
capaz de escribir esta tese no tempo no que o fixen se non tivese contado
coas s\'uas moitas  palabras de \'animo e a s\'ua confianza na mi\~na
capacidade para facelo. Neste senso, a s\'ua axuda e constante apoio persoal
foron, e son  a\'{\i}nda agora, moi importantes. Agrad\'ezolle tam\'en 
 a s\'ua interese no meu traballo, e 
a s\'ua axuda no desenvolvemento dalgunhas das partes desta tese, \'o
compartir comigo a s\'ua experiencia na programaci\'on en FORTRAN,
facendo as\'{\i} posible unha parte importante dos c\'alculos
num\'ericos que levei a cabo.

\vspace{0.15cm}

En segundo lugar, quero tam\'en mencionar a d\'uas persoas \'as que tam\'en
me une a amistade persoal: Dolores Sousa e Jaime \'Alvarez, ambos compa\~neiros
do departamento de F\'{\i}sica de Part\'{\i}culas da  Universidade de 
Santiago. Gracias a eles, e en xeral, a t\'odo-los meus compa\~neiros do
despacho 009: F\'elix del Moral, Carlos Lozano e, a xa mencionada,
  Dolores Sousa,
 puiden traballar nunha atmosfera inmellorable de compa\~neirismo.
Con Dolores Sousa compart\'{\i}n incontables conversaci\'ons, caf\'es
matutinos e horas de traballo. A s\'ua vitalidade e car\'acter optimista 
foron fundamentais para min en moitas ocasi\'ons, as\'{\i} coma os 
seus intelixentes consellos e continuo apoio persoal no tempo no que
compartimos despacho. A Jaime \'Alvarez, compa\~neiro do despacho 003, 
agrad\'ezolle tam\'en un mont\'on de conversaci\'ons que me axudaron
en momentos dif\'{\i}ciles, o ter sempre palabras de \'animo e amistade
para min, e formar parte de moitas das mi\~nas lembrazas m\'ais positivas
da  etapa de tese.

\vspace{0.15cm}

Ademais, debo mencionar tam\'en \'os veci\~nos do despacho 005,
M\'aximo Ave, Jose Camino, Marta G\'omez-Reino e C\'esar Seijas 
con quen compart\'{\i}n tam\'en moito do  meu tempo de doutorado, e mesmo
nalg\'uns casos, o tempo anterior da carreira. E falando de compa\~neiros de
carreira, debo mencionar especialmente a Jose Castro, 
Patricia Conde, Maite Flores, Carlos Garc\'{i}a
(Qui\~no), Xose Rodr\'{\i}guez ...e unha longa lista, da que de
seguro se me olvidan un mont\'on de persoas. Para todos eles vai o meu
agradecemento sinceiro por facer m\'ais levadeiros os meus anos de
estudios e doutorado.

\vspace{0.15cm}

Tam\'en merecen unha menci\'on especial, varias persoas \'as que
co\~nec\'{\i}n en diferentes congresos e escolas durante estes anos,
que permaneceron en contacto comigo, en moitos casos via e-mail, 
e que tam\'en me proporcionaron apoio e axuda en moitas ocasi\'ons.
Entre eles debo destacar a Miguel Aguado, Giovanni Feveratti, 
Chris Johnson, Ingo Runkel e Jon Urrestilla. 

\vspace{0.15cm}
Quixera tam\'en adicar un agradecemento moi especial a Vera Lesmeister
e \'a s\'ua filla Sangita Lesmeister, con quen tiven a enorme sorte
de compartir casa durante as mi\~nas diferentes estancias en Berl\'{\i}n.
A elas te\~no que agradecerlles a s\'ua enorme axuda nestes per\'{\i}odos,
especialmente a compa\~na que me proporcionaron, facendome sentir en familia,
e a amistade que a\'{\i}nda nos une e que segue sendo de gran importancia 
para min. 

\vspace{0.15cm}

Finalmente, a\'{\i}nda que, non menos importante foi e \'e o meu
amigo de moit\'{\i}simos anos, David Exp\'osito a quen lle debo agradecer
os moitos momentos agradables que compartimos e, en definitiva, o poder
contar coa s\'ua amistade, que segue sendo moi especial e importante para
min na actualidade.

Para rematar vai a lembranza e agradecemento \'a mi\~na familia, parte 
imprescindible da mi\~na vida e punto de referencia e apoio constante
dende sempre, en especial nestes anos, sen o cal non poder\'{\i}a ter
acadado moitos obxectivos. Por ensinarme a quererme e a querelos.

\vspace{0.15cm}

A todos, moit\'{\i}simas gracias.}

\vspace{0.5cm}

{\bf Olalla Castro Alvaredo.}

\vspace{0.5cm}

{\em \hspace{7cm}{En Berl{\'\i}n, a 4 de xu\~no do 2001.}}

\appendix{}
\chapter{Elementary symmetric polynomials.}
\indent \ \
In this appendix we assemble several properties of elementary symmetric
polynomials to which we wish to appeal from time to time. Most of them may
be found either in \cite{Don} or can be derived effortlessly. The elementary
symmetric polynomials are defined as 
\begin{equation}
\sigma _{k}(x_{1},\ldots ,x_{n})=\sum_{l_{1}<\ldots <l_{k}}x_{l_{1}}\ldots
x_{l_{k}}\,\,\,.
\end{equation}
\noindent Some examples are:
\begin{eqnarray}
\sigma_{1}(x_1, \cdots, x_n)&=&x_1 + x_2 + \cdots + x_n,\\
\sigma_2 (x_1, \cdots, x_n)&=& x_1 \, x_2 + x_1 \, x_3 + \cdots
\end{eqnarray}
\noindent The elementary symmetric polynomials  are generated by 
\begin{equation}
\prod_{k=1}^{n}(x+x_{k})=\sum_{k=0}^{n}x^{n-k}\sigma _{k}(x_{1},\ldots
,x_{n})\,\,.
\label{fund}
\end{equation}
and, as a consequence, can also be expressed in terms of an integral
representation 
\begin{equation}
\text{\thinspace }\sigma _{k}(x_{1},\ldots ,x_{n})\,=\frac{1}{2\pi i}%
\oint_{|z|=\varrho }\frac{dz}{z^{n-k+1}}\prod\limits_{k=1}^{n}(z+x_{k})\,,
\label{symint}
\end{equation}
which is convenient for various applications. Here $\varrho $ is an
arbitrary positive real number.

\noindent With the help of (\ref{symint}) we easily derive the identity 
\begin{equation}
\sigma _{k}(-x,x,x_{1},\ldots ,x_{n})=\sigma _{k}(x_{1},\ldots
,x_{n})-x^{2}\sigma _{k-2}(x_{1},\ldots ,x_{n})\,\,,  \label{symp}
\end{equation}
which will be central for us. Using the definition of the 
operators $\mathcal{T}_{a,b }^{\pm \lambda}$ introduced  in
subsection \ref{properties} of chapter \ref{ffs} for the analysis
of the momentum space cluster property, we derive the asymptotic behaviours 
\begin{equation}
\mathcal{T}_{1,\eta }^{\lambda }\,\sigma _{k}(x_{1},\ldots ,x_{n})\sim
\QATOPD\{ . {e^{\eta \lambda }\sigma _{\eta }(x_{1},\ldots ,x_{\eta })\sigma
_{k-\eta }(x_{\eta +1},\ldots ,x_{n})\,\,\,\quad \text{for }\eta
<k}{e^{k\lambda }\sigma _{k}(x_{1},\ldots ,x_{\eta
})\,\,\,\,\,\,\,\,\,\,\,\,\,\,\,\,\,\,\,\,\,\,\,\,\,\,\,\,\,\,\,\,\,\,\,\,\,%
\,\quad \quad \quad \text{for }\eta \geq k}  \label{as+}
\end{equation}
and 
\begin{equation}
\mathcal{T}_{1,\eta }^{-\lambda }\,\sigma _{k}(x_{1},\ldots ,x_{n})\sim
\QATOPD\{ . {\sigma _{k}(x_{\eta +1},\ldots ,x_{n})\,\,\,\quad \qquad \qquad
\quad \text{for }\eta \leq n-k}{\frac{\sigma _{k+\eta -n}(x_{1},\ldots
,x_{\eta })\,\sigma _{n-\eta }(x_{\eta +1},\ldots ,x_{n})}{e^{\lambda
(k+\eta -n)}}\,\,\,\,\,\,\,\,\,\,\,\,\,\,\text{for }\eta >n-k}  \label{as-}
\end{equation}
which may be obtained from (\ref{symint}) as well. 

\chapter{Explicit form factor formulae.}
\indent \ \
Having constructed the general solutions in terms of the parameterisation 
(\ref{fact}), it is simply a matter of collecting all the factors to get
explicit formulae. For the concrete computation of the correlation function,
it is convenient to have some of the evaluated expressions at hand in form
of hyperbolic functions. The following abbreviation
\begin{equation}
\tilde{F}_{\text{mim}}^{ij} (\theta)= e^{-\theta/4 }{F}_{\text{mim}}^{ij}(\theta), 
\end{equation}
\noindent will be used in what follows.

\subsubsection{One particle form factors}

\begin{equation}
F_{1}^{{\mathcal{O}}_{1,0}^{0,0}|+}=F_{1}^{{\mathcal{O}}%
_{0,1}^{0,0}|-}=F_{1}^{{\mathcal{O}}_{0,1}^{0,1}|-}=F_{1}^{{\mathcal{O}}%
_{1,0}^{1,0}|+}=H^{1,0}=H^{0,1}
\end{equation}

\subsubsection{Two particle form factors}

\begin{eqnarray}
F_{2}^{^{\!\!\Theta|\pm \pm }} &=&-2 \pi m^{2}
\sinh \tfrac{\theta }{2}, \\
F_{2}^{^{\!\!\mathcal{O}|\pm \pm }} &=&i\left\langle \mathcal{O}%
\right\rangle \tanh \tfrac{\theta }{2}\quad \qquad \text{for\quad }\mathcal{%
O}=\text{ }\mathcal{O}_{0,0}^{0,0}\text{, }\mathcal{O}_{0,2}^{0,1},\mathcal{O%
}_{2,0}^{1,0},\text{ } \nonumber \\
F_{2}^{{\mathcal{O}}_{1,1}^{0,1}|+-} &=&H^{1,1}e^{\theta _{21}/2}F_{\text{min%
}}^{+-}(\theta ),\quad \quad F_{2}^{{\mathcal{O}}_{1,1}^{1,0}|+-}=H^{1,1}F_{%
\text{min}}^{+-}(\theta ).
\label{twopp}
\end{eqnarray}

\subsubsection{Three particle form factors}

\begin{eqnarray}
F_{3}^{{\mathcal{O}}|\pm \pm \pm }&=&\frac{H^{0,1}\prod_{i<j}F_{\text{min}%
}^{\mu _{i}\mu _{j}}(\theta _{ij})}{\prod_{1\leq i<j\leq 3}\cosh (\theta
_{ij}/2)}\quad \text{for\quad }{\mathcal{O}}_{1,0}^{0,0}\text{,}{\mathcal{O}}%
_{0,1}^{0,0}\text{, }{\mathcal{O}}_{0,1}^{0,1}\text{, }{\mathcal{O}}%
_{1,0}^{1,0}\text{, } \\
F_{3}^{{\mathcal{O}}^{0,0}_{1,0}|+--}&=&\frac{i H^{1,0} 
e^{-\theta_{1}/2}{(\sigma_2^{-})}^{1/2}}
{2 \,\, \text{cosh}(\theta_{23}/2)} \prod_{i<j}
F_{\text{min}}^{\mu_i\mu_j}(\theta_{ij})\text{, } \\
F_{3}^{{\mathcal{O}}^{0,0}_{0,1}|++-}&=&\frac{-i H^{0,1}}
{2 \,\, \text{cosh}(\theta_{12}/2)} \prod_{i<j}
F_{\text{min}}^{\mu_i\mu_j}(\theta_{ij})\text{, }  \\
F_{3}^{{\mathcal{O}}^{1,0}_{1,0}|+--}&=&\frac{i H^{1,0}}
{2 \,\, \text{cosh}(\theta_{23}/2)} \prod_{i<j}
F_{\text{min}}^{\mu_i\mu_j}(\theta_{ij})\text{, }  \\
F_{3}^{{\mathcal{O}}^{0,1}_{0,1}|++-}&=&\frac{i H^{0,1}
e^{\theta_{3}/2}/{(\sigma_2^{+})}^{1/2}}
{2 \,\, \text{cosh}(\theta_{12}/2)} \prod_{i<j}
F_{\text{min}}^{\mu_i\mu_j}(\theta_{ij}).
\end{eqnarray}

\subsubsection{Four particle form factors}

\begin{eqnarray}
F_{4}^{\Theta |++--} &=&\frac{-\pi m^{2}(2+\sum_{i<j}\cosh (\theta _{ij}))}{%
2\cosh (\theta _{12}/2)\cosh (\theta _{34}/2)}\prod_{i<j}\tilde{F}_{\text{min%
}}^{\mu _{i}\mu _{j}}(\theta _{ij})\text{, } \\
F_{4}^{\mathcal{O}_{0,0}^{0,0}|++--} &=&\frac{-\left\langle \mathcal{O}%
_{0,0}^{0,0}\right\rangle \cosh (\theta _{13}/2+\theta _{24}/2)}{2\cosh
(\theta _{12}/2)\cosh (\theta _{34}/2)}\prod_{i<j}\tilde{F}_{\text{min}%
}^{\mu _{i}\mu _{j}}(\theta _{ij})\text{, }\\
F_{4}^{\mathcal{O}_{0,2}^{0,1}|++--} &=&\frac{-\left\langle \mathcal{O}%
_{0,2}^{0,1}\right\rangle}{2\cosh
(\theta _{12}/2)}\prod_{i<j}\tilde{F}_{\text{min}%
}^{\mu _{i}\mu _{j}}(\theta _{ij})\text{, }\\
F_{4}^{\mathcal{O}_{2,0}^{1,0}|++--} &=&\frac{-\left\langle \mathcal{O}%
_{2,0}^{1,0}\right\rangle}{2\cosh
(\theta _{34}/2)}\prod_{i<j}\tilde{F}_{\text{min}%
}^{\mu _{i}\mu _{j}}(\theta _{ij})\text{, }\\
F_{4}^{{\mathcal{O}}^{0,1}_{1,1}|+---}&=&
\frac{iH^{1,1} e^{-\theta_{1}/2}(\sigma_{3}^{-})^{1/2}}
{2\,\prod_{2\leq i<j\leq 4}\text{cosh}(\theta_{ij}/2)}
 \prod_{i<j}F_{\text{min}}^{\mu_i\mu_j}(\theta_{ij})\text{, }\\
F_{4}^{{\mathcal{O}}^{0,1}_{1,1}|+++-}&=&
\frac{-iH^{1,1} e^{\theta_{4}/2}/(\sigma_{3}^{+})^{1/2} }
{2\,\prod_{1\leq i<j\leq 3}\text{cosh}(\theta_{ij}/2)}
 \prod_{i<j}F_{\text{min}}^{\mu_i\mu_j}(\theta_{ij})\text{, }\\
F_{4}^{{\mathcal{O}}^{1,0}_{1,1}|+---}&=&
\frac{iH^{1,1}}
{2\,\prod_{2\leq i<j\leq 4}\text{cosh}(\theta_{ij}/2)}
 \prod_{i<j}F_{\text{min}}^{\mu_i\mu_j}(\theta_{ij})\text{, }\\
F_{4}^{{\mathcal{O}}^{1,0}_{1,1}|+++-}&=&
\frac{iH^{1,1}}
{2\,\prod_{1\leq i<j\leq 3}\text{cosh}(\theta_{ij}/2)}
 \prod_{i<j}F_{\text{min}}^{\mu_i\mu_j}(\theta_{ij}).
\end{eqnarray}

\subsubsection{Five particle form factors}

\begin{eqnarray}
F_{5}^{{\mathcal{O}}|\pm \pm \pm \pm \pm }&=&\frac{H^{0,1}\prod_{i<j}F_{\text{%
min}}^{\mu _{i}\mu _{j}}(\theta _{ij})}{\prod_{1\leq i<j\leq 5}\cosh (\theta
_{ij}/2)}\quad \text{for }{\mathcal{O}}_{1,0}^{0,0}\text{, }{\mathcal{O}}%
_{0,1}^{0,0}\text{, }{\mathcal{O}}_{1,0}^{1,0}\text{, }{\mathcal{O}}%
_{0,1}^{0,1}\text{, }\\
F_{5}^{{\mathcal{O}}^{0,0}_{1,0}|+----}&=&\frac{- H^{1,0} 
e^{-\theta_{1}/2}(\sigma_{4}^{-})^{1/2}}
{4\, \prod_{2 \leq i<j \leq 5}\text{cosh}(\theta_{ij}/2)} \prod_{i<j}
F_{\text{min}}^{\mu_i\mu_j}(\theta_{ij})\text{, }\\
F_{5}^{{\mathcal{O}}^{0,0}_{1,0}|+++--}&=&\frac{-iH^{1,0} 
(\sigma_2^{-})^{1/2}
(\sigma_2^{+}+\sigma_2^{-})/\sigma_3^{+}}
{8\, \text{cosh}(\theta_{45}/2)\, 
\prod_{1 \leq i<j \leq 3}\text{cosh}(\theta_{ij}/2)}\prod_{i<j}
F_{\text{min}}^{\mu_i\mu_j}(\theta_{ij})\text{, }\\
F_{5}^{{\mathcal{O}}^{0,0}_{0,1}|++---}&=&\frac{iH^{0,1} 
(\sigma_2^{+}+\sigma_2^{-})/\sigma_2^{+}}
{8\, \text{cosh}(\theta_{12}/2)\, 
\prod_{3 \leq i<j \leq 5}\text{cosh}(\theta_{ij}/2)}
\prod_{i<j}F_{\text{min}}^{\mu_i\mu_j}(\theta_{ij})\text{, }\\
F_{5}^{{\mathcal{O}}^{0,0}_{0,1}|++++-}&=&\frac{-H^{0,1}} 
{4\,\prod_{1 \leq i<j \leq 4}\text{cosh}(\theta_{ij}/2)}
\prod_{i<j}F_{\text{min}}^{\mu_i\mu_j}(\theta_{ij})\text{, }\\
F_{5}^{{\mathcal{O}}^{1,0}_{1,0}|+----}&=&\frac{-H^{1,0}} 
{4\,\prod_{2 \leq i<j \leq 5}\text{cosh}(\theta_{ij}/2)}
\prod_{i<j}F_{\text{min}}^{\mu_i\mu_j}(\theta_{ij})\text{, }\\
F_{5}^{{\mathcal{O}}^{1,0}_{1,0}|+++--}&=&\frac{-iH^{1,0}
(\sigma_3^{+}+\sigma_1^{+}\sigma_2^{-})/\sigma_3^{+}}
{8\,\text{cosh}(\theta_{45}/2)\,
\prod_{1 \leq i<j \leq 3}\text{cosh}(\theta_{ij}/2)}
\prod_{i<j}F_{\text{min}}^{\mu_i\mu_j}(\theta_{ij})\text{, }\\
F_{5}^{{\mathcal{O}}^{0,1}_{0,1}|++---}&=&\frac{-iH^{0,1}
 ( \sigma_3^{-}+\sigma_1^{-}\sigma_2^{+})/(\sigma_2^{+})^{3/2}}
{8\,\text{cosh}(\theta_{12}/2)\,
\prod_{3 \leq i<j \leq 5}\text{cosh}(\theta_{ij}/2)}
\prod_{i<j}F_{\text{min}}^{\mu_i\mu_j}(\theta_{ij})\text{, }\\
F_{5}^{{\mathcal{O}}^{0,1}_{0,1}|++++-}&=&\frac{-H^{0,1} 
e^{\theta_{5}/2}/(\sigma_4^{+})^{1/2}}
{4\,\prod_{1 \leq i<j \leq 4}\text{cosh}(\theta_{ij}/2)}
\prod_{i<j}F_{\text{min}}^{\mu_i\mu_j}(\theta_{ij}).
\end{eqnarray}

\subsubsection{Six particle form factors}

\begin{eqnarray}
F_{6}^{\Theta |++++--} &=&\frac{\pi m^{2}(3+\sum_{i<j}\cosh (\theta _{ij}))}{%
4\prod_{1\leq i<j\leq 4}\cosh (\theta _{ij}/2)}\prod_{i<j}\tilde{F}_{\text{%
min}}^{\mu _{i}\mu _{j}}(\theta _{ij}) \text{, }\\
F_{6}^{\mathcal{O}_{0,0}^{0,0}|++++--} &=&\frac{\left\langle \mathcal{O}%
_{0,0}^{0,0}\right\rangle \left( (\sigma _{2}^{-})^{2}+\sigma
_{4}^{+}+\sigma _{2}^{+}\sigma _{2}^{-}\right)/\sigma _{4}^{+}}{16\cosh
(\theta _{56}/2)\prod_{1\leq i<j\leq 4}\cosh (\theta _{ij}/2)}\prod_{i<j}F_{%
\text{min}}^{\mu _{i}\mu _{j}}(\theta _{ij}) \text{, }\\
F_{6}^{\mathcal{O}_{0,2}^{0,1}|++++--} &=&\frac{\left\langle \mathcal{O}%
_{0,2}^{0,1}\right\rangle \cosh (\theta _{56}/2)
 }{4 \prod_{1\leq i<j\leq 4}\cosh (\theta _{ij}/2)}\prod_{i<j}\tilde{F}_{%
\text{min}}^{\mu _{i}\mu _{j}}(\theta _{ij}) \text{, }\\
F_{6}^{\mathcal{O}_{2,0}^{1,0}|++++--} &=&\frac{\left\langle \mathcal{O}%
_{2,0}^{1,0}\right\rangle (\sigma _{2}^{-} \sigma _{4}^{+})^{-1/2}
 }{16\cosh (\theta _{56}/2) \prod_{1\leq i<j\leq 4}\cosh (\theta _{ij}/2)}
\prod_{i<j}\tilde{F}_{%
\text{min}}^{\mu _{i}\mu _{j}}(\theta _{ij})\text{, }\\
F_{6}^{{\mathcal{O}}^{0,1}_{1,1}|+-----}&=&
\frac{-H^{1,1}e^{-\theta_{1}/2}(\sigma _{5}^{-})^{1/2}}
{4\,\prod_{2\leq i<j\leq 6}\text{cosh}(\theta_{ij}/2)}
 \prod_{i<j}F_{\text{min}}^{\mu_i\mu_j}(\theta_{ij})\text{, }
\end{eqnarray}

\begin{equation}
F_{6}^{{\mathcal{O}}^{0,1}_{1,1}|+++---}=
\frac{-H^{1,1}(\sigma_3^{-})^{1/2}
(\sigma_3^{-}+\sigma_1^{-}\sigma_2^{+})/(\sigma_3^{+})^{3/2}}
{16\,\prod_{1\leq i<j\leq 3}\text{cosh}(\theta_{ij}/2)
\,\prod_{4\leq i<j\leq 6}\text{cosh}(\theta_{ij}/2)}
 \prod_{i<j}F_{\text{min}}^{\mu_i\mu_j}(\theta_{ij})\text{, }
\end{equation}
\begin{eqnarray}
F_{6}^{{\mathcal{O}}^{0,1}_{1,1}|+++++-}&=&
\frac{-H^{1,1}e^{\theta_{6}/2}/(\sigma _{5}^{+})^{1/2}}
{4\,\prod_{1\leq i<j\leq 5}\text{cosh}(\theta_{ij}/2)}
 \prod_{i<j}F_{\text{min}}^{\mu_i\mu_j}(\theta_{ij})\text{, }\\
F_{6}^{{\mathcal{O}}^{1,0}_{1,1}|+-----}&=&
\frac{-H^{1,1}}
{4\,\prod_{2\leq i<j\leq 6}\text{cosh}(\theta_{ij}/2)}
 \prod_{i<j}F_{\text{min}}^{\mu_i\mu_j}(\theta_{ij})\text{, }
\end{eqnarray}
\begin{equation}
F_{6}^{{\mathcal{O}}^{1,0}_{1,1}|+++---}=
\frac{H^{1,1}
(\sigma_3^{+}+\sigma_1^{+}\sigma_2^{-})/\sigma_3^{+}}
{16\,\prod_{1\leq i<j\leq 3}\text{cosh}(\theta_{ij}/2)
\,\prod_{4\leq i<j\leq 6}\text{cosh}(\theta_{ij}/2)}
 \prod_{i<j}F_{\text{min}}^{\mu_i\mu_j}(\theta_{ij})\text{, }
\end{equation}
\begin{equation}
F_{6}^{{\mathcal{O}}^{1,0}_{1,1}|+++++-}=
\frac{-H^{1,1}}
{4\,\prod_{1\leq i<j\leq 5}\text{cosh}(\theta_{ij}/2)}
\prod_{i<j}F_{\text{min}}^{\mu_i\mu_j}(\theta_{ij}).
\end{equation}

\subsubsection{Seven particle form factors}

\begin{eqnarray}
F_{7}^{{\mathcal{O}}|\pm \pm \pm \pm \pm \pm \pm }&=&\frac{H^{0,1}%
\prod_{i<j}F_{\text{min}}^{\mu _{i}\mu _{j}}(\theta _{ij})}{\,\prod_{1\leq
i<j\leq 7}\text{cosh}(\theta _{ij}/2)}\quad \text{for }{\mathcal{O}}%
_{1,0}^{0,0}\text{, }{\mathcal{O}}_{0,1}^{0,0}\text{, }{\mathcal{O}}%
_{1,0}^{1,0}\text{, }{\mathcal{O}}_{0,1}^{0,1}\text{, }\\
F_{7}^{{\mathcal{O}}^{0,0}_{1,0}|+------}&=&\frac{-iH^{1,0}
(\sigma_6^{-})^{1/2}/
(\sigma_1^{+})^3}
{8\prod_{2 \leq i<j \leq 7}\text{cosh}(\theta_{ij}/2)}
\prod_{i<j}F_{\text{min}}^{\mu_i\mu_j}(\theta_{ij}) \text{, }
\end{eqnarray}

\begin{equation}
F_{7}^{{\mathcal{O}}^{0,0}_{1,0}|+++----}=\frac{-H^{1,0}
(\sigma_4^{-})^{1/2}
(\sigma_4^{-}+\sigma_2^{+}\sigma_2^{-}+(\sigma_2^{+})^2)/
(\sigma_3^{+})^2}
{2^8\prod_{1 \leq i<j \leq 3}\text{cosh}(\theta_{ij}/2)
\prod_{4 \leq i<j \leq 7}\text{cosh}(\theta_{ij}/2)}
\prod_{i<j}F_{\text{min}}^{\mu_i\mu_j}(\theta_{ij})  \text{, }
\end{equation}
\begin{equation}
F_{7}^{{\mathcal{O}}^{0,0}_{1,0}|+++++--}=\frac{-H^{1,0}
(\sigma_2^{-})^{1/2}
(\sigma_4^{+}+\sigma_2^{+}\sigma_2^{-}+(\sigma_2^{-})^2)/
\sigma_5^{+}}
{2^5\,\text{cosh}(\theta_{67}/2)
\prod_{1 \leq i<j \leq 5}\text{cosh}(\theta_{ij}/2)}
\prod_{i<j}F_{\text{min}}^{\mu_i\mu_j}(\theta_{ij}) \text{, }
\end{equation}

\begin{equation}
F_{7}^{{\mathcal{O}}^{0,0}_{0,1}|++-----}=\frac{-iH^{0,1}
(\sigma_4^{-}+\sigma_2^{+}\sigma_2^{-}+(\sigma_2^{+})^2)/
(\sigma_2^{+})^2}
{2^5\,\text{cosh}(\theta_{12}/2)
\prod_{3 \leq i<j \leq 7}\text{cosh}(\theta_{ij}/2)}
\prod_{i<j}F_{\text{min}}^{\mu_i\mu_j}(\theta_{ij})\text{, }
\end{equation}
\begin{equation} 
F_{7}^{{\mathcal{O}}^{0,0}_{0,1}|++++---}=\frac{-H^{0,1}
(\sigma_4^{+}+\sigma_2^{+}\sigma_2^{-}+(\sigma_2^{-})^2)/
\sigma_4^{+}}
{2^8\,\prod_{1 \leq i<j \leq 4}\text{cosh}(\theta_{ij}/2)\,
\prod_{5 \leq i<j \leq 7}\text{cosh}(\theta_{ij}/2)}
\prod_{i<j}F_{\text{min}}^{\mu_i\mu_j}(\theta_{ij}) \text{, }
\end{equation}

\begin{eqnarray}
F_{7}^{{\mathcal{O}}^{0,0}_{0,1}|++++++-}&=&\frac{iH^{0,1}}
{8\,\prod_{1 \leq i<j \leq 6}\text{cosh}(\theta_{ij}/2)}
\prod_{i<j}F_{\text{min}}^{\mu_i\mu_j}(\theta_{ij}) \text{, }\\ 
F_{7}^{{\mathcal{O}}^{1,0}_{1,0}|+------}&=&\frac{-i H^{1,0}}
{8\prod_{1 \leq i<j \leq 6}\text{cosh}(\theta_{ij}/2)}
\prod_{i<j}F_{\text{min}}^{\mu_i\mu_j}(\theta_{ij}) \text{, }
\end{eqnarray}
 
\begin{equation} 
F_{7}^{{\mathcal{O}}^{1,0}_{1,0}|+++----}=\frac{- H^{1,0}
((\sigma_1^{+})^2\sigma_4^{-} +(\sigma_3^{+})^2+
\sigma_3^{+}\sigma_1^{+}\sigma_2^{-})/
(\sigma_3^{+})^2}
{2^8\prod_{1 \leq i<j \leq 6}\text{cosh}(\theta_{ij}/2)}
\prod_{i<j}F_{\text{min}}^{\mu_i\mu_j}(\theta_{ij}) \text{, }
\end{equation}
\begin{equation} 
F_{7}^{{\mathcal{O}}^{1,0}_{1,0}|+++++--}=\frac{i H^{1,0}
((\sigma_2^{-})^2\sigma_1^{+} +\sigma_5^{+}+
\sigma_3^{+}\sigma_2^{-})/
\sigma_5^{+}}
{2^5\text{cosh}(\theta_{67}/2)
\, \prod_{1 \leq i<j \leq 5}\text{cosh}(\theta_{ij}/2)}
\prod_{i<j}F_{\text{min}}^{\mu_i\mu_j}(\theta_{ij}) \text{, }
\end{equation}
\begin{equation} 
F_{7}^{{\mathcal{O}}^{0,1}_{0,1}|++-----}=\frac{i H^{0,1}
((\sigma_2^{+})^2 \sigma_1^{-} + \sigma_5^{-}+
\sigma_3^{-}\sigma_2^{+})/(\sigma_2^{+})^{5/2}}
{ 2^5 \text{cosh}(\theta_{12}/2)\,
\prod_{3 \leq i<j \leq 7}\text{cosh}(\theta_{ij}/2)}
\prod_{i<j}F_{\text{min}}^{\mu_i\mu_j}(\theta_{ij}) \text{, }
\end{equation}
\begin{equation}
F_{7}^{{\mathcal{O}}^{0,1}_{0,1}|++++---}=\frac{- H^{0,1}
((\sigma_3^{-})^2 + \sigma_4^{+}\sigma_1^{-}+
\sigma_3^{-}\sigma_1^{-}\sigma_2^{+})/(\sigma_4^{+})^{3/2}}
{ 2^8\prod_{1 \leq i<j \leq 3}\text{cosh}(\theta_{ij}/2) \,
\prod_{5 \leq i<j \leq 7}\text{cosh}(\theta_{ij}/2)}
\prod_{i<j}F_{\text{min}}^{\mu_i\mu_j}(\theta_{ij}) \text{, }
\end{equation}
\begin{equation}
F_{7}^{{\mathcal{O}}^{0,1}_{0,1}|++++++-}=\frac{-i H^{0,1}
(\sigma_1^{-})^3/(\sigma_6^{+})^{1/2}}
{8\prod_{1 \leq i<j \leq 6}\text{cosh}(\theta_{ij}/2)}
\prod_{i<j}F_{\text{min}}^{\mu_i\mu_j}(\theta_{ij}).
\end{equation}

\subsubsection{Eight particle form factors}

\begin{equation}
F_{8}^{\Theta |++------}=\frac{-\pi m^{2}(4+\sum_{i<j}\text{cosh}(\theta
_{ij}))\,\text{cosh}(\theta _{12}/2)}{8\,\prod_{3\leq i<j\leq 8}\text{cosh}%
(\theta _{ij}/2)}\prod_{i<j}\tilde{F}_{\text{min}}^{\mu _{i}\mu _{j}}(\theta
_{ij}),
\end{equation}

\begin{equation}
F_{8}^{\Theta |++++----}=\frac{\pi m^{2}(\sigma _{4}^{-})^{1/2}(\sigma
_{1}^{-}\sigma _{3}^{+}+\sigma _{1}^{+}\sigma _{3}^{-})(4+\sum_{i<j}\text{%
cosh}(\theta _{ij}))}{2^{7}\,(\sigma _{4}^{+})^{3/2}\prod_{1\leq i<j\leq 4}%
\text{cosh}(\theta _{ij}/2)\prod_{5\leq i<j\leq 8}\text{cosh}(\theta _{ij}/2)%
}\prod_{i<j}F_{\text{min}}^{\mu _{i}\mu _{j}}(\theta _{ij}),
\end{equation}

\begin{equation}
F_{8}^{\Theta |++++++--}=\frac{-\pi m^{2}(4+\sum_{i<j}\text{cosh}(\theta
_{ij}))\,\text{cosh}(\theta _{78}/2)}{8\,\prod_{1\leq i<j\leq 6}\text{cosh}%
(\theta _{ij}/2)}\prod_{i<j}\tilde{F}_{\text{min}}^{\mu _{i}\mu _{j}}(\theta
_{ij}),
\end{equation}

\begin{equation}
F_{8}^{{\mathcal{O}}^{0,0}_{0,0}|++------}=
\frac{\langle {\mathcal{O}}^{0,0}_{0,0}
\rangle (\sigma_6^{-}+ \sigma_2^{-} (\sigma_2^{+})^2+
\sigma_2^{+} \sigma_4^{-}+(\sigma_2^{+})^3)/(\sigma_2^{+})^3}{2^{6}\,
 \text{cosh}(\theta_{12}/2)\,
\prod_{3 \leq i<j \leq 8}\text{cosh}(\theta_{ij}/2)}
\prod_{i<j}F_{\text{min}}^{\mu_i\mu_j}(\theta_{ij}),
\end{equation}

\begin{equation}
F_{8}^{{\mathcal{O}}^{0,0}_{0,0}|++++----}=
\frac{-\langle {\mathcal{O}}^{0,0}_{0,0}
\rangle 
((\sigma_4^{+}+\sigma_4^{-})^2+ (\sigma_2^{-})^2\sigma_4^{+}+
 \sigma_2^{+}\sigma_2^{-}(\sigma_4^{-}+\sigma_4^{+}) +
(\sigma_2^{+})^2\sigma_4^{-})}
{2^{8}(\sigma_4^{+})^2 \prod_{1 \leq i<j \leq 4}\text{cosh}(\theta_{ij}/2)
 \prod_{5 \leq i<j \leq 8}\text{cosh}(\theta_{ij}/2)}
\prod_{i<j}F_{\text{min}}^{\mu_i\mu_j}(\theta_{ij}),
\end{equation}

\begin{equation}
F_{8}^{{\mathcal{O}}^{0,0}_{0,0}|++++++--}=
\frac{\langle {\mathcal{O}}^{0,0}_{0,0}
\rangle (\sigma_6^{+}+ \sigma_2^{+} (\sigma_2^{-})^2+
\sigma_2^{-} \sigma_4^{+}+(\sigma_2^{-})^3)/\sigma_6^{+}}
{2^{6}\, \text{cosh}(\theta_{78}/2)\,
\prod_{1 \leq i<j \leq 6}\text{cosh}(\theta_{ij}/2)}
\prod_{i<j}F_{\text{min}}^{\mu_i\mu_j}(\theta_{ij}),
\end{equation}

\begin{equation}
F_{8}^{{\mathcal{O}}^{0,1}_{0,2}|++------}=
\frac{-\langle {\mathcal{O}}^{0,1}_{0,2}
\rangle (\sigma_5^{-}+ \sigma_1^{-} (\sigma_2^{+})^2+
\sigma_3^{-} \sigma_2^{+})/(\sigma_2^{+})^{5/2}}
{2^{6}\, \text{cosh}(\theta_{12}/2)\,
\prod_{3 \leq i<j \leq 8}\text{cosh}(\theta_{ij}/2)}
\prod_{i<j}F_{\text{min}}^{\mu_i\mu_j}(\theta_{ij}),
\end{equation}

\begin{equation}
F_{8}^{{\mathcal{O}}^{0,1}_{0,2}|++++----}=
\frac{\langle {\mathcal{O}}^{0,1}_{0,2}
\rangle ((\sigma_3^{-})^2+ \sigma_1^{-}\sigma_3^{-}\sigma_2^{+}+
 \sigma_4^{+} (\sigma_1^{-})^2)/(\sigma_4^{-})^{3/2}}
{2^{8}\, \prod_{1 \leq i<j \leq 4}\text{cosh}(\theta_{ij}/2)
\prod_{5 \leq i<j \leq 8}\text{cosh}(\theta_{ij}/2)}
\prod_{i<j}F_{\text{min}}^{\mu_i\mu_j}(\theta_{ij}),
\end{equation}

\begin{equation}
F_{8}^{{\mathcal{O}}^{0,1}_{0,2}|++++++--}=
\frac{\langle {\mathcal{O}}^{0,1}_{0,2}
\rangle (\sigma_1^{-})^{3}/(\sigma_6^{+})^{1/2}}
{2^{6}\, \text{cosh}(\theta_{78}/2)\,
\prod_{1 \leq i<j \leq 6}\text{cosh}(\theta_{ij}/2)}
\prod_{i<j}F_{\text{min}}^{\mu_i\mu_j}(\theta_{ij}),
\end{equation}

\begin{equation}
F_{8}^{{\mathcal{O}}^{1,0}_{2,0}|++------}=
\frac{\langle {\mathcal{O}}^{1,0}_{2,0}
\rangle  (\sigma_1^{+})^{3}(\sigma_6^{-})^{1/2}/(\sigma_2^{+})^{3}}
{2^{6}\, \text{cosh}(\theta_{12}/2)\,
\prod_{3 \leq i<j \leq 8}\text{cosh}(\theta_{ij}/2)}
\prod_{i<j}F_{\text{min}}^{\mu_i\mu_j}(\theta_{ij}),
\end{equation}

\begin{equation}
F_{8}^{{\mathcal{O}}^{1,0}_{2,0}|++++----}=
\frac{\langle {\mathcal{O}}^{1,0}_{2,0}
\rangle (\sigma_4^{-})^{1/2}
((\sigma_3^{+})^2+ \sigma_1^{+}\sigma_3^{+}\sigma_2^{-}+
 \sigma_4^{-} (\sigma_1^{+})^2)/(\sigma_4^{+})^{2}}
{2^{8}\, \prod_{1 \leq i<j \leq 4}\text{cosh}(\theta_{ij}/2)
\prod_{5 \leq i<j \leq 8}\text{cosh}(\theta_{ij}/2)}
\prod_{i<j}F_{\text{min}}^{\mu_i\mu_j}(\theta_{ij}),
\end{equation}

\begin{equation}
F_{8}^{{\mathcal{O}}^{1,0}_{2,0}|++++++--}=
\frac{\langle {\mathcal{O}}^{1,0}_{2,0}
\rangle (\sigma_2^{-})^{1/2}(  
\sigma_5^{+}+\sigma_3^{+}\sigma_2^{-}+
 \sigma_1^{+} (\sigma_2^{-
})^2)/\sigma_6^{+}}
{2^{6}\, \text{cosh}(\theta_{78}/2)\,
\prod_{1 \leq i<j \leq 6}\text{cosh}(\theta_{ij}/2)}
\prod_{i<j}F_{\text{min}}^{\mu_i\mu_j}(\theta_{ij}),
\end{equation}

\begin{equation}
F_{8}^{{\mathcal{O}}^{1,0}_{1,1}|+-------}=\frac{-iH^{1,1}}
{8 \prod_{2 \leq i<j \leq 8}\text{cosh}(\theta_{ij}/2)}
\prod_{i<j}F_{\text{min}}^{\mu_i\mu_j}(\theta_{ij}),  
\end{equation}

\begin{equation}
F_{8}^{{\mathcal{O}}^{1,0}_{1,1}|+++-----}=
\frac{-iH^{1,1}((\sigma_1^{+})^2 \sigma_4^{-} + (\sigma_3^{+})^2+
\sigma_3^{+}\sigma_1^{+}\sigma_2^{-})/(\sigma_3^{+})^2}
{2^7 \prod_{1 \leq i<j \leq 3}\text{cosh}(\theta_{ij}/2)
 \prod_{4 \leq i<j \leq 8}\text{cosh}(\theta_{ij}/2)}
\prod_{i<j}F_{\text{min}}^{\mu_i\mu_j}(\theta_{ij}),  
\end{equation}

\begin{equation}
F_{8}^{{\mathcal{O}}^{1,0}_{1,1}|+++++---}=
\frac{-iH^{1,1}((\sigma_2^{-})^2 \sigma_1^{+} + \sigma_5^{+}+
\sigma_3^{+}\sigma_2^{-})/\sigma_5^{+}}
{2^7 \prod_{1 \leq i<j \leq 5}\text{cosh}(\theta_{ij}/2)
 \prod_{6 \leq i<j \leq 8}\text{cosh}(\theta_{ij}/2)}
\prod_{i<j}F_{\text{min}}^{\mu_i\mu_j}(\theta_{ij}),  
\end{equation}

\begin{equation}
F_{8}^{{\mathcal{O}}^{0,1}_{1,1}|+-------}=\frac{-iH^{1,1} (\sigma_7^{-})^{1/2}/
(\sigma_1^{+})^{7/2}}
{8 \prod_{2 \leq i<j \leq 8}\text{cosh}(\theta_{ij}/2)}
\prod_{i<j}F_{\text{min}}^{\mu_i\mu_j}(\theta_{ij}),  
\end{equation}

\begin{equation}
F_{8}^{{\mathcal{O}}^{0,1}_{1,1}|+++-----}=
\frac{iH^{1,1}(\sigma_5^{-})^{1/2}
((\sigma_2^{+})^2 \sigma_1^{-} + \sigma_5^{-}+
\sigma_3^{-}\sigma_2^{+})/(\sigma_3^{+})^{5/2}}
{2^7 \prod_{1 \leq i<j \leq 3}\text{cosh}(\theta_{ij}/2)
 \prod_{4 \leq i<j \leq 8}\text{cosh}(\theta_{ij}/2)}
\prod_{i<j}F_{\text{min}}^{\mu_i\mu_j}(\theta_{ij}),  
\end{equation}

\begin{equation}
F_{8}^{{\mathcal{O}}^{0,1}_{1,1}|+++++---}=
\frac{-iH^{1,1}(\sigma_3^{-})^{1/2}
 ((\sigma_1^{-})^2 \sigma_4^{+} + (\sigma_3^{-})^2+
\sigma_3^{-}\sigma_1^{-}\sigma_2^{+})/(\sigma_5^{+})^{3/2}}
{2^7 \prod_{1 \leq i<j \leq 5}\text{cosh}(\theta_{ij}/2)
 \prod_{6 \leq i<j \leq 8}\text{cosh}(\theta_{ij}/2)}
\prod_{i<j}F_{\text{min}}^{\mu_i\mu_j}(\theta_{ij}),  
\end{equation}

\begin{equation}
F_{8}^{{\mathcal{O}}^{0,1}_{1,1}|+++++++-}=\frac{iH^{1,1}(\sigma_1^{-})^{7/2}/
(\sigma_7^{+})^{1/2}}
{8 \prod_{1 \leq i<j \leq 7}\text{cosh}(\theta_{ij}/2)}
\prod_{i<j}F_{\text{min}}^{\mu_i\mu_j}(\theta_{ij}).  
\end{equation}

\begin{description}
\item  {\small \setlength{\baselineskip}{12pt}}
\end{description}

\end{document}